\documentclass[a4paper, twoside]{memoir}
\usepackage[british]{babel}
\usepackage[latin1]{inputenc}
\usepackage[T1]{fontenc}
\usepackage[babel]{microtype}
\usepackage{graphicx}
\usepackage{amsmath}
\usepackage{amsthm}
\usepackage[mathcalasscript, oldstylenums, widermath]{kpfonts}
\usepackage{tikz}
\usetikzlibrary{positioning, fit, decorations.pathmorphing, calc}
\tikzset{x=1em, y=1em,          
  >=latex}                      
\colorlet{shaded}{gray!60}

\usepackage[bookmarksnumbered, backref=page, colorlinks,
linkcolor=black, citecolor=black, urlcolor=black, filecolor=black]{hyperref}

\setsecnumdepth{subsection}
\openany
\indexintoc
\nouppercaseheads
\chapterstyle{ger}

\makeatletter \def\@pnumwidth{2em} \makeatother 

\setparaheadstyle{\normalfont\normalsize\bfseries\addperiod}
\setbeforesubparaskip{\parskip}
\setaftersubparaskip{-0.5em}
\setsubparaheadstyle{\normalfont\normalsize\itshape\addperiod}

\firmlists

\def\backref#1{}
\def\backrefalt#1#2#3#4{\ifcase #1%
  \or\small{Cited on page~#2.}%
  \else\small{Cited on pages #2.}\fi}

\def\mytheoremstyle#1#2{%
  \newtheoremstyle{#1}%
  {\topsep}{\topsep}
  {#2}
  {}
  {\bfseries\slshape}
  {.}
  {.5em plus .5em minus .25em}
  {}
}
\mytheoremstyle{definition}{}
\mytheoremstyle{plain}{\slshape}
\mytheoremstyle{remark}{}

\theoremstyle{definition}
\newtheorem{definition}{Definition}[chapter]
\theoremstyle{plain}
\newtheorem{theorem}[definition]{Theorem}
\newtheorem{lemma}[definition]{Lemma}
\newtheorem{corollary}[definition]{Corollary}
\theoremstyle{remark}

\makeindex
\def\idef#1{\index{#1}}
\def\intro#1{\idef{#1}\emph{#1}}
\newcommand{\introx}[2][]{\idef{#1}\emph{#2}}

\def\indexsplittwo#1 #2\relax{%
\index{#1!#1 #2@$\sim$ #2}\index{#2!#1 $\sim$}}
\def\indextwo#1{\indexsplittwo #1\relax}
\def\introtwo#1{\indextwo{#1}\emph{#1}}
  
\def\indexsplittwoshort#1 #2\relax{%
\index{#1 #2}\index{#2!#1 $\sim$}}
\def\indextwoshort#1{\indexsplittwoshort #1\relax}
\def\introtwoshort#1{\indextwoshort{#1}\emph{#1}}
  
\def\indexsplitthree#1 #2 #3\relax{%
\index{#1!#1 #2 #3@$\sim$ #2 #3}\index{#3!#1 #2 $\sim$}}
\def\indexthree#1{\indexsplitthree #1\relax}
\def\introthree#1{\indexthree{#1}\emph{#1}}
  
\def\indexsplitthreeshort#1 #2 #3\relax{%
\index{#1 #2 #3}\index{#3!#1 #2 $\sim$}}
\def\indexthreeshort#1{\indexsplitthreeshort #1\relax}
\def\introthreeshort#1{\indexthreeshort{#1}\emph{#1}}
  
\def\ci#1
{\protect\includegraphics{ci/#1}}

\def\etc{\textit{\&ct}}

\def\N{\mathbb N}
\def\Z{\mathbb Z}

\def\et{T} \def\ex{X}

\let\epsilon=\varepsilon
\let\phi=\varphi

\def\diag#1{\begin{smallmatrix} #1 \end{smallmatrix}} 

\def\op{\oplus}
\def\om{\ominus}
\def\opom{\mathbin{\oplus\,\ominus}}
\def\d[#1]{\mathbin{[#1]}}      

\def\cl{\mathop{\mathrm{cl}}\nolimits} 

\def\rea{\rightarrow}
\def\dom{\mathop{\mathrm{dom}}}  
\def\pr{\mathrm{pr}}             
\def\prx#1#2{\mathrm{pr}_{#1}(#2)} 
\def\comp{\mathrel\mathrm{comp}} 
\def\seq{\mathrel{\prec\succ}}   

\def\cproc#1{{\xymatrix@=0.5ex{#1}}} 

\def\ovl#1{\mathbin{\langle #1\rangle}}   
\def\leftend{\mathrel{{\backslash}\!{\backslash}}}
\def\rightend{\mathrel{{/}\!{/}}}
\def\set#1{\{\, #1 \,\}}

\def\abs#1{\mathopen|#1\mathclose|}

\def\Floor#1{\left\lfloor #1\right\rfloor}

\def\qqtext#1{\qquad\text{#1}\qquad}
\def\qtext#1{\quad\text{#1}\quad}

\begin{document}
\frontmatter


\begin{titlingpage}
  {\centering
    \textsc{Markus Redeker}\\[0.5ex]
    Faculty of Environment and Technology\\[7ex]
    {\HUGE Flexible Time and Ether\\
      \huge in One-dimensional Cellular Automata\\[4ex]}
    Thesis submitted in partial fulfilment
    of the requirements\\
    of the University of the West of England, Bristol\\
    for the degree of Doctor of Philosophy\\[0ex plus 1fill]
    \hspace*{3em}\ci{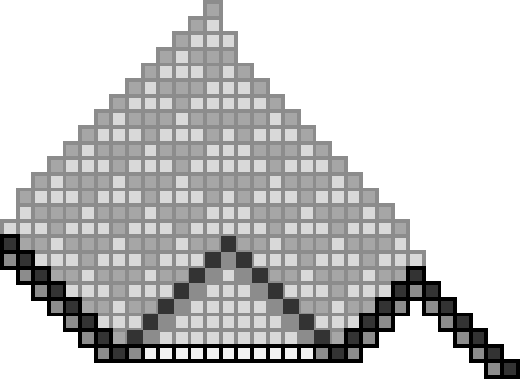}\\[0ex plus 1fill]
    December 2013\\}

  \newpage
  \hbox{}\vfill
  {
    \tabcolsep=0pt
    \noindent
    \begin{tabular*}{\textwidth}{c@{\,\,}p{0.85\textwidth}}
      \href{http://creativecommons.org/licenses/by-nc-sa/4.0}%
      {\raisebox{-2.25ex}{\includegraphics[height=3.5ex]{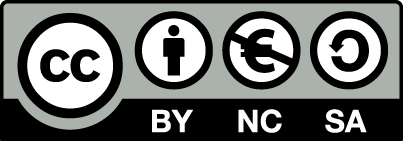}}} &
      \footnotesize
      This work is licensed under a
      Creative
      Commons Attribution--NonCommercial--ShareAlike 4.0 International
      License (BY-NC-SA).
    \end{tabular*}
  }
\end{titlingpage}

\cleardoublepage
\thispagestyle{empty}

\begin{abstract}
  \noindent
  A one-dimensional cellular automaton is an infinite row of identical
  machines---the cells---which depend for their behaviour only on the
  states of their direct neighbours.

  This thesis introduces a new way to think about one-dimensional
  cellular automata. The formalism of Flexible Time allows one to
  unify the states of of a finite number of cells into a single
  object, even if they occur at different times. This gives greater
  flexibility to handle the structures that occur in the development
  of a cellular automaton. Flexible Time makes it possible to
  calculate in an algebraic way the fate of a finite number of cells.

  In the first part of this thesis the formalism is developed in
  detail. Then it is applied to a specific problem of one-dimensional
  cellular automata, namely ether formation. The so-called ether is a
  periodic pattern of cells that occurs in some cellular automata: It
  arises from almost all randomly chosen initial configurations, and
  why this happens is not clear. For one of these cellular automata,
  the elementary cellular automaton with rule code 54, ether formation
  is expressed in the formalism of Flexible Time.

  Then a partial result about ether formation is proved: There is a
  certain fragment of the ether that arises with probability $1$ from
  every random initial configuration, and it is then propagated with
  probability $1$ to any later time. The persistence of the ether
  fragment is a strong argument that the ether under Rule 54 indeed
  arises from almost all input configurations. The result only
  requires that the states of the cells are chosen independently and
  with equal probability distributions, and that all cell states can
  occur. This is not yet a full proof of ether formation, but it is
  derived by formal means, not just by computer simulations.
\end{abstract}


\cleardoublepage
\tableofcontents*

\chapter*{List of Symbols}
\label{cha:list-symbols}

\def\symsection#1{\contentsline{chapter}{#1}{}{}}
\def\symbox#1{\hbox to 12ex{#1\hss}}

\makeatletter
\def\symline#1#2#3
{\@dottedtocline{1}{0ex}{0ex}%
  {\symbox{#1} \hyperref[#3]{#2}} {\pageref{#3}}}
\makeatother

\symsection{Sets and Functions}
\symline{$\N$}{Set of positive integers}{pg:integer-sets}
\symline{$\N_0$}{Set of nonnegative integers}{pg:integer-sets}
\symline{$B^A$}{Set of functions from $A$ to $B$}{pg:func-A-B}
\symline{$A \subset B$}{$A$ is proper subset of $B$}{pg:subset-notation}
\symline{$A \subseteq B$} {$A$ is subset or equal to $B$}{pg:subset-notation}
\symline{$\dom f$}{Domain of the function $f$}{pg:domain-func}

\symsection{Sequences}
\symline{$\lambda$}{Empty sequence}{pg:empty-sequence}
\symline{$A^*$}{Kleene closure of $A$}{eq:A-star}
\symline{$\abs{s}$} {Length of sequence $s$}{pg:length}

\symsection{Space-time}
\symline{$p_T$}{Time component of point $p$}{pg:space-time-point}
\symline{$p_X$}{Space component of point $p$}{pg:space-time-point}
\symline{$\et$}{Unit vector in temporal direction}{eq:unit-vectors}
\symline{$\ex$}{Unit vector in spatial direction}{eq:unit-vectors}
\symline{$N(p, r)$}{Neighbourhood domain of point $p$}{eq:nb-domain}
\symline{$I_t(i, j)$}{Interval domain at time $t$}{def:intervals}

\symsection{Processes}
\symline{$\mathcal{P}$}
{Set of cellular processes with states in $\Sigma$}{def:cellular-process}
\symline{$\dom \pi$}{Domain of process $\pi$}{eq:domain}
\symline{$\pi \comp \psi$}{$\pi$ is compatible with $\psi$}{def:compatible}
\symline{$\pi|_S$}
{Restriction of process $\pi$ to set $S$}{eq:process-restriction}
\symline{$\pi^{(t)}$}
{Time slice at time $t$ of process $\pi$}{def:cellular-process}
\symline{$\nu(p, w)$}
{Neighbourhood process for point $p$}{eq:neighbourhood}
\symline{$S(p, \pi)$}
{Set of possible states for the point $p$}{eq:values}
\symline{$\Delta \pi$}
{Events determined by $\pi$ under rule $\phi$}{pg:determined}
\symline{$\cl \pi$}
{Closure of process $\pi$ under rule $\phi$}{def:closure}
\symline{$\cl^{(t)} \pi$}
{Closure at time $t$ of process $\pi$}{def:closure}
\symline{$[p] \pi$}{Process $\pi$, shifted by $p$}{eq:shifted}
\symline{$\pi \seq \psi$}
{$\pi$ is left of $\psi$}{def:spatial-arrangement}
\symline{$\psi \supseteq_L \pi$}
{$\psi$ is left extension of $\pi$}{eq:left-extension}
\symline{$\pi \subseteq_R \psi$}
{$\psi$ is right extension of $\pi$}{eq:right-extension}

\symsection{Situations}
\symline{$\mathcal{S}$}
{Set of situations with state set $\Sigma$}{def:situation}
\symline{$\delta(s)$}{Size vector of situation $s$}{pg:size-vector}
\symline{$\pr(a)$}{Process of situation $a$}{def:process-of}
\symline{$\prx{a}{b}$}
{Process of situation $b$, shifted by $a$}{eq:relative-process}
\symline{$[p]$}{Displacement by distance $p$}{def:situation-notation}
\symline{$[0]$}{Empty situation}{pg:empty-situation}
\symline{$a \leftend x$}{$a$ is left factor of $x$}{def:end-factors}
\symline{$x \rightend a$}{$a$ is right factor of $x$}{def:end-factors}
\symline{$\ovl{b}$}{Overlap at situation $b$}{def:overlap}
\symline{$a \sim b$}{$a$ is equivalent to $b$}{def:equivalent}

\symsection{Reactions}
\symline{$a \rea b$}{$a$ reacts to $b$}{def:reaction}
\symline{$a \rea_R b$}
{The reaction $a \rea b$ is element of the set $R$}{pg:reaction}
\symline{$\dom R$}{Domain of the reaction set $R$}{pg:r-domain}
\symline{$\hat a$}
{Interval situation determined by $a$}{def:determined-situation}
\symline{$a_L$}
{Leftmost minimal separating interval in $a$}{def:boundary-interval}
\symline{$a_R$}
{Rightmost minimal separating interval in $a$}{def:boundary-interval}
\symline{$+_a$, $-_a$}{Slope operators for $a$}{def:slope-operators}
\symline{$\mathcal{A}_\phi$}
{Set of achronal situations for $\phi$}{def:achronal}
\symline{$\mathcal{A}_{\phi+}$, $\mathcal{A}_{\phi-}$}
{Slopes for $\phi$}{def:achronal}
\symline{$R_+$, $R_-$}
{Slope subsystems of the reaction system $R$}{def:slope-subsystem}


\clearpage
\listoffigures*
\clearpage
\listoftables*

\mainmatter
\chapter{Introduction}

Cellular automata came into being as an almost brutally simplified
model of information processing in a physical medium.

John von Neumann invented cellular automata (together with Stanislaw
Ulam) as part of his work on self-reproducing systems
\cite[p.~42]{Levy1993}. He needed a simplified physical universe in
which he could construct a model which captures the essential
properties of self-reproduction in a biological organism. This model
universe had to be simple enough that a single person could reason
about it while using only paper and pencil and no mechanical aid. It
also had to consist of simple components in order to make sure that
self-reproduction was a property of the simulated organism and not
already built into the physics of its universe. These requirements
lead to several simplifications. The first one is that only discrete
parameters could be used, especially no real numbers. Time in a
cellular automaton therefore runs in discrete steps, like the ticks of
a clock. Space is reduced to a rectangular grid. It consists of the
points of the $n$-dimensional grid $\Z^n$: we speak then of an
$n$-dimensional cellular automaton. The second simplification concerns
the interior of the universe. It must be possible to describe the
self-reproducing organism completely with a finite number of symbols.
The world simulated by the automaton consists of objects at the
lattice points, which are called cells. A cell can be in one of
several states, and there could be only finitely many of them if it
was possible to write down a configuration of the automaton. The cells
are thought as small information-processing machines, representing
atoms, electrical components or possibly biological cells in a tissue.
For the purpose of von Neumann, the simulated universe of the cellular
automaton would contain only a finite number of cells that simulated
the self-reproducing system. All other cells were in a special, quiet
state, which stood for a kind of vacuum that remained unchanged.
Activity was always caused by cells in other states.

The physics of this model universe is one of local interaction. The
purpose of the model universes of cellular automata is the simulation
of an object of moderate size. Therefore the world view of cellular
automata is based on Newtonian physics. It especially ignores General
Relativity. The physical laws in a cellular automaton are then the
same at every point and for every time step. In order to make it
possible that they can be described completely, they also have to
allow a finite description. Therefore the state of a cell in the next
time step cannot depend on the states of all the cells in the
automaton. This leads to the idea that the state of a cell in the next
time step should only depend on the states of the cells in its direct
neighbourhood. Such a neighbourhood contains only the cells at an
Euclidean distance less than or equal to a given constant~$r$. (Note
that with this definition a cell is always part of its neighbourhood.)
The number of cells in a neighbourhood is then always finite. The
number of combinations of states that these cells can have together is
also finite: therefore the behaviour of each cell can be described by
a finite table. It maps each state of the neighbourhood of a cell to
the state of the cell in the next time step. The neighbourhoods of all
points in $\Z^n$, and therefore those of the cells in the cellular
automaton, look the same: therefore it is possible to specify the
behaviour of the whole system of cells with a single finite rule. The
number $r$ is called the radius of the cellular automaton. A
beneficial side effect is that no point of $\Z^n$ is special, as it is
in Newtonian physics. Another side effect is that signal transmission
in the universe of cellular automata always has a finite speed. This
is what John von Neumann did, and he started to develop a
self-reproducing system in a specific two-dimensional cellular
automaton.

A second important step in the history of cellular automata was the
invention of the ``Game of Life'' by John H.~Conway in 1970
\cite[p.~49--53]{Levy1993}. It is a two-dimensional cellular automaton
with two states and an especially simple rule and became soon very
popular. This was the time when computers with a graphics display
started to become accessible to many people. A development that is
important in the context of the present work is that some of them did
run ``Life'' systematically with initial configurations that were
chosen at random. The usual method is to chose a probability $p$ and
then let the computer initialise independently every cell with
probability $p$ in state $1$ and with probability $1 - p$ in state
$0$. State $0$ is the quiet state in Life. Then after several time
steps of evolution\footnote{The word ``evolution'' means different
  things in different contexts. Here I use it in the wider sense of
  ``development over time'', not in the narrow sense of Darwinian
  evolution.

  There is also research on cellular automata where the subject is the
  Darwinian evolution of transition rules for a specific purpose. (See
  e.\,g.\ Mitchell, Crutchfield and Hraber \cite{Mitchell1994} or the
  review by Mitchell \cite{Mitchell1998}.) Nevertheless the use of
  ``evolution'' in the wider sense is also established in cellular
  automata theory. The word ``evolution'' has more specific
  associations than e.\,g.\ ``development'', therefore I use it here.}
stable patterns often emerge. Some of them stay unchanged, others
oscillate with a period of $2$, seldom more, time steps, and a few
move through the two-dimensional cellular space.

This set a precedent, and random initial configurations have become a
standard tool that is used when one wants to get an overview of the
behaviour of an unknown cellular automaton. So when in 1983 Stephen
Wolfram \cite{Wolfram1983} wanted to survey the possible behaviours of
cellular automata, he too tested them on random initial
configurations. The plan to survey the possible behaviours of cellular
automata had a side effect that influences cellular automata theory
until now. Wolfram worked with one-dimensional cellular automata, and
in order to have a subset of manageable size, he chose the set of
automata with two states and radius $1$, the simplest class at all
from which one can expect nontrivial behaviour. Wolfram called them
``elementary cellular automata''. There are $256$ of them, but if one
views automata as equivalent if they differ only by an interchange of
the states $0$ and $1$ or of left and right, only $88$ types of
behaviour remain. (See Li and Packard \cite[p.~284]{Li1990}.) There
has been earlier research on elementary cellular automata, but Wolfram
gave this class a name and introduced a system of code numbers with
which one can refer to their rules, and they have stayed in the centre
of research since then.

The choice of one-dimensional automata is also in another sense
advantageous, since their behaviour can be easily displayed in a
two-dimensional diagram with one space and one time dimension. This
makes communication about their behaviour much more direct than that
about two-dimensional automata.

In contrast to John von Neumann's rule and ``Life'', the set of
elementary cellular automata contains rules in which there is no state
that can be considered quiet. With some of the rules, the cellular
automata stay chaotic when started from a random initial
configuration. But with others, a similar phenomenon as with ``Life''
occurs. After some time, particles appear and move or stay on a simple
background, but here the background does not consist of a region of
cells all in the same state. Instead, the regions between the
particles consist of a spatially periodic pattern. After several steps
of evolution of the cellular automaton, the same pattern occurs again.
When the evolution of the cellular automaton is displayed as a
space-time diagram, the background looks like a wallpaper pattern.
Nowadays, such a periodic background pattern that arises from almost
every random initial configuration is often called an ``ether'', in
analogy to the ether concept of pre-relativistic physics. There, as in
cellular automata, the ether is a background in which particles and
signals move.

\paragraph{Examples} While this thesis focuses on ether formation
under Rule 54, we will now take a larger perspective and look for
examples of ether formation among the elementary cellular automata in
general.

{
  \def\rawfig#1#2#3{\begin{tabular}{l}\ci{eth_#1}\\
      #2{Rule #1\hfill #3}\end{tabular}}
  \def\fig#1{\rawfig{#1}\textrm{}}
  \def\figE#1{\rawfig{#1}\textit{E}}
  \def\capt#1{\caption{A survey of elementary cellular automata (Part
      #1).}}
  \newenvironment{tbl}{\begin{tabular}{c} \toprule}%
    {\bottomrule\end{tabular}}
  \tabcolsep=1ex
  \begin{table}[t]
    \centering
    \capt1
    \begin{tbl}
      \fig{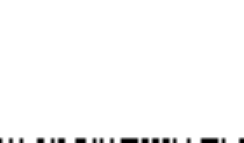} \fig{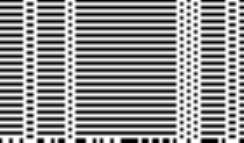} \fig{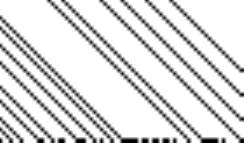} \fig{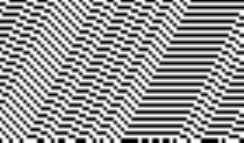} \\
      \fig{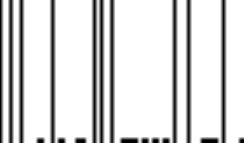} \fig{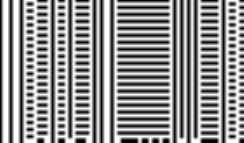} \fig{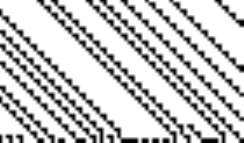} \fig{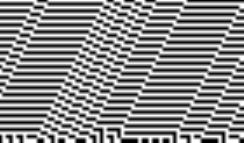} \\
      \fig{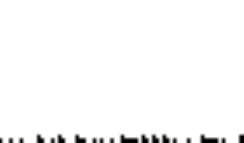} \figE{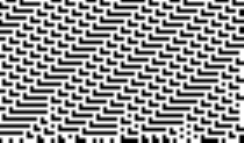} \fig{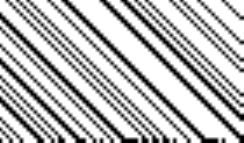} \fig{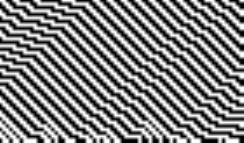} \\
      \fig{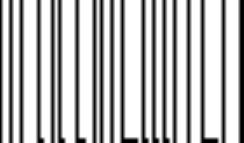} \fig{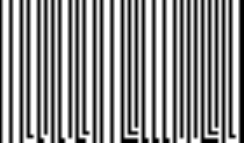} \figE{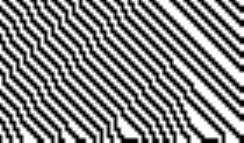} \fig{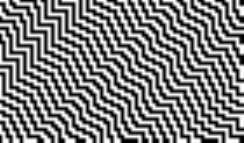} \\
      \fig{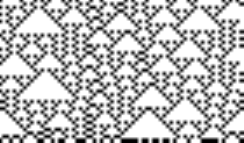} \fig{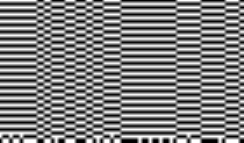} \fig{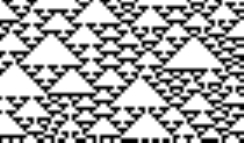} \fig{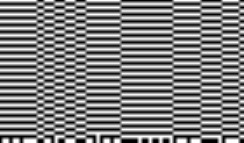} \\
      \fig{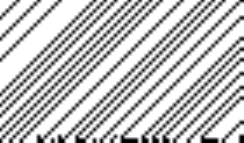} \figE{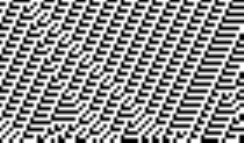} \fig{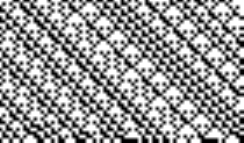} \fig{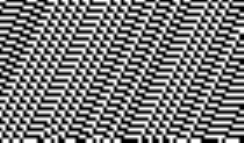} \\
      \fig{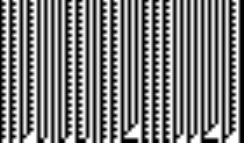} \fig{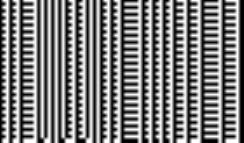} \fig{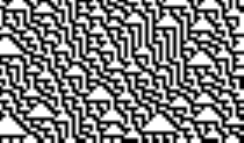} \fig{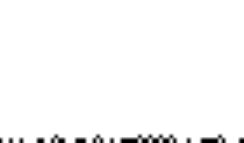} \\
      \fig{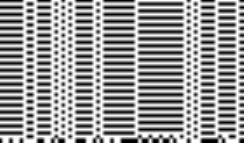} \fig{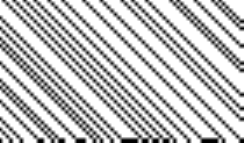} \fig{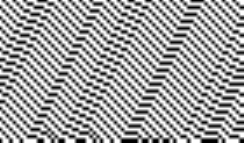} \fig{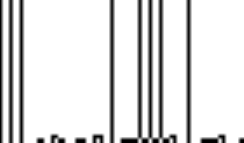} \\
    \end{tbl}
    \label{tab:survey-1}
  \end{table}
  \begin{table}[t]
    \centering
    \capt2
    \begin{tbl}
      \figE{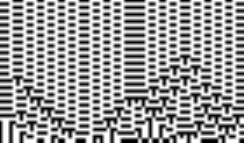} \fig{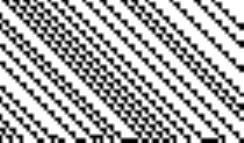} \fig{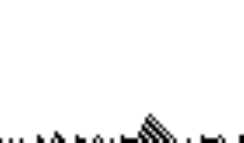} \fig{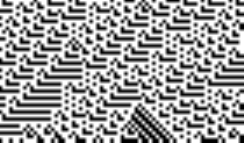} \\
      \fig{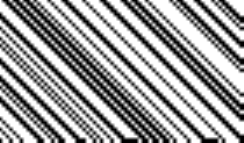} \fig{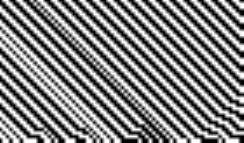} \fig{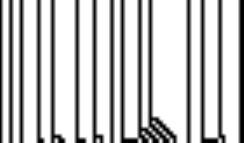} \fig{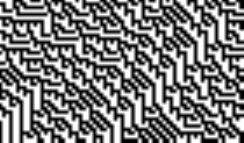} \\
      \fig{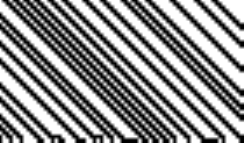} \fig{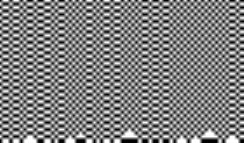} \fig{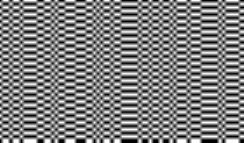} \figE{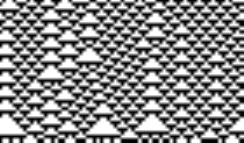} \\
      \fig{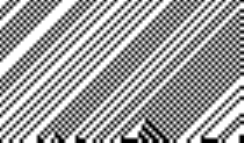} \figE{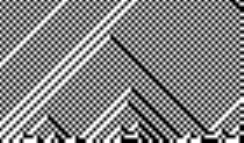} \fig{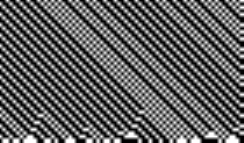} \fig{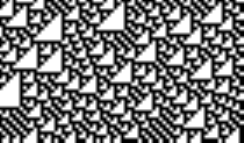} \\
      \figE{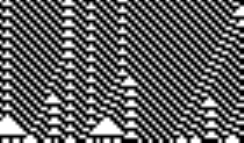} \fig{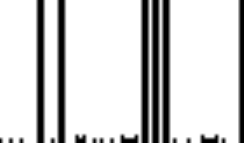} \fig{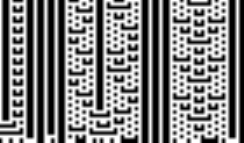} \fig{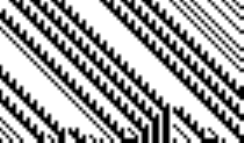} \\
      \fig{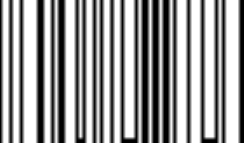} \fig{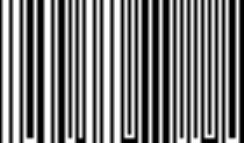} \fig{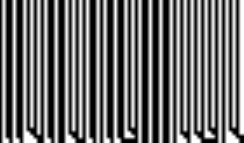} \fig{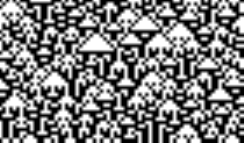} \\
      \fig{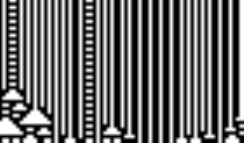} \fig{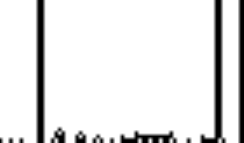} \fig{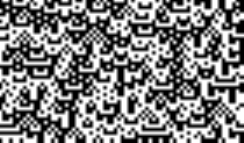} \fig{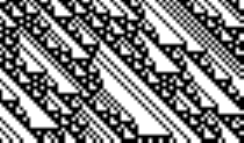} \\
      \fig{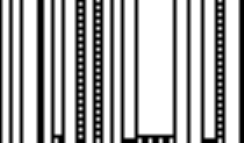} \figE{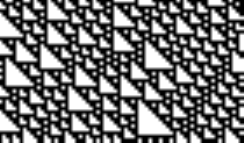} \fig{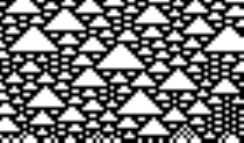} \fig{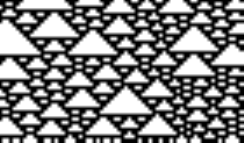} \\
    \end{tbl}
    \label{tab:survey-2}
  \end{table}
  \begin{table}[t]
    \centering
    \capt3
    \begin{tbl}
      \fig{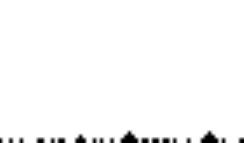} \fig{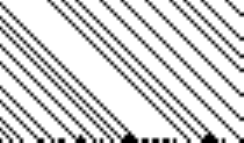} \fig{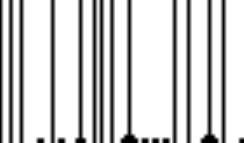} \fig{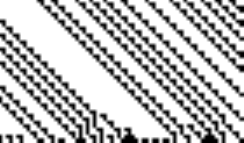} \\
      \fig{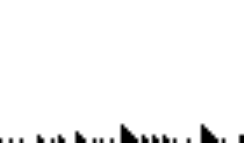} \fig{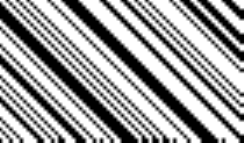} \fig{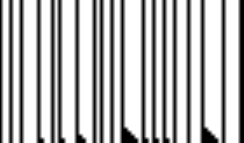} \figE{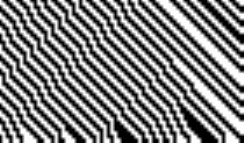} \\
      \fig{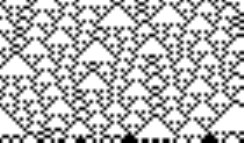} \fig{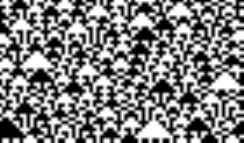} \fig{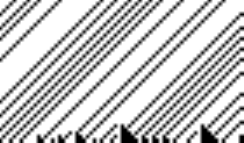} \fig{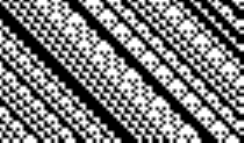} \\
      \fig{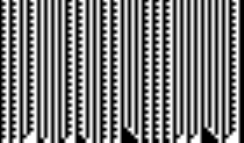} \fig{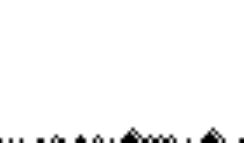} \fig{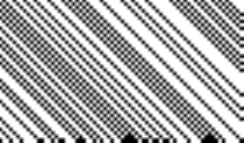} \fig{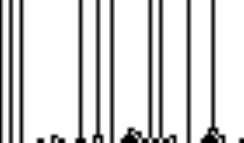} \\
      \fig{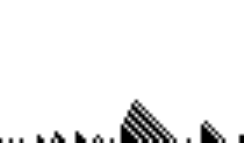} \fig{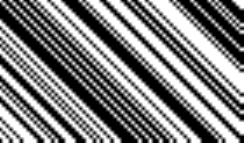} \fig{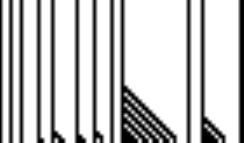} \fig{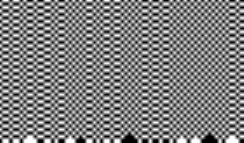} \\
      \figE{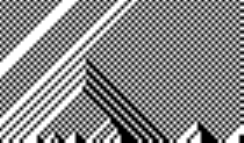} \fig{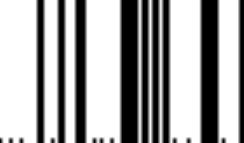} \fig{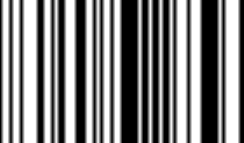} \fig{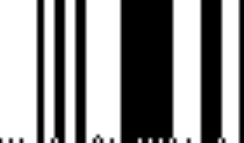} \\
    \end{tbl}
    \label{tab:survey-3}
  \end{table}
}

For this we need a criterion that tells us whether a cellular
automaton has an ether. To my knowledge there is however no general
definition in the literature under which conditions a cellular
automaton has an ether. As a working condition for the following small
survey, we will use the following definition, based on Rule 54:
\begin{quote}
  If the evolution diagram of a typical random initial configuration
  contains regions that are periodic in space and time, and these
  regions grow over time, then this cellular automaton has an
  ether.\footnote{Thus an ether may be an uniform pattern consisting
    of a single cell, like the quiet state under some rules. I do not
    exclude it because one of the properties of the ether is that it
    is the background on which particles move. In this aspect, an
    uniform ether is not different from other ethers. Not excluding
    the uniform case also makes the definition simpler.}
\end{quote}
With it we can check empirically for all elementary cellular automata
whether they have an ether. This is done in
Table~\ref{tab:survey-1}--\ref{tab:survey-3}. The diagrams in this
figure show the evolution of a random initial configuration under all
equivalence classes of elementary cellular automata. All evolution
diagrams use the same initial configuration, one in which each cell is
with probability $\frac12$ in state $0$ and with probability $\frac12$
in state $1$. For this probability distribution the behaviour of the
cellular automaton is unchanged when state $0$ is exchanged with $1$
in its transition rule, or left with right. We will therefore show
only the evolution of one of these related rules, namely that with the
lowest code number.

For some of the rules, e.\,g.\ Rule 110, a larger evolution diagram
than that one shown here is needed to clarify whether they have an
ether. But if we do this, we see that only the cellular automata with
the code numbers 9, 14, 25, 37, 54, 57, 62, 110, 142 and 184 (and
those equivalent to them) support an ether. Their evolution diagrams
are marked in Table~\ref{tab:survey-1}--\ref{tab:survey-3} with an
\emph{E}.

\paragraph{The Problem of Ether Formation} The formation of an ether
in some cellular automata has been mentioned by some authors as an
open question, but to my knowledge no one has published a solution.
Bruno Martin \cite[p.~3]{Martin2000} writes,
\begin{quote}
  ``By observing space-time diagrams of the rule 54 on a random
  configuration, we always see a kind of background with space and
  time period 4 [\dots]. The background apparition is usually very
  fast (less than ten iterations are enough) and still mysterious, we
  have no explanation of this phenomenon.''
\end{quote}
As a repeating pattern, the ether is the simplest structure in a
cellular automaton that can grow to an unlimited size. The problem of
ether formation is therefore the simplest question about an important
form of self-organisation in cellular automata, namely the autonomous
emergence of large structures.

Self-organisation and the construction of complex patterns and
machines in a cellular automaton have in common that they involve the
synchronised behaviour of many cells. To understand them one first
needs a language in which one can express the behaviour of a large
number of interacting cells over time. In an earlier
publication~\cite{Redeker2010} I have called this language ``Flexible
Time''.

\section{Flexible Time}

\paragraph{Motivation} A basic idea behind the formalism of Flexible
Time is that it generalises the way in which a finite part of a
configuration of a cellular automaton is written. For a
one-dimensional cellular automaton, which is an infinite sequence of
cells, a configuration is simply an infinite sequence of cell states.
A single cell state is usually written as a symbol, often a number,
and then one naturally writes a finite region of the cellular
automaton as a string of symbols. Then a sequence like $011101110111$
is a possible content of a region in a cellular automaton for which
the state set contains the values $0$ and $1$. Mathematics has already
developed notations and theories to work with such strings of symbols
efficiently. One simple notational device is the use of exponents to
express repetition, which allows us to express the previous string as
$(0111)^3$. There is also a whole theory of formal languages to handle
such strings. I want to be able to use such methods.

Another ingredient is selective interest. A natural way to understand
a complex system is to decompose it into subsystems and first try to
understand them first. However, a cellular automaton is a model of a
physical (or biological or information processing, \etc.) system. In
such a system there are lots of processes that start and end
independently of each other---lots of organisms that are born and die
at any moment in time, or lots of tasks that are completed
independently. There is no global synchronisation. In a world in which
information travels with finite speed, the starting time of a process
is only influenced by the processes in its direct neighbourhood. And
if it consists itself of subprocesses, which are also loosely coupled,
then they may finish at different times, and it makes no sense to
speak of the ``end time'' of the main process of which they are parts.
A notation for cellular automata that allows one to focus on the
behaviour of arbitrary subprocess has to take that into account.

\paragraph{The Formalism of Flexible Time} My idea to solve to the
questions implied here is influenced by the concepts of Relativity
theory. We will now give up the thought that there must be a globally
determined time. Instead, when a complex process consists of
subprocesses that end at different times in different places, then the
end times of the subprocesses together form the end time of the
process. They are, all together, viewed as a single moment in time.
With this concept of time it is no longer necessary to work with
configurations of infinite size. We will instead specify the content
of a finite region of the cellular automaton, namely that where a
specific process starts, and are then able to compute the generalised
end time of this process and the states of the cells at this time.
There is a mathematical object that specifies the location of some
cells, both in space and in time, together with their states. I call
it a \emph{situation}, and it is a generalisation of the finite
sequence of cell states that I mentioned before. Situations have in
common with cell state sequences that they are formally strings of
symbols; therefore the familiar concepts for words in a formal
language can be applied to them.

There is another kind of mathematical object, the \emph{reaction},
which we will use to specify such a process. It is a pair of
situations, one for the beginning and one for the end of the process.
The set of all reactions for a cellular automaton provides the same
information as the transition rule.

Reactions and situations are thus an alternative means to reason about
the behaviour of a cellular automaton. Together they form the
formalism of \indextwoshort{flexible time}Flexible Time. We have thus
the following (approximate) equivalences:
\begin{center}
  \begin{tabular}{cc}
    \toprule
    Flexible Time & Global Time \\
    \midrule
    situation & configuration \\
    reaction & step in evolution \\
    \bottomrule
  \end{tabular}
\end{center}

\section{Aims and Methods}

The intent of this thesis is to construct a general framework in which
one can solve questions about collective behaviour in one-dimensional
cellular automata. It should be a language that is adaptable to a wide
range of questions. This way I want to provide a step forward towards
the solution of the last one among Stephen Wolfram's ``Twenty Problems
in the Theory of Cellular Automata'' \cite{Wolfram1985},
\begin{quote}
  What higher-level descriptions of information processing in
  cellular automata can be given?
\end{quote}
My approach to address Wolfram's question is to invent a language in
which one can describe more easily the components of a cellular
automaton's information processing system. For a concrete problem this
allows to create a vocabulary of space-time patterns and their
interactions. With this language, increasingly larger structures could
be described and understood, until one would understand the behaviour
of a large and complex information-processing system.

In its current state the language is not so powerful. It does however
allow to express with situations the patterns that one can see in
space-time diagrams more or less directly, and then to express and
prove general theorems about them.

The development of such a tool is easier with a concrete application
in mind. Therefore, a second aim of this work is to find an
explanation why there is an ether in the elementary cellular automaton
with rule code 54.

I have chosen ether formation because it is a case of
self-organisation and therefore interesting in its own right.
Furthermore it is the simplest case of self-organisation in cellular
automata that I am aware of. Rule 54 has a relatively simple ether
that arises early in the evolution of a random initial configuration.
Nevertheless this cellular automaton is far from trivial: With Rule 54
one can e.\,g.\ compute arbitrary Boolean functions
\cite{Ju'arezMart'inez2006a}. This makes it more probable that the
results about ether formation in Rule 54 and the methods to derive
them carry over to other interesting cellular automata.


The thesis is a kind of sequel to my article \cite{Redeker2010}:
There, a description of Flexible Time in the context of a specific
cellular automaton was given, but its general theory was missing. Here
I provide a theoretical justification for the formalism, together with
a study of ether formation for a specific cellular automaton. A more
general theory of structure formation in one-dimensional cellular
automata is something I would like to be able to do at a later time.

\paragraph{Requirements on the Theory} At a very basic level it has
always been difficult to express the phenomena in a cellular automaton
in an understandable way. This is true especially if emphasis is
placed on concrete interactions, like the collision of two particles.
All authors use pictures in some way. This becomes however difficult
once larger structures are involved and the details of such diagrams
would become smaller and smaller.

The aim of this work is therefore to construct a formalism that
expresses the behaviour of a large mass of cells in a cellular
automaton. We need to express the evolution of a cellular automaton in
a way that corresponds to the structures that one can see in the
two-dimensional space-time diagrams, and which is at the same way able
to represent arbitrarily large structures. My goal was to find a
language in which one can describe the behaviour of cellular automata
in an algebraic notation---a kind of cellular automata evolution
program that is run by the human brain. The formalism is therefore
completely algebraic. No pictures are necessary to specify details.
They are however still useful for clarification and to get ideas.

It is an important requirement that the formalism is not bound to a
specific moment in time. A pattern that we see in a space-time diagram
usually extends over a longer period of time. We see it as a
two-dimensional form, and pieces that appear to us as connected may
belong to different times. A formalism that restricts us to snapshots
of the cellular automaton at specific moments in time can not display
this. I have therefore developed a formalism that allows us to jump
forward and backward in time.

\paragraph{Two Kinds of Mathematics} This work is also an attempt to
support a return, after years in which computer experiments dominated,
to the idea that the behaviour of cells in a cellular automaton should
be something that can be comprehended with the help of pencil and
paper alone. I will call these methods here ``traditional
mathematics''.

The methods of computer experiments and of traditional mathematics
have different aims and result in different kinds of understanding.
The result of a computer experiment is knowledge about a single case.
A result of traditional mathematics is a theorem about an infinity of
cases. One may say that, since the majority of the questions asked in
scientific research are about ``all cases'' of a certain kind,
traditional mathematics is the only way to answer them. However, the
requirement of traditional mathematics always to work with an infinity
of cases at once is a severe restriction. In contrast, computer
experiments can be done even in cases where there is not enough
understanding of the question to apply the methods of traditional
mathematics.

The restricted nature of traditional mathematics also holds a great
promise. If it works, traditional mathematics has results that
automatically apply to a great range of cases. This is because the
very restrictedness of its methods makes sure that its results have
only a small number of preconditions. This automatism suggests a
method that combines some of the advantages of both approaches: If one
takes a phenomenon that has been found empirically in a small number
of cases and finds a proof for it, then it will automatically tell us
something about an infinity of other cases. This is what I do here
with the ether in Rule 54.

\paragraph{Previous Work} This work is an extension of the ideas
presented in \cite{Redeker2010}, where the formalism of Flexible Time
was introduced for the case of Rule 110. An application of the
formalism to Rule 54 was presented in \cite{Redeker2010a}, where it
was also shown how the ether and particles were represented with
situations and reactions.

\paragraph{Thanks} I want to thank Andrew Adamatzky, Genaro J.\
Martínez and Lars Immisch for their help in getting me to Bristol.

I also want to thank my thesis supervisors Andrew Adamatzky, Rob
Laister and Tony Solomonides for general support and for comments on
earlier stages of this work. The comments on this thesis during the
various examinations also proved extremely helpful.


\section{Notation}

Since the purpose of this thesis is to introduce a new mathematical
language, I~have to introduce many new words and notations. When
reading the thesis, it may be difficult to keep an overview of all
these concepts. All newly introduced words are therefore listed in the
index at the end, and on page~\pageref{cha:list-symbols} there is a
list of all symbolic notations used in the text, together with short
explanations.

The new concepts themselves are introduced and explained step by step
in the following chapters. But first we have to clarify the notations
for some basic and well-known concepts which are written differently
by different authors.

\paragraph{Sets and Functions} \label{pg:integer-sets}The set of
positive integers is $\N = \{ 1, 2, 3, \dots \}$, and the set of
non-negative integers is $\N_0 = \N \cup \{ 0 \}$.

If $A$ and $B$ are two sets, then \label{pg:func-A-B}$B^A$ is the set
of functions from $A$ to $B$. Therefore we can say of a function $f
\colon A \to B$ that $f \in B^A$. The set $A$ is then called the
\intro{domain} of $f$, and we will write
it \label{pg:domain-func}$\dom f$. Thus for the specific function $f$
just mentioned we have $\dom f = A$.

\label{pg:subset-notation}There are two conventions in use for the
symbol of set inclusion, $\subset$. They differ in the case where the
two sets to be compared are equal. Here we will use the convention
that $A \subset B$ means that $A$ is a proper subset of $B$. If we
want to include the case that $A = B$, we write $A \subseteq B$.

A useful property of functions between sets is monotonity. Let $F$
be a function that maps subsets of $A$ to subsets of $B$. Then we will
say that $F$ is \intro{monotone} if $F(a) \subseteq F(a')$ whenever $a
\subseteq a'$. Similarly, a property $P$ of subsets of a set $A$ is
monotone if, when $a \subseteq A$ has property $P$ and $a'$ is a set
with $a \subseteq a' \subseteq A$, then $a'$ has property $P$.

\label{pg:closure-operator} Let now $F$ be a monotone function $F$
that maps subsets of $A$ to subsets of $A$. If it also has the
property that $a \subseteq F(a)$ and $F(F(a)) = F(a)$ for all $a
\subseteq A$, then $F$ is called a \introtwo{closure operator}. (The
last two concepts are taken from order theory
\cite[p.~145]{Davey2002}.)

\paragraph{Sequences} We will work very often with finite sequences of
arbitrary objects.

Let $A$ be a set. An \introx[sequence]{$A$-sequence} of length $\ell$
is then an $\ell$-tuple of elements of $A$. As it is usual in formal
language theory, we may write a sequence as a formal product of its
elements. So if $a = (\alpha_1, \dots, \alpha_\ell)$ is an element of
$A^\ell$, then it can also be written as $\alpha_1 \dots \alpha_\ell$.
This automatically leads to the notion of a product of $A$-sequences,
defined by concatenation. If $b = \beta_1 \dots \beta_m \in A^m$ is
another $A$-sequence, then their product is
\begin{equation}
  \label{eq:seq-product}
  a b = \alpha_1 \dots \alpha_\ell \beta_1 \dots \beta_m
  \in A^{\ell + m}\,.
\end{equation}

It is easier to work with sequences if one does not always have to
refer to its elements. We therefore introduce now a small arithmetic
for sequences, beginning with the product just defined. It introduces
a semigroup structure in the set of all sequences, therefore it is
natural to introduce an \label{pg:empty-sequence}\introtwo{empty
  sequence}. It is written $\lambda$ and will be used a lot, albeit
mostly under another name. Then, since it is useful to have a product
of an $A$-sequence and an element of $A$, we identify $A$ with $A^1$,
the set of $1$-tuples. There is also $A^0$, the set that only contains
$\lambda$.

With these notations we can introduce a name for the set of all
$A$-sequences. It is called the \introtwoshort{Kleene closure}
\cite[p.~29]{Hopcroft1993} of $A$ and has the algebraic structure of a
monoid,
\begin{equation}
  \label{eq:A-star}
  A^* = \bigcup_{\ell \geq 0} A^\ell\,.
\end{equation}

We will also use other notations and notions that are related to
products, without making much fuss about it. One example for this is
the use of exponents, another the concept of the \intro{decomposition}
of a sequence: If $a \in A^*$ and there are $b$, $c \in A^*$ with $a =
b c$, then we will speak of this equation as the decomposition of $a$
into $b$ and $c$. We will use this as a way to introduce the variables
$b$ and $c$ without explicitly mentioning that they are elements of
$A^*$. Note also that if e.\,g.\ $a = b c$ and of the two factors of
$a$ only $b$ is known, then this already determines $c$. We will use
this as a way to introduce $c$.

A concept that we have already used implicitly is
the \label{pg:length}\indexthree{length of sequence}\emph{length}
of a sequence. We will now introduce a notation for it: if $a \in
A^\ell$\label{pg:astar}, then its length is $\abs{a} = \ell$. We will
often use the fact that the length of a product $a b$ is $\abs{a b} =
\abs{a} + \abs{b}$.

\paragraph{Sequences and Functions} The identification of
$A$-sequences with tuples makes another simplification possible. A
function $f \colon A^n \to B$ can be viewed as taking $n$ parameters
from the set $A$ and mapping them to an element of the set $B$. In
this case we will encounter expressions of the form $f(\alpha_1,
\dots, \alpha_n)$, where $\alpha_1$, \dots, $\alpha_n$ are elements of
$A$. But with the definitions above, $f$ is also a function that maps
$A$-sequences of length $n$ to $B$. Then we can use expressions of the
form $f(a)$ instead, with an $a \in A^n$, for example with $a =
\alpha_1 \dots \alpha_n$.

This simplification becomes especially useful if $f$ is the transition
function of a one-dimensional cellular automaton. It is especially
convenient if $a$ is a product of sequences, say $a = b c$ with $b \in
A^\ell$ and $c \in A^{n-\ell}$. We can then write terms like $f(b c)$
instead of much more voluminous expressions like $f(\beta_1, \dots,
\beta_\ell, \gamma_1, \dots, \gamma_{n-\ell})$.


\chapter{Background}
\label{cha:background}

In this chapter I describe in greater detail how this thesis relates
to cellular automata research in general. I also describe the relation
of this work to other kinds of research that inspired it and how they
influenced it.

\section{Structures in Cellular Automata}
\label{sec:struct-cell-autom}

There are several approaches in use by with which researchers try to
get an understanding of the space-time structures that occur in the
evolution of one-dimensional cellular automata. To give an overview I
will now describe some of these works.

Besides the projects that are directly concerned with pattern
formation I will also describe research that has the description of
patterns as its main theme.

\paragraph{Turing's Work} The ancestor of all mathematical research
about pattern formation is certainly Alan Turing's paper on the
chemical basis of morphogenesis \cite{Turing1952}. From the viewpoint
of cellular automata, there are may similarities: Turing worked with a
ring of cells that have only knowledge of their direct neighbours, he
stressed the necessity of a randomised initial configuration, and he
found that his setup created periodic patterns. He even did a
computational (but not computer) experiment. On the other hand, his
time parameter was continuous, and the state of his cells was
characterised by two or three continuous parameter. The view that
cellular automata are a good model to study pattern formation still
had to wait for some time.

\paragraph{Triangles} For any research on pattern formation it is
useful to find a kind of structure that occurs in many cellular
automata. This enhances the probability that the results of the
research are applicable to many kinds of automata. Different
approaches on pattern formation can therefore be classified by the
kinds of patterns on which they concentrate.

Triangular structures appear in many one-dimensional cellular automata
when they are run from random initial configurations. It is therefore
natural to use them as the building block for the description of more
complex structures.

There is one such approach that uses triangles as building blocks for
larger structures \cite{Ju'arezMart'inez2001,McIntosh2001}. It
currently concentrates on Rule 110. This is one of the cellular
automata that have been studied in great detail. It has a very complex
behaviour and became even more interesting after Matthew Cook had
proved its support for universal computation \cite{Cook2004}. The
triangles in Rule 110 are the building blocks of larger structures.
They are therefore represented by ``tiles'', which are subsets of the
two-dimensional plane. The development of the cellular automaton can
therefore be understood by a covering of the two-dimensional plane
without a gap. From the work with these tiles one can therefore derive
the possible periodic patterns in a cellular automaton, especially
candidates for the ether and for particles
\cite{Ju'arezMart'inez2008}.

\paragraph{Tilings} A tiling approach somewhat similar to this is used
by Ollinger and Richard \cite{Ollinger2007,Richard2008} to express the
interactions of particles under Rule 110. It uses this approach to
express the behaviour of the cellular automaton in terms of particles
and collisions. There are ``tiles'' which represent pieces of the
ether, others which represent the movement of a particle over a finite
amount of time, and others that represent the collision of two or more
particles. A tiling of the two-dimensional plane that corresponds to
the space-time diagram of a cellular automaton is then represented in
an abstract form by ``a planar map whose vertices are labeled by
collisions and edges by particles''
\cite[Definition~1.3]{Richard2008}. These graphs are then used to
represent complex interactions between particles, especially by
Richard \cite{Richard2008} to understand Rule 110. The method is
however applicable to cellular automata in general.

\paragraph{Replicating Patterns} Another specialised approach to
express the large-scale structure for a specific class of cellular
automata concerns those rules which support \emph{replication}. This
class is a subclass of those rules that have a quiet state. In them
one can look at localised patterns that consist of a finite number of
cells in non-quiet states, while all the other cells are quiet.
Replication then occurs in rules under which a small localised pattern
in an initial configuration later reappears as several copies. These
too then replicate, and the evolution of such a pattern in a
one-dimensional cellular automaton generates a fractal-like structure,
a generalisation of the Sierpi\`nski triangle. Gravner and Griffeath
\cite{Gravner2011} give a formal definition for replication in
one-dimensional automata and then search among other things for
replicating patterns under Rule 22. In another paper, by Gravner,
Gliner and Pelfrey \cite{Gravner2011a}, several transition rules are
investigated for their replicating patterns.

\paragraph{Domains and Defects} In several articles the configuration
of the cellular automaton is decomposed into regions with a regular
structure and defects between them
\cite{Boccara1991a,Jen1990,Letourneau2010,Letourneau2010a}. When the
initial configuration is chosen at random, the defects usually take a
random walk. From time to time two of them collide and annihilate each
other, which enlarges the regular regions. This way the state of the
cellular automaton becomes more ordered over time, a phenomenon that
has some similarities with ether formation.

There is a more theoretical view of these phenomena, in a paper by
Eloranta \cite{Eloranta1993}, which yields rigorous results in a
simpler case. In it, the set of states of the cellular automaton is
divided into two subsets, $S$ and $T$, such that the next state of a
cell the neighbours of which are all elements of $S$ is another
element of $S$, and the same is true for $T$. The author then
investigates the behaviour of the boundary between a region of $S$
cells and a region of $T$ cells in which the cell states were chosen
at random. He finds that the boundary moves either deterministically
with maximal speed or it is a random walk, and that it is possible to
give explicit, albeit complicated expressions for the speed of the
walk.

A similar pattern of self-organisation occurs in cellular automata
with an ether and particles, as in Rule 54 \cite{Boccara1991} and Rule
110 \cite{Li1992}. In these automata an ether forms that is disrupted
by particles; the particles move and collide and sometimes destroy
each other. While the transition rule of these automata is
deterministic, the number of particles behaves nevertheless in these
automata as if the collisions and decays occurred at random. In both
papers a power law is found by the computer simulations. In the paper
by Li and Nordahl \cite{Li1992} it concerns the dependence of the
density of particles over time, while Boccara, Nasser and Roger
\cite{Boccara1991} measure the density of a specific particle.

How does one define particles and background? Mostly it was obvious to
the researchers, but there are systematic approaches. The method of
``computational mechanics'' by Crutchfield and Hanson
\cite{Crutchfield1994,Crutchfield1993,Hanson1995} is a systematic
approach that allows, among other things, to divide the configuration
of the cellular automaton into regular \emph{domains} and the
\emph{domain walls} between them. A domain wall may move, therefore
particles count as domain walls. A domain is in the simplest case a
spatially periodic pattern that is preserved by the transition rule,
so the ether counts as a domain. There also exist more complex
domains, and the authors have found a way to identify them
mechanically by a program. Then it is possible to create another
finite automaton that classifies the cells as belonging either to a
domain or one of the domain wall. This allows to show simplified
pictures of the often very complicated evolution diagrams.

Further research in this direction has been done by Marcus Pivato
\cite{Pivato2007,Pivato2007a}. Here, too, the aim is to divide the
cellular evolution into different regions with different behaviour,
again in the form of patterns and defects, but with finer
subdivisions.

\paragraph{Grouping and Supercells} Another method to describe
large-scale structures simply ignores the structures that arise in the
evolution of the cellular automaton. It uses ``grouping'' operations
for the classification of cellular automata
\cite{Delorme2011a,Delorme2011,Martin2000,Mazoyer1999,Moore1995}. In
it the cells of the automaton are arranged in blocks of $n$ cells, and
one then considers the cellular automaton that consists of these
``supercells''. One also considers transition rules that aggregate
several's time steps into one. This way one can establish equivalences
between automata and introduce a partial order between them in terms
of the complicatedness of their behaviour. Among these works the most
elaborate is the work of Delorme, Mazoyer, Ollinger and Theyssier
\cite{Delorme2011a,Delorme2011}. In it the authors give a formal
definition for the generalised grouping operations and then prove
theorems about them in an abstract way. They also define three
concrete grouping operations, find some equivalence classes of
one-dimensional automata under these operations and prove how they are
related in the partial order defined by the grouping operation.

\paragraph{Global Behaviour} All this work with local structures in
cellular evolution has also as its goal the understanding of cellular
automata and to classify them by their behaviour.

The first approach of this kind that found greater resonance was
Wolfram's \cite{Wolfram1984} classification. It divides the cellular
automata into four classes according to the behaviour they show when
starting from a random initial configuration---in other words, by
their ability for self-organisation. However, this classification
scheme is not decidable, as Culik and Yu \cite{CulikII1988} showed.

Another point is that only four classes provide only a very small
amount of information about the cellular automata---especially because
the automata with nontrivial behaviour end up in only two of them. For
this and other reasons, the business of finding classification schemes
for cellular automata is still going on actively. A recent survey
\cite{Martinez2013} lists 18 different classification schemes, just
for the elementary cellular automata.

\section{Physics as Metaphor and Model}

As we have seen in the introduction, a cellular automaton can be
understood as the simulation of a physical system. The nature of this
system is however the subject of some confusion: Is it Newtonian or is
it relativistic---and what role does such an old-fashioned concept as
the ether play?

\paragraph{Newtonian and Relativistic Physics} Since a cellular
automaton is only a rough approximation to a physical system, we have
a certain amount of freedom in our interpretation. We can choose what
kind of physics our cellular automaton should resemble. The formalism
of Flexible Time is an attempt to bring a relativistic interpretation
into the cellular automata, which have before mostly interpreted in a
Newtonian fashion.

A sign of the Newtonian viewpoint is the existence of an universal
clock. In the usual formulation, a one-dimensional cellular automaton
consists of an infinite line of cells, and they evolve in discrete
time steps. Time passes therefore at every point in the same way.

The central point of Relativity, on the other hand, is the finite
maximal speed with which signals can propagate. In a cellular
automaton we also have a finite maximal speed: It is given by the
radius of the transition rule. If the transition rule has radius $r$,
then the state of a cell can influence in the next time step only the
cells at most $r$ places to the left or the right. The analogy has
been known for a long time: In the context of the Game of Life, this
maximal speed has already been called by J.~H.\ Conway the ``speed of
light'' \cite[p.~217]{Gardner1983}. We can use the analogy to let the
cellular automaton play the role of the universe of Special
Relativity.

We can take this analogy a step further. 
As in Relativity, when there is no global concept of time, causality
becomes important. For cellular automata, causality can become the
question, ``If I change the state of one cell in the initial
configuration, which cells change their state in later time steps?''
This has been asked e.\,g.\ by Wolfram \cite[p.~171]{Wolfram1985}. In
this thesis, the dual question becomes important, ``If the states of
only a finite number of cells are known in a cellular automaton, the
states of which other cells can be determined from this knowledge?''
This question will lead to the concept of the closure in
Definition~\ref{def:closure}.

We can maintain the standpoint that the set of all cells at a given
time is not such an important concept. After all, each cell knows only
about a finite number of its direct neighbours. The concept of a
configuration, consisting of all cells at a time step, is therefore
nothing which one is forced to use. We have, as in Relativity, a
freedom to choose which events we consider as occurring at the same
time. In Relativity, they form a ``space-like'' set. In Flexible Time,
we will speak of achronal situations. We only have the requirement
that the events that can influence each other causally cannot be part
of the same time slice.\footnote{This requirement is broken a little
  bit in achronal situations, but it is correct in the large scale.}
Then we have the flexibility that allows us to follow more easily the
structures that occur in the evolution of a cellular automaton.

\paragraph{Space-time} Another important concept that became popular
through Relativity is that of space-time. We will used heavily the
freedom that it provides. The space-time viewpoint for cellular
automata is actually quite old. An early example occurs in Konrad
Zuse's article about Calculating Space \cite{Zuse1967}. Here, in
Figures 9 and 10, the author uses a mode of display in which events
from different times are displayed together. This way the movement of
a particle can be shown, even though it extends over several time
steps. This is however an informal use of a flexible time; I have not
seen diagrams of the same style elsewhere.

There is however an example where events from different times occur
naturally during a computation of a cellular automaton. William Gosper
\cite{Gosper1984} uses such a scheme to compute the evolution under
the Game of Life (or another two-dimensional cellular automaton) in a
faster way. One could view his scheme as a form of Flexible Time in
two dimensions---albeit one in which all situations are based on
squares with an edge length that is a power of~$2$. This work was an
important inspiration for me.

\paragraph{The Ether and Other Muddled Metaphors} There still remains
the question which role the old-fashioned concept of the ether plays
in such highly modern physics.

A part of the answer is that the word is already in use: The name
``ether'' has apparently been introduced by Matthew Cook
\cite{Cook2004} for the regular pattern in Rule 110, and it has been
used by other authors too.

We can however take the concept of the ether a bit more seriously, as
the physicists of the 19th century did. For them, the ether formed a
background on which signals travel. The ether was however specifically
invented to support the transmission of waves, for which there is no
analogue in the context of cellular automata. We do have particles
that move in the ether, but there is no ether at the place where the
particle is located. To have a true ether, we would need an analogy to
ether vibrations as they were thought to occur in the physical ether.
To my knowledge, nobody has attempted such an analogy. I therefore
believe that we should take the analogy to the ether---in contrast to
that with Relativity---not too seriously.

If one keeps this in mind, even an inexact metaphor can serve as
support for the intuition and help to find names for the phenomena
that occur in cellular automata. In case of the ether, I have done so
in Chapter~\ref{cha:ether}, where I speak of ``pure'' and
``disturbed'' ether (the disturbances being the particles), thus using
exactly that analogy I just have rejected as being not exact.

Another incongruent use of physical metaphors is the use of the word
``particle''. It is nowadays a common word for a localised structure
in a cellular automaton that moves with a constant speed. The name is
especially used for a localised structure that stays in its place,
like the static structures that occur under Rule 54
\cite{Boccara1991}. One of the earliest uses must be again Zuse
\cite{Zuse1967}, who explicitly set out to simulate physical particles
with cellular automata.

The particle metaphor is nowadays used by many authors (and also in
this work), but it is not a faithful image of, say, elementary
particles, or Newtonian idealised point particles. Among the features
that are generally missing are an analogy to mass or impulse, or to
any kind of conservation theorems. What remains is a kind of
``topological'' image of physical particles, in which the particles
move in straight lines and interact only when they collide, but there
are no general laws about that what is the result of the collision.
Once again this is a metaphor that should not be followed too far.

\section{Relation to Logic and Language}

As the subject of this thesis is the construction of a language for
easier mathematical reasoning, I have to name other projects that are
related to mathematical languages and their construction. Most of them
have provided context or direction for this project.

\paragraph{Combinatory Logic} The structure of the resulting reaction
system has some similarities with the systems used in Combinatory
Logic \cite{Curry1958,Hindley1986}. One of the motivation that lead to
the introduction of Combinatory Logic was the analysis of the
substitution process in formulas \cite[p.~1]{Curry1958}. The formal
process with which a term is substituted for a variable in a formula
is quite complicated, especially if the formula may contain free and
bound variables. In Combinatory Logic, the substitution process is
decomposed into elementary steps, which consists of purely textual
substitutions. (See e.\,g.\ \cite[p.~7]{Hindley1986}.)

This concentration on elementary, textual substitutions in Combinatory
Logic served as a model for the development of Flexible Time.
Especially the concepts of applying a reaction
(Definition~\ref{def:application}) and confluence
(Theorem~\ref{thm:church-rosser}) have their similarities in
Combinatory Logic.

\paragraph{Development of a Language} There are many predecessors for
the idea to develop a language that helps us to think more efficiently
about cellular automata. Among the first, and certainly the most
illustrious, was Gottfried Wilhelm Leibniz with his project of a
``universal characteristic''. This was to be an ideal language and a
general symbolic method or both, because there is a certain ambiguity
in Leibniz's writings \cite[p.~226--227]{Rutherford2006}. In the first
interpretation, the language should consist of ``signs which process a
determinate content and exactly correspond in their structure to the
analysis of thought'' \cite[p.~227]{Rutherford2006}, in the second it
would be a symbolic calculus, an ``instrument to reason''
\cite[p.~230]{Rutherford2006}.

In this second aspect, Leibniz's work is widely seen as a predecessor
to formal logic. (A rare concrete example for his thoughts about
formal reasoning looks to modern eyes like a formalisation of set
inclusion or propositional calculus \cite[p.~18]{Davis2000}.) The
first aspect emphasises the idea the \emph{signs} of the language
should correspond to the concepts of thought in a simple way. This is
an aspect that is mostly ignored in the theory of formal systems, but
not by Leibniz: His symbolism for integration was clearly designed
with the intention in mind to find symbols that aid thought. The
resulting mathematical language has always been seen as widely
superior to Newton's version
\cite[p.~12]{Davis2000}.\footnote{However, to my knowledge, Leibniz
  seems not to have understood the formalism of calculus as a part of
  his project of finding an universal characteristic.}

So far the current thesis could be seen as a part of Leibniz's
project, but there are differences. First, Leibniz had viewed the
signs of his universal characteristic as the most primitive concepts,
and believed that they could be found once and for all. Second, he
imagined his universal language as something complete, encompassing
all human knowledge. A growing language, intended for a small
subuniverse of mathematics, would not be his intention.

In order to find a model for this kind of project we need to look into
a direction that at first seems to be completely unrelated: the
construction of languages for the communication with
extraterrestrials. There is a program outlined by Lancelot Hogben
\cite{Hogben1952} on how to establish communication with an
extraterrestrial civilisation via radio signals. To establish a means
of communication, and a common vocabulary of concepts, ``lessons'' are
sent out to the extraterrestrials, starting with numbers and
arithmetic and then building up on this base increasingly complex
concepts. Hogben's paper is only a sketch of such a program. The most
elaborate implementation is certainly Hans Freudenthal's ``Lincos''
\cite{Freudenthal1960}, in which he introduces step by step the
concepts for mathematics, time, basic human behaviour and elementary
physics.

But this description is misleading in one point: Freudenthal's primary
interest was to create a logical language that was actually usable for
communication, and in order to do this he used interstellar
communication as an example problem. This is then the point where a
project like Freudenthal's becomes a model for works like this thesis.
We have here namely an example for a language that grows step by step
from examples, which is never complete, and which at every step of its
development can only access a limited set of concepts. It also sets an
example by requiring a concrete example to let the language grow.

\paragraph{Influence on this Work} As Freudenthal needed a
communication problem to develop a language for communication, we will
need a self-organisation problem to develop a language about
self-organisation and structures.

Another lesson from Freudenthal's work is to let the language evolve
step by step, from simple to complex concepts. For this thesis this
means that at the beginning the concepts are quite general and are
valid for every one-dimensional cellular automaton. Step by step, by
the amount that we learn about the theory and its abilities, the range
of the definitions and theorems becomes more restricted, but in
exchange they become more powerful, until finally they offer insight
into ether formation under Rule 54.


\chapter{Cellular Evolution}
\label{cha:cellular-evolution}

This chapter starts with the definition of the basic terms that are
needed to speak about one-dimensional automata, and then introduces
concepts that captures their development over time. It finishes with
theorems about that what can be said about the development of a
cellular automaton when only a part of its cells is known.

\section{One-dimensional Cellular Automata}

Imagine the cellular automaton as a physical object.

It consists of an infinite row of \emph{cells}. The positions of the
cells are integers. The cells are simple machines with a finite amount
of memory, and they are all equal. Two cells may differ only by the
content of their memory. The possible states of a cell are elements of
the finite \introtwo{state set} $\Sigma$. There is a function
\begin{equation}
  c \colon \Z \to \Sigma
\end{equation}
that maps the position of a cell to its state. Such a function is
called here a \intro{configuration} of the cellular automaton. The set
of all configurations of a cellular automaton is therefore the set
$\Sigma^\Z$.

Cellular automata evolve over time. Time for cellular automata is
discrete and the time coordinate takes integer values. We speak of
\indextwo{time step}\emph{time steps}. The behaviour of a cellular
automaton is given by a \indextwoshort{transition
  rule}\introthree{local transition rule} $\phi$ with \intro{radius}
$r$,
\begin{equation}
  \label{eq:local-rule}
  \phi \colon \Sigma^{2r+1} \to \Sigma\,.
\end{equation}
It maps the neighbourhood of a cell to the state of it one time step
later. To see in which way, we need a notion for the collection of all
configurations of a cellular automaton at all time steps. I call such
a collection the \introtwoshort{evolution sequence} of the cellular
automaton.\footnote{In the theory of dynamical systems this is often
  called the \intro{orbit}. But the definition of this term seems to
  vary between authors. For Alligood, Sauer and Yorke
  \cite[p.~5]{Alligood1996}, the orbit is a set, while for Strogatz
  \cite[p.~348]{Strogatz1994} it is a sequence. However, in the case
  of cellular automata a sequence is the more natural choice.} It is
an infinite sequence $(c_t)_{t \geq 0}$ of configurations. We will
only consider evolution sequences that belong to a specific local
transition rule $\phi$. In such an evolution the configuration $c_0$
must be specified in advance: it is the \introtwoshort{initial
  configuration} of the sequence. Each later configuration $c_t$, with
$t \geq 1$, depends by the \index{rule!global transition
  $\sim$}\introtwoshort{global transition rule}
\begin{equation}
  \label{eq:global-rule}
  c_t(x) = \phi(c_{t-1}(x - r), \dots, c_{t-1}(x + r))
  \qquad\text{for all $x \in \Z$}.
\end{equation}
on its predecessor configuration $c_{t-1}$.

\paragraph{Conventions} We can specify a one-dimensional cellular
automaton completely by specifying $\Sigma$, $r$ and $\phi$. Since we
will work almost always with one specific cellular automaton at a
time, we will from now on keep $\Sigma$, $\phi$ and $r$ fixed and not
refer to it in most of the notation.

Furthermore, since the subject of this thesis are one-dimensional
automata, we will from now on in most cases omit the adjective
``one-dimensional''.

\paragraph{Radius Invariance} It is possible that two different local
transition rules lead to the same global rule. A pair of such
equivalent transition rules is easy to construct: For a given local
transition rule $\phi$ with radius $r$, let $\phi'\colon \Sigma^{2r' +
  1} \to \Sigma$ be a rule with radius $r' >r$ such that
\begin{equation}
  \label{eq:sigma-extended}
  \phi'(\sigma_{-r'}, \dots, \sigma_{r'}) =
  \phi(\sigma_{-r}, \dots, \sigma_r)
  \qquad\text{for all $\sigma_{-r'}, \dots, \sigma_{r'} \in \Sigma$.}
\end{equation}
Then $\phi'$, which ignores the states of the additional cells, has
the same global transition rule as $\phi$. We call such a $\phi'$ the
\intro{extension} of $\phi$ to the radius $r'$. It is easy to see that
if two local transition rules lead to the same global rule, then one
must be the extension of the other one.

The centre of our interest in a cellular automaton is the behaviour of
its cells, not its local transition rule. Therefore we will view here
cellular automata with the same global transition rule as equivalent,
since they have the same evolutions. Nevertheless the local transition
rule provides an easy way to specify the properties of a cellular
automaton. So we will use it, but we will require that the properties
and functions defined for cellular automata are invariant of the
radius of its local transition rule, in the following sense:
\begin{definition}[Radius Invariance]
  \label{def:radius-invariance}
  A property of a cellular automaton is \intro{radius-invariant} when
  it is true for a local transition rule $\phi$ if and only if it is
  true for all its extensions.
\end{definition}

What is then the radius of the cellular automaton itself? We will here
allow that a cellular automaton has more than one radius: A number $r$
is a \emph{radius} for a cellular automaton if there exists a local
transition rule with radius $r$ that generates its global transition
function.

\section{Cellular Processes}

At this point, our only tool to analyse the concrete behaviour of a
cellular automaton---i.\,e.\ when its initial configuration $c_0$ is
given---is its evolution sequence $(c_t)_{t \geq 0}$. But this is for
many applications not enough. It requires the knowledge of infinitely
many cell states, which is too much when our interest is only on the
development of a specific localised pattern.

A better way of formalisation for cellular evolution is inspired by
the way cellular evolution is usually shown in pictures.

\begin{figure}[ht]
  \centering
  \ci{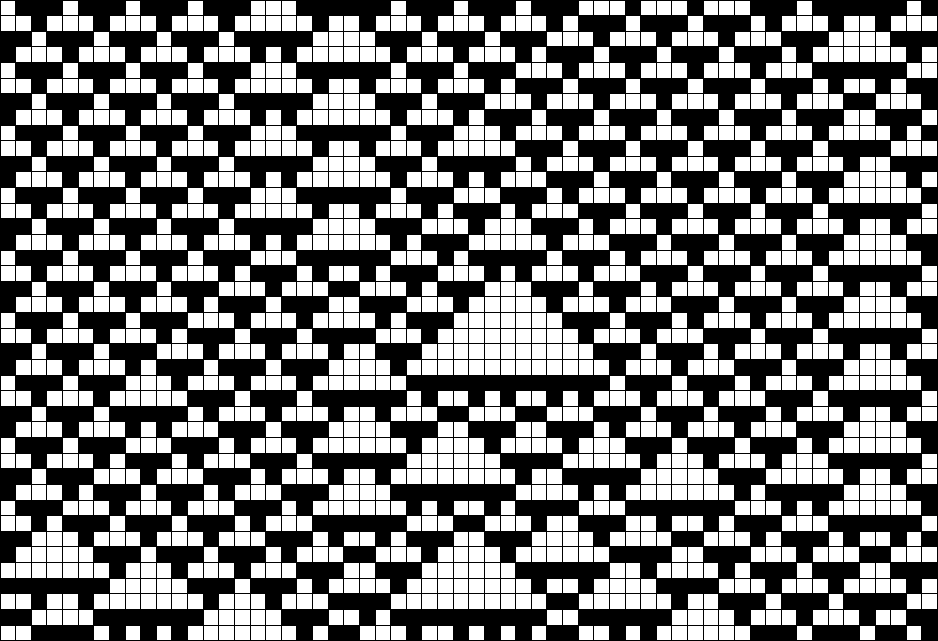}
  \caption{Space-time diagram of an evolution under Rule 54.}
  \label{fig:random_54}
\end{figure}

\paragraph{Space-time Diagrams} The evolution of a cellular automaton
is in general shown by a diagram like that in
Figure~\ref{fig:random_54}.
It is a rectangular array in which the states of the cells are shown
by squares of different colours. In this picture, which shows the
evolution of an elementary cellular automaton, the colours are white
for cells in state $0$ and black for cells in state $1$. This colour
convention is kept---sometimes in a modified way---in all the other
pictures of cellular evolution that occur in this thesis.

The place of the square in the diagram specifies its place in space
and time. The $x$-coordinate of a square determines its location in
space, and its horizontal position marks the point in time to which it
refers. In this thesis I use the physicist's convention for space-time
diagrams in which time runs
upward.\footnote{\label{fn:time-direction}There is also a strong
  tradition to draw the diagrams with time running downward. It
  probably has its origin in the time when one-dimensional cellular
  were simulated in a computer and then printed with a line printer,
  with one letter for every cell state and the cells of one time step
  in a line. Then the natural way to display the evolution is to print
  them in the way they are computed, and in the resulting diagram time
  runs downward. (For an example see Wolfram \cite{Wolfram1983}.)

  For some reason, the physicist's convention is also the preferred
  convention used by French authors on cellular automata.} Each row in
such a diagram is then part of a configuration, and the partial image
of configuration $c_{t+1}$ is directly above that of $c_t$.

Sometimes we will not need a whole rectangle and draw therefore only a
subset of its squares.

\paragraph{Explanation of Figure~\ref{fig:random_54}} The diagram in
Figure~\ref{fig:random_54} then displays the evolution of a random
initial configuration under Rule 54. Since it displays a the evolution
of an infinite line of cells, there is no wraparound and, in contrast
to many other space-time diagrams, the leftmost cell in each time step
is \emph{not} the right neighbour of its rightmost cell.

\paragraph{Events and Space-time} We will now introduce a mathematical
equivalent to the space-time diagrams, namely the concept of cellular
processes. It will be able to display any behaviour of the cells and
is not restricted to the case that the cells follow a transitions
rule.

The first step in defining a mathematical analog of an space-time
diagram is to find a representation for a single square. A square has
a position and a colour, and the colour represents a state. The
position has a space and a time component, and both are integers.
Therefore the following definition is reasonable.

\begin{definition}[Cellular Event]
  A \introtwo{cellular event} is a pair $(p, \sigma) \in \Z^2 \times
  \Sigma$.

  It consists of a \intro{position} $p$ and a \intro{state} $\sigma$.
  The first component of $p$ is its time coordinate and the second its
  space coordinate. The pair $(p, \sigma)$ will be usually written
  $[p]\sigma$.
\end{definition}

With this definition, negative time values are explicitly allowed. We
will need them later, when situations are introduced in
Definition~\ref{def:situation}, because for them it is natural to
refer to events with negative time values. Viewed from our current
standpoint, this gives us the choice to let the evolution of the
cellular automaton begin at an arbitrary time step, not just at time
$0$.

The name ``event'' has been taken from Relativity theory. There it
stands for a point of four-dimensional space-time (see Wald
\cite[p.~4]{Wald1984}). I have here extended it to mean
``space${}+{}$time${}+{}$physical conditions at this point'', first
because I have not found another word for this idea, and second
because it then harmonises with the use of ``event'' in everyday
language.

Using another word from Relativity \cite{Wald1984}, we will call the
position $p$ of an event $[p] \sigma$
a \label{pg:space-time-point}\index{point}\emph{space-time point}. The
convention that the first component of a space-time point is the time
and the second the space coordinate is also from Relativity. It
extends in a natural way to all other cases where an element of $\Z^2$
is used for the same purpose in the context of cellular automata. I
use here concepts from Relativity because Relativity theory has
already well-developed concepts to treat space and time in an unified
way.

With the notations
\begin{equation}
  p_T = t \qqtext{and} p_X = x
\end{equation}
we will refer to the components of a space-time point $p \in \Z^2$
with $p = (t, x)$. The use of capital letters for this purpose in
unusual, but lower case $t$ is already used as time variable and
occurs also as index letter.

As a kind of inverse to the component notation we will need the
\indextwoshort{unit vector}\emph{unit vectors} of space-time,
\begin{equation}
  \label{eq:unit-vectors}
  \et = (1, 0) \qqtext{and} \ex = (0, 1),
\end{equation}
especially to refer to differences between space-time points in a more
abstract way.

A final set of conventions refer to the ``point'' component of an
event $[p] \sigma$. If $p = (t,x)$, we may write $[p]$ as $[t, x]$. If
$t = 0$, we may abbreviate $[t, x]$ to $[x]$. We also write $[p][q]$
for $[p + q]$. I list them here only for reasons of completeness. They
will become useful later in the context of situations.

\paragraph{Processes} Next we need a mathematical object that
resembles a whole space-time diagram. There are two possibilities. We
may interpret the diagram as the picture of a function that maps a
space-time point to a cell state. Then the correct way to represent a
space-time diagram would be a map from a subset of $\Z^2$ to $\Sigma$.
We will however also need to express unions and intersections of
cellular processes, something that is easier to express if a cellular
process were a set of events. Therefore I will introduce now a concept
that intends to unite the good properties of functions and
sets.\footnote{There is a viewpoint in mathematics that functions
  \emph{are} sets, but it is apparently not shared by everyone.
  Therefore I do here the unification explicitly. (I had used the
  other approach in my previous paper \cite{Redeker2010}.)}

Let $A$ and $B$ be two sets. I call a set $F \subseteq A \times B$
\indextwoshort{function-like set}\emph{function-like} if there is a
set $D \subseteq A$ and a function $f\colon D \to B$ such that
\begin{equation}
  \label{eq:function-like}
  F = \set{ (a, f(a)) \colon a \in D}\,.
\end{equation}
In notation we will treat function-like sets like functions. The term
$F(a)$ stands for the element $b \in B$ for which $(a, b) \in F$.
There is exactly one such $b$ because $F$ is function-like. The
\indexthree{domain of process}\emph{domain} of $F$ is the set
\begin{equation}
  \label{eq:domain}
  \dom F = \set{ a \colon \exists b \in B \colon (a, b) \in F}\,.
\end{equation}
For the $F$ of equation~\eqref{eq:function-like} we have $\dom F = D$.
The \intro{restriction} of $F$ to a set $A' \subseteq A$ is the set
\begin{equation}
  \label{eq:process-restriction}
  F|_{A'} = \set{ (a, b) \in F \colon a \in A' },
\end{equation}
which is also function-like.

Then we can define cellular processes as a special kind of
function-like sets. Together with the cellular processes we define
also a short notation for the subset of all events at a certain time.
\begin{definition}[Cellular Process]
  \label{def:cellular-process}
  A \introtwo{cellular process} is a function-like set of cellular
  events. The set of all cellular processes is called $\mathcal{P}$.

  If $\pi \in \mathcal{P}$ and $t \in \Z$, then its restriction to
  events at time $t$, its \introtwo{time slice}, is the cellular
  process
  \begin{equation}
    \label{eq:time-slice}
    \pi^{(t)} = \set{ ([t, x] \sigma \in \pi \colon x \in \Z }\,.
  \end{equation}
\end{definition}

\paragraph{Compatibility} Next we consider the set-theoretic
operations for cellular processes. Here we must know whether the
result of a set-theoretic operation applied to one or more cellular
processes is again a cellular process.

This is no problem with subset formation and intersection: since the
subset of a cellular process is again a cellular process, the
intersection of two processes is a process too. The only exception is
the union of cellular processes. It is not always a function-like set.

An exception may occur when two cellular processes $\pi$, $\theta \in
\mathcal{P}$ have domains that overlap in a point $p$. It is then
possible that there are events $[p]\sigma \in \pi$ and $[p]\tau \in
\theta$ with $\sigma \neq \tau$. Then the set $\pi \cup \theta$
exists, but it is no longer function-like. If it were, there would be
a function $f \colon \dom \pi \cup \dom \theta \to \Sigma$ with $f(p)
= \sigma$ and $f(p) = \tau$ at the same time, which is impossible.

If this does not happen, we say that $\pi$ and $\theta$ are
compatible:
\begin{definition}[Compatibility]
  \label{def:compatible}
  Two cellular processes $\pi$, $\theta \in \mathcal{P}$ are
  \intro{compatible} if
  \begin{equation}
    \pi(p) = \theta(p)
    \qquad\text{for all  $p \in \dom \pi \cap \dom \theta$.}
  \end{equation}
  We write this as $\pi \comp \theta$.
\end{definition}
The question of compatibility plays an important role in this theory
of cellular processes. In the rest of this text we must check very
often whether a certain construction is possible, and if it is not,
the cause is almost always incompatibility.

\section{Evolution}
\label{sec:cellular-evolution}

We will now define what it means when a cellular process follows a
transition rule. The construction that will be defined at the end must
generalise the way in which an evolution sequence depends on its
initial configuration. This is because currently the only way the
behaviour of a cellular automaton is formally defined is via the
evolution sequence.

As an intermediate step and to verify later the definition, we will
translate now evolution sequences and configurations into the language
of cellular processes. For this, let $(c_t)_{t \geq 0}$ be an
evolution sequence of a cellular automaton. Then there exists a
cellular process
\begin{equation}
  \label{eq:gamma-whole}
  \gamma = \set{[t, x]c_t(x) \colon t, x \in \Z}
\end{equation}
that contains all the information in $(c_t)_{t \geq 0}$. The
information of every configuration $c_t$ in the sequence is contained
in the time slice $\gamma^{(t)} = \set{[t, x]c_t(x) \colon x \in \Z}$
of $\gamma$. Our task is then to find a construction that, among other
things, extends the initial time slice $\gamma^{(0)}$ to the whole
process $\gamma$, in the same way as the global transition
rule~\eqref{eq:global-rule} extends the initial configuration $c_0$ to
$(c_t)_{t \geq 0}$.

We will call this construction the \emph{closure} of a process,
because it will turn out to be a closure operator as defined on
page~\pageref{pg:closure-operator}.

The closure will be defined in two steps. First we consider the case
of a single event. Given a process $\pi$ and a point $p$, what does it
mean that we can reconstruct the state of the event at $p$ from $\pi$?
If this is the case, we say that the event at $p$ is determined by
$\pi$. What this means exactly will be described in
Definition~\ref{def:determined}.

As a second step we consider the events that are determined by $\pi$,
together with the events that are determined by $\pi$ and them, and so
on: together they form the closure of $\pi$. It will turn out that not
every process has a closure. The result of the second step is
Definition~\ref{def:closure}.\footnote{The construction introduced
  here has some similarity with the use of ``tiling constraints'' to
  specify the space-time pattern of the cell states in a
  one-dimensional cellular automaton by Ollinger and Richard
  \cite[p.~4]{Ollinger2007}.}

\paragraph{Determined Events} Let $p = (t, x)$ be a space-time point.
Given a cellular process $\pi \in \mathcal{P}$ and a transition rule
$\phi$ of radius $r$, what could be the state of the event at $p$?

We will answer this question first for the case of $\pi = \gamma$,
with $\gamma$ as in~\eqref{eq:gamma-whole}. In $\gamma$, the states of
an event at time $t > 0$ depend, by the global transition rule, on the
events at time $t - 1$. We will then say that the events of $\gamma
\setminus \gamma^{(0)}$ are \emph{determined} by $\gamma$. From this
we will now distill concepts that tell us how a transition rule $\phi$
acts on a cellular process. The first goal is then to express the
global transition rule for evolution sequences in a form that is
meaningful for processes like $\gamma$.

The point $p = (t, x)$ is the coordinate of the cell at position $x$
and time $t$. The state of a cell at time $t$ depends on the states of
the cells in its neighbourhood at time $t - 1$. So we must consider
the neighbourhood of the point $p - \et = (t - 1, x)$ to compute the
state of the event at $p$.

\paragraph{Neighbourhoods} For easier notation we will now first
describe the neighbourhood of the point $p$ instead of that of $p -
\et$. We begin with the neighbourhood of a cell as a set of space-time
points, without reference to a cellular process. Since the transition
rule has radius $r$, the cell in the cellular automaton at position
$x$ has a neighbourhood that consists of the cells at positions $x -
r$, \dots, $x + r$. At a time $t$, these cells are located at the
space-time points $(t, x - r)$, \dots, $(t, x + r)$. The central cell
itself is located at $(t, x)$, or $p$. Therefore we can say that the
neighbourhood of the point $p$ is consists of the points $p - r \ex$,
\dots, $p + r \ex$. To refer to it we introduce the following
definition.
\begin{definition}[Neighbourhood Domain]
  Let $p \in \Z^2$ and $r \in \N_0$. The \introtwo{neighbourhood
    domain} of $p$ with radius $r$ is the set
  \begin{equation}
    \label{eq:nb-domain}
    N(p, r) = \{ p - r \ex, \dots, p + r \ex \}\,.
  \end{equation}
\end{definition}

Next we must find an expression for the states of those events in
$\gamma$ that are located at the points of $N(p, r)$. We need them not
just as a set, but also in their natural order. Therefore we express
them as the cellular process $\nu(p, w)$, defined below. In the same
way that we can write $[p] \sigma \in \gamma$ to express the fact that
in the process $\gamma$ the event at point $p$ has state $\sigma$, we
will write $\nu(p, \omega_{-r} \dots \omega_r) \subseteq \gamma$ to
express the fact that in $\gamma$ the events at the points $p - r
\ex$, \dots, $p + r \ex$ have, respectively, the states $\omega_{-r}$,
\dots, $\omega_r$.
\begin{definition}[Neighbourhood Process]
  Let $w = \omega_{-r} \dots \omega_r \in \Sigma^{2r+1}$. The
  \introtwo{neighbourhood process} for $w$ at $p$ is the cellular
  process
  \begin{equation}
    \label{eq:neighbourhood}
    \nu(p, w) =
    \{ [p - r \ex]\omega_{-r}, \dots, [p + r \ex]\omega_r \}\,.
  \end{equation}
\end{definition}

\paragraph{The Transition Rule} We now return to the computation of
the state of $\gamma$ at $p$. With neighbourhood processes we can
express the global transition rule~\eqref{eq:global-rule} for
evolution sequences in a new way for cellular processes like for
$\gamma$. A direct translation of~\eqref{eq:global-rule} uses the fact
that $\gamma(t, x) = c_t(x)$ for all $t \in \N_0$ and $x \in \Z$. We
now replace all terms like $c_t(x)$ with terms of the form $\gamma(t,
x)$ and get the formula
\begin{equation}
  \gamma(t, x) = \phi(\gamma(t-1, x-r), \dots, \gamma(t-1, x+r)),
\end{equation}
which is valid for all $t > 0$ and $x \in \Z$. With neighbourhood
processes this becomes the condition,
\begin{equation}
  \text{if}\quad
  \nu((t-1, x), \omega_{-r} \dots \omega_r) \subseteq \gamma,
  \qtext{then}
  \gamma(t, x) = \phi(\omega_{-r}, \dots, \omega_r),
\end{equation}
which can be shortened by using $p = (t, x)$ again and by setting $w =
\omega_{-r} \dots \omega_r$. Then it becomes the requirement that
\begin{equation}
  \label{eq:gamma-global-rule}
  \text{if}\quad
  w \in \Sigma^{2r+1}
  \qtext{and}
  \nu(p - \et, w) \subseteq \gamma,
  \qtext{then}
  \gamma(p) = \phi(w)\,.
\end{equation}
This is is a formulation of the global transition rule for the process
$\gamma$. It is valid for all $p \in \Z^2$ with $p_T > 0$.

\paragraph{Arbitrary Processes} Now we return to an arbitrary process
$\pi$. We ask which state we should expect for the event at $(t, x)$,
given the information in $\pi$. To do this we will view $\pi$ as a
window into the evolution of a cellular automaton that follows rule
$\phi$.

If $w = \omega_{-r} \dots \omega_r$ and $\nu((t - 1, x), w) \subseteq
\pi$, then at the time $t - 1$ the states of the cells in the
neighbourhood of the cell at $x$ are $\omega_{-r}$, \dots, $\omega_r$.
At time $t$, the state of the cell at $x$ must then be $\phi(w)$. This
is then the expected state for the event at $(t, x)$.

\begin{figure}[ht]
  \centering
  \begin{tikzpicture}[
    direct/.style={line width=.75ex},
    somewhere/.style={direct, draw=shaded, decorate},
    decoration={random steps, amplitude=1.5, segment length=5},
    every pin edge/.style={black}]
    \fill[shaded, decorate]
    (0,0) coordinate (A)
        -- (6, 6.5) coordinate (B)
        -- (13, 8) coordinate (midtop)
        -- (23, 1)
        -- (15, -1) coordinate (bottom)
        -- cycle;
    \node at ($(bottom)!0.6!(midtop)$) {$\pi$};
    \path ($(A)!0.7!(B)$)
        ++(.75, 0)  coordinate [label=right:{$p$}] (p)
        ++(0, -0.8) coordinate (p')
        let \n{rad} = 3 in
            +(-\n{rad}, 0) coordinate (left)
            +(\n{rad}, 0)  coordinate (right);
    \fill (p) circle (.375ex);
    \node[fit=(left) (right), draw, dashed,
        pin=175:{$\nu(p - \et, w)$}] {};
    \draw[direct] ($(left)!0.3!(right)$) -- (right);
    \coordinate[pin=below right:{$\pi \cap\nu(p - \et, w)$}] (x)
       at ($(left)!0.65!(right)$);
  \end{tikzpicture}
  \caption{Determining the possible states for the event at $p$.}
  \label{fig:possible-states}
\end{figure}
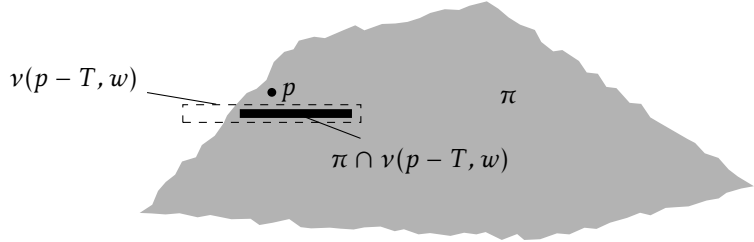

When however $N(p, r) \not\subseteq \dom \pi$, then we have not enough
information about the evolution to find the state for $p$ in this way.
Instead we can find a set of possible states: If there is a process
$\pi' \supseteq \pi$ such that $\nu(p - \et, w) \subseteq \pi'$, then
$\nu(p - \et, w)$ is a possible neighbourhood for $p - \et$, and
$\phi(w)$ is a possible state for the event at $p$. The set of
possible states for the event at $p$ is therefore
\begin{equation}
  \set{\phi(w) \colon
    w \in \Sigma^{2r+1},
    \exists \pi' \subseteq \mathcal{P} \colon
    \nu(p - \et, w) \subseteq \pi' \supseteq \pi}\,.
\end{equation}
However, as stated here this definition involves an infinite number of
processes $\pi'$, which is bad for actual computations. We avoid this
by choosing only those $\pi'$ that contain only as many additional
points that $\nu(p - \et, w) \subseteq \pi$. Then $\nu(p, - \et, w)$
is compatible with $\pi$ (Figure~\ref{fig:possible-states}). This then
leads to the following definition, in which $\pi'$ does no longer
occur explicitly.

\begin{definition}[Set of Possible States]
  Let $\pi \in \mathcal{P}$ be a cellular process and $\phi$ be a
  transition rule for $\Sigma$ of radius $r$. The \introthree{set of
    possible states} for the event at $p$ is
  \begin{equation}
    \label{eq:values}
    S(p, \pi) =
    \set{ \phi(w) \colon w \in \Sigma^{2r + 1},
      \nu(p - \et, w) \comp \pi}\,.
  \end{equation}
\end{definition}

If there is only one possible state for the event at $p$, then it is
determined. This is expressed in the following definition.

\begin{definition}[Determined Events]
  \label{def:determined}
  Let $\phi\colon \Sigma^{2r+1} \to \Sigma$ be a transition rule and
  $\pi \in \mathcal{P}$ a process. If
  \begin{equation}
    S(p, \pi) = \{ \sigma \},
  \end{equation}
  then both the event $[p] \sigma$ and the point $p$ are
  \indextwo{determined event}\emph{determined} by $\pi$.
  \label{pg:determined}The set of all events that are determined by
  $\pi$ is $\Delta \pi$.
\end{definition}

\paragraph{Non-constant Transition Rules} Note that if $\phi$ is a
constant function, then every point of $\Z^2$ is determined. We will
therefore restrict the following definitions and theorems to
non-constant transition functions, in order to avoid this unintuitive
property.

The following lemma summarises useful properties of determinedness
that are only true for a transition rule that is non-constant. The
statements of this lemma are also the reason why the set $N(p, r)$
gets a special name.
\begin{lemma}[Events Determined by a Time Slice]
  \label{thm:timeslice-determines}
  Let $\phi$ be a non-constant transition rule for $\Sigma$ and $\pi
  \in \mathcal{P}$ a process. Then:

  \begin{enumerate}
  \item A point $p$ is determined by $\pi$ if and only if it is
    determined by $\pi|_{N(p - \et, r)}$.

  \item If $p$ is determined by $\pi$, the set $\pi|_{N(p - \et, r)}$
    is nonempty.

  \item For every time $t$, the set $\Delta \pi^{(t)}$ consists
    only of events at time $t + 1$.
  \end{enumerate}
\end{lemma}

The first statement of this lemma expresses again the fact that
determinedness is a local property and relies only on a finite number
of events.

The second statement is about causality. For it we need to have the
view of a cellular automaton as a physical system governed by the
``physics'' $\phi$. Then the events of $\pi|_{N(p - \et, r)}$ can be
understood as the ``cause'' of the event at $p$. The second statement
of the theorem then states that an event at time $t$ which is
determined by $\pi$ is always caused by an event at time $t - 1$. It
is also a statement about the maximal speed with which information is
transmitted: the state of a cell at time $t$ can only be caused by the
cells at most $r$ positions to its left or right.

The third statement is tailored for its use in connection with the
closure of a process, which will be defined next.

\begin{proof}[Proof of the lemma]
  For the proof of the first statement we note that for every $w \in
  \Sigma^{2r+1}$ the domain of the neighbourhood process $\nu(p, w)$
  is $N(p, r)$. Therefore the set $S(p, \pi)$ does actually depend
  only on $\pi|_{N(p - \et, r)}$. The knowledge of this part of $\pi$
  is therefore also enough to find out whether $p$ is determined.

  To show the second statement we prove its converse. Assume that
  $\pi|_{N(p - \et, r)}$ is empty. Then every neighbourhood process
  $\nu(p - \et, w)$ in~\eqref{eq:values} is compatible to $\pi$. Since
  $\phi$ is non-constant, the set $S(p, \pi)$ has more than one
  element. Therefore the point $p$ is then not determined by $\pi$.

  The third statement then follows from the second.
\end{proof}

\paragraph{The Closure} Now we can extend the global transition
rule~\eqref{eq:gamma-global-rule} from $\gamma$ to arbitrary cellular
processes. Similar to the way an evolution sequence is generated by
always computing the configuration for time $t$ from the configuration
for time $t - 1$, the closure of a cellular process is created from
time slices, each of them depending on the previous one, that are
finally put together. There is however no direct analog to the initial
configuration.

To understand what is meant with the closure of a cellular process,
imagine that cells are multicoloured lights that can be switched on or
off. If a light is switched on, it has one of a finite set of colours.
A cellular automaton is then an infinite line of such lights, and the
colours represent the states of its cells. A cellular process $\pi$ is
then a certain light pattern, a rule when to switch the lights on and
with which colours. The closure of $\pi$ is another light pattern,
where the lights are switched on not only when it is required by $\pi$
but also depending on the lights that were switched on at the previous
time step. When the lights switched on at the previous time step
determine the state of a cell in the current time step, then this cell
is also switched on, besides the cells that are required by $\pi$ to
be switched on. A conflict between these two rules is possible: It can
happen that the light pattern prescribes one colour for a cell at a
certain time and $\pi$ describes another. Then we will say that for
this $\pi$ the closure does not exist.

To be practical this procedure must have a starting time. We will
require that there was a time when no event of $\pi$ happened;
otherwise $\pi$ will have no closure. In the following definition we
therefore call a cellular process $\pi$ \indexthreeshort{quiet {before
    a} time}\emph{quiet before $t_0$} if $\pi^{(t)} = \emptyset$ for
all $t < t_0$.

\begin{definition}[Closure]
  \label{def:closure}
  Let $\phi$ be a non-constant transition rule for $\Sigma$ and $\pi
  \in \mathcal{P}$ be a cellular process that is quiet before
  $t_0$.

  The \indexthree{closure {at a} time}\emph{closure of $\pi$ at time
    $t$} under $\phi$ is the process $\cl^{(t)} \pi$. It exists always
  when $t \leq t_0$. When $t > t_0$, the process $\cl^{(t)} \pi$
  exists if $\cl^{(t-1)} \pi$ exists and $\pi^{(t)}$ is compatible
  with $\Delta \cl^{(t-1)} \pi$. It is defined by the recursion
  \begin{equation}
    \label{eq:closure-recursion}
    \cl^{(t)} \pi =
    \begin{cases}
      \pi^{(t)} & \text{if $t \leq t_0$,} \\
      \pi^{(t)} \cup \Delta \cl^{(t-1)} \pi
       & \text{if $t > t_0$.}
    \end{cases}
  \end{equation}
  The \introx[closure]{closure of $\pi$} under $\phi$ exists if all
  time slices $\cl^{(t)} \pi$ exist. It is the process
  \begin{equation}
    \label{eq:closure}
    \cl \pi = \bigcup_{t \in \Z} \cl^{(t)} \pi\,.
  \end{equation}
\end{definition}
An incidental result of equation~\eqref{eq:closure} is that always
$(\cl \pi)^{(t)} = \cl^{(t)} \pi$. This is the way the
operator $\cl^{(t)}$ fits into the formalism of time
slices of Definition~\ref{def:cellular-process}.

There is one act of choice in this definition. If a process $\pi$ is
quiet before $t_0$, then it is also quiet before any time $t'_0 <
t_0$. Therefore one can also compute the closure of $\pi$ using $t'_0$
as starting time. But this has no influence on $\cl \pi$. To see
this, assume that $\psi$ is the closure of $\pi$ as computed with
$t_0$ and $\psi'$ the closure of $\pi$ as computed with $t'_0$. Then
for $t \leq t_0$, the first case of~\eqref{eq:closure-recursion}
applies to the computation of $\psi$, and we have $\psi^{(t)} =
\pi^{(t)} = \emptyset$. For $t \leq t'_0$ it also applies to the
computation of $\psi'$, so we have then $\psi'^{(t)} = \pi^{(t)} =
\emptyset$. For $t'_0 < t \leq t_0$, the second case
of~\eqref{eq:closure-recursion} applies and we can see by induction
that then $\psi'^{(t)} = \emptyset$: If $\psi'^{(t-1)} = \emptyset$,
then
\begin{equation}
  \label{eq:psi-prime-calc}
  \psi'^{(t)}
  = \pi^{(t)} \cup \Delta \psi'^{(t-1)}
  = \emptyset
  \cup \Delta \emptyset
  = \emptyset\,.
\end{equation}
Therefore for $t \leq t_0$, $\psi^{(t)} = \psi'^{(t)}$. At later
times, the second part of~\eqref{eq:closure-recursion} comes into play
for both $\psi$ and $\psi'$ to construct the next time slice.
Therefore for $t > t_0$ the time slice $\psi^{(t)}$ exists if and only
if $\psi'^{(t)}$ exists, and when one of them exists, then $\psi^{(t)}
= \psi'^{(t)}$. This shows then that $\psi$ exists if and only $\psi'$
exists, and if they exist, they are equal. In other words, the choice
of $t_0$ has no influence on the closure.

Most properties of the closure are proved by inductions in the style
of this proof. In it one can also see why the definition of the
closure was restricted to non-constant transition rules: This
restriction ensured in~\eqref{eq:psi-prime-calc} that $\Delta
\emptyset = \emptyset$ and thus removed unnecessary complexity.

\section{Properties of the Closure}

We now will prove some properties of the closure that either become
useful later or will provide insight about this concept.

\paragraph{Shift Invariance} Like the physical laws which they
imitate, the laws of a cellular automaton are independent of an
absolute location in space and time. In the next chapter we will use
this property for a simplification of the formalism; at this point we
are mainly concerned with a way to express it for cellular automata.

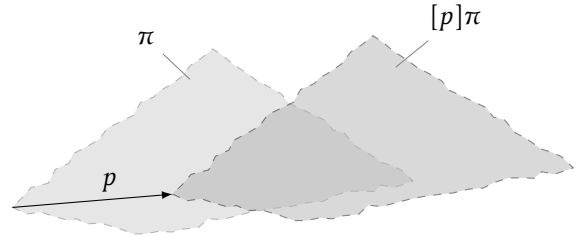
\begin{figure}[ht]
  \centering
  \begin{tikzpicture}[
    direct/.style={line width=.75ex},
    somewhere/.style={direct, draw=shaded, decorate},
    decoration={random steps, amplitude=1.5, segment length=5}]
    \path coordinate (s1)
          ++(5, -1) coordinate (s2)
          ++(5, 1)  coordinate (s3)
          ++(5, 1)  coordinate (s4);
    \coordinate (s5) at ($(s1)!0.5!(s4) + (0, 5.5)$);
    \foreach \p in {s1,s2,s3,s4,s5}
        \coordinate (\p') at ($(\p)+(6,.5)$);
    \pgfmathsetseed{999}
    \fill[gray!20, decorate, dashed, thin, draw=shaded]
         (s1) -- (s2) -- (s3) -- (s4) -- (s5) -- (s1);
    \pgfmathsetseed{999}
    \fill[gray!60, decorate, dashed, thin, draw=black, opacity=0.5]
         (s1') -- (s2') -- (s3') -- (s4') -- (s5') -- (s1');
    \node at ($(s5)-(.5,1.5)$) [pin=135:{$\pi$}] {};
    \node at ($(s5')+(.5,-1.5)$) [pin=45:{$[p]\pi$}] {};
    \draw[->] (s1) -- node [above right] {$p$} (s1');
  \end{tikzpicture}
  \caption{A shifted cellular process.}
  \label{fig:shifted}
\end{figure}
We can express it by the concept of a space-time shifted process
(Figure~\ref{fig:shifted}). The notation $[p] \pi$ has been chosen in
harmony to the other uses of square brackets in this text.
\begin{definition}[Space-time Shift]
  Let $\pi \in \mathcal{P}$ be a process and $p \in \Z^2$ be a
  space-time point. We write for the copy of $\pi$ that is
  \indextwoshort{shifted process}\emph{shifted} by $p$,
  \begin{equation}
    \label{eq:shifted}
    [p] \pi = \set{[p + q] \sigma \colon [q] \sigma \in \pi}\,.
  \end{equation}

  A property of cellular processes is \intro{shift-invariant} if it is
  true for a process $\pi$ if and only if it is true for $[p] \pi$. A
  function $F \colon \mathcal{P} \to \mathcal{P}$
  between cellular processes is shift-invariant if $F([p] \pi) = [p]
  F(\pi)$ for all $p$ and $\pi$.
\end{definition}

Then we can say that $\Delta$ and $\cl$ are two shift-invariant
functions. In a more informal way we will also say that determinedness
is a shift-invariant notion, meaning that $\pi$ determines $[p]
\sigma$ if and only if $[q] \pi$ determines $[q + p] \sigma$. I do not
give here a formal proof for these facts: they can easily be verified
from their definitions, by checking that they use only differences
between space-time points and no absolute coordinates.

\paragraph{Radius Invariance} As requested at the beginning of this
chapter, the concepts introduced here are radius-invariant. Neither
the closure nor the set of determined events of a process are
dependent of the radius of the transition rule. To verify this, the
following lemma is sufficient, since the closure is defined with help
of the set of determined events, and determined events are defined
with help of the set of possible states for a point.

In the proof we have to distinguish between the sets $S(p, \pi)$ for
the transition rules $\phi$ and $\phi$; we will therefore write the
two sets of possible states as $S_\phi(p, \pi)$ and $S_{\phi'}(p,
\pi)$.

\begin{lemma}[Possible States are Radius Invariant]
  Let $\phi \colon \Sigma^{2r+1} \to \Sigma$ be a transition rule and
  $\phi' \colon \Sigma^{2r'+1} \to \Sigma$ an extension of $\phi$ with
  radius $r' > r$.

  Let $\pi \in \mathcal{P}$ be a cellular process and $p \in
  \Z^2$. Then $S_\phi(p, \pi) = S_{\phi'}(p, \pi)$.
\end{lemma}
\begin{proof}
  Assume that $\sigma \in S_\phi(p, \pi)$. Then there are states
  $\omega_{-r}$, \dots, $\omega_r \in \Sigma$ such that
  $\phi(\omega_{-r}, \dots, \omega_r) = \sigma$ and $\nu(p - \et,
  \omega_{-r} \dots \omega_r) \comp \pi$. Now choose the cell states
  $\omega_{-r'}$, \dots, $\omega_{-r-1}$ and $\omega_{r+1}$, \dots
  $\omega_{r'} \in \Sigma$ in the following way: If $p - \et + i \ex
  \in \dom \pi$, then $\omega_i = \pi(p - \et + i \ex)$; otherwise
  $\omega_i$ is arbitrary. Then $\nu(p - \et, \omega_{-r'} \dots
  \omega_{r'})$ is compatible with $\pi$. Since $\phi'$ is an
  extension of $\phi$, we have also $\phi'(\omega_{-r'} \dots
  \omega_{r'}) = \phi(\omega_{-r}, \dots, \omega_r) = \sigma$.
  Therefore $\sigma \in S_{\phi'}(p, \pi)$, which in turn proves that
  $S_\phi(p, \pi) \subseteq S_{\phi'}(p, \pi)$.

  Assume that $\sigma \in S_{\phi'}(p, \pi)$. This means that there
  are states $\omega_{-r'}$, \dots, $\omega_{r'} \in \Sigma$ such that
  $\phi'(\omega_{-r'}, \dots, \omega_{r'}) = \sigma$ and $\nu(p - \et,
  \omega_{-r'} \dots \omega_{r'}) \comp \pi$. Then $\phi(\omega_{-r},
  \dots, \omega_r) = \sigma$ because $\phi'$ is an extension of
  $\phi$, and $\nu(p - \et, \omega_{-r} \dots \omega_r)$ is compatible
  with $\pi$ because $\nu(p - \et, \omega_{-r} \dots \omega_r)$ is a
  subset of $\nu(p - \et, \omega_{-r'} \dots \omega_{r'})$. This then
  proves that $\sigma \in S_\phi(p, \pi)$, and therefore that
  $S_{\phi'}(p, \pi) \subseteq S_\phi(p, \pi)$.
\end{proof}

\paragraph{Monotony} We return for a moment to the view of a cellular
process as a partial description for an evolution of a cellular
automaton. Finding the closure can then be seen as reconstructing an
evolution from incomplete information. More information should then
result in a larger reconstruction. So we will expect that the closure
of the superset of a process is a superset of its closure, or, in
other words, that the closure operator defines a monotone function.
This property is used very often.

Its proof begins with the proof of the same property for
determinateness.
\begin{lemma}[Determinateness is Monotone]
  \label{thm:det-monotone}
  Let $\pi \subseteq \psi \in \mathcal{P}$ be two processes
  and $\phi$ be a non-constant transition rule. Then $\Delta \pi
  \subseteq \Delta \psi$.
\end{lemma}
\begin{proof}
  Let $[p]\sigma$ be determined by $\pi$. Then $S(p, \pi) = \{
  \sigma \}$. Since $\psi \supseteq \pi$, the requirement that $\nu(p
  - \et, w) \comp \psi$ is a stronger restriction on $w$ than the
  requirement that $\nu(p - \et, w) \comp \pi$. So we must have
  $S(p, \psi) \subseteq S(p, \pi)$ and therefore $S(p,
  \psi) \subseteq \{p\}$. On the other hand, $S(p, \psi)$ has at
  least one element. So $[p]\sigma$ is determined also by $\psi$.
\end{proof}

All properties of the closure involve questions of its existence,
therefore also this one. The theorem below expresses the intuitive
notion that more requirements on the behaviour of a cellular automaton
make it more likely that they are inconsistent and cannot be satisfied
by the evolution of a cellular automaton.

\begin{theorem}[Closure is Monotone]
  \label{thm:closure-monotone}
  Let $\pi \subseteq \psi \in \mathcal{P}$ be two processes
  and $\phi$ be a non-constant transition rule.

  If $\cl \psi$ exists, then $\cl \pi$ exists and $\cl
  \pi \subseteq \cl \psi$.
\end{theorem}
\begin{proof}
  Assume that $\cl \psi$ exists. We will say that the theorem is
  \emph{true for time $t$} if $\cl^{(t)} \pi$ exists and
  $\cl^{(t)} \pi \subseteq \cl^{(t)} \psi$.

  Since $\cl \psi$ exists, there must be a $t_0 \in \Z$ such that
  $\psi$ is quiet before $t_0$. Then $\pi$ is quiet before $t_0$ too,
  because $\pi \subseteq \psi$. Therefore $\cl^{(t)} \pi =
  \pi^{(t)} \subseteq \psi^{(t)} = \cl^{(t)} \psi$ for all $t
  \leq t_0$. So the theorem is true for every $t \leq t_0$.

  Let now $t > t_0$ and assume that the theorem is true for $t - 1$.
  Then $\cl^{(t-1)} \pi$ exists. Because $\cl^{(t-1)} \pi
  \subseteq \cl^{(t-1)} \psi$, we have $\Delta
  \cl^{(t-1)} \pi \subseteq \Delta \cl^{(t-1)} \psi$
  with Lemma~\ref{thm:det-monotone}. Because $\pi^{(t)} \subseteq
  \psi^{(t)}$, we have
  \begin{equation}
    \label{eq:monotone-inclusion}
    \pi^{(t)} \cup \Delta \cl^{(t-1)} \pi
    \subseteq \psi^{(t)} \cup \Delta \cl^{(t-1)} \psi,
  \end{equation}
  but only as an inclusion between sets of events. We have not yet
  proved that these sets are cellular processes. The right side
  of~\eqref{eq:monotone-inclusion} is however the cellular process
  $\cl^{(t)} \psi$, and therefore the left side, as its subset,
  must also be a process. This then means that $\pi^{(t)}$ is
  compatible with $\Delta \cl^{(t-1)} \pi$ and that
  therefore $\cl^{(t)} \pi$ exists. Since the left side
  of~\eqref{eq:monotone-inclusion} is then $\cl^{(t)} \pi$, while
  its right side is $\cl^{(t)} \psi$, we have proved that
  $\cl^{(t)} \pi \subseteq \cl^{(t)} \psi$. Therefore the
  theorem is true for time $t$ if it is true for time $t - 1$ when $t
  > t_0$.

  So we have shown by induction that the theorem is true for all times
  $t$, and therefore true in general.
\end{proof}

\paragraph{Closure and Evolution Sequence} As the final task of this
chapter we now verify what we required of the closure at the
beginning, when we motivated its construction: The process $\gamma$
of~\eqref{eq:gamma-whole}, the translation of the evolution sequence
$(c_t)_{t \geq 0}$, is the closure of the cellular process for its
initial configuration, $\gamma^{(0)}$.

We will prove a bit more: The following lemma shows that for every
time $t \geq 0$, the time slice $\gamma^{(t)}$ determines the
following time slice $\gamma^{(t+1)}$ in the same way that the
configuration $c_t$ determines the following configuration $c_{t+1}$.
The lemma is then the analog of the transition
rule~\eqref{eq:global-rule} for cellular processes.

\begin{lemma}[Global Transition Rule for Processes]
  \label{thm:gamma-t-determines}
  Let $\phi$ be a non-constant transition rule and let $\gamma$ be as
  in~\eqref{eq:gamma-whole}. Then for all $t \geq 0$,
  \begin{equation}
    \label{eq:gamma-global}
    \gamma^{(t+1)} = \Delta \gamma^{(t)}\,.
  \end{equation}
\end{lemma}

\begin{proof}
  By Lemma~\ref{thm:timeslice-determines}, the only events that can
  possibly be determined by $\gamma^{(t)}$ have a time coordinate of
  $t + 1$. It only remains to prove that $\gamma^{(t)}$ determines all
  events $[t+1, x] c_{t+1}(x)$ with $x \in \Z$.

  Let $p = (t + 1, x)$. To know whether this point is determined we
  have to find $S(p, \gamma^{(t)})$.

  Because $\dom \nu(p - \et, w) \subset \dom \gamma^{(t)}$ for all $w \in
  \Sigma^{2r+1}$, the process $\nu(p - \et, w)$ is compatible with
  $\gamma^{(t)}$ if and only if it is a subset of $\gamma^{(t)}$. So we must
  find all $w \in \Sigma^{2r+1}$ that satisfy
  \begin{equation}
    \nu(p - \et, w) =
    \{ [t, x - r]c_t(x - r), \dots,  [t, x + r]c_t(x + r) \},
  \end{equation}
  where $r$ is the radius of $\phi$. This equation has one solution,
  $w = c_t(x - r) \dots c_t(x + r)$.

  So the set $S(p, \gamma^{(t)})$ has exactly one element, which
  means that the event at $p$ is determined. Its state is $\phi(w)$,
  which is equal to $c_{t+1}(t + 1, x)$ by the global transition
  rule~\eqref{eq:global-rule}. This proves the lemma.
\end{proof}

Using this lemma we can then easily see that $\cl^{(t)}
\gamma^{(0)} = \gamma^{(t)}$ for $t \geq 0$ and that $\cl^{(t)}
\gamma^{(0)}$ is empty for $t < 0$. This then shows that $\gamma$ is
indeed generated by its initial time slice $\gamma^{(0)}$.

\section{Summary}

In this chapter we have formalised the concept of the evolution of a
cellular automaton in a new way, in order to be able to understand the
fate of a localised arrangement of cells. The goal was to have a
notation that treats events at different times on an equal footing.

The starting point was the description of cellular automata with
configurations and evolution sequences. This is a natural way to
understand the cellular automaton as a machine that evolves over time.
The transition rule then is a description of the law that governs the
behaviour of the cells.

This method to describe the behaviour of cellular automata was then
decomposed into its components. The configurations became sets of
cellular events. The transition rule was expressed in a
radius-invariant way as the function that generates the set of
determined events for a process. The evolution sequence became the
closure of a process.

The concept of closure helps us to express how information propagates
in a cellular automaton. We have seen that the closure operator is
monotone, which will help us to reason in an abstract way about
cellular processes. We have seen how the evolution of a configuration
is expressed with cellular processes. We have introduced the concept
of radius-invariance. The closure operator, as the new form of the
transition rule, is radius-invariant even for a cellular process of
finite size. We have therefore extended the concept of the initial
configuration to an arbitrary set of cells at arbitrary times.

Some cellular processes however have no closure and it is not yet
clear how to construct processes that have a closure. This question
will be answered in Chapter~\ref{cha:local-reactions} with the concept
of achronal situations.


\chapter{Reaction Systems}

One of the goals of this thesis is to find a way in which we can
express the laws of large-scale behaviour in a cellular automaton. A
``law of large-scale behaviour'' is here any statement that involves
an arbitrarily large number of cells. The triangles below in
Figure~\ref{fig:triangle-computation} are an intuitive example: One
knows that the exact number of white cells in the triangle's base does
not matter. It could become arbitrarily large, and the same kind of
triangular shape would result. We need to express this kind of
intuitive law---and much more complex laws---in a formal
way.\footnote{The specific law that is expressed in
  Figure~\ref{fig:triangle-computation} is expressed in
  Table~\ref{tab:large-scale}. Other examples are the laws of ether
  formation, like Lemma~\ref{thm:fragment-propagation}.}
\begin{figure}[ht]
  \centering
  \tabcolsep2em
  \def\arraystretch{1.5}
  \begin{tabular}{cc}
    \ci{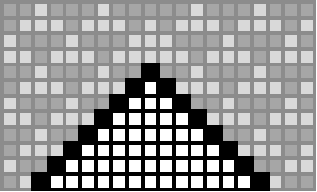} & \ci{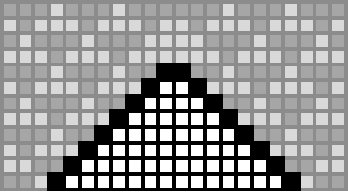}
  \end{tabular}
  \caption{Triangles as computations under Rule 54.}
  \label{fig:triangle-computation}
\end{figure}

In this chapter we will define a formalism with which we can express
instances of such large-scale laws in terms of the input and result of
a computation. The computation is then represented by the ordered pair
of input and output; intermediate steps are ignored.\footnote{The
  concepts introduced here rely to a great extent on the \emph{virtual
    state machine} (VSM) introduced by Christopher G.~Langton
  \cite{Langton1986}. In contrast to the VSM, which is understood by
  Langton as a long-lived entity, a reaction (defined below) always
  refers to a finite time span. One could then understand a reaction
  as the description of a single computational step in the existence
  of a VSM.}

\section{Situations}

We will take here the viewpoint that we build the computer inside the
universe represented by the cellular automaton. The transition rule is
kept fixed; it represents the physics of that world.

In a computer, we distinguish between the data and the computing
machine. In the cellular automaton, the data and the machine are both
cell states. A very simple example for the way data and computation
interact are the triangular structures found under Rule 54 and other
cellular automata (Figure~\ref{fig:triangle-computation}). Here we may
view the cell configuration at the initial time as the input; it
consists of two cells in state $1$ that surround a sequence of zeros.
The highlighted cells in the figure are then the computation initiated
by this input. How long the computation lasts depends directly on the
number of zeros in the input. We can view therefore it as a kind of
counter, or a loop that counts downward. Another viewpoint, since the
triangles have a different shape depending on whether the number of
zeros is odd or even, is to understand the triangles as programs that
test for parity.

When we now look at these triangular processes in terms of the cells
involved, we see that the computation takes a different amount of time
at different locations. It lasts longer at the center of the triangle
than at its margins. Since we have viewed the initial interval of
cells as the input of the computation, we will take the other sides of
the triangle as its result. It then consists roughly of the black
cells at the boundaries of the triangles, together with their direct
neighbours. Later, in Figure~\ref{fig:triangles}, the input and result
of this kind of computation will be shown explicitly.

So the formalism of Flexible Time makes no difference between the data
and the machine and represents them in a single mathematical object.
In a similar vein, input and result of a computation are the same kind
of mathematical object. This is so because it must be possible that
the result of one computation is the input of another computation.

\paragraph{Properties of Situations} The mathematical objects that
represent the input or result of a cellular computation are called
here \emph{situations}; they are defined below. But before we can
write down the formal definition, we will collect the properties a
situation must have.

\emph{(a) A situation specifies a finite cellular process.} To specify
the input or the result of a computation in a cellular automaton means
to express requirements on the states of cells: at a specified time a
cell must be in a specified state. Put together, these requirements
are the events of a cellular process. It is a finite process because
the input and output of a computation always have a finite number of
bits.

\emph{(b) A situation may specify events at different times.} With the
finite speed at which signals travel in a cellular automaton, exact
synchronisation of the components in the computer is difficult, and
different parts of a computation may end at a different time. We must
therefore allow that the events of a situation belong to different
times.

\emph{(c) A situation specifies a sequence of events.} The cells in a
one-dimensional cellular automaton have a natural order from left to
right. Situations generalise finite sets of adjacent cells together
with their states. Therefore it is good for our intuition if the
events in a situation also form a sequence. It especially allows us to
use the formalism of finite sequences for situations.

I will now sketch the way in which we get a meaningful left-to-right
arrangement for the events of a situation, starting from the order in
a set of adjacent cells. Take a situation that specifies the states of
such a set of cells and use it as the input of a computation. In
general, the output situation of this computation specifies events at
different times. They have no longer a natural left-to-right order,
but if the duration of the computation is short enough, there remains
an approximate order. As information travels with finite speed in a
cellular automaton, each event in the result has only a few events in
the input that are near enough to have caused it. So we can arrange
the events in the output approximately from left to right with help of
the input events. Later we will make this idea precise; we will then
have a correspondence between the order of the input of any
computation and that of its output.

\emph{(d) A situation has a size vector.} This vector is an analog to
the length $\abs{u}$ of a finite sequence $u$. Imagine that the
sequence $u$ consists of letters written on grid paper, with the
$\abs{u}$ letters of $u$ on the squares numbered $0$, \dots, $\abs{u}
- 1$. Now consider $u v$, the product of $u$ with some other sequence
$v$. In it, when it is written down the same way, the letters of $v$
will occupy the positions $\abs{u}$, \dots, $\abs{u} + \abs{v} - 1$,
instead of $0$, \dots, $\abs{v} - 1$, as it would have been if $v$
alone were written. So the length of a sequence $u$ marks the point
behind $u$, and this point is the starting point of the second factor
in a product involving $u$. The same happens with situations, except
that its size vector is an element of $\Z^2$ and that it can be chosen
arbitrarily. We then can define a notion of product for situations
similar to that for sequences; the arbitrariness of the size vector
makes it possible that there are gaps between the event sequences of
the factors.

The following is then a situation: a sequence of cellular events
together with a size vector.
\begin{definition}[Situations]
  \label{def:situation}
  A \indexthree{length of situation}\emph{situation of length $\ell$}
  with states in $\Sigma$ is a pair
  \begin{equation}
    \label{eq:situation-tuple-short}
    ((e_0, \dots, e_{\ell-1}), p) \in (\Z^2 \times \Sigma)^\ell \times \Z^2
  \end{equation}
  for which the set $\set{ e_i \colon 0 \leq i < \ell }$ is a cellular
  process. The set of all situations of length $\ell$ is
  $\mathcal{S}_\ell$. A \intro{situation} in general is an
  element of the set
  \begin{equation}
    \mathcal{S} = \bigcup_{\ell \geq 0} \mathcal{S}_\ell\,.
  \end{equation}
\end{definition}
The \introtwo{event sequence} of the situation $((e_0, \dots,
e_{\ell-1}), p)$ is then the tuple of events, $(e_0, \dots,
e_{\ell-1})$, and its size vector is the point $p$. A third property
of the situation that we required before, namely a cellular process
associated to it, is introduced in the following definition. It also
specifies a notation for the size vector that does not require to
spell out a situation as a pair.
\begin{definition}[Components of a Situation]
  \label{def:process-of}
  Let $a = ((e_0, \dots, e_{\ell-1}), p)$ be a situation. The
  \indexthree{process of situation}\emph{process} of $a$ is the cellular
  process
  \begin{equation}
    \label{eq:pi_a}
    \pr(a) = \set{ e_i \colon 0 \leq i < \ell }\,.
  \end{equation}
  For $p$, the \label{pg:size-vector}\introtwoshort{size vector} of
  $a$, we write $\delta(a)$.
\end{definition}
The symbol for the size vector, $\delta$, should remind of another way
to view a situation. We can view a situation $a$ as having a left end
at the coordinate origin an a left end at $\delta(a)$; then
$\delta(a)$ is the difference between the two ends of a situation.
Therefore the symbol.

With the notations of Definition~\ref{def:process-of} we can already
express a convenient shorthand notation. Let $a$ and $b$ be two
situations. The \indextwoshort{shifted process}\emph{process of $b$,
  shifted by $a$} is then
\begin{equation}
  \label{eq:relative-process}
  \prx{a}{b} = [\delta(a)]\pr(b)\,.
\end{equation}
It will become useful once we have defined the product of situations.

\paragraph{Path Notation} Let $a$ be a situation of length $\ell$. If
we want to express it in full detail, we currently have to write it in
the form
\begin{equation}
  \label{eq:situation-tuple}
  a = (([p_0]\alpha_0, \dots, [p_{\ell-1}]\alpha_{\ell-1}), p_\ell),
\end{equation}
with $p_i \in \Z^2$ and $\alpha_i \in \Sigma$ for all $i$. The use of
expressions of this kind for longer calculations and proofs would
however soon become quite cumbersome. Therefore we now introduce a
shorter form. It will fulfil the remaining requirement on situations
and provide a way to treat situations in the same manner as finite
sequences. In the full-developed formalism we will then refer to the
properties of a situation $a$ only with help of the new notation, the
process $\pr(a)$ and the size vector $\delta(a)$ and no longer refer
to terms of the form~\eqref{eq:situation-tuple} directly.

The new notation uses a relative notation for the locations of the
events in a situation.
\begin{definition}[Path Notation]
  \label{def:situation-notation}
  Let $a$ be a situation written in the
  form~\eqref{eq:situation-tuple}. Let $\tilde p_0 = p_0$ and $\tilde
  p_i = p_i - (p_{i-1} + \ex)$ for $i > 0$. The \index{notation!long
    path $\sim$}\indextwoshort{path notation}\emph{long path notation} for
  $a$ is then
  \begin{equation}
    \label{eq:situation-write}
    a = [\tilde p_0]\alpha_0[\tilde p_1]\alpha_1 \dots
    [\tilde p_{\ell-1}]\alpha_{\ell-1}[\tilde p_\ell],
  \end{equation}
  The terms $[\tilde p_i]$ are the
  \introx[displacement]{displacements} of $a$.

  The \index{notation!short path $\sim$}\intro{short path notation}
  of $a$ is similar to this, but all terms $[\tilde p_i]$ with $\tilde
  p_i = (0, 0)$ are removed from it. An exception is the case of $\ell
  = 0$: a situation $a = [\tilde p_0]$ cannot be shortened. It is by
  definition already in short path notation.
\end{definition}
One can understand the path notation as the description of a writing
process. In it, symbols for the cell states are written into a square
grid similar to a space-time diagram. After writing a symbol into the
square at point $p_{i-1}$ the cursor is at $p_{i-1} + \ex$. The
displacement $[\tilde p_i]$ is then the amount of \emph{extra}
movement before the next symbol can be written down. This explains the
occurrence of the unit vector $\ex$ in the definition of $\tilde p_i$
above. It also explains the formula
\begin{equation}
  \label{eq:displacements-back}
  p_i = \sum_{j = 0}^i \tilde p_j + j \ex
  \qquad\text{for $i = 0$, \dots, $\ell$}
\end{equation}
that converts the displacements of the long path notations back into
absolute positions. (Note that $p_0$, \dots, $p_{\ell - 1}$ are
locations of events, while $p_\ell$ is the size vector!)

The abbreviations for $[\tilde p_i]$ that were defined before in the
context of cellular events are also allowed for situations. So we can
write $[(t, x)]$ as $[t, x]$, and $[0, x]$ as $[x]$. Sometimes we will
use a ``mixed'' path notation, with not all $[0]$-terms omitted.

As an example of how this works, let us look at a cellular automaton
with state set $\Sigma = \{ 0, 1 \}$. We assume that at time $t = 0$
the cells at position $x = 0$, $1$, $2$ and $3$ are in the states $1$,
$0$, $0$ and $1$. We now want to express this information with a
situation. For this we start with a cellular process. We know the
cellular events $[0, 0]1$, $[0, 1]0$, $[0, 2] 0$ and $[0, 3] 1$. Using
the abbreviation convention for the positions of cellular events, we
can write the process that contains them as
\begin{equation}
  \pi = \{ [0] 1, [1] 0, [2] 0, [3] 1 \}\,.
\end{equation}
A natural way to write $\pi$ as a situation in the
form~\eqref{eq:situation-tuple} is
\begin{equation}
  \label{eq:tuple-notation}
  a = (( [0] 1, [1] 0, [2] 0, [3] 1), (0, 4))\,.
\end{equation}
In a situation, the events of $\pi$ must be arranged in a sequence; we
have here chosen the most natural one, an arrangement from left to
right by their $x$-positions. For the size vector of the situation we
have chosen the point $(0, 4)$, one position to the right of the last
event in the event sequence of $a$. If then $a$ is written in the long
path notation~\eqref{eq:situation-write}, it becomes
\begin{equation}
  \label{eq:long-path-notation}
  a = [0]1 [0]0 [0]0 [0]1 [0]\,.
\end{equation}
We now see that the choice of $a$ in~\eqref{eq:tuple-notation} was
natural: all displacements in the new notation become $[0]$. We can
remove them all, and this leads to the short path notation for $a$,
namely $1001$. (The similarity between the short path notation and the
notation for finite sequences is intended.)

\section{Products}

The path notation leads to a natural definition for the product of two
situations. For the concatenation of situations in the following
definition we employ the convention that $[p][q] = [p + q]$ that was
already introduced for cells: Here is where it becomes useful.
\begin{definition}[Product]
  \label{def:product}
  Let $a$, $b \in \mathcal{S}$ be two situations. We get their
  \indexthreeshort{product of situations}\emph{product} by concatenating
  the long path notation for $a$ with the long path notation for $b$.
  The product exists if the resulting expression is a situation.

  The product of $a$ and $b$ is written $a b$.
\end{definition}

To understand this definition, let
\begin{equation}
  a = [\tilde p_0]\alpha_0 \dots \alpha_{\ell-1}[\tilde p_\ell]
  \qqtext{and}
  b = [\tilde q_0]\beta_0 \dots \beta_{m-1} [\tilde q_m]
\end{equation}
be two processes in long path notation. Then their product, if it
exists, has the long path notation
\begin{equation}
  \label{eq:product}
  \begin{aligned}[b]
    ab
    &= [\tilde p_0]\alpha_0 \dots \alpha_{\ell-1}[\tilde p_\ell]
       [\tilde q_0]\beta_0   \dots \beta_{m-1} [\tilde q_m]\\
    &= [\tilde p_0]\alpha_0 \dots \alpha_{\ell-1}[\tilde p_\ell +
        \tilde q_0]\beta_0   \dots \beta_{m-1} [\tilde q_m]\,.
  \end{aligned}
\end{equation}
The first line in this equation is that what we get when we simply
concatenate the path notations for $a$ and $b$. The second line is
that what we get after applying the convention.

We now return to the pair notation for situations to find the process
of $a b$. We assume that $a$ is as in~\eqref{eq:situation-tuple}, and
$b$ is similar, so that we have
\begin{align}
  a &= (([p_0]\alpha_0, \dots, [p_{\ell-1}]\alpha_{\ell-1}), p_\ell),\\
  b &= (([q_0]\beta_0, \dots, [q_{m-1}]\beta_{m-1}), q_m)\,.
\end{align}
In the second equation the $q_i$ are related to the $\tilde q_i$ in
the same way as the $p_i$ to the $\tilde p_i$ in
Definition~\ref{def:situation-notation}. Then the product of $a$ and
$b$ has the form
\begin{equation}
  \label{eq:product-tuple}
  \begin{aligned}[b]
    ab &=
    (([p_0]\alpha_0, \dots, [p_{\ell-1}]\alpha_{\ell-1},\\
    &\qquad
    [p_\ell + q_0]\beta_0, \dots, [p_\ell + q_{m-1}]\beta_{m-1}),
    p_\ell + q_m)\,.
  \end{aligned}
\end{equation}
We can then translate this formula into a lemma that describes the
product in a more abstract form.

\begin{lemma}[Properties of the Product]
  Let $a$, $b \in \mathcal{S}$ be two situations. If $\pr(a)$
  is compatible to $\prx{a}{b}$, then the product $a b$ exists. Its
  process and size vector are
  \begin{equation}
    \label{eq:product-as-union}
    \pr(a b) = \pr(a) \cup \prx{a}{b}
    \qqtext{and}
    \delta(a b) = \delta(a) + \delta(b)\,.
  \end{equation}
\end{lemma}

\begin{proof}
  The event sequence in~\eqref{eq:product-tuple} contains the events
  in $\pr(a)$ together with the events in $\pr(b)$, but the latter
  shifted by the size vector $p_\ell$ of $a$. The set of these events
  is therefore $\pr(a) \cup [\delta(a)]\pr(b)$, or $\pr(a) \cup
  \prx{a}{b}$ in the notation of~\eqref{eq:relative-process}. It is a
  cellular process if $\pr(a)$ is compatible to $\prx{a}{b}$, and if
  this is true, then $a b$ is a situation.

  The equations in~\eqref{eq:product-as-union} can then be read
  directly from~\eqref{eq:product-tuple}.
\end{proof}

The left equation in~\eqref{eq:product-as-union} is the \intro{chain
  rule} for situations. It is the reason why the notation $\prx{a}{b}$
was introduced in~\eqref{eq:relative-process}.

Together with the product for situations we get the usual notations
that are related to it. Among them are exponentiation, Kleene closure
and other constructions that were already described for sequences. In
contrast to ordinary sequences we must however be careful whether a
product actually exists. The set of situations therefore does not form
a semigroup. Nevertheless it has a neutral element of multiplication.
As we can infer from~\eqref{eq:product}, it is the
situation \label{pg:empty-situation}$[0]$. We will speak of it as the
\introtwo{empty situation}.

A very important subset of $\mathcal{P}$ is the set $\set{[0] \sigma
  \colon \sigma \in \Sigma}^*$. It contains all those situations that,
when written down in short path notation, look like elements of
$\Sigma^*$. Therefore we will introduce no special symbol for them but
call this set of situation also $\Sigma^*$. It will be always clear
from the context which set is meant.

As a means to distinguish the elements of $\Sigma^*$ from other
situations we will use for them, and only for them, the length
notation $\abs{\,\cdot\,}$ of finite sequences.

\paragraph{Induction Proofs with Situations} The product of situations
is also important because it allows induction proofs. Assume e.\,g.\
that a property $P$ of situations is ``closed under non-empty
multiplication'': This shall mean that if two non-empty situations
$a$, $b \in \mathcal{S}$ have the property $P$ and $a b$ exist, then
it has property $P$. Then if all situations in $\Sigma^*$ have
property $P$, all elements of $\mathcal{S}$ have it. This is in fact
an induction over the length of the situations.

This specific form of induction requires however that the factors $a$
and $b$ are nonempty, which is not always easy to check. Therefore we
will use another, more complex induction principle.

Before we can express it, we have to handle an ambiguity of the
notation for situations. We will have to split a situation $s$ into
the product of the three terms $a$, $[p]$ and $b$. But because of the
convention that $[p_1][p_2]$ is equal to $[p_1 + p_2]$, we could also
split $s$ into $a[p - q]$, $[q]$ and $b$, for any $q \in \Z$. So we
cannot unambiguously say that the displacement $[p]$ is a factor of
the situation $s$. This problem is solved with the help of the long
path notation~\eqref{eq:situation-write}: in it the terms $[\tilde
p_i]$ are unambiguous.
\begin{definition}[Honest Decomposition]
  \label{def:honest}
  Let $s \in \mathcal{S}$ be a situation with the long path notation
  $s = [p_0] \sigma_0 \dots \sigma_{\ell - 1} [p_\ell]$. A
  decomposition $s = a [p] b$ of $s$ is \indextwoshort{honest
    decomposition}\emph{honest} if there is an index $i$ such that
  \begin{align}
    \label{eq:honest}
    a &= [p_0] \sigma_0 \dots \sigma_{i - 1} [0],
    &p &= p_i,
    &b &= [0] \sigma_{i} \dots \sigma_{\ell - 1} [p_\ell]\,.
  \end{align}
\end{definition}
The cases of $a = [0]$ or $b = [0]$ are here explicitly allowed. They
refer to the one-sided decompositions $s = [p] b$ and $s = a [p]$.

With this definition we can now describe the new induction principle.
\begin{theorem}[Induction over Displacements]
  \label{thm:induction-displacement}
  Let $P$ be a property of the elements of $\mathcal{S}$.
  Assume that for all $s \in \mathcal{S}$,
  \begin{enumerate}
  \item if $s \in \Sigma^*$, then $P$ is true for $s$,
  \item if $s = a [p] b$ is a honest decomposition with $p \neq (0,
    0)$ and $P$ is true for $a$ and $b$, then $P$ is true for $s$.
  \end{enumerate}
  Then $P$ is true for all elements of $\mathcal{S}$.
\end{theorem}
\begin{proof}
  For this proof we define the \emph{number of nontrivial
    displacements} in a situation $s$ as the number of displacements
  $[p_i]$ with $p_i \neq (0, 0)$ that occur in its long path notation.
  We write this number as $d(s)$. The proof is then an induction over
  $d(s)$.

  If $d(s) = 0$, then $s \in \Sigma^*$, and $P$ is true for $s$. If
  $d(s) > 0$, then a term $[p]$ with $p \neq (0, 0)$ occurs in the
  long path notation of $s$. So there is a honest decomposition $s = a
  [p] b$. We then have $d(s) > d(a) + d(b)$, so $P$ is true for $a$
  and $b$ by induction. Therefore $P$ is also true for $s$.
\end{proof}

\section{Reactions}

A reaction represents a computation in a cellular automaton. It
consists of two situations that are related to each other by a
cellular process. The first situation is the \indexthreeshort{input of
  reaction}\emph{input} of the computation. Its events start the
activity of the cellular automaton; the activity itself is represented
by the cellular process that is the closure of the input situation;
the \indexthreeshort{result of reaction}\emph{result} of the computation is
represented by the second situation of the reaction. Its process must
lie completely inside the closure of the input process: This means
that all events of the output are determined by the input via the
transition rule. Its size vector must be the same as that of the input
situation: This will allow us to replace the input of a reaction with
its output when the input is part of a larger situation.\footnote{See
  below at Theorem~\ref{thm:application} for more details.} Otherwise
the choice of the second situation is arbitrary.

Reactions were introduced in \cite{Redeker2010}. They have their name
from the arrow with which reactions are written here, because it
reminds of chemical reactions.

\begin{definition}[Reactions]
  \label{def:reaction}
  Let $\phi$ be a transition rule for $\Sigma$ and $a$, $b \in
  \mathcal{S}$ be two situations. If
  \begin{equation}
    \pr(b) \subseteq \cl \pr(a)
    \qqtext{and}
    \delta(a) = \delta(b),
  \end{equation}
  then the pair $(a, b)$ is a \indexthree{reaction {for a} transition
    rule}\emph{reaction} for $\phi$. For $(a, b)$ we will usually
  write $a \rea b$.
  \begin{figure}[ht]
    \centering
    \begin{tikzpicture}[
      direct/.style={line width=.75ex},
      somewhere/.style={direct, draw=shaded, decorate},
      decoration={random steps, amplitude=1.5, segment length=5}]
      \path coordinate (a1)
            ++(5, -1) coordinate (a2) +(0, 2.5) coordinate (b2)
            ++(5, 1)  coordinate (a3) +(0, 2) coordinate (b3)
            ++(5, 0)  coordinate (a4);
      \coordinate (top) at ($(a1)!0.5!(a4) + (0,5)$);
      \fill[gray!20, decorate, dashed, thin, draw=black]
           (a1) -- (a2) -- (a3) -- (a4) -- (top) -- (a1);
      \draw[direct, draw=shaded]
           (a1) -- (b2) -- node [below left] {$\pr(b)$} (b3) -- (a4);
      \draw[direct]
           (a1) -- (a2) -- (a3) node [below] {$\pr(a)$} -- (a4);
      \node at ($(b3)+(-1,1)$) [pin=45:{$\cl \pr(a)$}] {};
    \end{tikzpicture}
    \caption{The processes involved in the reaction $a \rea b$.}
    \label{fig:reaction}
  \end{figure}
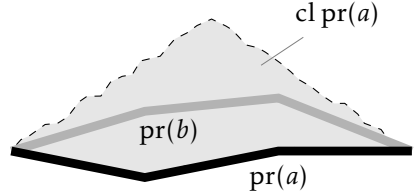

  Sometimes we will use the more general term of an \introtwo{abstract
    reaction}. This is a pair $(a, b)$ with $\delta(a) = \delta(b)$ in
  which $a$ is compatible with $b$.

  We will also use the formula $a \rea b$ as a proposition. Then it
  expresses the fact that there is a reaction $(a, b)$. It may be a
  reaction for $\phi$ or an abstract reaction, depending on the
  context.
\end{definition}
The processes that belong to this reaction are shown in
Figure~\ref{fig:reaction}. Often, when the situations that are part of
a reaction are complex, they will be drawn separately, as in
Figure~\ref{fig:reaction-diagram} below. This diagram is annotated
with the names of the situations and not the processes, as in the
previous figure: we will choose whichever is appropriate.
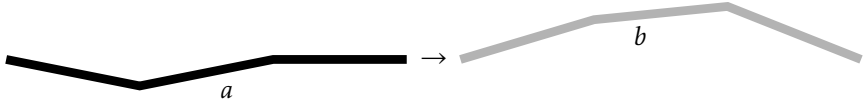
\begin{figure}[ht]
  \centering
  \begin{tikzpicture}[
    direct/.style={line width=.75ex},
    somewhere/.style={direct, draw=shaded, decorate},
    decoration={random steps, amplitude=1.5, segment length=5}]
    \def\backgroundpath{
    \path coordinate (a1)
          ++(5, -1) coordinate (a2) +(0, 2.5) coordinate (b2)
          ++(5, 1)  coordinate (a3) +(0, 2) coordinate (b3)
          ++(5, 0)  coordinate (a4);
    }
    \matrix [cells=midway] {
    \backgroundpath
    \draw[direct]
         (a1) -- (a2) -- node [below right] {$a$} (a3) -- (a4);
    & \node {$\rea$}; &
    \backgroundpath
    \draw[direct, draw=shaded]
         (a1) -- (b2) -- node [below left] {$b$} (b3) -- (a4);
    \\ };
  \end{tikzpicture}
  \caption{Another way to display the reaction of
    Figure~\ref{fig:reaction}.}
  \label{fig:reaction-diagram}
\end{figure}

\paragraph{Sets of Reactions} In order to be able to calculate with
them, we will now consider reactions that belong to a set. For this
let $S \subseteq \mathcal{S}$ be a set of situations and $R \subseteq
S \times S$ a set of reactions between its elements. The set $S$ is
then the \label{pg:r-domain}\introthree{domain of reaction
  set}\emph{domain} of $R$.

We use a special notation for reactions that belong to a set. If $(a,
b) \in R$, we write this as \label{pg:reaction}$a \rea_R b$. If $R$ is
known from context, we may write it even as $a \rea b$. As before, an
expression $a \rea_R b$ may be used as the proposition. It then means
that the pair $(a, b)$ is an element of $R$.

As a set of pairs, $R$ is a binary relation on $S$. The reaction sets
that we use for the understanding of cellular automata are mainly
pre-orders. In order theory, a binary relation is called a
\intro{pre-order} if it is transitive and reflexive
\cite[p.~2]{Davey2002}. We recapitulate what this means:
\begin{enumerate}
\item $R$ is transitive if $a \rea_R c$ whenever $a \rea_R b$
  and $b \rea_R c$.
\item $R$ is reflexive if $a \rea_R a$ for all $a \in R$.
\end{enumerate}
In a set of reaction that is a pre-order, reflexivity allows to
reconstruct the domain by
\begin{equation}
  \dom R =
  \set{ a \in \mathcal{S} \colon a \rea_R a}\,.
\end{equation}
Therefore for a set of reactions that is a pre-order it is not
necessary to specify the domain separately from the reaction set.

There is another small fact that is useful in its own right: If a set
$R$ of reactions is a pre-order, then every situation in its domain
has a closure. This is because for every $a \in \dom R$ there is a
reaction $a \rea_R a$, by the transitivity of $R$, and the definition
of reactions requires that $a$ then has a closure.

\paragraph{Reaction Systems} Transitivity of a reaction set allows to
form a chain of reactions, each using as input the result of the
previous one, and combine them into a single reaction that computes
the result of the last reaction in the chain from the input of the
first one. We now introduce another way to create new reactions from
old ones, one that is specific to cellular automata. It reflects the
local nature of the interactions between the cells.

\begin{definition}[Application of a Reaction]
  \label{def:application}
  We call the reaction $xay \rea xby$, where $a$, $b$, $x$ and $y$ are
  situations, the \indexthree{application of
    reaction}\emph{application} of $a \rea b$ on $xay$.
\end{definition}

\begin{figure}[ht]
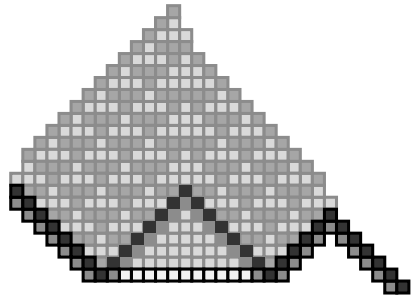

  \centering
  \ci{situations}
  \caption{Application of a triangle reaction under Rule 54.}
  \label{fig:appliction}
\end{figure}
One example for the application of the reaction $a \rea b$ to the
situation $x a y$ is shown in Figure~\ref{fig:appliction}. The events
of $\pr(x a y)$ are displayed as squares with thick frames, like
\ci{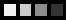}, and the events of $\pr(x b y)$ are displayed in darker
colours, like \ci{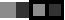}. The events displayed in lighter colours
belong to the closure of $\pr(x a y)$. In both colour sets, the
lighter and the darker, the brighter squares represent cells in state
$0$ and the darker squares, cells in state $1$. One can also see from
this diagram that the process $\pr(x b y)$ is a subset of the closure
of $\pr(x a y)$.

The reaction $a \rea b$, displayed in the centre of the diagram, is a
triangle reaction.\footnote{See also Figure~\ref{fig:triangles}. This
  kind of reaction will be formally defined later, in
  Definition~\ref{def:triangle}.} The base line of its triangle
consists of the squares with frames but in lighter colours at the
bottom of the diagram; the events belonging to it are all in state
$0$. The other two sides of the triangle are part of $\prx{x}{b}$; they
are shown as squares with frames, but with lightly coloured interior,
and they are partially in the states $0$ and $1$.

There is also an operation of \indexthree{application of
  reaction}\emph{applying a reaction} to a situation: It generates a
reaction $x a y \rea x b y$ from a reaction $a \rea b$ and a situation
$x a y$. We will show now that under reasonable conditions on $x$,
$y$, $a$ and $b$ application is always possible. This is done in two
steps, because the definition of application in the form it was stated
above uses too many variables at once. Instead of working with a
reaction between situations that consist of three factors, we will
first work with reactions between products of two factors.

\begin{lemma}[Parallel Processing]
  \label{thm:parallel-reactions}
  Let $a \rea a'$ and $b \rea b'$ be reactions for $\phi$. Assume that
  $\cl \pr(a b)$ exists. Then $a b \rea a' b'$ is a reaction for
  $\phi$.
\end{lemma}

\begin{proof}
  It is clear that $\delta(a b) = \delta(a' b')$. So it remains to
  prove that $\pr(a' b') \subseteq \cl \pr(a b)$. We have $\pr(a'
  b') = \pr(a') \cup \prx{a'}{b'}$ by the chain
  rule~\eqref{eq:product-as-union}. Therefore the proof of the lemma
  is complete if we show that $\pr(a') \subseteq \cl \pr(a b)$
  and $\prx{a'}{b'} \subseteq \cl \pr(a b)$.

  Because $a \rea a'$ is a reaction, we have $\pr(a') \subseteq
  \cl \pr(a)$. Since $\pr(a) \subseteq \pr(a b)$, we have
  $\cl \pr(a) \subseteq \cl \pr(a b)$ by monotony of the
  closure (Theorem~\ref{thm:closure-monotone}). Therefore $\pr(a')
  \subseteq \cl \pr(a b)$.

  Because $b \rea b'$ is a reaction, we have $\pr(b') \subseteq
  \cl \pr(b)$. The closure is shift-invariant, therefore
  $\prx{a'}{b'} \subseteq \cl \prx{a'}{b}$. Now $\delta(a) =
  \delta(a')$ because $a \rea a'$ is a reaction, so we have
  $\prx{a'}{b'} \subseteq \cl \prx{a}{b}$. Since $\prx{a}{b}
  \subseteq \pr(a b)$, we have $\cl \prx{a}{b} \subseteq \cl
  \pr(a b)$, again by monotony of the closure. Therefore $\prx{a'}{b'}
  \subseteq \cl \pr(a b)$.
\end{proof}

In this proof the condition $\delta(a) = \delta(a')$ played a crucial
role in keeping the processes $\prx{a}{b}$ and $\prx{a'}{b'}$ at the
same position. This is why it appeared in the definition of reactions.

\begin{theorem}[Applying Creates a Reaction]
  \label{thm:application}
  Let $\phi$ be a transition rule for $\Sigma$. If $a \rea b$ is a
  reaction for $\phi$ and there are $x$, $y \in \mathcal{S}$ for which
  $\cl(x a y)$ exists, then $x a y \rea x b y$ is a reaction for
  $\phi$. (Figure~\ref{fig:application-diagram}.)
\end{theorem}

\begin{figure}[ht]
  \centering
  \begin{tikzpicture}[
    direct/.style={line width=.75ex},
    somewhere/.style={direct, draw=shaded, decorate},
    decoration={random steps, amplitude=1.5, segment length=5}]
    \def\backgroundpath{
    \pgfmathsetseed{999}
    \path (0,0) coordinate (x)
        ++(3, 0) coordinate (x')  +(2.5, 2.5) coordinate (bmid)
        ++(5, 0) coordinate (y)
        ++(3, 0) coordinate (y');
    \draw[somewhere]
        (x) -- node[below] {$x$} (x')
        (y) -- node[below] {$y$} (y');
    }
    \matrix [cells=midway] {
    \backgroundpath
    \draw[somewhere, black]
        (x') -- node[below] {$a$} (y);
    & \node {$\rea_R$}; &
    \backgroundpath
    \draw[somewhere, black]
        (x') ..controls (bmid).. node[above] {$b$} (y);
    \\ };
  \end{tikzpicture}
  \caption{Applying $a \rea b$ to $x a y$.}
  \label{fig:application-diagram}
\end{figure}
\begin{proof}
  Since $\cl \pr(xay)$ exists, the processes $\cl \pr(x)$
  and $\cl \prx{a x}{y}$ exist by
  Theorem~\ref{thm:closure-monotone}. This means that $x \rea x$ and
  $y \rea y$ are reactions for $\phi$.

  By Lemma~\ref{thm:parallel-reactions}, $x a \rea x b$ is a reaction
  for $\phi$ because $x \rea x$ and $a \rea b$ are, and $x a y \rea x
  b y$ is a reaction for $\phi$ because $x a \rea x b$ and $y \rea y$
  are.
\end{proof}

Now we will introduce a name for the property of a set of reactions
that the operation of application in it is freely possible. Note that
in its definition there are no explicit restrictions on the situations
$x$ and $y$ at the sides of $a$; there is however the implicit
restriction that $x a y$ must be an element of $R$.
\begin{definition}[Closed under Application]
  Let $\Sigma$ be a set. Let $R$ be a set of reactions with $\dom R
  \subseteq \mathcal{S}$.

  If for all reactions $a \rea_R b$ and for all situations $x$, $y \in
  \mathcal{S}$, with $x a y \in R$ there is a reaction $x a y \rea_R x
  b y$, then $R$ is \introthreeshort{closed under application}.
\end{definition}
Theorem~\ref{thm:application} also expresses that the reactions in $R$
are local in scope: When $R$ is closed under application, it depends
only on the initial situation $a$ and not on the situations $x$ and
$y$ around it, whether a reaction $x a y \rea_R x b y$ is possible.

Now we can finally introduce the central concept of this thesis.
Reaction systems will serve as a replacement of the evolution sequence
defined in Chapter~\ref{cha:cellular-evolution} for the understanding
of cellular automata.
\begin{definition}[Reaction System]
  Let $\phi$ be a transition rule. A \intro{reaction system} for
  $\phi$ is a set of reactions for $\phi$ that is a pre-order and
  closed under application.

  Similarly, an \introtwo{abstract reaction system} is a set of
  abstract reactions that is a pre-order and closed under application.
\end{definition}

The set of all reactions for a given transition rule is obviously a
reaction system, but we will usually need smaller ones. The operation
of applying a reaction to a situation will allow us to define a large
set of reactions with the help of a small set of local reactions.
Therefore we define now how a small set of reactions and a set of
situations together generate a reaction system. As it is common with
generated sets in mathematics, a large part of the work with generated
reaction systems is about deriving properties of the whole system from
those of the set of generators.

\begin{definition}[Generated Reaction System]
  \label{def:generated-system}
  Let $S \subseteq \mathcal{S}$ be a set of situations and $G$ a set
  of reactions.

  Let $R$ be the smallest reaction system with $S \subseteq \dom R$
  and $G \subseteq R$ (i.\,e.\ no proper subset of $R$ has this
  property). Then $R$ is the \indextwoshort{generated reaction
    system}\emph{reaction system generated by $G$} from $S$. The set
  $S$ is the set of \indextwo{generating situation}\emph{generating
    situations} for $R$, and $G$ is the set of \indextwo{generating
    reaction}\emph{generating reactions} for $R$.
\end{definition}

We are mainly interested in non-abstract reaction systems. An abstract
reaction system is usually created from a reaction system for a rule
$\phi$ in order to have a system that is easier to handle.

With Theorem~\ref{thm:application} we see that a reaction system for
$\phi$ can be generated from an arbitrary reaction set $G$ and a set
$S$ of situations for which the only requirement is that all its
elements must have a closure under $\phi$. If we know this, we can
work with the reaction system in a quasi-algebraic way, without
referring to the closure again.

\section{Summary}

In this chapter we have introduced situations and reactions. They are,
in a manner of speaking, the substantives and basic propositions of
the new language. Much effort has been done to establish an intuitive
notation for situations.

We have then seen how to construct larger situations from smaller
situations by multiplication, and larger reactions from smaller
reactions by the concatenation of applications. This made it possible
to define a reaction system in terms of a small number of situations
and reactions. A small set of generating reactions then defines the
reactions of a large set of situations, in the same way as the local
transition rule defines the behaviour of a cellular automaton.


\chapter{Interval-preserving Automata}

In this and the next chapter we will show how to construct a reaction
system for a one-dimensional cellular automata from its transition
rule. But since the behaviour of cellular automata varies greatly, we
will consider here a subclass for which it is not too complex in a
geometrical sense. This subclass of \emph{interval-preserving} rules
contains however the complex elementary cellular automata rule 54 and
the computationally universal rule 110 \cite{Cook2004}; therefore no
restriction on the computational complexity of cellular automata is
apparent if one restricts one's view to interval-preserving transition
rules.

\section{Intervals}

Before we can describe what interval preservation shall mean, we must
define intervals and develop a notation for them and their arrangement
in space-time. We will develop it first in the context of cellular
processes and then, a bit later, for situations.

An interval consists of a finite number of cells that are positioned
without a gap. In the space-time viewpoint of cellular processes it is
also bound to a specific moment in time. We define intervals together
with a notation for their domain.\footnote{We do not use a square
  bracket notation analogous to the notation $[i, j]$ for intervals on
  the real line: This would lead to too much optical confusion with
  the other uses of square brackets in this text.}
\begin{definition}[Intervals]
  \label{def:intervals}
  An \introtwo{interval domain} at time $t \in \Z$ is a set of points
  of the form
  \begin{equation}
    I_t(i, j) = \set{ (t, x): i \leq x < j}
  \end{equation}
  with $i$, $j \in \Z$ and $i \leq j$
  (Figure~\ref{fig:interval-notation}). An \introtwo{interval process}
  at time $t$ is a cellular process whose domain is an interval
  domain.
\end{definition}
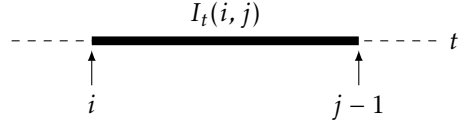
\begin{figure}[ht]
  \centering
  \begin{tikzpicture}[
    direct/.style={line width=.75ex},
    somewhere/.style={direct, draw=shaded, decorate},
    decoration={random steps, amplitude=1.5, segment length=5},
    every pin edge/.style={<-}]
    \path ++(3,0) coordinate (a)
          ++(10,0) coordinate (a')
          ++(3,0) coordinate (end0);
    \draw[dashed] (0,0)  -- (end0)  node[right] {$t$};
    \draw[direct] (a) -- node [above] {$I_t(i, j)$} (a');
    \node at (a) [pin=-90:{$i$}] {};
    \node at (a') [pin=-90:{$j-1$}] {};
  \end{tikzpicture}
  \caption{An interval process at time $t$.}
  \label{fig:interval-notation}
\end{figure}
Thus the set $I_t(i, j)$ stands for the interval domain at time $t$
that reaches from the cell position $i$ to the cell position $j$ (but
excludes it). We allow that an interval is empty: This happens if an
interval has a domain of the form $I_t(i, j)$ with $i = j$. In this
case the time $t$ is no longer determined by the set $I_t(i, j)$. We
will then use the convention that the empty set is an interval at any
time.

Now consider two compatible intervals that belong to the same time. A
nice property of intervals is that there is only a limited number of
ways in which these intervals can lie with respect to each other.
Three kinds of spatial arrangement are especially important. In the
following definition we will introduce notations for them. One
arrangement occurs if neither of the two intervals is a proper subset
of the other one: then one of them must be at the left of the other
one, if we allow overlap. The other two arrangements occur when one of
the intervals is the left or the right end of the other interval.

While the three notations below are intended especially for the use
with intervals, their definitions are meaningful for any cellular
process.
\begin{definition}[Spatial Arrangement of Processes]
  \label{def:spatial-arrangement}
  Let $\pi, \psi \in \mathcal{P}$ be two processes.


  \begin{enumerate}
  \item $\pi$ is \intro{left of} $\psi$, written $\pi \seq \psi$, if
    \begin{enumerate}[(i)]
    \item for all $p \in \dom \pi$ there is a $\xi \geq 0$ such that
      $p + \xi \ex \in \dom \psi$,
    \item for all $q \in \dom \psi$ there is a $\xi \geq 0$ such that
      $q - \xi \ex \in \dom \pi$, and
    \item $\pi$ is compatible with $\psi$.
    \end{enumerate}

  \item $\psi$ is a \introtwo{left extension} of $\pi$, written $\psi
    \supseteq_L \pi$, if
    \begin{equation}
      \label{eq:left-extension}
      \psi \supseteq \pi
      \qqtext{and}
      \psi \seq \pi\,.
    \end{equation}

  \item $\psi$ is a \introtwo{right extension} of $\pi$, written $\pi
    \subseteq_R \psi$, if
    \begin{equation}
      \label{eq:right-extension}
      \pi \subseteq \psi
      \qqtext{and}
      \pi \seq \psi\,.
    \end{equation}
  \end{enumerate}
\end{definition}
The expression ``left of'' is here used in an inclusive sense, such
that always $\pi \seq \pi$. The relation $\pi \seq \psi$ is always
true when $\pi$ or $\psi$ are empty, and $\psi \supseteq_L \pi$ and
$\pi \subseteq_R \psi$ are always true when $\pi$ is empty. If two
nonempty interval processes are related by $\seq$, $\supseteq_L$ or
$\subseteq_R$, they always occur at the same time.

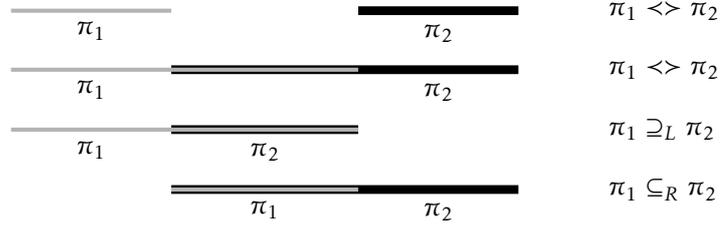
\begin{figure}[ht]
  \centering
  \begin{tikzpicture}[
    direct/.style={line width=.75ex},
    other/.style={direct, draw=shaded, line width=.35ex},
    somewhere/.style={direct, draw=shaded, decorate},
    decoration={random steps, amplitude=1.5, segment length=5}]
    \def\background{
      \path coordinate (a)
          ++(6,0) coordinate (b)
          ++(7,0) coordinate (c)
          ++(6,0) coordinate (d)
          ++(3,0) coordinate (end);}
    \matrix [cells=midway] {
      \background
      \draw[direct] (c) -- node[below] {$\pi_2$} (d);
      \draw[other]  (a) -- node[below] {$\pi_1$} (b);
      \path (end) node[right] {$\pi_1 \seq \pi_2$};\\
      \background
      \draw[direct] (b) -- (c)
                        -- node[below] {$\pi_2$} (d);
      \draw[other]  (a) -- node[below] {$\pi_1$} (b)
                        -- (c);
      \path (end) node[right] {$\pi_1 \seq \pi_2$};\\
      \draw[direct] (b) -- node[below] {$\pi_2$} (c);
      \draw[other]  (a) -- node[below] {$\pi_1$} (b)
                        -- (c);
      \path (end) node[right] {$\pi_1 \supseteq_L \pi_2$};\\
      \draw[direct] (b) -- (c)
                        -- node[below] {$\pi_2$} (d);
      \draw[other]  (b) -- node[below] {$\pi_1$} (c);
      \path (end) node[right] {$\pi_1 \subseteq_R \pi_2$};\\
      };
  \end{tikzpicture}
  \caption{Spatial orientation of intervals.}
  \label{fig:seq-interval}
\end{figure}
In the case of two intervals the relations of
definition~\ref{def:spatial-arrangement} have an especially simple
form (Figure~\ref{fig:seq-interval}). To see this, let $\pi_1$ and
$\pi_2$ be two intervals at time $t$ with $\dom \pi_1 = I_t(i_1, j_1)$
and $\dom \pi_2 = I_t(i_2, j_2)$. Then,
\begin{subequations}
  \label{eq:seq-interval}
  \begin{alignat}{2}
    \label{eq:seq-interval-1}
    \pi_1 \seq \pi_2 & \qqtext{iff}
    \pi_1 \comp \pi_2 && \qtext{and}
    i_1 \leq i_2,\ j_1 \leq j_2, \\
    \pi_1 \supseteq_L \pi_2 & \qqtext{iff}
    \pi_1 \supseteq \pi_2 && \qtext{and}
    i_1 = i_2,\ j_1 \leq j_2, \\
    \label{eq:sub-interval}
    \pi_1 \subseteq_R \pi_2 & \qqtext{iff}
    \pi_1 \subseteq \pi_2 &&\qtext{and}
    i_1 \leq i_2,\ j_1 = j_2\,.
  \end{alignat}
\end{subequations}

Another connection between the relations in
Definition~\ref{def:spatial-arrangement} is the following lemma. It
describes $\seq$ in terms of $\subseteq_R$ and $\supseteq_L$.
\begin{lemma}
  \label{thm:equivalence}
  Let $\pi$, $\psi \in \mathcal{P}$ be two cellular processes.
  Then $\pi \seq \psi$ is equivalent to $\pi \subseteq_R \pi \cup \psi
  \supseteq_L \psi$.
\end{lemma}
\begin{proof}
  It can be seen directly from the definition that $\pi \subseteq_R
  \pi \cup \psi$ is equivalent to $\pi \seq \pi \cup \psi$, and that
  $\pi \cup \psi \supseteq_L \psi$ is equivalent to $\pi \cup \psi
  \seq \psi$. Therefore $\pi \subseteq_R \pi \cup \psi \supseteq_L
  \psi$ is true if and only if $\pi \seq \pi \cup \psi \seq \psi$.

  Assume $\pi \seq \pi \cup \psi \seq \psi$: If $p \in \dom \pi$, then
  there is a $\xi \geq 0$ with $p + \xi \ex \in \dom \psi$ because
  $\pi \cup \psi \seq \psi$, and if $p \in \dom \psi$, then there is a
  $\xi \geq 0$ with $p - \xi \ex \in \dom \pi$ because $\pi \seq \pi
  \cup \psi$; together this shows $\pi \seq \psi$.

  Assume $\pi \seq \psi$: Then $\pi$ and $\psi$ are compatible, and
  therefore the process $\pi \cup \psi$ exists. To check whether $\pi
  \seq \pi \cup \psi$ is true, we only have to check that for $p \in
  \dom \psi$ there is a $\xi \geq 0$ such that $p - \xi \ex \in \dom
  \pi$, but that is true because $\pi \seq \psi$. The same way we can
  show that $\pi \cup \psi \seq \psi$. Together this proves that $\pi
  \seq \pi \cup \psi \seq \psi$.
\end{proof}

\paragraph{Interval Situations} Since a sequence $u \in \Sigma^*$ of
cell states is interpreted as a situation, its process $\pr(u)$ is an
interval process. More general, the process $\pr([t, x]u)$ has the
domain $I_t(x, x + \abs{u})$. Therefore every interval process can be
written as $\pr([p]u)$ with an appropriate $p \in \Z^2$ and $u \in
\Sigma^*$. This leads to the following definition for the set of
situations that represent interval processes.
\begin{definition}[Interval Situations]
  \label{def:interval-situations}
  An \introtwo{interval situation} with states in $\Sigma$ is a
  situation $[p]u \in \mathcal{S}$ with $p \in \Z^2$ and $u \in
  \Sigma^*$.
\end{definition}

For the following calculation we will need a notation that mirrors the
notations for interval processes in the language of situations.

First we introduce a notation for the left and right ends of a
situation, in analogy to $\subseteq_R$ and $\supseteq_L$. If $a$ is
the left or right end of $x$, then it is a factor of it; therefore I
have chosen symbols for these concepts that remind of division
operators.\footnote{This notation is also influenced by the
  alternative notation $m \mathrel\backslash n$ for ``$m$ divides
  $n$'' by Knuth, Graham and Patashnik \cite[p.\ 102]{Graham1989}.}

\begin{definition}[Left and Right Factors]
  \label{def:end-factors}
  Let $a$ and $x \in \mathcal{S}$ be situations.

  If there is a situation $x'$ such that $a x' = x$, then $a$ is a
  \introtwo{left factor} of $x$. We will write this as $a \leftend x$.

  If there is a situation $x'$ such that $x = x' a$, then $a$ is a
  \introtwo{right factor} of $x$. We will write this as $x \rightend
  a$.
\end{definition}

Next we need a notation for overlapping situations. Here I have chosen
the symbol $\ovl{b}$, in analogy to the symbol $\diamond$ for the
overlapping of two strings that is used by Harold V.\ McIntosh
\cite[p.\ 216]{McIntosh2009}.
\begin{definition}[Overlap]
  \label{def:overlap}
  Let $b \in \mathcal{S}$ be a situation. Then the displacement
  $\ovl{b} = [-\delta(b)]$ is the \introtwoshort{overlap operator} for
  $b$.
\end{definition}
This notation is subject to a convention: We will use $\ovl{b}$ only
for a situation that contains $b$ as a factor. This means that:
\begin{enumerate}
\item If $x \ovl{b} y \in \mathcal{S}$, then $x \rightend b$ and $b
  \leftend y$,
\item if $x \ovl{b} \in \mathcal{S}$, then $x \rightend b$, and
\item if $\ovl{b} y \in \mathcal{S}$, then $b \leftend y$.
\item The term $\ovl{b}$ will never be used at its own to indicate a
  situation.
\end{enumerate}

With the convention for overlap operators we can state a \intro{chain
  rule} for situations with overlap, a special case of the chain rule
in~\eqref{eq:product-as-union}. In contrast to the general case, this
equation has also a kind of converse, with intersection instead of
union:
\begin{subequations}
  \begin{align}
    \label{eq:ovl-chain}
    \pr(x \ovl{b} y) &= \pr(x) \cup \prx{x \ovl{b}}{y}, \\
    \prx{x \ovl{b}}{b} &=\pr(x) \cap \prx{x \ovl{b}}{y}\,.
  \end{align}
\end{subequations}

In almost all cases where we use $\ovl{b}$, the situation $b$ will be
an interval. If the situations $x$ and $y$ in the term $x \ovl{b} y$
are also intervals, then the relations $\seq$, $\supseteq_L$ and
$\subseteq_R$ between their processes can be expressed by the overlap
operator.
\begin{lemma}[Overlap and Spatial Arrangement of Intervals]
  \label{thm:overlap}
  Let $x$, $y$ and $b \in \Sigma^*$ be intervals. Then
  \begin{subequations}
    \begin{alignat}{2}
      \label{eq:seq-ovl}
      \pr(x) &\seq \prx{x \ovl{b}}{y} && \qtext{iff}
      x \rightend b \leftend y, \\
      \label{eq:sup-ovl}
      \pr(x) &\supseteq_L \prx{x \ovl{b}}{b} && \qtext{iff}
      x \rightend b, \\
      \label{eq:sub-ovl}
      \pr(b) &\subseteq_R \pr(y) && \qtext{iff}
      b \leftend y\,.
    \end{alignat}
  \end{subequations}
\end{lemma}
\begin{proof}
  These relations can be derived with the help of the equivalences
  in~\eqref{eq:seq-interval}.

  We will prove~\eqref{eq:seq-ovl} first. Assume that $\pr(x) \seq
  \prx{x \ovl{b}}{y}$. Then $\pr(x)$ is compatible with $\prx{x
    \ovl{b}}{y}$, and therefore the union of these processes exists.
  This union is according to~\eqref{eq:ovl-chain} the process $\pr(x
  \ovl{b} y)$. Therefore the situation $x \ovl{b} y$ exists, and this
  is, according to our convention, equivalent to $x \rightend b
  \leftend y$.

  For the opposite direction, assume that $x \rightend b \leftend y$
  is true and therefore $x \ovl{b} y$ exists. We will use here the
  equivalence~\eqref{eq:seq-interval-1}. We have
  \begin{equation}
    \dom \pr(x) = I_0(0, \abs{x}), \qquad
    \dom \prx{x \ovl{b}}{y}
    = I_0(\abs{x} - \abs{b}, \abs{x} - \abs{b} + \abs{y}).
  \end{equation}
  To apply~\eqref{eq:seq-interval-1}, we have to show that $\pr(x)$ is
  compatible with $\prx{x \ovl{b}}{y}$ and that $0 \leq \abs{x} -
  \abs{b}$ and $\abs{x} \leq \abs{x} - \abs{b} + \abs{y}$. The first
  condition is true because $\pr(x) \cup \prx{\ovl{b}}{y}$ exists. The
  second condition is equivalent to $\abs{x} \geq \abs{b} \leq
  \abs{y}$. It is true because $b$ is, according to the convention for
  $\ovl{b}$, the common part of $x$ and $y$. Therefore $\pr(x) \seq
  \prx{x \ovl{b}}{y}$.

  The other two equivalences are special cases of the first one, with
  $y = b$ or $x = b$ and are proved in a similar way.
\end{proof}

\paragraph{Borrowing an Interval} We will use the overlap operator for
making situations and reactions more readable. We will split a
situation into an equivalent situation that consists of overlapping
parts; then we apply reactions to the parts and put the parts together
again.

For this procedure we need a notion of equivalence under which two
equivalent situations initiate the same computation. The following
definition does this.
\begin{definition}[Equivalent Situations]
  \label{def:equivalent}
  Let $a$, $b \in \mathcal{S}$ be two situations. If
  \begin{equation}
    \pr(a) = \pr(b)
    \qqtext{and}
    \delta(a) = \delta(b),
  \end{equation}
  we say that $a$ is \intro{equivalent} to $b$ and write it as $a \sim
  b$.
\end{definition}
With this definition we can express $x \rightend b$ as $x \sim x
\ovl{b} b$ and $b \leftend y$ as $y \sim b \ovl{b} y$. We will use
this equivalence from time to time to split situations into
overlapping parts.

To see that equivalent situations cause the same reactions we note
that if $a \sim b$ and the closure of $\pr(a)$ exists, then there is a
reaction $a \rea b$, with no other requirements on the transition
rule. So if $b \rea x$ is a reaction for $\phi$, then $a \rea x$ is
also a reaction for $\phi$. Equivalence is symmetric, therefore the
converse is also true and the set of situations that start from $a$ is
the same as the set of reactions that start from $b$.

The following derivation then illustrates the work with overlapping
situations: Assume that the reaction system $R$ contains a reaction $b
y \rea_R b y'$ and that there is a situation $x \in \dom R$ that ends
with $b$. If also $x y \in \dom R$, then there is also a reaction $x y
\rea_R x y'$. We could prove this by introducing a situation $x'$ such
that $x = x' b$ and then applying the reaction $b y \rea_R b y'$ to
$x' b y$, but there is a notationally shorter way: We will then
instead say that $x \rightend b$ and write the following chain of
reactions, withot the need to introduce $x'$.
\begin{equation}
  \label{eq:equivalence-example}
  x y \sim x \ovl{b} b y
  \rea x \ovl{b} b y'
  \sim x y'\,.
\end{equation}
With this technique the descriptions of longer chains of reactions
become considerably shorter. An example for it occurs
in~\eqref{eq:intervals-twice}.

Note that this derivation only shows that $x y \rea x y'$ is a
reaction for $\phi$, not that it belongs to $R$. This must be verified
separately. Finding a reaction result is however often the more
difficult part, especially if the derivation is long.

\section{Interval Preservation}

Like many concepts we need the concept of interval preservation in two
forms. One is a global form that applies to a transition rule, the
other a localised form that applies to a single interval. The
localised form is defined for situations and not processes, because
that is the form where we need it in Lemma~\ref{thm:preserving-local}.

\begin{definition}[Interval Preservation]
  A transition rule $\phi$ for $\Sigma$ is \intro{interval-preserving}
  if for all interval processes $\pi \in \mathcal{P}$ the process
  $\Delta \pi$ is an interval.

  Let $u \in \Sigma^*$ be an interval situation. If $\Delta \pr(u)$ is
  an interval, then $\phi$ is \index{interval-preserving!$\sim$ for
    an interval}\emph{interval-preserving for $u$}.
\end{definition}

Interval-preserving rules are never constant functions: If $\phi$ is
constant, then the domain of $\Delta \pi$ is $\Z^2$ for every process
$\pi$, and it is therefore never an interval. So we do not specify
explicitly for interval-preserving rules that they are non-constant.

\begin{figure}[ht]
  \centering
  \ci{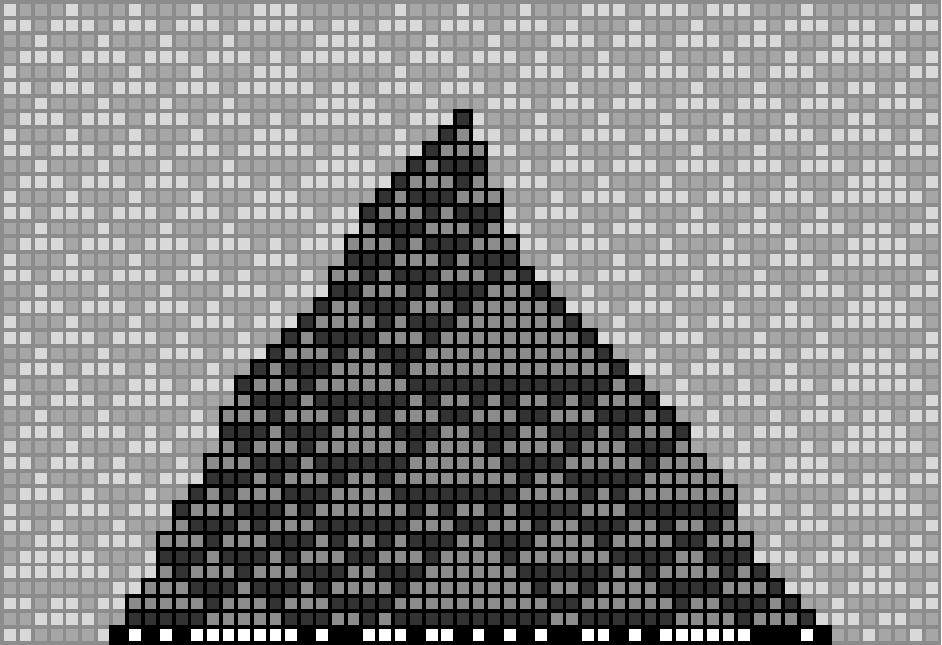}
  \caption{The closure of an interval under Rule 54 as part of the
    evolution of a random initial configuration.}
  \label{fig:dependent}
\end{figure}

The simple behaviour of interval-preserving rules, announced at the
beginning of this chapter, becomes visible when we look at the closure
of an interval process.
\begin{lemma}[Closure of an Interval]
  \label{thm:closure-interval}
  If a process is an interval, then its closure exists under an
  interval-preserving transition rule and all its time slices are then
  intervals.
\end{lemma}
\begin{proof}
  Let $\pi$ be an interval process at time $t_0$ and $\phi$ the
  transition rule. Then $\pi$ is quiet before $t_0$, and we have
  $\cl^{(t)} \pi = \emptyset$ for $t < t_0$ and $\cl^{(t_0)}
  \pi = \pi^{(t_0)}$: these time slices are intervals.

  If $t > t_0$, then $\pi^{(t)} = \emptyset$ and
  equation~\eqref{eq:closure-recursion} becomes $\cl^{(t)} \pi =
  \Delta \cl^{(t-1)} \pi$. This means that if
  $\cl^{(t-1)} \pi$ is an interval, then $\cl^{(t)} \pi$ is
  also an interval because $\phi$ is interval-preserving. Therefore
  all $\cl^{(t)} \pi$ with $t > t_0$ are intervals by induction,
  and $\cl \pi$ exists.
\end{proof}
We can see an example for such a closure in
Figure~\ref{fig:dependent}. It is part of a larger background process,
the same process as in Figure~\ref{fig:random_54}. The initial
interval is shown by the squares \ci{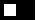}, its closure consists
of \ci{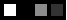}, and the whole background process of all kinds
of squares, \ci{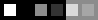}\,.

\paragraph{Testing for Interval Preservation} The following theorem
shows that it is possible to determine in finite time whether a
transition rule is interval-preserving.

\begin{theorem}[Interval Preservation is Local]
  \label{thm:preserving-local}
  A transition rule $\phi$ of radius $r$ is interval-preserving if and
  only if it is interval-preserving for all $u \in \Sigma^*$ with
  $\abs{u} \leq 2r + 1$.
\end{theorem}

{
  \newenvironment{tab}{\begin{tabular}{rcl}
      \toprule
      Rule & Evolution & IP \\
      \midrule
    }{\bottomrule\end{tabular}}
  \def\fig#1{#1 & \ci{intp_#1}}
  \def\n{no}
  \def\y{yes}
  \def\capt#1{\caption{Tests for interval preservation (Part #1).}}
  \begin{table}[t]
    \centering
    \capt1
    \begin{tab}
      \fig{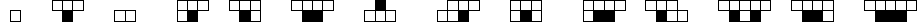} & \\
      \fig{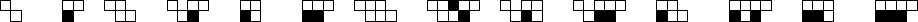} & \\
      \fig{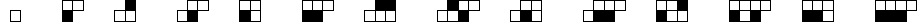} & \\
      \fig{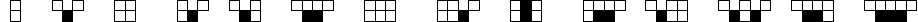} & \n\\
      \fig{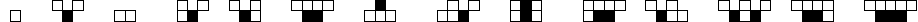} & \n\\
      \fig{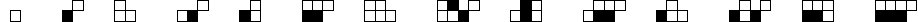} & \n\\
      \fig{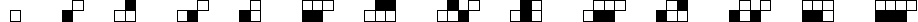} & \n\\
      \fig{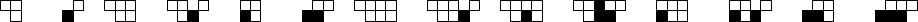} & \n\\
      \fig{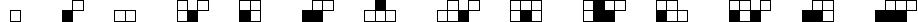} & \n\\
      \fig{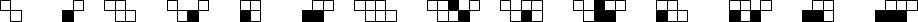} & \n\\
      \fig{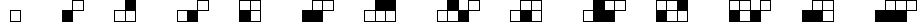} & \n\\
      \fig{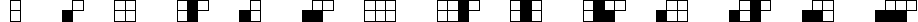} & \\
      \fig{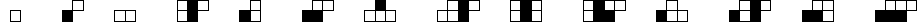} & \\
      \fig{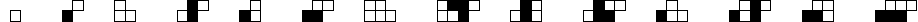} & \\
      \fig{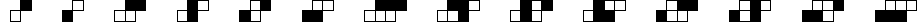} & \\
      \fig{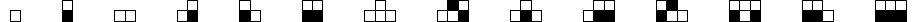} & \\
      \fig{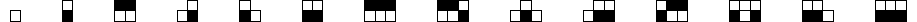} & \\
      \fig{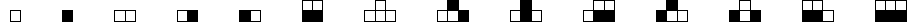} & \\
      \fig{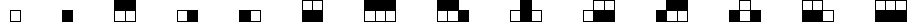} & \\
      \fig{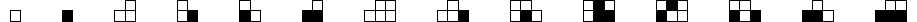} & \\
      \fig{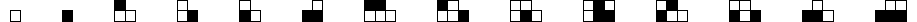} & \\
      \fig{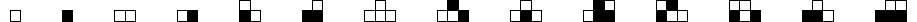} & \\
      \fig{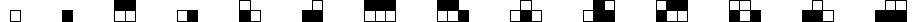} & \\
      \fig{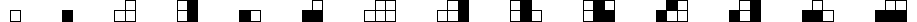} & \\
      \fig{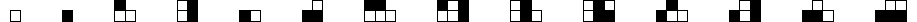} & \\
      \fig{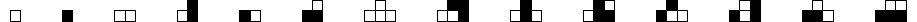} & \\
      \fig{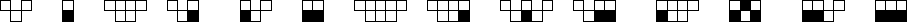} & \n\\
      \fig{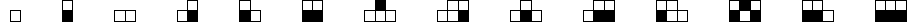} & \\
      \fig{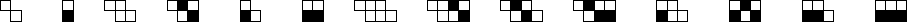} & \\
      \fig{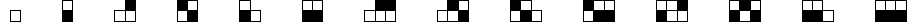} & \\
      \fig{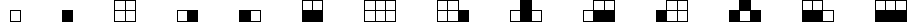} & \\
    \end{tab}
    \label{tab:interval-preserving-1}
  \end{table}
  \begin{table}[t]
    \centering
    \capt2
    \begin{tab}
      \fig{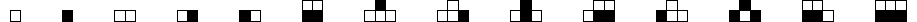} & \\
      \fig{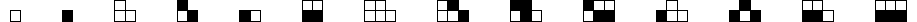} & \\
      \fig{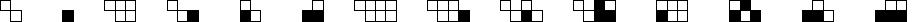} & \n\\
      \fig{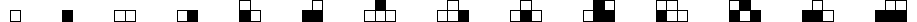} & \\
      \fig{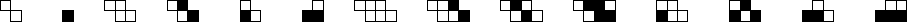} & \\
      \fig{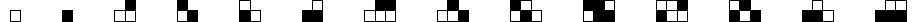} & \\
      \fig{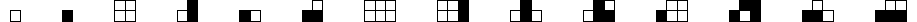} & \\
      \fig{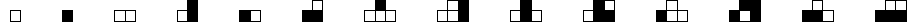} & \\
      \fig{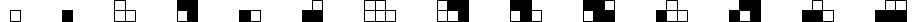} & \\
      \fig{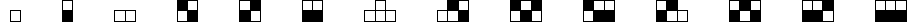} & \\
      \fig{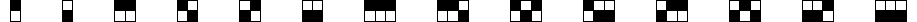} & \\
      \fig{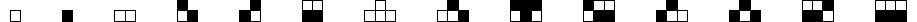} & \\
      \fig{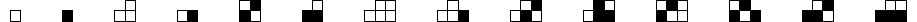} & \\
      \fig{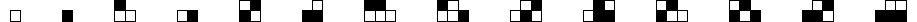} & \\
      \fig{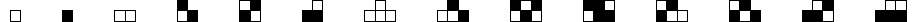} & \\
      \fig{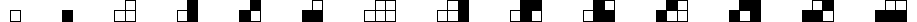} & \\
      \fig{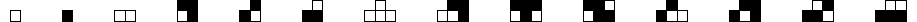} & \\
      \fig{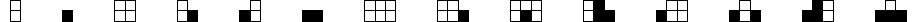} & \\
      \fig{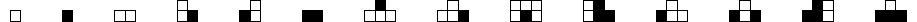} & \\
      \fig{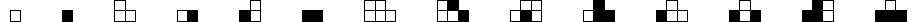} & \\
      \fig{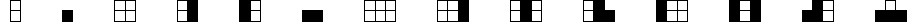} & \\
      \fig{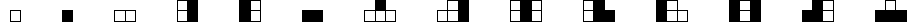} & \\
      \fig{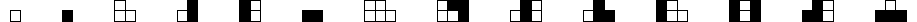} & \\
      \fig{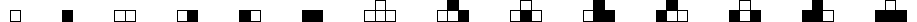} & \\
      \fig{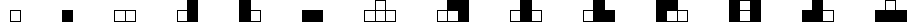} & \\
      \fig{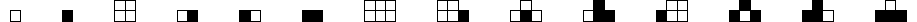} & \\
      \fig{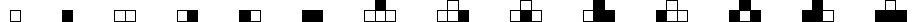} & \\
      \fig{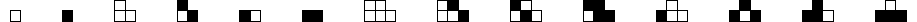} & \\
      \fig{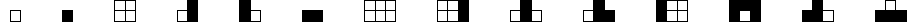} & \\
      \fig{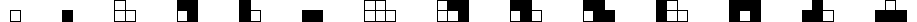} & \\
      \fig{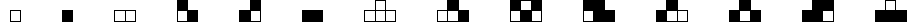} & \\
      \fig{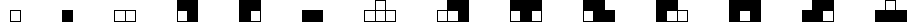} & \\
    \end{tab}
    \label{tab:interval-preserving-2}
  \end{table}
  \begin{table}[t]
    \centering
    \capt3
    \begin{tab}
      \fig{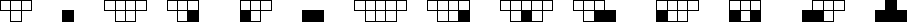} & \\
      \fig{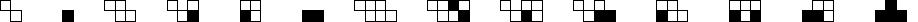} & \n\\
      \fig{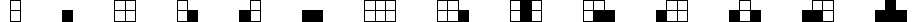} & \\
      \fig{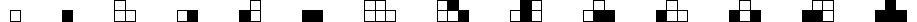} & \\
      \fig{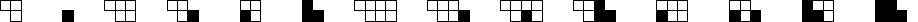} & \\
      \fig{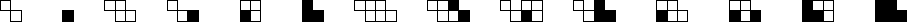} & \n\\
      \fig{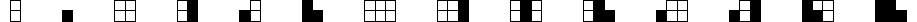} & \\
      \fig{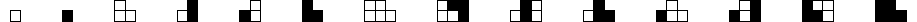} & \\
      \fig{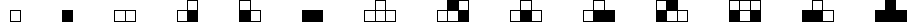} & \\
      \fig{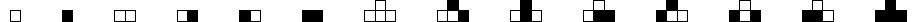} & \\
      \fig{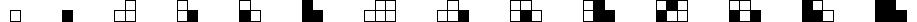} & \\
      \fig{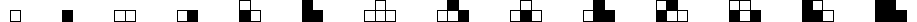} & \\
      \fig{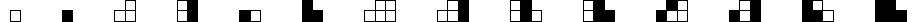} & \\
      \fig{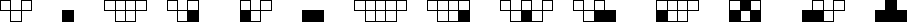} & \n\\
      \fig{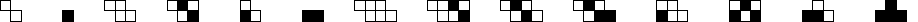} & \\
      \fig{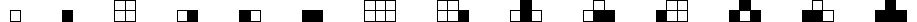} & \\
      \fig{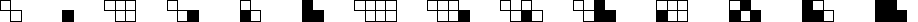} & \n\\
      \fig{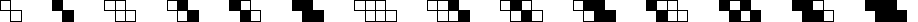} & \\
      \fig{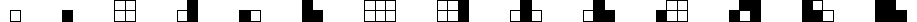} & \\
      \fig{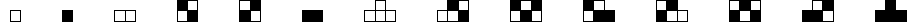} & \\
      \fig{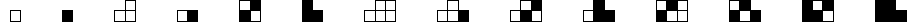} & \\
      \fig{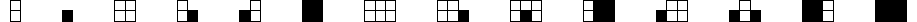} & \\
      \fig{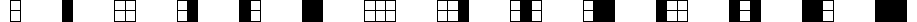} & \\
      \fig{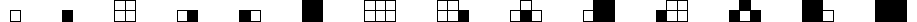} & \\
    \end{tab}
    \label{tab:interval-preserving-3}
  \end{table}
}
With this theorem we can easily find out which of the elementary
cellular automata are interval-preserving. This is done in
Table~\ref{tab:interval-preserving-1}--\ref{tab:interval-preserving-3}.
Each row in this table describes one transition rule. It contains a
list of evolution diagrams, one for each interval of maximal length
$3$. If a transition rule is interval-preserving, the top rows of all
its diagrams must be intervals.

As before, only one of the maximally four equivalent transition rules
is shown. Rule 0 is omitted because it is constant and therefore
cannot be interval-preserving. We then find that most of the
elementary cellular automata are interval-preserving: Therefore, for
better visual recognition only the non-preserving rules are marked
with a ``no'' in the last column. We can then conclude that all
elementary cellular automata except 0, 4--11, 32, 40, 130, 138, 160
and 168 are interval-preserving.

The tables here were generated by a program; a more detailed
description how one can check by hand whether a transition rule is
interval-preserving appears in Section~\ref{sec:basic-prop-rule54}.

\begin{proof}[Proof of Theorem~\ref{thm:preserving-local}]
  We need only to prove that if $\phi$ is interval-preserving for all
  $u \in \Sigma^*$ with $\abs{u} \leq 2r + 1$, then it is also
  interval-preserving for every $v \in \Sigma^*$ with $\abs{v} > 2r +
  1$.

  Let us therefore write $v = \nu_0 \dots \nu_{\ell-1} \in
  \Sigma^\ell$ with $\ell \geq 2r + 1$ and let
  \begin{equation}
    \psi_i = \pr(v)|_{I_0(i - r, i + r)}
  \end{equation}
  be, for $-r \leq i < \ell + r$, a sequence of maximally $2r + 1$
  events of $\pr(v)$, centred at $(0, i)$. Then every point that is
  determined by $\pr(v)$ is determined by some of the $\psi_i$, since
  interval-preserving rules are non-constant. There are three
  different shapes of $\psi_i$, depending on $i$:
  \begin{equation}
    \psi_i =
    \begin{cases}
      \pr([0]\nu_0 \dots \nu_{i+r})
      & \text{if $-r \leq i < r$,} \\
      \pr([i-r]\nu_{i-r} \dots \nu_{i+r})
      & \text{if $r \leq i < \ell - r$,} \\
      \pr([i-r]\nu_{i-r} \dots \nu_{\ell-1})
      & \text{if $\ell - r \leq i < \ell + r$.}
    \end{cases}
  \end{equation}
  Let us call them \emph{left}, \emph{central} and \emph{right}
  $\psi_i$ (Figure~\ref{fig:psi-i}). If $-r \leq i < r$, then $\psi_i
  \subseteq \psi_r$, so all events that are determined by a left
  $\psi_i$ are already determined by $\psi_r$. If $\ell - r \leq i <
  \ell + r$, then $\psi_i \subseteq \psi_{\ell - 1 - r}$, so all
  events determined by a right $\psi_i$ are already determined by
  $\psi_{\ell - 1 - r}$. Therefore all events determined by $\pr(v)$
  are determined by at least one of the central $\psi_i$.
  \begin{figure}[ht]
    \centering
    \begin{tikzpicture}[
      direct/.style={line width=.75ex},
      other/.style={direct, draw=shaded, line width=.35ex},
      somewhere/.style={direct, draw=shaded, decorate},
      decoration={random steps, amplitude=1.5, segment length=5}]
      \def\background{
        \path coordinate (a)
            ++(20,0) coordinate (b)
            ++(2,0) coordinate (end);}
      \path let \n{r}=3 in node[matrix,cells=midway] {
        \background
        \path (1, 0) coordinate (p) node[below] {$\psi_i$};
        \draw[direct] (a) -- ($(p)+(\n{r},0)$);
        \draw[other] ( a) -- node[below] {$\psi$} (b);
        \path (end) node[right] {left};\\
        \background
        \path (10, 0) coordinate (p) node[below] {$\psi_i$};
        \draw[direct] ($(p)+(-\n{r},0)$) -- ($(p)+(\n{r},0)$);
        \draw[other]  (a) -- node[below, pos=0.2] {$\psi$} (b);
        \path (end) node[right] {central};\\
        \background
        \path (19, 0) coordinate (p) node[below] {$\psi_i$};
        \draw[direct] ($(p)+(-\n{r},0)$) -- (b);
        \draw[other] (a) -- node[below] {$\psi$} (b);
        \path (end) node[right] {right};\\
        };
    \end{tikzpicture}
    \caption{The different shapes of $\psi_i$.}
    \label{fig:psi-i}
  \end{figure}
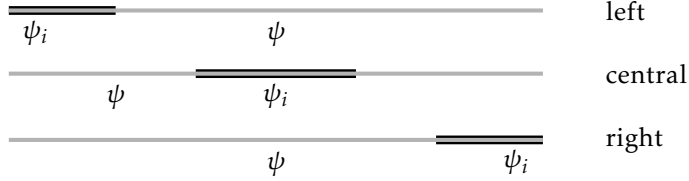

  All central $\psi_i$ are intervals of length $2r + 1$. Under a rule
  with radius $r$, a central $\psi_i$ therefore determines the point
  $(1, i)$. Since $\phi$ is interval-preserving, the interval
  determined by $\psi_i$ must therefore contain an event at $(1, i)$.
  The set of events determined by $\pr(v)$ is a union of such
  intervals, one for each $i$ with $r \leq i < \ell -r$, and is
  therefore itself an interval. Therefore $\phi$ is also
  interval-preserving for all $u$ with $\abs{u} \geq 2r + 1$.
\end{proof}

\section{Separating Intervals}

Let now $\phi$ be interval-preserving. We will now introduce a method
to compute the states of cells at the next time step only for parts of
an interval. For this we need as a technical tool a special kind of
intervals, the ``separating intervals'', that serve as a boundary
between the different parts of an interval.
\begin{definition}[Separating Intervals]
  \label{def:separating}
  An interval process $\pi \in \mathcal{P}$ is \indextwoshort{separating
    interval}\emph{separating} if for all processes $\psi$, $\psi_1$,
  $\psi_2 \in \mathcal{P}$ with $\psi = \psi_1 \cup \psi_2$ and
  $\psi_1 \supseteq_L \pi \subseteq_R \psi_2$,
  \begin{equation}
    \label{eq:separating}
    \Delta \psi_1 \cup \Delta \psi_2 = \Delta \psi
    \qqtext{and}
    \Delta \psi_1 \cap \Delta \psi_2 = \Delta \pi\,.
  \end{equation}
  An interval situation $a \in \mathcal{S}$ is separating if
  $\pr(a)$ is separating.
\end{definition}
Note here that if $\psi_1 \supseteq_L \pi$ and $\pi \subseteq_R
\psi_2$, then $\psi_1 \cap \psi_2 = \pi$. So an equivalent formulation
of~\eqref{eq:separating} is that under the conditions of
Definition~\ref{def:separating},
\begin{equation}
  \label{eq:separating2}
  \Delta \psi_1 \cup \Delta \psi_2 =
  \Delta (\psi_1 \cup \psi_2)
  \qtext{and}
  \Delta \psi_1 \cap \Delta \psi_2 =
  \Delta (\psi_1 \cap \psi_2)\,.
\end{equation}

With separating intervals we can split an interval process into parts
and evolve them independently. They solve the following question:
Assume that we have an interval $\psi$ that consists of two parts,
$\psi_1$ and $\psi_2$, such that $\psi = \psi_1 \cup \psi_2$. (The
parts may overlap.) Is it then possible to compute $\Delta
\psi_1$ and $\Delta \psi_2$ independently such that their union
is $\Delta \psi$, or does something get lost? The answer is: Yes,
it is possible, if there is a separating interval $\pi$ with $\psi_1
\supseteq_L \pi \subseteq_R \psi_2$.

By the first part of~\eqref{eq:separating}, no element of $\Delta
\psi$ is omitted by the separate computation from $\psi_1$ and
$\psi_2$, while the second part requests that the contents of
$\Delta \psi_1$ and $\Delta \psi_2$ are independent of each
other. It is actually a restricted form of independence because the
process $\pi$ is subset of $\psi_1$ and of $\psi_2$. The elements of
$\Delta \pi$ can therefore be computed from both processes, but
no other element of $\Delta \psi_1$ can be computed from $\psi_2$
and no other element of $\Delta \psi_2$ can be computed from
$\psi_1$ because of the second part of~\eqref{eq:separating}.

In Definition~\ref{def:separating}, the processes $\psi_1$ and
$\psi_2$ are arranged from left to right. We naturally should then
expect that then $\Delta \psi_1$ too is at the left of $\Delta
\psi_2$. The following lemma proves this for $\psi_1$ and $\psi_2$
that are intervals. It also shows that if $\Delta \psi_1$ and $\Delta
\psi_2$ overlap and $\psi_1 \seq \psi_2$, then the right boundary of
$\Delta \psi_1$ and the left boundary of $\Delta \psi_2$ depend only
on~$\pi$.

\begin{lemma}[Order Preservation for Overlapping Intervals]
  \label{thm:order-preservation}
  Let $\phi$ be an interval-preserving transition rule for $\Sigma$
  and let $\pi$, $\psi_1$, $\psi_2 \in \mathcal{P}$ be three intervals
  with $\psi_1 \supseteq_L \pi \subseteq_R \psi_2$, where $\pi$ is a
  separating interval. Then $\Delta \psi_1 \seq \Delta \psi_2$.

  Moreover, if $\Delta \psi_1$ and $\Delta \psi_2$ are nonempty and
  \begin{align}
    \label{eq:order-boundaries-1}
    \dom \Delta \psi_1 &= I_t(i'_1, j'_1), &
    \dom \Delta \psi_2 &= I_t(i'_2, j'_2),
  \end{align}
  then $i'_1 \leq i'_2 \leq j'_1 \leq j'_2$, and the numbers $i'_2$
  and $j'_1$ depend only on $\pi$.
\end{lemma}

The difficulty in the following proof is that $\Delta \pi$ may be
empty, so we cannot prove the lemma by proving that $\Delta
\psi_1 \supseteq_L \Delta \pi \subseteq_R \Delta \psi_2$.
Instead we must work with the interval boundaries.
\begin{proof}
  We will assume that the four processes $\psi_1$, $\psi_2$,
  $\Delta \psi_1$ and $\Delta \psi_2$ are all nonempty,
  because otherwise the lemma would be trivially true.

  We will now view $\pi$ as a constant and $\psi_1$ and $\psi_2$ as
  variables. For all the other quantities in the proof we will keep
  track whether they depend on $\psi_1$ or $\psi_2$. Let now
  \begin{align}
    \label{eq:order-boundaries-0}
    \dom \psi_1 &= I_t(i_1, j_1), &
    \dom \psi_2 &= I_t(i_2, j_2)\,.
  \end{align}
  Then the numbers $i_1$, $j_1$, \dots, $i'_2$, $j'_2$ in
  \eqref{eq:order-boundaries-0} and \eqref{eq:order-boundaries-1} are
  uniquely determined by $\psi_1$ respectively $\psi_2$.

  Because we have $\psi_1 \supseteq_L \pi \subseteq_R \psi_2$, we must
  have $i_1 \leq i_2 \leq j_1 \leq j_2$. The intersection of $\psi_1$
  and $\psi_2$ is $\pi$, therefore $\dom \pi = I_0(i_2, j_1)$, so
  $i_2$ and $j_1$ are constants. This also means that if $\psi_1$ is a
  large set, $i_1$ must be a small number, and if $\psi_2$ is large,
  then $j_2$ is a large number. The set $\Delta \pi$ may be empty, so
  we cannot give an expression for $\dom \Delta \pi$ similar to those
  in~\eqref{eq:order-boundaries-1}. But the number of elements in
  $\Delta \pi$ is a well-defined constant, and we will call it~$\ell$.

  We have always
  \begin{align}
    \dom \Delta \psi_1 &\supseteq I_{t+1}(i_1 - r, j_1 + r),&
    \dom \Delta \psi_2 &\subseteq I_{t+1}(i_2 + r, j_2 - r)\,.
  \end{align}
  Assume now that $j_2 > j_1 + 2r$, where $r$ is the radius of $\phi$.
  Then the rightmost elements of $\Delta \psi_1 \cup \Delta
  \psi_2$ must belong to $\Delta \psi_2$. This means that
  $\Delta \psi_1$ is at the left of $\Delta \psi_2$ and they
  overlap at $\ell$ events. So we must have $i'_1 \leq i'_2 \leq j'_1
  \leq j'_2$ and $i'_2 + \ell = j'_1$. If we keep $\psi_2$ fixed but
  let $\psi_1$ vary arbitrarily, the condition $i'_2 + \ell = j'_1$
  must always be true. This means that $j'_1$ stays the same for all
  values of $\psi_1$, and because it only depends on $\psi_1$, it must
  be a constant.

  The same way, by assuming $i_1 < i_2 - 2r$, we can see that $i'_2$
  is a constant.

  So we have always $i'_2 \leq j'_1$, and $i'_2$ and $j'_1$ are
  constants. For the validity of $\Delta \psi_1 \cap \Delta
  \psi_2 = \Delta \pi$ it is then necessary that $i'_1 \leq i'_2
  \leq j'_1 \leq j'_2$ is true in general, which means that $\psi_1
  \seq \psi_2$.
\end{proof}

A corollary of this lemma shows what happens if we split $\pi$ into
two intervals.
\begin{corollary}[Separation by Bounded Intervals]
  \label{thm:separation-bounded}
  Let $\phi$ be an interval-preserving transition rule for $\Sigma$.

  Let $\pi_1$, $\pi_2 \in \mathcal{P}$ be separating intervals
  and $\psi_1$, $\psi_2 \in \mathcal{P}$ be intervals with
  \begin{equation}
    \label{eq:separation-bounded}
    \psi_1 \supseteq_L \pi_1 \seq \pi_2 \subseteq_R \psi_2\,.
  \end{equation}
  Then $\Delta \psi_1 \seq \Delta \psi_2$.
\end{corollary}
\begin{proof}
  Let $\pi$ be an arbitrary interval process that reaches from the
  left end of $\pi_1$ to the right end of $\pi_2$, such that we have
  $\pi_1 \subseteq_R \pi \supseteq_L \pi_2$.

  Then $\pi \cup \psi_2$ is an interval, and we have $\psi_1
  \supseteq_L \pi_1 \subseteq_R \pi \cup \psi_2$. So we can apply
  Lemma~\ref{thm:order-preservation} and get $\Delta \psi_1 \seq
  \Delta (\pi \cup \psi_2)$. Because $\pi_2$ is separating and $\pi
  \supseteq_L \pi_2 \subseteq_R \psi_2$, we have $\Delta (\pi
  \cup \psi_2) = \Delta \pi \cup \Delta \psi_2$. We derive
  from the resulting relation $\Delta \psi_1 \seq \Delta \pi
  \cup \Delta \psi_2$ with Lemma~\ref{thm:equivalence} the
  relation $\Delta \psi_1 \subseteq_R \Delta \psi_1 \cup
  \Delta \pi \cup \Delta \psi_2$, and this leads to
  $\Delta \psi_1 \subseteq_R \Delta \psi_1 \cup \Delta
  \psi_2$.

  The same way we can also prove $\Delta \psi_1 \cup \Delta
  \psi_2 \supseteq_L \Delta \psi_2$. These two relations together
  are equivalent to $\Delta \psi_1 \seq \Delta \psi_2$,
  again by Lemma~\ref{thm:equivalence}.
\end{proof}

\paragraph{The Set of Separating Intervals} All this assumes that
separating intervals exist. We need to make that certain and would
also like to get an overview about which intervals are separating.
This we will do in two steps. First we will prove that being a
separating interval is a monotone property: an interval that contains
a separating interval as a subset is itself separating. Then we will
show that under a rule of radius $r$, every interval of exactly $2r$
cells is separating. From this we can then conclude that every
interval of at least $2r$ cells is separating.

\begin{lemma}[Being Separating is Monotone]
  \label{thm:separating-monotone}
  Let $\phi$ be an interval-preserving transition rule and let $\pi
  \subseteq \pi' \in \mathcal{P}$ be two intervals. If $\pi$ is
  separating, then $\pi'$ is separating too.
\end{lemma}
\begin{proof}
  We will first prove the lemma for the case that $\pi \subseteq_R
  \pi'$. For this, let $\psi'_1$, $\psi'_2$, $\psi' \in
  \mathcal{P}$ be any processes with $\psi'_1 \supseteq_L \pi'
  \subseteq_R \psi'_2$ and $\psi'_1 \cup \psi'_2 = \psi'$. Let $\psi_1
  = (\psi'_1 \setminus \pi') \cup \pi$; then $\psi_1 \supseteq_L \pi
  \subseteq_R \psi'_2$ and $\psi_1 \cup \psi'_2 = \psi'$.

  We have $\Delta \psi'_1 = \Delta (\psi_1 \cup \pi') =
  \Delta \psi_1 \cup \Delta \pi'$ because $\pi$ is separating.
  Therefore,
  \begin{equation}
    \begin{alignedat}[b]{2}
      \Delta \psi'_1 \cup \Delta \psi'_2
      &= \Delta \psi_1 \cup \Delta \pi' \cup \Delta
      \psi'_2 \\
      &= \Delta \psi_1 \cup \Delta \psi'_2
      &&\qquad\text{because $\Delta \pi' \subseteq \Delta \psi'_2$} \\
      &= \Delta (\psi_1 \cup \psi')
      &&\qquad\text{because $\psi_1 \supseteq_L \pi \subseteq_R \psi'_2$} \\
      &= \Delta \psi'\,.
      &&\qquad\text{because $\psi_1 \subseteq \psi$}
    \end{alignedat}
  \end{equation}
  On the other hand,
  \begin{equation}
    \begin{alignedat}[b]{2}
      \Delta \psi'_1 \cap \Delta \psi'_2
      &= (\Delta \psi_1 \cup \Delta \pi')
      \cap \Delta \psi'_2 \\
      &= (\Delta \psi_1 \cap \Delta \psi'_2)
      \cup \Delta \pi'
      &&\qquad\text{because $\Delta \pi' \subseteq \Delta \psi'_2$} \\
      &= \Delta \pi \cup \Delta \pi'
      &&\qquad\text{because $\psi_1 \supseteq_L \pi \subseteq_R \psi'_2$} \\
      &= \Delta \pi'\,.
      &&\qquad\text{because $\Delta \pi \subseteq \Delta \pi'$}\\
    \end{alignedat}
  \end{equation}
  This proves the lemma in the case that $\pi \subseteq_R \pi'$.

  The same kind of argument works when $\pi' \supseteq_L \pi$. In the
  general case we note that if $\pi \subseteq \pi'$, then there is
  always an interval process $\pi''$ such that $\pi \subseteq_R \pi''$
  and $\pi' \supseteq_L \pi''$. This reduces the general case to the
  two other cases.
\end{proof}

\begin{lemma}[Existence of Separating Intervals]
  \label{thm:separating-exists}
  Under a transition rule with radius $r$, every interval that
  consists of at least $2r$ events is separating.
\end{lemma}

The proof makes use of the fact that determinedness is a local
property: if $\theta$ is a process, then a point $p$ is determined by
$\theta$ if and only if it is determined by $\theta|_{N(p - \et, r)}$.
We have seen this in Lemma~\ref{thm:timeslice-determines}. The set
$N(p - \et, r)$ is the neighbourhood domain of $p$ for the previous
time step, defined in~\eqref{eq:nb-domain}.

\begin{proof}
  Let $\pi$ be the separating interval and let $\psi$, $\psi_1$ and
  $\psi_2$ be as in Definition~\ref{def:separating}.

  We know already that $\Delta \psi_1 \cup \Delta \psi_2
  \subseteq \Delta \psi$ and $\Delta \pi \subseteq
  \Delta \psi_1 \cap \Delta \psi_2$ because $\Delta$ is
  monotone (Lemma~\ref{thm:det-monotone}). So it remains to prove
  \begin{equation}
    \label{eq:separate-inclusions}
    \Delta \psi
    \subseteq \Delta \psi_1 \cup \Delta \psi_2
    \qqtext{and}
    \Delta \psi_1 \cap \Delta \psi_2
    \subseteq \Delta \pi\,.
  \end{equation}

  Let now $p\in \dom \Delta \psi$ be an arbitrary point and let
  $N_p$ stand for $N(p - \et, r)$. The proof
  of~\eqref{eq:separate-inclusions} then relies on the fact that $N_p$
  is an interval domain of length $2r + 1$, but $\dom \pi$ has at
  least $2r$ points. The points of ${N_p} \setminus \dom \pi$ must
  therefore be either completely at the left or completely at the
  right of $\pi$. In the first case we have $(\psi_2 \setminus
  \pi)|_{N_p} = \emptyset$, in the second, $(\psi_1 \setminus
  \pi)|_{N_p} = \emptyset$. Therefore,
  \begin{equation}
    \label{eq:N-alternative}
    \psi_2|_{N_p} = \pi|_{N_p}
    \qqtext{or}
    \psi_1|_{N_p} = \pi|_{N_p}\,.
  \end{equation}

  Assume first that $p \in \dom \Delta \psi$. Because
  $\psi|_{N_p}$ is equal to $\psi_1|_{N_p} \cup \psi_2|_{N_p}$, and
  with equation~\eqref{eq:N-alternative}, the process $\psi|_{N_p}$
  must either be equal to $\pi|_{N_p} \cup \psi_2|_{N_p} =
  \psi_2|_{N_p}$ or $\psi_1|_{N_p} \cup \pi|_{N_p} = \psi_1|_{N_p}$ or
  both. In the first case, $p \in \dom \Delta \psi_1$, in the
  second case, $p \in \dom \Delta \psi_2$. This proves
  $\Delta \psi \subseteq \Delta \psi_1 \cup \Delta
  \psi_2$.

  Assume now that $p \in \dom (\Delta \psi_1 \cap \Delta
  \psi_2)$. Then $p$ depends on $\psi_1|_{N_p}$ and on
  $\psi_2|_{N_p}$. One of these processes is equal to $\pi|_{N_p}$
  by~\eqref{eq:N-alternative}, therefore $p \in \dom \Delta
  (\pi|_{N_p})$. This proves $\Delta \psi_1 \cap \Delta
  \psi_2 \subseteq \Delta \pi$.
\end{proof}

A transition rule may also have separating intervals of less than $2r$
elements. So to get an overview about the separating intervals of a
specific transition rule, we should know the set of its \emph{minimal
  separating intervals}. We will actually need three kinds of minimal
intervals, defined below.
\begin{definition}[Minimal Separating Interval]
  \label{def:minimal-separating}
  A separating interval process is \indextwoshort{{left minimal}
    {separating interval}}\emph{left minimal} if no events can be
  removed from its right side without making it non-separating. It is
  \indextwoshort{{right minimal} {separating interval}}\emph{right
    minimal} if no events can be removed from its left side without
  making it non-separating. It is \indextwoshort{minimal separating
    interval}\emph{minimal} if it is both left minimal and right
  minimal.
\end{definition}

Left and right minimal intervals occur as the boundaries of a separating
interval: If $\psi$ is a separating interval, then there are intervals
$\pi_1$ and $\pi_2$ such that $\pi_1 \subseteq_R \psi \supseteq_L
\pi_2$. The processes $\pi_1$ and $\pi_2$ can be viewed as the ``left
and right end'' of $\psi$. The shortest interval $\pi_1$ that is still
separating is then a left minimal interval, and the shortest separating
interval $\pi_2$ is a right minimal interval.

\paragraph{Example: The Elementary Cellular Automata} For the
elementary cellular automata there exists a simple test to find the
minimal separating intervals. The main reason for this is the small
radius of their transition rules. This means, with
Lemma~\ref{thm:separating-exists}, that we only need to check whether
there are intervals of length $0$ and $1$ that are separating.

The case of a separating interval of length $0$ does indeed occur, and
it means that the cells never interact. There are two rules that have
this property: Rule 51, which lets the cells alternate between the
states $0$ and $1$, and the identity function, Rule 204.

For the other rules we must check whether there are intervals that
consist of a single event and are separating. We will call such an
interval $\pi$ and place it at the origin; it consists of the single
event $[0, 0] \sigma$.

As we have seen in the proof of Lemma~\ref{thm:separating-exists}, we
need only to verify that for all cellular processes $\psi_1$, $\psi_2$
with $\psi_1 \supseteq_L \pi \subseteq_R \psi_2$ we have $\Delta
(\psi_1 \cup \psi_2) \subseteq \Delta \psi_1 \cup \Delta \psi_2$ and
$\Delta \psi_1 \cap \Delta \psi_2 \subseteq \Delta \pi$ in order to
prove that $\pi$ is a separating interval.

We will use the convention that at time $0$, the cell at position $i$
is in state $\xi_i$; we have then $\sigma = \xi_0$. For an arbitrary
event at time $1$ we will write $e$: its state is $\eta$ and its
location, $x$, such that we have $e = [1, x]\eta$. Since $\phi$ has
radius $1$, the state of the event $e$ can only depend on $\xi_{x -
  1}$, $\xi_x$ and $\xi_{x + 1}$: This will be important in the
following derivation.

Let now $e$ be element of $\Delta (\psi_1 \cup \psi_2)$. Then, if $x <
0$, we have $e \in \Delta \psi_1$ and if $x > 1$, we have $e \in
\Delta \psi_2$. The remaining case, $x = 0$, is the key to finding a
necessary condition for $\pi$: If the state of $e$ depends both on
events in $\psi_1$ and $\psi_2$, then $e$ cannot be an element of
$\Delta \psi_1 \cup \Delta \psi_2$, and $\pi$ cannot be separating.
This can only occur when $\phi(\xi_{-1}, \sigma, \xi_1)$, the state of
$e$, depends on both $\xi_{-1}$ and $\xi_1$. Therefore a necessary
condition for $\pi$ being separating is that such a dependency does
not happen.

We can ensure that by requiring that at least one of the following two
equations is true:
\begin{subequations}
  \label{eq:elementary-separating}
  \begin{alignat}{2}
    \label{eq:elementary-separating-left}
    \forall \xi_{-1} \in \Sigma &\colon&
    \phi(\xi_{-1}, \sigma, 0) &= \phi(\xi_{-1}, \sigma, 1), \\
    \label{eq:elementary-separating-right}
    \forall \xi_1 \in \Sigma &\colon&
    \phi(0, \sigma, \xi_1) &= \phi(1, \sigma, \xi_1)\,.
  \end{alignat}
\end{subequations}
Here the first equation means that the cell to the right has no
influence on the next state of a cell in state $\sigma$, and the
second, that the cell at the left has no influence on the next state.

The conditions~\eqref{eq:elementary-separating} are also sufficient.
To show this, we will first prove that $\Delta (\psi_1 \cup \psi_2)
\subseteq \Delta \psi_1 \cup \Delta \psi_2$. In the proof we will
assume that $e \in \Delta(\psi_1 \cup \psi_2)$. As we have seen
before, if $x < 0$, then $e \in \Delta \psi_1$ and if $x > 1$, then $e
\in \Delta \psi_2$. In the remaining case of $x = 0$, at least one of
the two conditions in~\eqref{eq:elementary-separating} must be true.
If~\eqref{eq:elementary-separating-left} is true, then $\eta$ depends
only on $\xi_{-1}$ and $\sigma$, and therefore $e \in \Delta \psi_1$.
If~\eqref{eq:elementary-separating-right} is true, then $\eta$ depends
only on $\sigma$ and $\xi_1$, and $e \in \Delta \psi_2$. This shows
that $\Delta (\psi_1 \cup \psi_2) \subseteq \Delta \psi_1 \cup \Delta
\psi_2$.

Next we prove that $\Delta \psi_1 \cap \Delta \psi_2 \subseteq \Delta
\pi$. We will assume that $e \in \Delta \psi_1 \cap \Delta \psi_2$.
Since $\phi$ has radius $1$, the only possible values for $x$ are then
$-1$, $0$ and $1$. In the case of $x = -1$ we use the fact that $e$ is
an element of $\Delta \psi_2$. This means that we have $\phi(\xi_{-2},
\xi_{-1}, \sigma) = \eta$ for all $\xi_{-2}$, $\xi_{-1} \in \Sigma$
and that therefore $e \in \Delta \pi$. In the case of $x = 0$, the
condition $e \in \Delta \psi_1$ requires
that~\eqref{eq:elementary-separating-left} is true, and $e \in \Delta
\psi_2$ requires that~\eqref{eq:elementary-separating-right} is true.
Taken together, the two conditions imply that $\phi(\xi_{-1}, \sigma,
\xi_1) = \eta$ for all $\xi_{-1}$, $\xi_1 \in \Sigma$ and that we have
here again $e \in \Delta \pi$. The case $x = 1$ can be handled in a
similar way as $x = -1$. This shows that $\Delta \psi_1 \cap \Delta
\psi_2 \subseteq \Delta \pi$.

We therefore have now proved that~\eqref{eq:elementary-separating} are
necessary and sufficient conditions that the interval $\pi$ is
separating.

The tests~\eqref{eq:elementary-separating} have been done by a program
for all interval-preserving elementary cellular automata different
from 51 and 204. The results are shown in Table~\ref{tab:separating}.
If none of the length $1$ intervals are separating, then the intervals
of length $2$ are the minimal separating intervals. If an interval
$\pi$ of length $1$ is minimally separating, then only those intervals
of length $2$ are separating that do not contain $\pi$. So if, e.\,g.,
$0$ is a separating interval but $1$ is not separating, then the only
minimal separating interval is $11$. This argument explains the second
row of Table~\ref{tab:separating}; the other rows are explained
similarly.

\begin{table}[t]
  \centering
  \caption{Separating intervals of the interval-preserving elementary
    cellular automata.}
  \begin{tabular}{lp{24em}}
    \toprule
    Intervals & Rules \\
    \midrule
    \ci{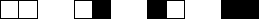} & 22, 23, 24, 26, 27, 36, 37, 41, 43, 54, 57, 58,
    73, 74, 77, 78, 90, 94, 104, 105, 108, 122, 126, 134, 142, 146,
    150, 156, 164, 172, 178, 232. \\
    \ci{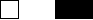} & 25, 38, 42, 56, 72, 76, 106, 110, 128, 132, 140,
    162, 200. \\
    \ci{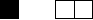} & 1, 2, 13, 14, 18, 19, 28, 30, 33, 35, 44, 45, 50,
    62, 152, 154. \\
    \ci{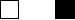} & 3, 12, 15, 29, 34, 46, 60, 136, 170, 184 \\
    \small $[0]$ & 51, 204. \\
    \bottomrule
  \end{tabular}
  \label{tab:separating}
\end{table}

In Section~\ref{sec:basic-prop-rule54} and in the context of Rule 54,
we will see a related method to find the minimal separating intervals
of a transition rule. It uses the characteristic reactions, which are
defined next.

\section{Characteristic Reactions}

We will now introduce the concept of \emph{characteristic reactions}
as a way to express the relation between an interval $\pi$ and the set
$\Delta \pi$ in the language of situations and reactions. The
characteristic reaction specifies the location and the dimensions of
$\Delta \pi$ in relation to $\pi$ in a kind of shorthand.

We will introduce it in two steps. The first step is for the case that
$\Delta \pi$ is nonempty. Then the notion of the location of $\Delta
\pi$ has an obvious meaning. If $\pi$ is separating and it is part of
a longer interval, then the cells of $\Delta \pi$ serve as a separator
between the events determined by the cells left of $\pi$ from those
determined by the cells right of $\pi$. This property is important for
proofs about the behaviour of reaction systems, so we would like to
have it for all of the separating intervals. We will use it in the
second step to extend the notion of the characteristic reaction such
that for a separating interval $\pi$ we can speak of the ``location''
of $\Delta \pi$ even then when $\Delta \pi$ is empty.

The characteristic reactions themselves will however not become part
of the final reaction system; they are only a tool to define it.

\paragraph{Construction of the Characteristic Reactions} We can
express the property of a rule $\phi$ to be interval-preserving in the
following way: If $a \in \Sigma^*$, then there are $i \in \Z$ and
$\hat a \in \Sigma^*$ such that $\Delta \pr(a) = \pr([1, i]\hat
a)$. The interval situation $\hat a$ is then always uniquely
determined by $a$. So we define
\begin{definition}[Determined Interval]
  \label{def:determined-situation}
  Let $\phi$ be an interval-preserving transition rule for $\Sigma$
  and let $a \in \Sigma^*$. Then the situation $\hat a \in \Sigma^*$
  for which there is an $i \in \Z$ such that
  \begin{equation}
    \label{eq:determined}
    \Delta \pr(a) = \pr([1, i] \hat a),
  \end{equation}
  is the \introtwo{determined interval} of $a$ under $\phi$.
\end{definition}
The ``hat'' accent of $\hat a$ should remind of the operator
$\Delta$.

In contrast to $\hat a$, the number $i$ is only then uniquely
determined when $\hat a \neq [0]$, because only then $\Delta
\pr(a) \neq \emptyset$. In this case we can express the relation
between $\pr(a)$ and $\Delta \pr(a)$ by a reaction. This
definition is important because it specifies the numbers $i$ and $j$,
which will be needed later to define the actual reaction system.
\begin{figure}[ht]
  \centering
  \begin{tikzpicture}[
    direct/.style={line width=.75ex},
    somewhere/.style={direct, draw=shaded, decorate},
    decoration={random steps, amplitude=1.5, segment length=5},
    displace/.style={->, >=latex, shorten <=.5ex, shorten >=1ex}]
    \path ++(5,0) coordinate (ahat) +(-3,-1) coordinate (a)
          ++(5,0) coordinate (ahat') +(3,-1) coordinate (a')
          ++(5,0) coordinate (end0)
          ++(0,-1) coordinate (end-1);
    \draw[dashed] (0,0)  -- (end0)  node[right] {$t=1$}
                  (0,-1) -- (end-1) node[right] {$t=0$};
    \draw[direct]
                  (ahat) -- node [above] {$\hat a$} (ahat');
    \draw[direct, draw=shaded] (a)    -- node [below] {$a$} (a');
    \draw[->, shorten <=.5ex, shorten >=.7ex]
        (a) -- coordinate (mid) (ahat);
    \node at (mid) [pin=120:{\small$[1, i]$}] {};
    \draw[->, shorten <=.7ex, shorten >=.5ex]
        (ahat') -- coordinate (mid') (a');
    \node at (mid') [pin=80:{\small$[-1, j]$}] {};
  \end{tikzpicture}
  \caption{The two sides of the characteristic
    reaction~\eqref{eq:characteristic-reaction}, overlayed.}
  \label{fig:characteristic-reaction}
\end{figure}
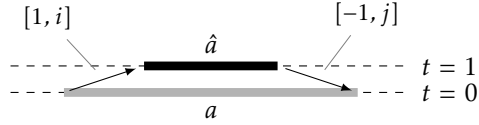
\begin{definition}[Characteristic Reactions, Preliminary Form]
  \label{def:characteristic}
  Let $\phi$ be an interval-preserving transition rule for $\Sigma$.
  Let $a \in \Sigma^*$ be an interval with $\hat a \neq [0]$ and let
  $\Delta \pr(a) = \pr([1, i] \hat a)$. Then the reaction
  \begin{equation}
    \label{eq:characteristic-reaction}
    a \rea [1, i] \hat a [-1, j]
  \end{equation}
  with $j = \abs{a} - i - \abs{\hat a}$, is the
  \introtwoshort{characteristic reaction} for $a$ under $\phi$.
  (Figure~\ref{fig:characteristic-reaction}.)
\end{definition}

Now we extend this definition to interval situations $a$ for which
$\hat a$ can be empty---but in this case $a$ must be a separating
interval. We can do this by using other separating intervals as ``test
functions''. 

The following lemma is there to show that the final definition of
separating intervals extends the preliminary definition.
\begin{theorem}[Intervals as Reactions]
  \label{thm:characteristic-reaction}
  Let $\phi$ be an interval-preserving transition rule for $\Sigma$
  and $a \in \Sigma^*$ a separating interval for $\phi$.

  Then there exist $i$, $j \in \Z$ such that for all $x \in \Sigma^*$
  with $\widehat{x a} \neq [0]$ there is an $i' \in \Z$ such that the
  characteristic reaction for $x a$ is
  \begin{subequations}
    \label{eq:characteristic-extension}
    \begin{equation}
      xa \rea [1, i'] \widehat{x a} [-1, j],
    \end{equation}
    and for all $y \in \Sigma^*$ with $\widehat{a y} \neq [0]$ there
    is a $j' \in\Z$ such that the characteristic reaction for $a y$ is
    \begin{equation}
      ay \rea [1, i] \widehat{a y} [-1, j'].
    \end{equation}
  \end{subequations}
  If also $\hat a \neq [0]$, then $a$ has the characteristic reaction
  $a \rea [1, i] \hat a [-1, j]$, the same as
  in~\eqref{eq:characteristic-reaction}.
\end{theorem}

We will use in this proof that if $u \rea [1, i] \hat u [-1, j]$
is a characteristic reaction, then $\dom \Delta \pr(u) = I_1(i,
\abs{u} - j)$: The process $\Delta \pr(u)$ reaches from $i$ cells
to the left of $\pr(u)$ to $j$ cells to the right of $\pr(u)$, one
time step later.

\begin{proof}
  To apply Lemma~\ref{thm:order-preservation} we first define
  processes $\pi$, $\psi_1$ and $\psi_2$ with $\psi_1 \supseteq_L \pi
  \subseteq_R \psi_2$, namely
  \begin{equation}
    \label{eq:psi-definitions}
    \psi_1 = \prx{[-\abs{x}]}{x a}, \qquad
    \pi = \pr(a), \qquad
    \psi_2 = \pr(ay)\,.
  \end{equation}
  In order to verify the inclusion $\psi_1 \supseteq_L \pi$ we note
  that we have $\prx{[-\abs{x}]}{x a} = \prx{[-\abs{x}]}{x} \cup
  \prx{[-\abs{x}]x}{a}$ with the chain rule, and $\prx{[-\abs{x}]x}{a}
  = \pr(a)$ because $\delta([-\abs{x}]x) = (0, 0)$. Therefore
  $\prx{[-\abs{x}]}{x a} = \pr(x) \cup \pr(a) \supseteq_L \pr(a)$,
  which means that $\psi_1 \supseteq_L \pi$. This explains also why
  the shift of $\pr(xa)$ to the left with $[-\abs{x}]$ is necessary.
  The other inclusion, $\pi \subseteq_R \psi_2$, can also be verified
  by an application of the chain rule.

  The domains of the processes $\psi_1$ and $\psi_2$ are then
  \begin{equation}
    \label{eq:psi-domains}
    \dom \psi_1 = I_0(-\abs{x}, \abs{a}), \qquad
    \dom \psi_2 = I_0(0, \abs{ay}),
  \end{equation}
  and the events determined by them are located at
  \begin{equation}
    \label{eq:delta-psi-domains}
    \dom \Delta \psi_1 = I_1(-\abs{x} + i', \abs{a} - j), \qquad
    \dom \Delta \psi_2 = I_1(i, \abs{ay} - j')\,.
  \end{equation}
  as we can see from the characteristic
  reactions~\eqref{eq:characteristic-extension} and from the remark
  before the proof. For the process $\psi_1$ we have to keep in mind
  that it, and therefore also $\Delta \psi_1$, is shifted to the
  left by $\abs{x}$ cells. Then we know by
  Lemma~\ref{thm:order-preservation} that the boundaries of the
  intervals in~\eqref{eq:delta-psi-domains} are arranged in the form
  $-\abs{x} + i' \leq i \leq \abs{a} - j \leq \abs{ay} - j'$ and that
  $i$ and $\abs{a} - j$ only depend on $a$. Therefore the variables
  $i$ and $j$ only depend on $a$.

  Assume now that $\hat a \neq [0]$ and therefore $\Delta \pi
  \neq \emptyset$. Then $\dom \pi = I_0(0, \abs{a})$ and $\dom
  \Delta \pi = I_1( i, \abs{a} - j)$, which means that the
  characteristic reaction for $a$ (in the preliminary form of
  Definition~\ref{def:characteristic}) must be $a \rea [1, i] \hat a
  [-1, j]$.
\end{proof}

This theorem then justifies the following definition of the
characteristic reaction for separating intervals.
\begin{definition}[Characteristic Reactions, Final Form]
  \label{def:characteristic-separating}
  Let $a \in \Sigma^*$ be a separating interval situation.

  If there are intervals $x$, $y \in \Sigma^*$ with $\widehat{x a}
  \neq [0] \neq \widehat{a y}$, and they have the characteristic
  reactions $x a \rea [1, i'] \widehat{x a} [-1, j]$ and $a y \rea
  [1, i] \widehat{a y} \, [-1, j']$, then the reaction
  \begin{equation}
    \label{eq:characteristic-separating}
    a \rea [1, i] \hat a [-1, j]
  \end{equation}
  is the \introtwoshort{characteristic reaction} for $a$ under $\phi$.
\end{definition}

The following lemma expresses the monotony of the closure in terms of
characteristic reaction. It shows that in the previous definition
$\hat a$ is always a part of $\widehat{x a}$ and $\widehat{a y}$, and
how to recover it.
\begin{lemma}[Reactions of Separating Intervals]
  \label{thm:separating-reactions}
  Let $a \in \Sigma^*$ be a separating interval under an
  interval-preserving transition rule, and let $x$, $y\in \Sigma^*$.

  Then $\widehat{x a} \rightend \hat a$ and $\hat a \leftend
  \widehat{a y}$.
\end{lemma}

\begin{proof}
  Let $\phi$ be the transition rule. We will prove only the first
  equivalence, the second one is its mirror image and the proof is
  similar.

  Assume that the characteristic reaction of $a$ and $x a$ are
  \begin{equation}
    \label{eq:separating-1}
    a \rea [1, i] \hat a [-1, j]
    \qqtext{and}
    x a \rea [1, i'] \widehat{x a} [-1, j]\,.
  \end{equation}
  We will show first that $\prx{x [1, i]}{\hat a} \subseteq \prx{[1,
    i']}{\widehat{x a}}$. We see from~\eqref{eq:separating-1} that
  $\prx{[1, i]}{\hat a} = \Delta \pr(a)$ and $\Delta \pr(x a) =
  \prx{[1, i']}{\widehat{x a}}$. With the chain rule we get
  $\prx{x}{a} \subseteq \pr(xa)$, by the monotony of the $\Delta$
  operator, $\Delta \prx{x}{a} \subseteq \Delta \pr(xa)$, and by
  putting these relations together we get
  \begin{equation}
    \prx{x [1, i]}{\hat a} = \Delta \prx{x}{a} \subseteq
    \Delta \pr(xa) = \prx{[1, i']}{\widehat{x a}}\,.
  \end{equation}

  Next we show that $\delta(x [1, i]) = \delta([1, i'] \widehat{x a}
  \ovl{\hat a})$. We see from the characteristic reaction for $a$ that
  $i = \abs{a} - \abs{\hat a} - j$ and from the characteristic reaction
  for $x a$ that $j = \abs{x a} - i' - \abs{\widehat{x a}}$. Therefore
  $i = \abs{a} - \abs{\hat a} - \abs{x a} + i' + \abs{\widehat{x a}} =
  i' + \abs{\widehat{x a}} - \abs{x} - \abs{\hat a}$, and $\delta(x
  [1, i]) = (1, \abs{x} + i) = (1, i' + \abs{\widehat{x a}} -
  \abs{\hat a}) = \delta([1, i'] \widehat{x a} \ovl{\hat a})$. This
  means that
  \begin{equation}
    \begin{aligned}[b]
      \prx{[1, i']}{\widehat{x a}}
      &= \prx{[1, i']}{\widehat{x a}} \cup \prx{x [1, i]}{\hat a} \\
      &= \prx{[1, i']}{\widehat{x a}} \cup
      \prx{[1, i'] \widehat{x a}}{\ovl{\hat a} \hat{a}}
      = \prx{[1, i']}{\widehat{x a} \ovl{\hat a} \hat{a}},
    \end{aligned}
  \end{equation}
  which also means that $\widehat{x a} \rightend \hat{a}$.
\end{proof}

\section{Summary}

In this chapter we were concerned with interval behaviour and the
left-to-right arrangement of intervals. The goal of this was to find a
subset of the transition rules that harmonise with the definition of
reaction systems in the previous chapter.

We have seen that interval-preserving transition rules are such a
subset. Interval preservation is a useful property because intervals
already play an important role in cellular automata: The transition
rule is expressed in terms of intervals. Intervals are conceptually
simple cellular processes: Their domain can be expressed with three
numbers, and their content can be expressed in a natural way as a
sequence of cell states. They are therefore easy to express with
situations. Interval \emph{preservation} then also puts a limit on the
complicatedness of the closure of an interval. Every time slice of it
is an interval, so it does not become more complicated over time. The
complexity of the behaviour of an interval-preserving cellular
automaton is therefore confined to the interior of this closure.

A specific result of this chapter was the usefulness of \emph{separating
  intervals}. They form the boundaries between different regions in a
cellular automaton that do not influence each other in the next time
step. This makes them useful for the selective evolution of an initial
configuration of cells that we define in the next chapter.

We have learned how to express intervals and their interactions both
in terms of processes and of situations. We have seen that it is
possible to test for interval preservation of a transition rule in a
finite number of steps. It is also possible to find the minimal
separating intervals of a transition rule in a finite number of steps.
As a technical tool to express the properties of separating intervals we
have introduced characteristic reactions. They express which events
are determined by a separating interval and where the zones of influence
are for the cells at the left and the right of the separating interval.


\chapter{A Local Reaction System}
\label{cha:local-reactions}

Although they do characterise interval-preserving transition rules,
characteristic reactions cannot be used as generators of a reaction
system without unpleasant side effects.

I mean the following: Let $a_1 \rea [1, i_1] \hat a_1 [-1, j_1]$
and $a_2 \rea [1, i_2] \hat a_2 [-1, j_2]$ be the characteristic
reactions of separating intervals. If we apply them in sequence to the
situation $a_1 a_2$, then we get
\begin{equation}
  \label{eq:characteristic-twice}
  \begin{aligned}[b]
    a_1 a_2
    &\rea [1, i_1] \hat a_1 [-1, j_1] a_2 \\
    &\rea [1, i_1] \hat a_1 [-1, j_1]
    [1, i_2] \hat a_2 [-1, j_2] \\
    &= [1, i_1] \hat a_1 [j_1 + i_2] \hat a_2 [-1, j_2]\,.
  \end{aligned}
\end{equation}
So, unless $j_1 + i_2 \leq 0$, which is usually not the case, the
result of this reaction is not an interval but contains a gap. If we
had used the characteristic reaction of $a_1 a_2$ directly, we would
have encountered no gap:
\begin{equation}
  a_1 a_2
  \rea  [1, i_1] \widehat{a_1 a_2} [-1, j_2]\,.
\end{equation}
(In this reaction the coefficients $i_1$ and $j_2$ occur because of
Theorem~\ref{thm:characteristic-reaction}.)

No characteristic reaction could have recovered the missing piece,
even when starting from $[1, i_1] \hat a_1 [-1, j_1] a_2$: if we had
applied a characteristic reaction to $\hat a_1$ or parts of it, it
would only yield events at time step $2$.

But if we had preserved in the first step
of~\eqref{eq:characteristic-twice} some events at the right end of
$a_1$, then the gap in the interval could have been avoided. We will
now define a new kind of reaction that accomplishes this.

\section{Reactions for Separating Intervals}

The parts of a separating interval that must be preserved in a reaction
are the minimal separating intervals of
Definition~\ref{def:minimal-separating}. We now introduce a notation for
the situations that correspond to them.
\begin{definition}[Leftmost and Rightmost Intervals]
  \label{def:boundary-interval}
  Let $a \in \mathcal{S}$ be a separating interval.

  Let $a_L$ be the shortest separating interval for which $a_L
  \leftend a$.

  Let $a_R$ be the shortest separating interval for which $a \rightend
  a_R$.

  Then $a_L$ and $a_R$ are the \indextwoshort{{leftmost minimal} separating
    interval}{leftmost} and \indextwoshort{{rightmost minimal} separating
    interval}\emph{rightmost minimal separating intervals} of~$a$.
\end{definition}

If an interval $a$ is minimally separating, then $a_L = a_R = a$. On the
other hand, if $a$ is separating, then $(a_L)_R$ and $(a_R)_L$ are
minimal separating intervals, but not necessarily identical. Other
properties of $a_L$ and $a_R$ are the subject of the following lemma.

\begin{lemma}[Extending a Separating Interval]
  \label{thm:separating-extensions}
  Let $a$, $x \in \Sigma^*$ be interval situations, of which $a$ is
  separating. Then
  \begin{align}
    (a x)_L &= a_L, & (a x)_R &= (a_R x)_R, \\
    (x a)_R &= a_R, & (x a)_L &= (x a_L)_L\,.
  \end{align}
\end{lemma}
\begin{proof}
  Only the equations of the first line need to be proved. We get $(a
  x)_L$ by removing events from the right part of $a x$ as long as the
  result is still separating. Since $a$ is separating, we can remove $x$
  completely. This proves $(a x)_L = a_L$.

  The rightmost separating interval in $a x$ of which we know must is
  $a_R x$ because we do not know whether $x$ is separating. Therefore
  the rightmost separating part of $a x$ must be the rightmost separating
  part of $a_R x$. This proves $(a x)_R = (a_R x)_R$.
\end{proof}

The second piece of the definition is a notation for a specific kind
of displacement terms.
\begin{definition}[Slope Operators]
  \label{def:slope-operators}
  \index{slope operator}Let $a \in \Sigma^*$ be a separating interval
  with characteristic reaction $a \rea [1, i] \hat a [-1, j]$. Then
  the \indextwo{left slope operator}\emph{left} and \indextwo{right
    slope operator}\emph{right slope operators} of $a$ are
  \begin{equation}
    {+_a}  = [1,  i -\abs{a_L}]
    \qqtext{and}
    {-_a} = [-1, j -\abs{a_R}]\,.
  \end{equation}
\end{definition}
The slope operators are defined in this way because then we have
\begin{equation}
  \label{eq:slope-operator-2}
  \delta(a) = \delta(a_L +_a a -_a a_R)
  \qqtext{and}
  \Delta \pr(a) = \prx{a +_a}{\hat a}\,.
\end{equation}
Taken together these equations imply that $a \rea a_L +_a a -_a a_R$
is a reaction for $\phi$. The two conditions can be verified with the
help of Definitions~\ref{def:characteristic}
and~\ref{def:characteristic-separating}.

The slope operators are no addition operators, and $+_a$ is not the
inverse of $-_a$. Their symbols have been chosen to stay in harmony
with the operators $\oplus_k$ and $\ominus_k$ that were already
introduced in \cite{Redeker2010}; here they will reappear in
Definition~\ref{def:concrete-boundaries}. Other notations that have no
relations to addition, like $\uparrow_a$ and $\downarrow_a$, were
considered but rejected. In the case of an arrow notation, the main
reason was that there is no complete agreement whether the future or
the past is ``up'' (see p.~\pageref{fn:time-direction}) and a notation
that is agnostic in this aspect is therefore preferable. Another
point is that we will need to distinguish between two kinds of slope
operators; the existence of both encircled and not encircled plus and
minus symbols in \LaTeX\ is therefore another reason to use them as
the notation for slopes.

\paragraph{A System of Interval Reactions} Now we can define the new
reaction system. It will later become a part of the ``local reaction
system'' of Definition~\ref{def:slow-system}, but is conceptually a
bit simpler. It is however complex enough to show essential features
and motivate the extensions.

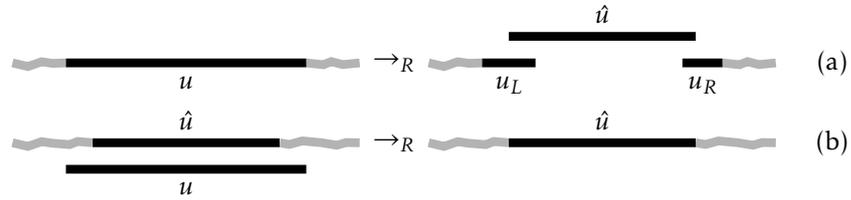
\begin{figure}[ht]
  \centering
  \begin{tikzpicture}[
    direct/.style={line width=.75ex},
    other/.style={direct, draw=shaded, line width=.35ex},
    somewhere/.style={direct, draw=shaded, decorate},
    decoration={random steps, amplitude=1.5, segment length=5}]
    \def\background(#1){
      \pgfmathsetseed{999}
      \path (#1)   coordinate (u)  +(2, 0)    coordinate (uL')
          ++(9,0) coordinate (u')  +(-1.5, 0) coordinate (uR);
      \path (u)  +(1, 1)  coordinate (uhat);
      \path (u') +(-1, 1) coordinate (uhat');
        }
    \matrix[cells=midway] {
      \background(2,0)
      \draw[somewhere] (0, 0) -- (u)  (u') -- +(2, 0);
      \draw[direct] (u) -- node[below] {$u$} (u');
      & \node{$\rea_R$}; &
      \background(2,0)
      \draw[somewhere] (0, 0) -- (u)  (u') -- +(2, 0);
      \draw[direct] (u) -- node[below] {$u_L$} (uL')
          (uR) -- node[below] {$u_R$} (u');
      \draw[direct] (uhat) -- node[above] {$\hat u$} (uhat');
      & \node {\quad(a)};
      \\
      \background(2,-1)
      \draw[somewhere] (0, 0) -- (uhat)  (uhat') -- +(3, 0);
      \draw[direct] (uhat) -- node[above] {$\hat u$} (uhat')
          (u) -- node[below] {$u$} (u');
      & \node{$\rea_R$}; &
      \background(2,-1)
      \draw[somewhere] (0, 0) -- (uhat)  (uhat') -- +(3, 0);
      \draw[direct] (uhat) -- node[above] {$\hat u$} (uhat');
      & \node {\quad(b)};
      \\
      };
  \end{tikzpicture}
  \caption{A system of interval reactions.}
  \label{fig:interval-system}
\end{figure}

For this system let $\phi$ be an interval-preserving transition rule
for $\Sigma$. Let $R$ be the reaction system that is generated by
$\Sigma^*$ and the following reactions, where $u \in \Sigma^*$ may be
any separating interval,
\begin{subequations}
  \label{eq:interval-system}
  \begin{alignat}{2}
    \label{eq:interval-generator}
    u &\rea_R u_L +_u \hat u -_u u_R, \\
    \label{eq:interval-destructor}
    \hat u -_u u +_u \hat u &\rea_R \hat u
    &&\quad\text{if $u$ is minimally separating.}
  \end{alignat}
\end{subequations}
The diagrams for these reactions can be seen in
Figure~\ref{fig:interval-system}. The reactions are shown as parts of
a larger situation, which is displayed in grey.
In the first diagram, which shows
reaction~\eqref{eq:interval-generator}, the interval $u$ is replaced
by the interval $\hat u$ that is determined by it, but the left and
right ends of $u$ are kept for the use in later reactions. The second
diagram shows reaction~\eqref{eq:interval-destructor}. Its left side
is somewhat difficult to display: the situation $\hat u -_u u +_u \hat
u$ begins with $\hat u$, followed by $u$, and then, because
$\delta(\hat u -_u u +_u) = (0, 0)$, the same interval $\hat u$ occurs
again. The reaction then eliminates the $u$ interval.

To prove that $\delta(\hat u -_u u +_u {}) = (0, 0)$, we only have to
notice that $u$ is a minimal separating interval and that therefore $u_L
= u_R = u$. Then the first condition in~\eqref{eq:slope-operator-2}
becomes $\delta(u +_u \hat u -_u u) = \delta(u)$. From this follows
$\delta(u +_u \hat u -_u {}) = (0, 0)$, which is equivalent to the
assertion.

\paragraph{How it works} With this reaction system we can avoid the
problems we had with characteristic reactions. To see how this works,
let $b$ be a minimal separating interval. Instead of $a_1$ and $a_2$ as
before, we now consider the separating intervals $a_1 b$ and $b a_2$,
which overlap in $b$. Because of this we have $(a_1 b)_L = (b a_2)_R =
b$ by Lemma~\ref{thm:separating-extensions}. We can then apply a
reaction of the form~\eqref{eq:interval-generator} on $a_1 b$; the
result is
\begin{equation}
  \label{eq:intervals-twice}
  \begin{aligned}[b]
    a_1 b
    &\rea_R (a_1 b)_L +_{a_1 b} \widehat{a_1 b} -_{a_1
      b} (a_1 b)_R \\
    &= (a_1 b)_L +_{a_1} \widehat{a_1 b} -_b b\,.
  \end{aligned}
\end{equation}
In the same way $b a_2$ reacts to $b +_b \widehat{b a_2} -_{a_2} (b
a_2)_R$.

We now evolve the left part of the interval $a_1 b a_2$ first, as we
did in~\eqref{eq:characteristic-twice}. In the following computation
the parts of the formulas that change in the next step are underlined,
to make it more readable. Then we get,
\begin{equation}
  \begin{aligned}[b]
    \underline{a_1 b} a_2
    & \rea_R (a_1 b)_L +_{a_1} \widehat{a_1 b} -_b
    \underline{b a_2} \\
    & \rea_R (a_1 b)_L +_{a_1} \underline{\widehat{a_1 b}} -_b b
    +_b \underline{\widehat{b a_2}} -_{a_2} (b a_2)_R \\
    & \sim (a_1 b)_L +_{a_1} \widehat{a_1 b} \ovl{\hat b}
    \underline{\hat b -_b b +_b \hat b} \ovl{\hat b}
    \widehat{b a_2} -_{a_2} (b a_2)_R \\
    & \rea_R (a_1 b)_L +_b
    \underline{\widehat{a_1 b} \ovl{\hat b} \hat b} \ovl{\hat b}
    \widehat{b a_2} -_{a_2} (b a_2)_R \\
    & \sim (a_1 b)_L +_b \widehat{a_1 b} \ovl{\hat b}
    \widehat{b a_2} -_{a_2} (b a_2)_R\,.
  \end{aligned}
\end{equation}
We can then see that $\widehat{a_1 b} \ovl{\hat b} \widehat{b a_2}$,
the part of the reaction result that belongs to time step 1, is now an
interval.

If instead we apply rule~\eqref{eq:interval-generator} directly to
$a_1 b a_2$, then we get the reaction $a_1 b a_2 \rea_R (a_1 b)_L +_b
\widehat{a_1 b a_2} -_{a_2} (b a_2)_R$. By comparing its result with
the result of the previous computation we see also that $\widehat{a_1
  b} \ovl{\hat b} \widehat{b a_2} \sim \widehat{a_1 b a_2}$.

\section{Well-Behaved Transition Rules}

There were two ideas in the previous section that motivated the jump
from characteristic reactions to the reaction
system~\eqref{eq:interval-system}: It should be possible to reach all
elements of the closure of a situation with reactions, and one should
be able to do it by applying reactions to this situation in any order.
We will make these vague concepts later precise as \emph{covering
  property} and \emph{confluence} and prove them at the end if this
chapter. The proofs however are valid only for a subclass of the
interval-preserving transition rules.

This class of \emph{well-behaved} transition rules, which is defined
next, is introduced mainly for convenience. It was found by trial and
error, trying to exclude special cases that would make proofs and
concepts too complex, while keeping the theory applicable for Rules 54
and 110.

\begin{definition}[Well-Behaved]
  \label{def:well-behaved}
  A transition rule $\phi$ on $\Sigma$ is \intro{well-behaved} if
  \begin{enumerate}
  \item $\phi$ is interval-preserving,
  \item if $\pi \in \mathcal{P}$ is a non-separating interval, then
    $\Delta \pi = \emptyset$,
  \item if $\pi \in \mathcal{P}$ is a minimal separating interval, then
    $\Delta \pi$ is either a minimal separating interval or
    non-separating, and
  \item the empty interval is not separating.
  \end{enumerate}
\end{definition}

Condition 2 in this definition is necessary for the proof of
Lemma~\ref{thm:covering-the-closure}. It ensures that all reactions
that start from intervals and compute new events, i.\,e.\ those of the
form~\eqref{eq:interval-generator}, do indeed start from separating
intervals. We do not need to consider very short intervals as special
cases. Condition 2 is a completeness property for reaction
system~\eqref{eq:interval-system} and for the systems that will be
later derived from it.

The condition is always true for elementary cellular automata, but it
can become false for radii greater than $1$. We will now construct a
counterexample for $r = 2$ and $\Sigma = \{0, 1, 2\}$. Its transition
rule is
\begin{equation}
  \label{eq:counterexample-2}
  \phi(\sigma_{-2}, \sigma_{-1}, \sigma_0, \sigma_1, \sigma_2) = 
  \left\{
    \begin{array}{l@{\quad}l}
      0 & \text{if $\sigma_0 = 0$,} \\
      \max\{\sigma_{-2}, \dots, \sigma_2\}
      & \text{otherwise.}
    \end{array}
  \right.
\end{equation}
Then the interval $\pi = \pr(0)$ provides a contradiction. It is not
separating, since the state of the cell at location $-1$ can influence
the next state of the cell at location $1$ and \emph{vice versa}, but
we also have $\Delta \pi = \pr([1, 0] 0)$. It is this crossover
influence between cells that Condition 2 prevents.

Condition 3 is necessary in the context of achronal situations
(Definition~\ref{def:achronal} below). It concerns situations of the
form $a +_a \hat a$ or $\hat b -_b b$, with $a$ and $b$ minimal
separating intervals. These situations arise frequently in reaction
system~\eqref{eq:interval-system} and other systems that have a
reaction $u \rea u_L +_u \hat u -_u u_R$. In the result of this
reaction, the interval $\hat u$ has $\widehat{u_L}$ as its left end
and $\widehat{u_R}$ as its right end; so with $a = u_L$ and $b = u_R$
we can say that $u_L +_u \hat u -_u u_R$ begins with $a +_a \hat a$
and ends with $\hat b -_b b$. The condition then ensures that a
reaction of the same type as before, when applied to $\hat u$, does
not destroy $\hat a$ and $\hat b$. This is because such a reaction,
when applied to an interval, leaves its left and right minimal
separating intervals intact: The interval $\hat a$ is by condition~3 not
longer than a minimal separating interval and is therefore part of $\hat
u_L$, and $\hat b$ is for the same reason a part of $\hat u_R$.

Here a counterexample occurs with an elementary cellular automaton,
Rule~1. We have found in Table~\ref{tab:separating} that the interval
$1$ is minimally separating for this rule, but $\Delta \pr(1)$ is the
interval $\pr([1, -1] 000)$, as we can see in
Table~\ref{tab:interval-preserving-1}. We have thus an interval $\pi$
for which $\Delta \pi$ consists of three events; such intervals are
never minimally separating for elementary cellular automata. The
purpose of Condition 3 is to exclude rules with separating intervals
that have such excessive influence.

Condition 4 is an intuitively obvious requirement on separating
intervals, but it is violated by transition rules of radius $0$. These
are rules in which the state of a cell does only depend on the state
of a single cell at the previous time step. Excluding them from
consideration therefore is no loss.

\section{Achronal Situations}

The reaction products in the system~\eqref{eq:interval-system} have a
specific form, a generalisation of intervals, for which we will now
give a definition. The set is called ``achronal'' because these
situations, like the intervals, consist of events that belong almost
to the same time. We think of the events in them as arranged from left
to right, not in a temporal sequence.

Achronal situation also have in common with intervals that every
achronal situation has a closure and can therefore be the starting
point of a reaction. This will be proved later, in
Theorem~\ref{thm:achronal-closure}.

\begin{definition}[Achronal Situations]
  \label{def:achronal}
  The set of \indextwoshort{achronal situation}\emph{achronal situations}
  for an interval-preserving transition rule $\phi$ is the set
  $\mathcal{A}_\phi \subset \mathcal{S}$.

  It is defined recursively in the following way: A situation $s \in
  \mathcal{S}$ is an element of $\mathcal{A}_\phi$ if and only if
  \begin{enumerate}
  \item $s \in \Sigma^*$, or
  \item $s = y b +_b \hat b x$, with $y b$, $\hat b x \in
    \mathcal{A}_\phi$ and $b \in \Sigma^*$ minimally separating,
    or
  \item $s = x \hat b -_b b y$, with $x \hat b$, $b y \in
    \mathcal{A}_\phi$ and $b \in \Sigma^*$ minimally separating.
  \end{enumerate}
\end{definition}

We will also use two subsets of $\mathcal{A}_\phi$. The set
$\mathcal{A}_{\phi+}$ consists of those elements of $\mathcal{A}_\phi$
that are constructed only with the $+$ operators, and the set
$\mathcal{A}_{\phi-}$ consists of those elements of $\mathcal{A}_\phi$
that are constructed only with the $-$ operators. These sets are
called the \indextwo{positive slope}\emph{positive} and
\indextwo{negative slope}\emph{negative slopes}.

Similarly, the terms $\hat b -_b b$ and $b +_b \hat b$ in
Definition~\ref{def:achronal} are called \indextwo{generating
  slope}\emph{generating slopes}. 

\paragraph{Use of Slopes} The positive and negative slopes provide a
notation with which we can name the different parts of a situation.
Later, in Lemma~\ref{thm:slope-decomposition}, we will see that every
situation can react into a situation that is the product of a positive
and negative slope.
\begin{figure}[ht]
  \centering
  \tabcolsep1em
  \def\arraystretch{1.5}
  \begin{tabular}{cc}
    \ci{triangle} & \ci{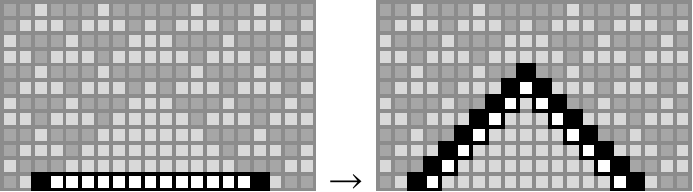} \\
    (a) Process &
    (b) Reaction
  \end{tabular}
  \caption{A triangle and its slopes under Rule 54.}
  \label{fig:slopes}
\end{figure}
Figure~\ref{fig:slopes} is an example. Here the left triangle of
Figure~\ref{fig:triangle-computation} is expressed as a reaction. It
begins with an interval situation and ends with a situation that
consists of a positive and a negative slope. (This example will be
continued with Figure~\ref{fig:triangles}.)

The generating slopes are important because the reactions that
transform positive generating slopes into positive, or negative
generating slopes into negative generating slopes, are among the
building blocks for the reaction system associated to a transition
rule, which is described in Definition~\ref{def:slow-system}.

\paragraph{Induction} If we view the recursive construction of the
achronal situation as a sequential process, then the intervals are
created at its beginning, and every other achronal situation $s$ has
either a decomposition $s = y b +_b \hat b x$ with $y b$ and $\hat b
x$ constructed earlier, or a decomposition $s = x \hat b -_b b y$ with
$x \hat b$ and $b y$ constructed earlier. We have therefore an
induction principle for achronal situations:
\begin{lemma}[Achronal Induction]
  \label{thm:achronal-induction}
  Let $\phi$ be an interval-preserving transition rule.
  Let $S \subseteq \mathcal{A}_\phi$ be a set of situations where
  \begin{enumerate}
  \item $\Sigma^* \subseteq S$,
  \item if $y b$, $\hat b x \in S$, where $b$ is a minimal separating
    interval, then $y b +_b \hat b x \in S$, and
  \item if $x \hat b$, $b y \in S$, where $b$ is a minimal separating
    interval, then $x \hat b -_b b y \in S$.
  \end{enumerate}
  Then $S = \mathcal{A}_\phi$. \qed
\end{lemma}

This induction principle uses the operators $+_b$ and $-_b$.
Situations are however defined in terms of displacements, not in terms
of slope operators. Therefore it is not yet clear whether we can, when
given an achronal situation, reconstruct the slope operators with
which it was constructed. The following lemma shows that the answer is
``yes, but it is not completely trivial''.

\begin{lemma}[Slope Operators]
  \label{thm:slope-operators}
  Let $\phi$ be a well-behaved transition rule and let $s \in
  \mathcal{A}_\phi$ be an achronal situation with a honest
  decomposition $s = x[p]y$ in which $p \neq (0, 0)$. Then there is
  either
  \begin{enumerate}
  \item $[p] = +_b$ with $x \rightend b$, where $b$ is a minimal
    separating interval, or
  \item $[p] = -_b$ with $b \leftend y$, where $b$ is a minimal
    separating interval, or
  \item $[p] = +_{b_1} -_{b_2}$ with $x \rightend b_1$ and $b_2
    \leftend y$, where $b_1$ and $b_2$ are minimal separating
    intervals and $\hat b_1 = \hat b_2 = [0]$.
  \end{enumerate}
\end{lemma}
\begin{proof}
  We see from the definition of $\mathcal{A}_\phi$ that $s$ can be
  written as a sequence of cell states and slope operators. In this
  proof we will call this sequence the \emph{symbol sequence} for $s$.

  In the symbol sequence for $s$, every one of the symbols either
  contributes to $x$, to $[p]$ or to $y$. The symbols that contribute
  to $[p]$ can only be slope operators. They form a subsequence of
  maximal length in the symbol sequence; it is maximal because the
  decomposition is honest.

  If an operator $+_b$ contributes to $[p]$, then the interval $b$
  must appear at the left of it in the symbol sequence. Because $\phi$
  is well-behaved, $b$ is never empty. Therefore $+_b$ can only appear
  at the left end of the sequence of slope operators that
  contribute to $[p]$. At its right side it must be followed by $\hat
  b$, but only if $\hat b \neq [0]$; in that case $+_b$ is the only
  factor of $[p]$. For the same reason $-_b$ can only appear at the
  right end of $[p]$; and if $\hat b \neq [0]$, then $-_b$ is the only
  factor of $[p]$.

  Therefore $[p]$ is a product of at most two slope operators in a
  prescribed order. Since $p \neq (0, 0)$, at least one of them must
  appear. This leads to the three cases of the lemma.
\end{proof}
It is clear that all of these three cases can occur. They can easily
be distinguished: we have either $\delta(p)_T = +1$, $-1$ or $0$.
Therefore ``the number of slope operators''\footnote{More exactly,
  this number is the minimal number of slope operators with which a
  situation can be written. Ambiguous cases are possible: if $+_{b_1}
  = [1, 0]$ and $-_{b_2} = [-1, 0]$, then $b_1 b_2 = b_1 +_{b_1}
  -_{b_2} b_2$, and this is in fact an equality, not just an
  equivalence.} in a situation is a well-defined concept, and
induction over this number is possible. It will be the most common
form of induction used in this text.

\paragraph{Achronal Situation Occur Naturally} In the proof of
Lemma~\ref{thm:slope-operators} we have seen that every achronal
situation can be written as a sequence of elements of $\Sigma$
together with slope operators: Every $-_b$ must be surrounded by $\hat
b$ at the left and $b$ at the right and every $+_b$ must be surrounded
by $b$ at the left and $\hat b$ at the right. Whether a situation is
achronal therefore depends only on the terms next to the slope
operators. This means that if the two situations $s_1 x$ and $x s_2$
are achronal, their ``overlapping product'' $s_1 x s_2$ is also
achronal.

The converse is not always true, but at least when the common part of
the two situations is a separating interval:
\begin{lemma}[Splitting at Separating Intervals]
  \label{thm:achronal-decomposition}
  Let $\phi$ be a well-behaved transition rule for $\Sigma$. Let
  $s_1$, $b$, $s_2 \in \mathcal{S}$, where $b$ is a separating
  interval for $\phi$.

  Then if $s_1 b s_2 \in \mathcal{A}_\phi$, then $s_1 b \in
  \mathcal{A}_\phi$ and $b s_2 \in \mathcal{A}_\phi$.
\end{lemma}

\begin{proof}
  Let $s = s_1 b s_2$. We perform an induction over the number of
  slope operators in $s$.

  If $s \in \Sigma^*$, the lemma is obviously true. Otherwise $s$
  has at least one slope operator, either $+_a$ or $-_a$.

  Assume now that $s = x \hat a -_a a y$ is a decomposition of $s$
  with $x \hat a$, $a y \in \mathcal{A}_\phi$, where $a$ is minimally
  separating. Since $b$ is a nonempty interval, it must be part of
  either $x \hat a$ or of $a y$.

  If $b$ is part of $ x \hat a$, then there is a situation $s'_2$ such
  that $x \hat a = s_1 b s'_2$. Since $s_1 b s'_2$ has fewer slope
  operators than $s$, the induction hypothesis can be applied to it,
  and therefore $s_1 b \in \mathcal{A}_\phi$ and $b s'_2 \in
  \mathcal{A}_\phi$. Because $\phi$ is well-behaved, the interval
  $\hat a$ is not longer than a separating interval; therefore $b s'_2,$
  which contains the separating $b$ interval as its factor, cannot be
  just a part of $\hat a$. So there must be a situation $z \in
  \mathcal{S}$ such that $b s'_2 = z \hat a$. Then $b s'_2 -_a
  a y = z \hat a -_a a y$ and therefore $b s'_2 -_a a y \in
  \mathcal{A}_\phi$.

  If $b$ is part of $a y$, then there is a situation $s'_1$ such that
  $a y = s'_1 b s_2$. Then $s'_1 b \in\mathcal{A}_\phi$ and $b s_2 \in
  \mathcal{A}_\phi$ by induction. When dividing up $a y$, the
  situation $a$ must become a part of $s'_1 b$ because it is the
  leftmost minimal separating interval of this situation and $s'_1 b$
  contains already the separating interval $b$: the interval $a$ could
  not have been cut into pieces. Therefore $x \hat a -_a s'_1 b \in
  \mathcal{A}_\phi$.

  So if $s = x \hat a -_a a y$, then we could divide $s$ either into
  $s_1 b$ and $b s'_2 -_a a y$ or into $x \hat a -_a s'_1 b$ and $b
  s_2$. If $s = y a +_a \hat a x$, then there are similar
  decompositions for it; they can be found by a mirror image of this
  argument.
\end{proof}

With the methods developed so far we can now show that achronal
situations occur naturally in the reaction
system~\eqref{eq:interval-system}.
\begin{lemma}[Achronal Domain]
  Let $R$ be the interval reaction system
  of~\eqref{eq:interval-system}. Then $\dom R \subseteq
  \mathcal{A}_\phi$.
\end{lemma}
\begin{proof}
  We will show that the generating reactions~\eqref{eq:interval-system}
  transform achronal situations into achronal situations. As the
  initial situations of $R$ are intervals and therefore obviously
  achronal, this will prove that all situations in $\dom R$ are
  achronal.

  Let $xuy \in \mathcal{A}_\phi$, where $u$ is a separating interval.
  Then $x u_L$, $u$ and $u_R y$ are achronal by
  Lemma~\ref{thm:achronal-decomposition}. One can see directly that
  $u_L +_u \hat u -_u u_R$ is achronal. Therefore $x u_L +_u \hat u
  -_u u_R y \in \mathcal{A}_\phi$, again by
  Lemma~\ref{thm:achronal-decomposition}. This proves that the
  reaction $u \rea_R u_L +_u \hat u -_u u_R$ preserves achronality.

  Let now $x \hat u -_u u +_u \hat u y \in \mathcal{A}_\phi$, where
  $u$ is a minimal separating interval. Then $x \hat u$ and $\hat u y$
  are achronal by by Lemma~\ref{thm:achronal-decomposition}. Therefore
  we have $x \hat u y \in \mathcal{A}_\phi$. This proves that the
  reaction $\hat u -_u u +_u \hat u \rea_R \hat u$ preserves
  achronality.
\end{proof}

\section{Closure}

A preference for symmetry now leads to another question: If all
reactions results are achronal situations, can we then also extend the
set of input situations of the reaction system
in~\eqref{eq:interval-system} from $\Sigma^*$ to $\mathcal{A}_\phi$?
The following theorem shows that this is possible for a subset of the
achronal situations.

\begin{figure}[ht]
  \centering
  \begin{tikzpicture}[
    direct/.style={line width=.5ex},
    somewhere/.style={direct, draw=shaded, decorate},
    decoration={random steps, amplitude=1.5, segment length=5}]
    \draw[somewhere] (0,0)
        -- node [above] {$s_1$} ++(4,-1) coordinate (a);
    \draw[direct] (a)
        -- node [below] {$u_1$} ++(3, 0) coordinate (b);
    \draw[somewhere] (b)
        -- ++(2, -1)
        -- node [below] {$s_2$} ++(3,0)
        -- ++(2, 1) coordinate (c);
    \draw[direct] (c)
        -- node [below] {$u_2$} ++(2, 0) coordinate (d);
    \draw[somewhere] (d)
        -- node [above] {$s_3$} ++(3,1);
    \draw[dashed] ($(a)-(2,0)$)
        -- ($(d)+(4,0)$) node [right] {$t$};
  \end{tikzpicture}
  \caption{The situation $s$ in \eqref{eq:order-decomposition}. The
    processes belonging to $u_1$ and $u_2$ occur at the same time.}
  \label{fig:achronal-closure}
\end{figure}
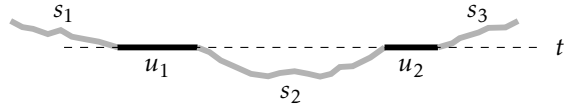

For this we have to introduce a new concept. It represents the
intuitive notion that the events of a situation are arranged
approximately from left to right. To express the concept for a
situation $s$ we consider the honest decompositions of $s$ of the form
\begin{equation}
  \label{eq:order-decomposition}
  s = s_1 [p_1] u_1 [q_1] s_2 [p_2] u_2 [q_2] s_3,
\end{equation}
in which $u_1$ and $u_2$ are intervals and $p_1$, $q_1$, $p_2$, $q_2
\neq (0, 0)$. We then write $\pi_1 = \prx{s_1[p_1]}{u_1}$ and $\pi_2 =
\prx{s_1[p_1]u_1[q_1]s_2[p_2]}{u_2}$ for the processes that belong to
$u_1$ and $u_2$ (Figure~\ref{fig:achronal-closure}).
Now consider the decompositions of the
form~\eqref{eq:order-decomposition} the processes $\pi_1$ and $\pi_2$
belong to the same time. If for every decomposition of $s$ of this
form we have $\pi_1 \seq \pi_2$, then the situation $s$ is
\indextwoshort{ordered situation}\emph{ordered}.

\begin{theorem}[Closure of Achronal Situations]
  \label{thm:achronal-closure}
  Let $\phi$ be a well-behaved transition rule for $\Sigma$ and $s \in
  \mathcal{A}_\phi$ be ordered. Then $\cl \pr(s)$
  exists.\footnote{It is this theorem for which we need the fact that
    achronal situations are ordered: A counterexample in the reaction
    system for Rule 54 is the situation $000 \oplus 01 \oplus_2 1
    \ominus_2 10$, written in the notation~\eqref{eq:osymbols}. It is
    not ordered but achronal, and it has no closure.}
\end{theorem}

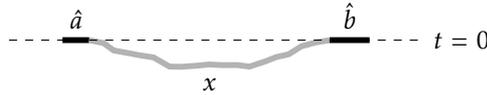
\begin{figure}[ht]
  \centering
  \begin{tikzpicture}[
    direct/.style={line width=.5ex},
    somewhere/.style={direct, draw=shaded, decorate},
    decoration={random steps, amplitude=1.5, segment length=5}]
    \draw[direct] (0,0)
        -- node [above] {$\hat a$} ++(1, 0) coordinate (b);
    \draw[somewhere] (b)
        -- ++(3, -1)
        -- node [below] {$x$} ++(3,0)
        -- ++(3, 1) coordinate (c);
    \draw[direct] (c)
        -- node [above] {$\hat b$} ++(1.5, 0) coordinate (end);
    \draw[dashed] (-2,0)
        -- ($(end)+(2,0)$) node [right] {$t = 0$};
  \end{tikzpicture}
  \caption{A balanced situation.}
  \label{fig:balanced}
\end{figure}

It is enough if we restrict the proof of the theorem to the case where
$s$ is a \emph{balanced} situation. This shall mean that $\delta(s)_T
= 0$, that $\pr(s)^{(t)} = \emptyset$ for all $t \geq 1$ and that $s$
is either an interval or there is a decomposition $s = \hat a -_a x
+_b \hat b$ with minimal separating intervals $a$ and $b$. (See
Figure~\ref{fig:balanced}. Note that $\hat a$ or $\hat b$ may be
empty.)
Every situation can be extended to a balanced situation; if that
situation has a closure, then the original situation has a closure
too.

First we will however prove the existence of the closure for the
simplest nontrivial balanced situations. This result is then used as a
stepping stone for the proof of Theorem~\ref{thm:achronal-closure}.

\begin{lemma}[Closure of Simple Balanced Situation]
  \label{thm:balanced-closure}
  Let $\phi$ be a well-behaved transition rule for $\Sigma$ and let $s
  = \hat a -_a u +_b \hat b$ be an achronal situation for $\Sigma$ in
  which $u$ is an interval and $a$, $b$ are minimal separating
  intervals.

  Then $s \rea \hat u$ is a reaction for $\phi$, and $\cl^{(0)}
  \pr(s) = \pr(\hat u)$.
\end{lemma}

\begin{figure}[ht]
  \centering
  \begin{tikzpicture}[
    direct/.style={line width=.75ex},
    somewhere/.style={direct, draw=shaded, decorate},
    decoration={random steps, amplitude=1.5, segment length=5}]
    \path ++(3,0) coordinate (ahat1) +(-1,-1) coordinate (u1)
          ++(2,0) coordinate (ahat2)
          ++(6,0) coordinate (bhat1)
          ++(3,0) coordinate (bhat2) +(1,-1) coordinate (u2)
          ++(3,0) coordinate (end0)
          ++(0,-1) coordinate (end-1);
    \draw[dashed] (0,0) -- (end0) node[right] {$t=0$}
                  (0,-1) -- (end-1) node[right] {$t=-1$};
    \draw[direct] (ahat1) -- node [above] {$\hat a$} (ahat2)
                  (bhat1) -- node [above] {$\hat b$} (bhat2)
                  (u1)    -- node [below] {$u$} (u2);
    \draw[direct, draw=shaded, line width=.35ex]
                  (ahat1) -- node [above] {$\hat u$} (bhat2);
  \end{tikzpicture}
  \caption{The situations in Lemma~\ref{thm:balanced-closure}.}
  \label{fig:balanced-closure}
\end{figure}
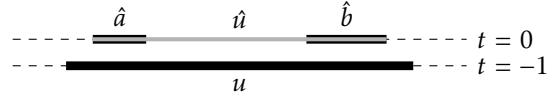
A diagram of the processes for $s$ and $\hat u$ can be seen in
Figure~\ref{fig:balanced-closure}. The process belonging to $\hat u$
overlaps with those of $\hat a$ and $\hat b$.

\begin{proof}
  To compute the closure of $\pr(s)$ we must express $s$ in the
  language of cellular processes. Let therefore $\pi = \pr(s)$; its
  components are then
  \begin{equation}
    \alpha' = \pr(\hat a),
    \qquad
    \mu = \prx{\hat a -_a}{u},
    \qquad
    \beta' = \prx{\hat a -_a u +_b}{b},
  \end{equation}
  such that $\pi = \alpha' \cup \mu \cup \beta'$. The process $\pi$
  then consists of the time slices $\pi^{(-1)} = \mu$ and $\pi^{(0)} =
  \alpha' \cup \beta'$. Since $s$ is an achronal situation, we must
  have $a \leftend u \rightend b$. We will therefore also need names
  for the end intervals of $\mu$. They are
  \begin{equation}
    \alpha = \prx{\hat a -_a}{a}
    \qqtext{and}
    \beta = \prx{\hat a -_a u \ovl{b}}{b}.
  \end{equation}
  Then we can say that $\mu$ begins with $\alpha$ and ends with
  $\beta$, such that we have $\alpha \subseteq_R \mu \supseteq_L
  \beta$ (Figure~\ref{fig:balanced-closure-proof}).
  \begin{figure}[ht]
    \centering
    \begin{tikzpicture}[
      direct/.style={line width=.75ex},
      somewhere/.style={direct, draw=shaded, decorate},
      decoration={random steps, amplitude=1.5, segment length=5}]
      \path ++(3,0) coordinate (ahat1)  +(-1,-1) coordinate (a1)
            ++(2,0) coordinate (ahat2)   +(1,-1) coordinate (a2)
            ++(6,0) coordinate (bhat1)  +(-1,-1) coordinate (b1)
            ++(3,0) coordinate (bhat2)   +(1,-1) coordinate (b2)
            ++(3,0) coordinate (end0)
            ++(0,-1) coordinate (end-1);
      \draw[dashed] (0,0) -- (end0) node[right] {$t=0$};
      \draw[dashed] (0,-1) -- (end-1) node[right] {$t=-1$};
      \draw[direct] (ahat1) -- node [above] {$\alpha'$} (ahat2)
                    (bhat1) -- node [above] {$\beta'$} (bhat2)
                    (a1) -- node [below] {$\alpha$} (a2)
                    (b1) -- node [below] {$\beta$} (b2);
      \draw[direct, draw=shaded, line width=.35ex]
                    (ahat1) -- node [above] {$\Delta \mu$} (bhat2)
                    (a1)    -- node [below] {$\mu$} (b2);
    \end{tikzpicture}
    \caption{The processes related to the situations in
      Figure~\ref{fig:balanced-closure}.}
    \label{fig:balanced-closure-proof}
  \end{figure}
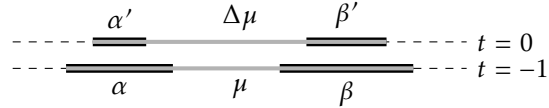

  The main part of the proof then consists of a computation of the
  space-time locations of all these processes. For this we let the
  characteristic reactions of $a$ and $b$ be
  \begin{equation}
    a \rea [1, i] \hat a [-1, j']
    \qqtext{and}
    b \rea [1, i'] \hat b [-1, j]\,.
  \end{equation}
  By Theorem~\ref{thm:characteristic-reaction}, we must then have $u
  \rea [1, i] \hat u [-1, j]$ as characteristic reaction for $u$. This
  is because $u$ begins with $a$, and therefore its characteristic
  reaction shares its left displacement term $[1, i]$ with that of
  $a$, and $u$ ends with $b$ and therefore its characteristic reaction
  shares its right displacement term $[-1, j]$ with that of $b$. This
  ``bounding'' of the location of $\hat u$ by $a$ and $b$ is the core
  of the proof. Because of the left-oriented structure of the
  formalism it however does not become directly visible in the
  following calculations.

  Before we start with the calculations proper, we will determine
  short expressions for the values of $\delta(\hat a -_a)$ and
  $\delta(\hat a -_a u)$, terms that will occur at several places. For
  the first term we begin with the equation $-_a = [-1, j' -
  \abs{a}]$, which follows from
  Definition~\ref{def:slope-operators}. Then we can see that
  $\delta(\hat a -_a) = (0, \abs{\hat a}) + (-1, j' - \abs{a}) = (-1,
  \abs{\hat a} + j' - \abs{a})$. We now use a relation derived from
  the characteristic reaction for $a$ to simplify that term. The left
  and the right side of a reaction must have the same size vector, and
  this means for the characteristic reaction for $a$ that $\abs{a} = i
  + \abs{\hat a} + j'$. Using that we see that $\delta(\hat a -_a) =
  (-1, -i)$. With this result we get an expression for the second size
  vector, $\delta(\hat a -_a u) = (-1, -i) + (0, \abs{u}) = (-1,
  \abs{u} - i)$. It too can be brought into a form that is more useful
  later, this time with the equation $\abs{u} = i + \abs{\hat u} + j$
  that is derived from the characteristic reaction of $u$. The result
  is $(\abs{\hat u} + j)$, such that we have
  \begin{equation}
    \label{eq:initial-offsets}
    \delta(\hat a -_a) = (-1, -i)
    \qqtext{and}
    \delta(\hat a -_a u) = (0, \abs{\hat u} + j)\,.
  \end{equation}

  Next we will show that $\alpha' = \Delta \alpha$ and $\beta' =
  \Delta \beta$. To find a term that expresses $\Delta
  \alpha$ in terms of situations, we use the characteristic reaction
  for $a$. We can read it as saying that the set of events determined
  by the process $\pr(a)$ is $\pr([1, i] \hat a)$. (See
  Definition~\ref{def:determined-situation}.) This is also valid for
  shifted versions of $\pr(a)$, so the set of events determined by
  $\alpha = \prx{\hat a -_a}{a}$ must be $\Delta \alpha =
  \prx{\hat a -_a}{[1, i] \hat a}$. The position of $\hat a$ in this
  term is the sum of two displacements, $\delta(\hat a -_a)$ and $(1,
  i)$. Since $\delta(\hat a -_a) + (1, i) = (0, 0)$, we have therefore
  \begin{equation}
    \label{eq:pos-alpha-prime}
    \Delta \alpha = \pr(\hat a)= \alpha'\,.
  \end{equation}
  For the same reason, this time with the characteristic reaction of
  $\hat b$, the set of events determined by $\beta = \prx{\hat a -_a u
    \ovl{b}}{\hat b}$ must be $\Delta \beta = \prx{\hat a -_a u
    \ovl{b}}{[1, i']\hat b}$. So we must compute $\delta(\hat a -_a u
  \ovl{b}) + (1, i')$ to find the position of $\hat b$ in this term:
  Then we get $\delta(\hat a -_a u \ovl{b}) + (1, i') = (-1, \abs{\hat
    u} + j) + (0, -\abs{b}) + (1, i') = (0, \abs{\hat u} + j - \abs{b}
  + i')$. We use the equation $\abs{b} = i' + \abs{\hat b} + j$, which
  is derived from the characteristic reaction for $b$, to simplify the
  result of this computation to $(0, \abs{\hat u} - \abs{\hat b})$.
  Therefore we have
  \begin{equation}
    \Delta \beta = \pr([0, \abs{\hat u} - \abs{\hat b}] \hat b)\,.
  \end{equation}
  To find the location of $\beta' = \prx{\hat a -_a u +_b}{\hat b}$ we
  use the fact that $+_b = [1, i' - \abs{b}]$. (See
  Definition~\ref{def:slope-operators}). Then we can calculate
  $\delta(\hat a -_a u +_b) = (-1, \abs{\hat u} + j) + (1, i' -
  \abs{b}) = (0, \abs{\hat u} + j + i' - \abs{b})$ and simplify the
  result in the same way as before to $(0, \abs{\hat u} - \abs{\hat
    b})$. Therefore we get
  \begin{equation}
    \label{eq:pos-beta-prime}
    \beta' = \pr([0, \abs{\hat u} - \abs{\hat b}] \hat b),
  \end{equation}
  which shows that
  $\Delta \beta = \beta'$.

  With this data we can compute the time slices of $\cl \pi$ and
  therefore show that $\pi$ actually has a closure. We have already
  seen that $\pi^{(-1)} = \mu$, $\pi^{(0)} = \alpha' \cup \beta'$, and
  that $\pi^{(t)} = \emptyset$ for all other values of $t$. Therefore,
  applying the definition~\eqref{eq:closure-recursion} of the closure
  we get
  \begin{subequations}
    \allowdisplaybreaks[3]
    \begin{alignat}{3}
      \cl^{(t)} \pi &= \pi^{(t)} &&= \emptyset
      &\qquad&\text{for $t < -1$,}\\
      \cl^{(-1)} \pi &= \pi^{(-1)} &&= \mu, \\
      \label{eq:closure-zero}
      \cl^{(0)} \pi
      &= \pi^{(0)} \cup \Delta \cl^{(-1)} \pi \notag \\
      &= (\alpha' \cup \beta') \cup \Delta \mu
      &&= \Delta \mu, \\
      \cl^{(t)} \pi
      &= \pi^{(t)} \cup \Delta \cl^{(t - 1)} \pi
      &&= \Delta \cl^{(t - 1)} \pi
      &&\text{for $t \geq 1$.}
    \end{alignat}
  \end{subequations}
  Only the third equation must be explained. It is true because
  $\alpha' = \Delta \alpha$ and $\beta' = \Delta \beta$. Since
  $\alpha$ and $\beta$ are subsets of $\mu$, we must then have
  $\alpha' \subseteq \Delta \mu \supseteq \beta'$ by the monotony of
  the $\Delta$ operator. This then proves that $\alpha' \cup \beta'$
  is compatible with $\Delta \mu$ and that $\cl^{(0)} \pi$ actually
  exists. Together these equations show that $\cl \pi$ exists.

  It remains to prove that $s \rea \hat u$ is a reaction for $\phi$.
  We have to show that $\delta(s) = \delta(\hat u)$ and that $\pr(\hat
  u) \subseteq \cl \pr(s)$. We know already that $\delta(\hat a -_a u
  +_b) = (0, \abs{\hat u} - \abs{\hat b})$; therefore $\delta(s) =
  \delta(\hat a -_a u +_b \hat b) = (0, \abs{\hat u}) = \delta(\hat
  u)$. To prove that $\pr(\hat u) \subseteq \cl \pr(s)$ we will now
  show that $\pr(\hat u) = \Delta \mu$. For this we use the fact that
  the set of events determined by $\mu = \prx{\hat a -_a}{u}$ is the
  process $\Delta \mu = \prx{\hat a -_a}{[-1, -i] u}$. Then, since
  $\delta(\hat a -_a) + (-1, -i) = (0, 0)$, we must have $\pr(\hat u)
  = \Delta \mu$. Now we can apply the result
  of~\eqref{eq:closure-zero} that $\Delta \mu = \cl^{(0)} \pi$ and see
  that $\pr(\hat u) \subseteq \cl \pr(s)$. This then concludes the
  proof that $s \rea \hat u$ is a reaction; it also proves that
  $\cl^{(0)} \pi = \pr(\hat u)$.
\end{proof}

\begin{proof}[Proof of Theorem~\ref{thm:achronal-closure}]
  As explained above, we restrict our case to balanced intervals.

  For a situation $a$, we will call the first time $t$ for which
  $\pr(a)^{(t)} \neq \emptyset$ the \emph{starting time} of $a$. A
  balanced situation has then a starting time $t \leq 0$, and the
  balanced situations with starting time $0$ are the intervals. Since
  intervals have a closure, it is therefore enough to show that if $t
  < 0$ and every balanced situation with starting time $t + 1$ has
  closure, then every situation with starting time $t$ has a closure
  too.

  We do this in the following way. A process $s$ with starting time
  $t_0$ has by definition the time slices of the closure
  $\cl^{(t)} \pr(s)$ for every $t \leq t_0$, with
  $\cl^{(t_0)} \pr(s) = \pr(s)^{(t_0)}$. We will then show that
  for every such $s$ exists another situation $s'$ with starting time
  $t_0 + 1$ such that $\cl^{(t_0 + 1)} \pr(s) = \pr(s')^{(t_0 +
    1)}$. The closure of $s'$ exists by induction, and we have, as
  before, $\cl^{(t_0 + 1)} \pr(s') = \pr(s')^{(t_0 + 1)}$.
  Therefore,
  \begin{equation}
    \cl \pr(s) = \bigcup_{t \leq t_0} \pr(s)^{(t)} \cup
    \bigcup_{t > t_0} \cl \pr(s')^{(t)},
  \end{equation}
  so the closure of $s$ exists then.

  Now we must isolate in $s$ the factors that contribute to
  $\pr(s)^{(t_0)}$. For this we will use the decomposition
  \begin{equation}
    \label{eq:balanced-decomposition}
    s = s_0 -_{a_1} u_1 +_{b_1} s_1 \dots
    s_{\ell-1} -_{a_\ell} u_\ell +_{b_\ell} s_\ell\,.
  \end{equation}
  \begin{figure}[ht]
    \centering
    \begin{tikzpicture}[
      direct/.style={line width=.5ex},
      somewhere/.style={direct, draw=shaded, decorate},
      decoration={random steps, amplitude=1.5, segment length=5}]
      \path (2,0) coordinate (s)
            ++(4, -3) coordinate (u1)  ++(1, 0)   coordinate (u1')
            ++(2, 0) coordinate (u2)   ++(1.5, 0) coordinate (u2')
            ++(3, 0) coordinate (u3)   ++(1, 0)   coordinate (u3')
            ++(5, 0) coordinate (uell) ++(2, 0)   coordinate (uell')
            ++(4, 3) coordinate (s')
            ++(2,0) coordinate (end);
      \draw[dashed]
           (0,0)  -- (end)    node[right] {$t=0$}
           let \p1 = (end) in
           (0,-3) -- (\x1,-3) node[right] {$t=t_0$};
      \draw[direct]
           (u1) -- node[below] {$u_1$} (u1')
           (u2) -- node[below] {$u_2$} (u2')
           (u3) -- node[below] {$u_3$} (u3')
           (uell) -- node[below] {$u_\ell$} (uell');
      \draw[somewhere]
           (s) to[out=-10] node[below] {$s_0$} (u1)
           (u1') to[out=40, in=140] node[above] {$s_1$} (u2)
           (u2') to[out=40, in=140] node[above] {$s_2$} (u3)
           (u3') to[out=40, in=140] node[pos=0.15,above] {$s_3$} (uell)
           (uell') to[in=-170] node[below] {$s_\ell$} (s');
    \end{tikzpicture}
    \caption{The situation $s$ in \eqref{eq:balanced-decomposition}.}
    \label{fig:balanced-decomposition}
  \end{figure}
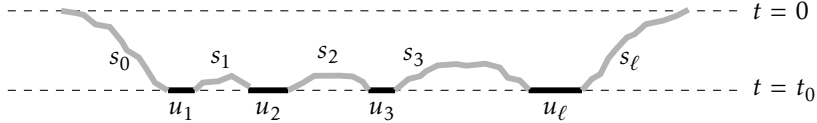
  in which the $u_i$ are the intervals that belong to time $t_0$. More
  precisely, we write $\pi_i = \prx{s_0 \dots s_{i_1} +_{a_i}}{u_i}$
  for the process that belongs to $u_i$ and require that
  $\pr(s)^{(t_0)} = \bigcup_{i=1}^\ell \pi_\ell$.

  The situations $s_i$ are arbitrary and need not be intervals.
  Nevertheless, since $s$ is achronal, every situation $s_i$ with $i <
  \ell$ ends with $\hat a_{i+1}$, and every $s_i$ with $i > 0$ begins
  with $\hat b_i$. Therefore
  \begin{equation}
    \label{eq:spread-decomposition}
    s \sim s_0 \ovl{\hat a_1}
    \underline{\hat a_1 -_{a_1} u_1 +_{b_1} \hat b_1}
    \ovl{\hat b_1} s_1 \dots s_{\ell-1} \ovl{\hat a_\ell}
    \underline{\hat a_\ell -_{a_\ell} u_\ell +_{b_\ell} \hat b_\ell}
    \ovl{\hat b_\ell} s_\ell\,.    
  \end{equation}
  We can now apply Lemma~\ref{thm:balanced-closure} to the underlined
  factors in this equation. In the current context it says that
  $\Delta \pi_i = \prx{s_0 \dots s_{i-1} \ovl{\hat a_i}}{\hat u_i}$
  for every $i$. We therefore get the situation $s'$ by replacing the
  underlined factors in $s$ with the intervals~$\hat u_i$,
  \begin{equation}
    \label{eq:s-prime}
    s' \sim s_0 \ovl{\hat a_1}
    \hat u_1
    \ovl{\hat b_1} s_1 \dots s_{\ell-1} \ovl{\hat a_\ell}
    \hat u_\ell
    \ovl{\hat b_\ell} s_\ell\,.    
  \end{equation}
  In fact the situation $s'$ is what we get when we resolve the
  overlaps at the right side of the previous equation. This is always
  possible because every $\hat u_i$ begins with $\hat a_i$ and ends
  with $\hat b_i$.
  \begin{figure}[ht]
    \centering
    \begin{tikzpicture}[
      direct/.style={line width=.75ex},
      somewhere/.style={direct, draw=shaded, decorate},
      decoration={random steps, amplitude=1.5, segment length=5}]
      \path ++(8,0) coordinate (ahat1) +(-1,-1) coordinate (u1)
            ++(2,0) coordinate (ahat2)
            ++(6,0) coordinate (bhat1)
            ++(3,0) coordinate (bhat2) +(1,-1) coordinate (u2)
            ++(6,0) coordinate (end0)
            ++(0,-1) coordinate (end-1);
      \draw[dashed] (0,0) -- (end0) node[right] {$t=t_0 + 1$}
                    (0,-1) -- (end-1) node[right] {$t=t_0$};
      \draw[direct] (ahat1) -- node [above] {$\hat a_1$} (ahat2)
                    (bhat1) -- node [above] {$\hat b_1$} (bhat2)
                    (u1)    -- node [below] {$u_1$} (u2);
      \draw[direct, draw=shaded, line width=.35ex]
                    (ahat1) -- node [above] {$\hat u_1$} (bhat2);
      \draw[somewhere] 
          (1, 6) to[out=-40, in=180] node [right] {$s_0$} (ahat1)
          let \p1 = (end0) in
          (bhat2) to[out=0, in=180] node [above] {$s_1$} (\x1, 2);
    \end{tikzpicture}
    \caption{The left end of Figure~\ref{fig:balanced-decomposition},
      with the factors of $s$ and $s'$ overlayed.}
    \label{fig:balanced-decomposition-detail}
  \end{figure}
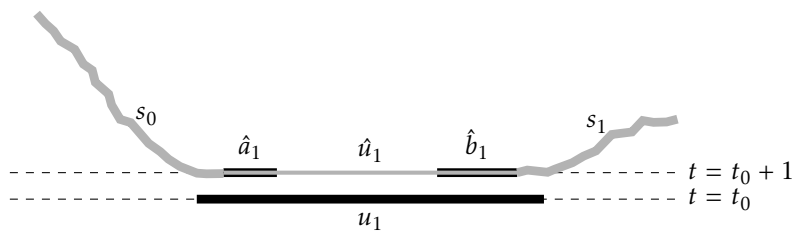

  We also have to check whether one of the processes $\Delta \pi_i$
  intersects with other parts of $\pr(s)^{(t_0+1)}$. But $\Delta
  \pi_i$ may have a non-empty intersection only with the processes
  belonging to $s_{i-1}$ and $s_i$. This is because $s$ is ordered:
  The intervals of $\pr(s)^{(t_0+1)}$ to the left of $\hat a_i$ must
  all belong to a situation $s_k$ or $u_k$ with $k \leq i - 1$.
  Therefore such an interval may extend to the right at most as far as
  the right end of $\hat a_i$. Similarly, the intervals of
  $\pr(s)^{(t_0+1)}$ to the right of $\hat b_i$ must all belong to a
  situation $s_k$ or $u_k$ with $i \leq k$ and therefore extend to the
  left at most to the left end of $\hat b_i$. The left-to-right
  arrangement of the intervals of $s$ is therefore preserved in $s'$,
  with the intervals $\hat u_i$ inserted in the gaps between the
  $s_i$. This shows that $s'$ is ordered.

  So $\Delta \pr(s)^{(t_0)}$ is compatible with $\pr(s)^{(t_0 + 1)}$,
  and $\cl^{(t_0 +1)} \pr(s)$ exists and is equal to $\pr(s')^{(t_0 +
    1)}$.
\end{proof}

Therefore we can define now a reaction system that has
$\mathcal{A}_\phi$ as its domain.

\section{The Local Reaction System}

The preliminary reaction system~\eqref{eq:interval-system} has the
disadvantage that its set of generating reactions is infinite. We
cannot specify them in a list in the same way as we can do this with a
transition rule.

We will now make the local nature of the interactions in a cellular
automaton more visible by decomposing the generating reactions
of~\eqref{eq:interval-system} into a finite number of reactions that
involve only a finite number of events. The following lemma specifies
these reactions and shows how they generate the reactions
of~\eqref{eq:interval-system}.

\begin{figure}[ht]
  \centering
  \begin{tikzpicture}[
    direct/.style={line width=.75ex},
    other/.style={direct, draw=shaded, line width=.35ex},
    hollow/.style={preaction={draw, direct}, draw=white, line width=.55ex},
    somewhere/.style={direct, draw=shaded, decorate},
    decoration={random steps, amplitude=1.5, segment length=5}]
    \def\background(#1){
      \pgfmathsetseed{999}
      \path (#1)  coordinate (u)   +(2, 0)    coordinate (uL')
          ++(5,0) coordinate (u')  +(-1.5, 0) coordinate (uR);
      \path (u)  +(1, 1)  coordinate (uhat);
      \path (u') +(-1, 1) coordinate (uhat');
        }
      \def\toleft{
        \path (u)    +(-1,0) coordinate (sigma);
        \path (u')   +(-1,0) coordinate (usigmaL');
        \path (uhat) +(-1,0) coordinate (sigmauhat);
      }
      \def\toright{
        \path (u')    +(1,0) coordinate (sigma');
        \path (u)     +(1,0) coordinate (usigmaR);
        \path (uhat') +(1,0) coordinate (usigmahat');
      }
    \matrix[cells=midway] {
      \background(3,0)
      \draw[somewhere] (0, 0) -- (u)  (u') -- +(3, 0);
      \draw[direct] (u) -- node[below] {$u$} (u');
      & \node{$\rea_R$}; &
      \background(3,0)
      \draw[somewhere] (0, 0) -- (u)  (u') -- +(3, 0);
      \draw[direct] (u) -- node[below] {$u_L$} (uL')
          (uR) -- node[below] {$u_R$} (u');
      \draw[direct] (uhat) -- node[above] {$\hat u$} (uhat');
      & \node {\quad(a)};
      \\
      \background(3,-0) \toleft
      \draw[somewhere] (0, 0) -- (sigma)  (uhat') -- +(4, 0);
      \draw[hollow] (sigma) -- node[below] {$\sigma$} (u);
      \draw[direct] (uhat) -- node[above] {$\hat u$} (uhat')
          (u) -- node[below] {$u$} (u');
      & \node{$\rea_R$}; &
      \background(3,-0) \toleft
      \draw[somewhere] (0, 0) -- (sigma)  (uhat') -- +(4, 0);
      \draw[direct]
          (sigmauhat) -- node[above] {$\widehat{\sigma u}$} (uhat')
          (sigma)     -- node[below] {$(\sigma u)_L$} (usigmaL');
      & \node {\quad(b)};
      \\
      \background(3,0) \toright
      \draw[somewhere] (0, 1) -- (uhat)  (sigma') -- +(2, 0);
      \draw[direct]
          (uhat) -- node[above] {$\hat u$} (uhat')
          (u) -- node[below] {$u$} (u');
      \draw[hollow] (u') -- node[below] {$\sigma$} (sigma');
      & \node{$\rea_R$}; &
      \background(3,0) \toright
      \draw[somewhere] (0, 1) -- (uhat)  (sigma') -- +(2, 0);
      \draw[direct]
          (uhat) -- node[above] {$\widehat{u \sigma}$} (usigmahat')
          (u) -- node[below] {$(u \sigma)_R$} (sigma');
      & \node {\quad(c)};
      \\
      };
  \end{tikzpicture}
  \caption{Reactions to generate an interval.}
  \label{fig:local-generators}
\end{figure}
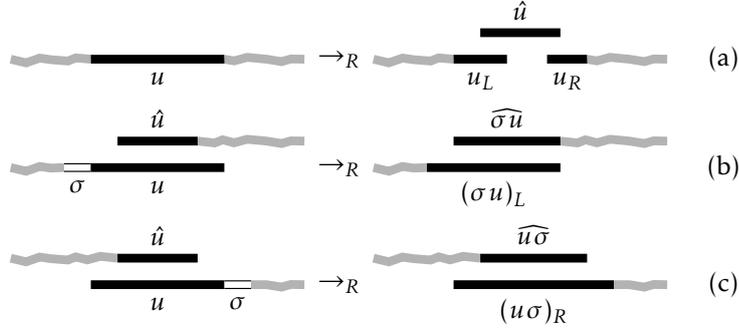

\begin{lemma}[Local Generators]
  \label{thm:local-generators}
  Let $R$ be a reaction system that contains for all separating
  intervals $u$ and all $\sigma \in \Sigma$ the reactions
  \begin{subequations}
    \begin{align}
      \label{eq:local-expand}
      u &\rea_R u_L +_u \hat u -_u u_R,
      &&\text{if $u$ is minimal,}\\
      \label{eq:local-grow-left}
      \hat u -_u u\sigma
      &\rea_R \widehat{u\sigma} -_{u\sigma} (u\sigma)_R,
      &&\text{if $u$ is right minimal,} \\
      \label{eq:local-grow-right}
      \sigma u +_u \hat u
      & \rea_R (\sigma u)_L +_{\sigma u} \widehat{\sigma u},
      &&\text{if $u$ is left minimal.}
    \end{align}
  \end{subequations}
  Then $R$ contains for all separating intervals $v$, not just those
  that are minimally separating, the reaction $v \rea_R v_L +_v
  \hat v -_v v_R$.
\end{lemma}
The diagrams for these reactions are shown in
Figure~\ref{fig:local-generators}; it uses the same conventions as
Figure~\ref{fig:interval-system}.
The intention behind the definitions is that
reaction~\eqref{eq:local-expand} is used to generate a new interval
one time step in the future---but this time one of minimal
length---and that~\eqref{eq:local-grow-left}
and~\eqref{eq:local-grow-right} then are used to expand it to the left
and the right.

\begin{proof}
  We prove the lemma by induction over the length of $v$: A separating
  interval $v$ is either minimally separating or there exist a separating
  interval $w \in \Sigma^*$ and a state $\sigma \in \Sigma$ such that
  $v = w \sigma$ or $v = \sigma w$.

  If $v$ is minimal, then there is by~\eqref{eq:local-expand} a
  reaction $v \rea_R v_L +_v \hat v -_v v_R$.

  If $v = w \sigma$, then there is by induction a reaction $w \rea_R
  w_L +_w \hat w -_w w_R$. We apply it to $v$ and get $v
  \rea_R w_L +_w \hat w -_w w_R \sigma$. Now let $x \in
  \Sigma^*$ such that $w = x w_R$. Since $w_R$ is separating, there is
  by Lemma~\ref{thm:separating-reactions} an $x' \in \Sigma^*$ such that
  $\widehat{x w_R} = x' \widehat{w_R}$. Therefore
  \begin{equation}
    \label{eq:w-sigma-reaction}
    v \rea_R
    w_L +_w x' \widehat{w_R} -_w w_R \sigma
  \end{equation}
  is a reaction in $R$. Then, since $w_R$ is right minimal and ${-_w}
  = {-_{w_R}}$, we can apply the reaction~\eqref{eq:local-grow-right}
  with $u = w_R$ to the result of~\eqref{eq:w-sigma-reaction} and get
  \begin{equation}
    \label{eq:w-sigma-reaction-2}
    w_L +_w x' \widehat{w_R} -_w w_R \sigma
    \rea_R w_L +_w x' \widehat{w_R \sigma}
    -_{w_R \sigma} (w_R \sigma)_R\,.
  \end{equation}
  We must now interpret the result of this reaction in terms of $v$.
  We have $w_L = v_L$ and $(w_R \sigma)_R = (w \sigma)_R = v_R$ by
  Lemma~\ref{thm:separating-extensions}, which also means that
  ${+_w} = {+_v}$ and ${-_{w_R \sigma}} =
  {-_v}$. For the middle term of the reaction result we apply
  again Lemma~\ref{thm:separating-reactions}: Since $w_R \sigma$ is
  separating and $x w_R \sigma = w \sigma = v$, we must have $x'
  \widehat{w_R \sigma} = \hat v$. Therefore the result
  of~\eqref{eq:w-sigma-reaction-2} is $v_L +_v \hat v -_v
  v_R$. Putting everything together we show this way that $v \rea_R
  v_L +_v \hat v -_v v_R$ if $v = w \sigma$.

  If $v = w \sigma$, a similar argument can be used.
\end{proof}

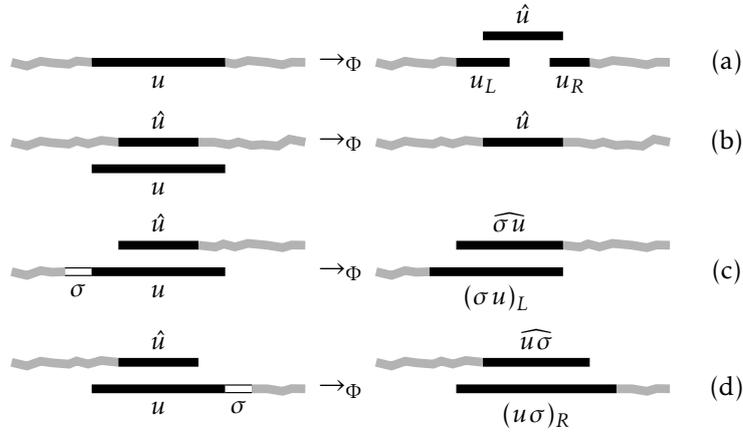
\begin{figure}[ht]
  \centering
  \begin{tikzpicture}[
    direct/.style={line width=.75ex},
    other/.style={direct, draw=shaded, line width=.35ex},
    hollow/.style={preaction={draw, direct}, draw=white, line width=.55ex},
    somewhere/.style={direct, draw=shaded, decorate},
    decoration={random steps, amplitude=1.5, segment length=5}]
    \def\background(#1){
      \pgfmathsetseed{999}
      \path (#1)  coordinate (u)   +(2, 0)    coordinate (uL')
          ++(5,0) coordinate (u')  +(-1.5, 0) coordinate (uR);
      \path (u)  +(1, 1)  coordinate (uhat);
      \path (u') +(-1, 1) coordinate (uhat');
        }
      \def\toleft{
        \path (u)    +(-1,0) coordinate (sigma);
        \path (u')   +(-1,0) coordinate (usigmaL');
        \path (uhat) +(-1,0) coordinate (sigmauhat);
      }
      \def\toright{
        \path (u')    +(1,0) coordinate (sigma');
        \path (u)     +(1,0) coordinate (usigmaR);
        \path (uhat') +(1,0) coordinate (usigmahat');
      }
    \matrix[cells=midway] {
      \background(3,0)
      \draw[somewhere] (0, 0) -- (u)  (u') -- +(3, 0);
      \draw[direct] (u) -- node[below] {$u$} (u');
      & \node{$\rea_\Phi$}; &
      \background(3,0)
      \draw[somewhere] (0, 0) -- (u)  (u') -- +(3, 0);
      \draw[direct] (u) -- node[below] {$u_L$} (uL')
          (uR) -- node[below] {$u_R$} (u');
      \draw[direct] (uhat) -- node[above] {$\hat u$} (uhat');
      & \node {\quad(a)};
      \\
      \background(3,-1)
      \draw[somewhere] (0, 0) -- (uhat)  (uhat') -- +(4, 0);
      \draw[direct] (uhat) -- node[above] {$\hat u$} (uhat')
          (u) -- node[below] {$u$} (u');
      & \node{$\rea_\Phi$}; &
      \background(3,-1)
      \draw[somewhere] (0, 0) -- (uhat)  (uhat') -- +(4, 0);
      \draw[direct] (uhat) -- node[above] {$\hat u$} (uhat');
      & \node {\quad(b)};
      \\
      \background(3,-0) \toleft
      \draw[somewhere] (0, 0) -- (sigma)  (uhat') -- +(4, 0);
      \draw[hollow] (sigma) -- node[below] {$\sigma$} (u);
      \draw[direct] (uhat) -- node[above] {$\hat u$} (uhat')
          (u) -- node[below] {$u$} (u');
      & \node{$\rea_\Phi$}; &
      \background(3,-0) \toleft
      \draw[somewhere] (0, 0) -- (sigma)  (uhat') -- +(4, 0);
      \draw[direct]
          (sigmauhat) -- node[above] {$\widehat{\sigma u}$} (uhat')
          (sigma)     -- node[below] {$(\sigma u)_L$} (usigmaL');
      & \node {\quad(c)};
      \\
      \background(3,0) \toright
      \draw[somewhere] (0, 1) -- (uhat)  (sigma') -- +(2, 0);
      \draw[direct]
          (uhat) -- node[above] {$\hat u$} (uhat')
          (u) -- node[below] {$u$} (u');
      \draw[hollow] (u') -- node[below] {$\sigma$} (sigma');
      & \node{$\rea_\Phi$}; &
      \background(3,0) \toright
      \draw[somewhere] (0, 1) -- (uhat)  (sigma') -- +(2, 0);
      \draw[direct]
          (uhat) -- node[above] {$\widehat{u \sigma}$} (usigmahat')
          (u) -- node[below] {$(u \sigma)_R$} (sigma');
      & \node {\quad(d)};
      \\
      };
  \end{tikzpicture}
  \caption{Generators of the local reaction system.}
  \label{fig:slow-system}
\end{figure}

\begin{definition}[Local Reaction System]
  \label{def:slow-system}
  Let $\phi$ be a well-behaved transition rule. Let $\Phi$ be the
  reaction system generated by the ordered situations in
  $\mathcal{A}_\phi$ and the following reactions, for all separating $u
  \in \Sigma^*$ and $\sigma \in \Sigma$,
  \begin{subequations}
    \label{eq:slow-system}
    \begin{align}
      \label{eq:slow-create}
      u &\rea_\Phi u +_u \hat u -_u u,
      &&\text{if $u$ is minimal,} \\
      \label{eq:slow-destroy}
      \hat u -_u u +_u \hat u &\rea_\Phi \hat u,
      &&\text{if $u$ is minimal,} \\
      \label{eq:oplus-reaction}
      \sigma u +_u \hat u
      & \rea_\Phi (\sigma u)_L +_{\sigma u} \widehat{\sigma u},
      &&\text{if $u$ is left minimal,} \\
      \label{eq:ominus-reaction}
      \hat u -_u u\sigma
      &\rea_\Phi \widehat{u\sigma} -_{u\sigma} (u\sigma)_R,
      &&\text{if $u$ is right minimal.}
    \end{align}
  \end{subequations}
  This reaction system is called the \introtwo{local reaction system}
  for $\phi$.
\end{definition}
For completeness, the diagrams for these reactions are also shown, in
Figure~\ref{fig:slow-system}.

In the rest of this chapter we will prove the following properties of
the local reaction system.
\begin{theorem}[Properties of Local Reaction Systems]
  Let $\phi$ be a well-behaved transition rule and $\Phi$ the local
  reaction system for $\phi$. Then $\Phi$ has the following
  properties:
  \begin{enumerate}
  \item \emph{(Covering, Figure~\ref{fig:covering})}. If $s \in \dom
    \Phi$ and $[p] \sigma \in \cl \pr(s)$, then there is a reaction $s
    \rea_\Phi v$ such that $[p] \sigma \in \pr(v)$.

  \item \emph{(Confluence, Figure~\ref{fig:confluence})}. If there are
    reactions $a \rea_\Phi b_1$ and $a \rea_\Phi b_2$, then there is a
    situation $c \in \dom \Phi$ such that $b_1 \rea_\Phi c$ and $b_2
    \rea_\Phi c$.
  \end{enumerate}
\end{theorem}
\begin{proof}
  The first property is proved in Theorem~\ref{thm:achronal-covering},
  the second property in Theorem~\ref{thm:church-rosser}.
\end{proof}

\section{Covering}

The property that is the subject of this section is a kind of converse
to the definition of reactions with help of the closure: Given a
reaction system $R$, do we have
\begin{equation}
  \label{eq:covering}
  \cl \pr(a) = \bigcup \set{ \pr(b): a \rea_R b}
\end{equation}
for a situation $a \in \dom R$? If this is true, then we say that $R$
\introx[covering]{covers} the closure of $a$. If $R$ covers the
closure of every $a \in \dom R$, then no information about the
cellular automaton gets lost when switching from the work with
closures to the work with reaction systems. The most important case is
of course the local reaction system $\Phi$.

We will now prove the closure property in a slightly different form,
by asking whether a specific event belongs to the closure of $a$.

The simplest case occurs when the initial situation $a$ itself an
interval. We can then express Lemma~\ref{thm:closure-interval} for
well-behaved transition rules in terms of reactions.
\begin{lemma}[Intervals are Covering]
  \label{thm:covering-the-closure}
  Let $\phi$ be a well-behaved transition rule. Let $R$ be a reaction
  system for $\phi$ where for every separating interval $a \in \Sigma^*$
  with characteristic reaction $a \rea [1, i] \hat a [-1, j]$ there is
  a reaction
  \begin{equation}
    \label{eq:extend-characteristic}
    a \rea_R a_+ \hat a a_-
  \end{equation}
  with $\delta(a_+) = (1, i)$ and $\delta(a_-) = (-1, j)$.

  Then for every interval $u \in \dom R$ and every event $[p]\sigma
  \in \cl \pr(u)$ there is a reaction $u \rea_R v$ with
  $[p]\sigma\in \pr(v)$.
\end{lemma}

\begin{proof}
  We will prove the following assertion for every $t \geq 0$: If there
  is a reaction $u \rea_R u_+ a u_-$ with $\cl^{(t)} \pr(u) =
  \prx{u_+}{a}$, then there is a reaction $u \rea_R u'_+ a' u'_-$ with
  $\cl^{(t + 1)} \pr(u) = \prx{u'_+}{a'}$.

  Since $cl_\phi^{(0)} \pr(u) = \pr(u)$, we know then by induction
  that for all $t \geq 0$ there is a reaction $u \rea_R v$ with
  $\cl^{(t)} \pr(u) \subseteq \pr(v)$, which proves the lemma.

  Assume now that $u \rea_R u_+ a u_-$ with $\cl^{(t)} \pr(u) =
  \prx{u_+}{a}$. If $a$ is separating, then there is a
  reaction~\eqref{eq:extend-characteristic} for it. Then $\Delta
  \pr(a) = \pr([1, i] \hat a)$, and therefore $\cl^{(t + 1)}
  \pr(u) = \Delta (\cl^{(t)} \pr(u)) = \Delta
  \prx{u_+}{a} = \prx{u_+}{[1, i] \hat a}$. So if we set $u'_+ = u_+
  a_+$, $a' = \hat a$ and $u_- = a_- u_-$, the assertion is true for a
  separating $a$.

  If $a$ is non-separating, especially empty, then $\Delta \pr(a) =
  \emptyset$ by assumption. This also means that $\cl^{(t + 1)}
  \pr(u) = \Delta (\cl^{(t)} \pr(u)) = \Delta
  \prx{u_+}{a} = \emptyset$. So we may choose $u'_+ = u_+ a$, $a' =
  [0]$, and $u'_- = u_-$ to fulfil the initial assertion of this
  proof.
\end{proof}

The local reaction system is then a specific case of the reaction
system in the previous lemma, so we get:
\begin{lemma}[Covering]
  Let $\phi$ be a well-behaved transition rule. Let $R$ be a reaction
  system for $\phi$ which has for every separating interval $a \in
  \Sigma^*$ a reaction
  \begin{equation}
    \label{eq:interval-covering}
    a \rea_R a_L +_a \hat a -_a a_R\,.
  \end{equation}
  Then for every interval $u \in \dom R$ and every event $[p]\sigma
  \in \cl \pr(u)$ there is a reaction $u \rea_R v$ with
  $[p]\sigma\in \pr(v)$.
\end{lemma}
\begin{proof}
  Let $a \rea [1, i] \hat a [-1, j]$ be the characteristic
  reaction for $a$ in~\eqref{eq:interval-covering}. Then $\delta(a_L
  +_a) = (0, \abs{a_L}) + (1, i - \abs{a_L}) = (1, i)$ and
  $\delta(-_a a_R) = (-1, j - \abs{a_R}) + (0, \abs{a_R}) = (-1,
  j)$. Therefore we can apply Lemma~\ref{thm:covering-the-closure},
  which finishes the proof.
\end{proof}

With this lemma we can prove covering for the general case
(Figure~\ref{fig:covering}).
\begin{theorem}[Covering by Achronal Situations]
  \label{thm:achronal-covering}
  Let $\Phi$ be the local reaction system for a well-behaved
  transition rule $\phi$ and let $s \in \dom \Phi$. Then for all
  events $[p] \sigma \in \cl \pr(s)$ there is a reaction $s
  \rea_\Phi v$ with $[p]\sigma \in \pr(v)$.
\end{theorem}
\begin{figure}[ht]
  \centering
  \begin{tikzpicture}[
    direct/.style={line width=.75ex},
    somewhere/.style={direct, draw=shaded, decorate},
    decoration={random steps, amplitude=1.5, segment length=5}]
    \path coordinate (s1)
          ++(5, -1) coordinate (s2) +(0, 2.5) coordinate (v2)
          ++(5, 1)  coordinate (s3) +(0, 2) coordinate (v3)
          ++(5, 0)  coordinate (s4);
    \coordinate (p) at ($(v2)!0.6!(v3)$);
    \coordinate (top) at ($(s1)!0.5!(s4) + (0, 5.5)$);
    \fill[gray!20, decorate, dashed, thin, draw=black]
         (s1) -- (s2) -- (s3) -- (s4) -- (top) -- (s1);
    \draw[direct, draw=shaded]
         (s1) -- (v2) node [below] {$v$} -- (v3) -- (s4);
    \fill (p) circle [radius=.375ex];
    \node at (p) [above] {$p$};
    \draw[direct]
         (s1) -- (s2) -- (s3) node [below] {$s$} -- (s4);
    \node at ($(v3)+(-.5,1)$) [pin=45:{$\cl \pr(s)$}] {};
  \end{tikzpicture}
  \caption{A reaction that covers the point $p$.}
  \label{fig:covering}
\end{figure}
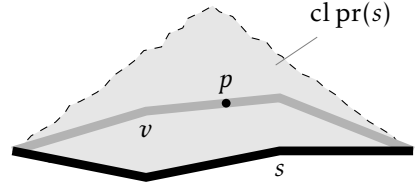
\begin{proof}
  Assume that $a = a_1 u a_2$ with $u \in \Sigma^*$ separating. Then
  there is a reaction $a \rea_\Phi a_1 +_u \hat u -_u a_2$, and if
  $[p]\sigma \notin \cl \pr(a_1 +_u \hat u -_u a_2)$, we must
  have $[p] \sigma \in pr_{a_1}(u)$.

  Assume that $a = a_1 \hat u -_u u +_u \hat u a_2$ with $u \in
  \Sigma^*$ minimally separating. Then there is a reaction $a \rea_\Phi
  a_1 \hat u a_2$, and if $[p]\sigma \notin \cl \pr(a_1 \hat u
  a_2)$, we must have $[p] \sigma \in pr_{a_1 \hat u -_u}(u)$.

  These two types of reactions generate all reactions in $\Phi$, so we
  see: If $a \rea_\Phi b$ and $[p] \sigma \notin \cl \pr(b)$,
  then there must be a reaction $a \rea_\Phi b'$ with $[p] \sigma \in
  \pr(b')$.

  Now there is for every $a \in \dom \Phi$ a reaction $a \rea_\Phi b$
  with
  \begin{equation}
    b =
    u_1 +_{u_1} \dots +_{u_{k-1}} u_k +_{u_k}
    v -_{w_\ell} w_\ell  -_{w_{\ell -1}} \dots -_{w_1} w_1\,. 
  \end{equation}
  in which $v$ is an interval and the $u_i$ and $w_i$ are minimal
  separating intervals. This can shown in an analogous way to the
  proof of Lemma~\ref{thm:slope-decomposition} below. With this
  definition we have $\cl \pr(b) = \pr(b) \cup \cl \prx{u_1 +_{u_1}
    \dots +_{u_\ell}}{v}$. This is so because $\phi$ is well-behaved
  and therefore the $u_i$ and $v_i$ provide no additional events to
  the closure of $b$.\footnote{We have e.\,g.\ $\Delta \pr(u_1)
    \subseteq \prx{u_1 +_{u_1}}{u_2}$ by the third property of
    Definition~\ref{def:well-behaved}.}

  Now $v$, as an interval, is covering, so there is either $[p] \sigma
  \in \cl \pr(b)$; then a reaction $v \rea_\Phi c$ can be applied
  to $b$ in order to cover $[p] \sigma$. Or, by the argument outlined
  above, there is directly a reaction $a \rea_\phi b'$ with $[p]
  \sigma \in \pr(b')$. In either case $[p] \sigma$ is covered.
\end{proof}

\section{Confluence}

I have borrowed the notion of confluence from the theory of term
rewriting systems, especially the lambda calculus \cite[p.\
4--5]{Abramsky1992}. If a term rewriting system is \intro{confluent} and
a term $a$ can be transformed by one rule of that system to a term
$b_1$ and by another rule to a term $b_2$, then there is a term $c$ in
that system that serves as a unifying target for $b_1$ and $b_2$:
there is a rule that transforms $b_1$ to $c$ and another rule that
transforms $b_2$ to $c$.

\begin{figure}[ht]
  \centering
  \begin{tikzpicture}[
    direct/.style={line width=.75ex},
    somewhere/.style={direct, draw=shaded, decorate},
    decoration={random steps, amplitude=1.5, segment length=5}]
    \path coordinate (a1)
          ++(6, -2) coordinate (a2)
              +(0, 2) coordinate (b12)
              +(0, 4) coordinate (b22)
              +(1, 6) coordinate (c2)
          ++(7, -.25)  coordinate (a3)
              +(-1.5, 4.5) coordinate (b13)
              +(0, 2) coordinate (b23)
              +(-1, 6) coordinate (c3)
          ++(6, 2)  coordinate (a4);
    \coordinate (top) at ($(a1)!0.5!(a4) + (0, 7.5)$);
    \fill[gray!20, decorate, dashed, thin, draw=black]
         (a1) -- (a2) -- (a3) -- (a4) -- (top) -- (a1);
    \draw[direct, draw=gray!45]
         (a1) -- (b12) node [below] {$b_1$} -- (b13) -- (a4);
    \draw[direct, draw=gray!60]
         (a1) -- (b22) -- (b23) node [below] {$b_2$} -- (a4);
    \draw[direct, draw=gray!90]
         (a1) -- (c2) -- node [above] {$c$} (c3) -- (a4);
    \draw[direct]
         (a1) -- (a2) -- node [below] {$a$} (a3) -- (a4);
  \end{tikzpicture}
  \caption{Two confluent reactions.}
  \label{fig:confluence}
\end{figure}
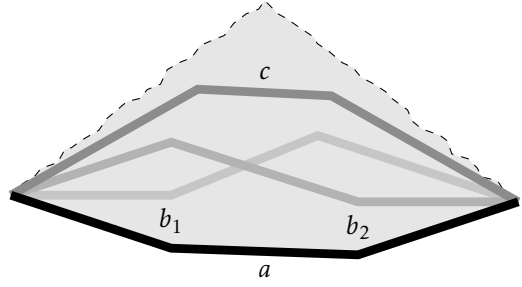
In the language of reaction systems this means: A reaction system $R$
is confluent if for every pair of reactions $a \rea_R b_1$ and $a
\rea_R b_2$ there is a situation $c \in \dom R$ such that $b_1 \rea_R
c$ and $b_2\rea_R c$ (Figure~\ref{fig:confluence}).

We will now show that this is also true for the local reaction system
$\Phi$. The situation $c$ will then have a specific form, which we
will specify with help of a subset of $\Phi$.

\begin{definition}[Slope Subsystems]
  \label{def:slope-subsystem}
  Let $R \subseteq \mathcal{A}_\phi \times \mathcal{A}_\phi$ be a
  reaction system for the transition rule $\phi$. Then the subsystems
  \begin{equation}
    R_+ = R \cap (\mathcal{A}_{\phi+} \times \mathcal{A}_{\phi+})
    \qqtext{and}
    R_- = R \cap (\mathcal{A}_{\phi-} \times \mathcal{A}_{\phi-})
  \end{equation}
  of $R$ are the systems of \indextwo{positive slope
    reaction}\emph{positive} and \indextwo{negative slope
    reaction}\emph{negative slope reactions}. The elements of $\dom
  R_-$ and $\dom R_+$ are the \indextwo{positive slope}\emph{positive}
  and \indextwo{negative slope}\emph{negative slopes}.
\end{definition}
The slope subsystems of $R$ therefore consist only of reactions
between slopes of the same kind. The system $R_+$ consists of all
reactions in $R$ that transform a positive slope into a positive
slope, while $R_-$ consists of all the reactions that transform
negative slopes into negative slopes. (Note that every interval
situation is a negative and a positive slope and that therefore the
reactions between intervals belong to both subsystems.)

The following lemma about slope subsystems is also important in its
own right.

\begin{lemma}[Slope Decomposition]
  \label{thm:slope-decomposition}
  Let $\Phi$ be a local reaction system and $a \in \Phi$. Then there
  exist situations $a_+ \in \dom \Phi_+$ and $a_- \in \dom \Phi_-$
  such that $a \rea_\Phi a_+ a_-$.
\end{lemma}

\begin{proof}
  If $a$ is not the product of an element of $\dom \Phi_+$ with an
  element of $\dom \Phi_-$, then it must contain a $-_b$ operator left
  of an $+_c$ operator. Especially there must be a pair of $-_b$ and
  $+_c$ operator that are separated only by an interval.

  This means that there must be a decomposition $a = a_1 \hat b_1
  -_{b_1} u +_{b_2} \hat b_2 a_2$, where $u$ is an interval, $b_1$ and
  $b_2$ are minimal separating intervals and $b_1 = u_L$ and $b_2 =
  u_R$. Then, since $u$ is a separating interval, there is a reaction
  $u \rea_\Phi b_1 +_{b_1} \hat u -_{b_2} b_2$. The application of
  this reaction to $a$ results in
  \begin{equation}
    \label{eq:slope-destroy-1}
    a \rea_\Phi
    a_1 \underline{\hat b_1 -_{b_1}
      b_1 +_{b_1} \hat u} -_{b_2} b_2
    +_{b_2} \hat b_2 a_2
  \end{equation}
  We concentrate now on the underlined part of the reaction result.
  Because the result is achronal, there must be an interval $u'$ such
  that $\hat u = \hat b_1 u'$. This means that we can apply a reaction
  of the form~\eqref{eq:slow-destroy} to $\hat b_1 -_{b_1} b_1 +_{b_1}
  \hat u$: We get then the reaction $\hat b_1 -_{b_1} b_1 +_{b_1} \hat
  u = \hat b_1 -_{b_1} b_1 +_{b_1} \hat b_1 u' \rea_\Phi \hat b_1 u' =
  \hat u$. Applying this reaction to the result
  of~\eqref{eq:slope-destroy-1} results in
  \begin{equation}
    a \rea_\Phi
    a_1  \underline{\hat u -_{b_2} b_2
      +_{b_2} \hat b_2} a_2
  \end{equation}
  With the same kind of argument we can show that there is a reaction
  that replaces the underlined part of this reaction with $\hat u$,
  resulting in $a \rea_\Phi a_1 \hat u a_2$.

  This reaction has removed one $-_{b_1}$ and one $+_{b_2}$ operator
  from $a$. Repeating this removes all pairs of $-_b$ and $+_{b'}$
  from $a$. The result is a reaction $a \rea_\Phi a_+ a_-$ to a
  situation of the required form.
\end{proof}

The situation $c$ that makes the two reactions $a \rea_\Phi b_1$ and
$a \rea_\phi b_2$ confluent will be constructed step by step in an
induction proof. The following proposition is a technical lemma that
will be used in Lemma~\ref{thm:inductive-confluence} to transform
$b_1$ and $b_2$ into successively better approximations of $c$.

\begin{lemma}[Creation of a Minimal Separating Boundary]
  \label{thm:minimal-separating}
  Let $a \in \dom \Phi$, where $\Phi$ is a local reaction system. Then
  at least one of the following cases occurs:
  \begin{enumerate}
  \item $\delta(a)_T \leq 0$ and there is a reaction $a \rea_\Phi a'
    -_u u$, where $u$ is a right minimal separating interval,
  \item $\delta(a)_T \geq 0$ and there is a reaction $a \rea_\Phi u
    +_u a'$, where $u$ is a left minimal separating interval, or
  \item $\delta(a)_T = 0$, there is a reaction $a \rea_\Phi v$ to a
    non-separating interval $v$, and $\cl^{(t)} \pr(a) = \emptyset$ for
    all $t \geq 1$.
  \end{enumerate}
\end{lemma}

\begin{proof}
  Let $\delta(a)_T \leq 0$. Because of
  Lemma~\ref{thm:slope-decomposition} there is a reaction $a \rea_\Phi
  a_+ a_-$ with $a_+ \in \dom \Phi_+$ and $a_- \in \dom \Phi_-$.

  If $\delta(a_-)_T < 0$, there must be a decomposition $a_- = a'_-
  -_v v x$, where $v$ is a minimal separating interval and $x$ an
  interval. Since $v x$ is separating, there is by
  Lemma~\ref{thm:local-generators} a reaction $v x \rea_\Phi (v x)_L
  +_{v x} \widehat{v x} -_{v x} (v x)_R$. So we have a
  reaction $a \rea_\Phi a_+ a'_- -_v (v x)_L +_{v x}
  \widehat{v x} -_{v x} (v x)_R$ and case~1 occurs with $u = (v
  x)_R$.

  If $\delta(a_-)_T = 0$, we must also have $\delta(a_+)_T = 0$
  because $\delta(a_+)_T + \delta(a_-)_T = \delta(a)_T \leq 0$ while
  $\delta(a_+) \geq 0$. Then $a_+ a_-$ is an interval. If
  $\cl^{(1)} \pr(a) \neq \emptyset$, then there must be a
  reaction $a_+ a_- \rea_\Phi (a_+ a_-)_L +_{a_+ a_-}
  \widehat{a_+ a_-} -_{a_+ a_-} (a_+ a_-)_R$; then case~1 occurs
  with $u = (a_+ a_-)_R$. If $\cl^{(1)} \pr(a) = \emptyset$, then
  $a_+ a_-$ must be a non-separating interval because $\phi$ is
  well-behaved; then case~3 occurs with $v = a_+ a_-$.

  The case of $\delta(a)_T \geq 0$ is handled in a mirror-symmetric
  way.
\end{proof}

We will now show a slightly stronger form of confluence, in order to
get a good induction proof. In the following lemma we will say that
two situations $x_1$ and $x_2$ are \emph{equal until time $t$} if for
all $\tau \leq t$ we have $\pr(x_1)^{(\tau)} = pr(x_2)^{(\tau)}$.
\begin{lemma}[Approximated Confluence]
  \label{thm:inductive-confluence}
  Let $\Phi$ be a local reaction system.

  If there are reactions $a \rea_\Phi b_1$ and $a \rea_\Phi b_2$ and
  the situations $b_1$ and $b_2$ are equal before time $t$, then there
  are situations $c_1$, $c_2 \in \dom \Phi$ that are equal until time
  $t +1 $, and reactions $a \rea_\Phi c_1$, $b_1 \rea_\Phi c_1$, $a
  \rea_\Phi c_2$ and $b_2 \rea_\Phi c_2$.
\end{lemma}

In the following proof, $\mathcal{S}_{+-}$ is the set $\set{a_+
  a_-\colon a_+ \in \Phi_+, a_- \in \Phi_-}$.

\begin{proof}
  Let $b_1$ and $b_2$ be equal before time $t$. If both
  $\pr(b_1)^{(t)}$ and $\pr(b_1)^{(t)}$ are empty, the lemma is
  trivially true, so we assume from now on that this is not the case.

  We know already that $\delta(b_1) = \delta(b_2)$. With
  Lemma~\ref{thm:slope-decomposition} we can also assume that $b_1$
  and $b_2$ are elements of $\mathcal{S}_{+-}$.

  If $b_1$ and $b_2$ are equal until time $t$, then there are
  situations $x$, $y$, $b'_1$ and $b'_2 \in \dom \Phi$ such that $b_1
  = x b'_1 y$ and $b_2 = x b_2 y$, and $t$ is the minimum of
  $\delta(x)$ and $\delta(x b_1)$. Since $\delta(b_1)$ and
  $\delta(b_2)$ are equal, $\delta(b'_1)$ and $\delta(b'_2)$ are equal
  too and also elements of $\mathcal{S}_{+-}$.

  If $\delta(b'_1)_T = \delta(b'_2)_T > 0$, then there are by
  Lemma~\ref{thm:minimal-separating} two reactions $b'_1 \rea u_1
  +_{u_1} b''_1$ and $b'_2 \rea u_2 +_{u_2} b''_2$, with intervals
  $u_1$ and $u_2$. We may assume without loss of generality that
  $\abs{u_1} \leq \abs{u_2}$. Then $\prx{x}{u_1} \subseteq \cl \pr(a)$
  and $\prx{x}{u_2} \subseteq \cl \pr(a)$, therefore $\prx{x}{u_2}|_{\dom
    \prx{x}{u_1}} = \prx{x}{u_1}$. So $\prx{x}{u_1} \subseteq \prx{x}{u_2}$,
  and if $u_1 \neq u_2$, then $u_1$ is an initial segment of $u_2$,
  which is impossible because $u_2$ is already a right minimal
  separating interval. So we must have $u_1 = u_2$. Then we can set $c_1
  = x u_1 +_{u_1} b''_1 y$ and $c_2 = x u_1 +_{u_1} b''_2 y$; these
  situations are equal until time $t + 1$. The same kind of argument
  works if $\delta(b'_1) = \delta(b'_2) < 0$.

  Now we assume that $\delta(b'_1)_T = \delta(b'_2)_T = 0$. If $b'_1$
  and $b_2$ are intervals, then they must be equal, by an argument
  similar to that in the previous paragraph. We can then set $c_1 =
  c_2 = b_2$.

  Otherwise, if $b_1$ is not an interval, it must still be an element
  of $\mathcal{S}_{+-}$, so there must be a reaction $b'_1 \rea u_+
  +_{u_+} b''_1 -_{u_-} u_-$, with separating intervals $u_+$ and $u_-$.
  But this means that $b_2$ cannot react to a non-separating interval
  $v$: If this were the case, the process $\prx{x}{u_+}$ would be a part
  of $\prx{x}{v}$, but $\prx{x}{u_+}$ is a separating interval and therefore
  cannot be part of a non-separating interval. So there must be a
  reaction $b'_2 \rea u_+ +_{u_+} b''_2 -_{u_-} u_-$, where the
  ``re-use'' of $u_+$ and $u_-$ can be justified as in the previous
  paragraphs. In this case we have $c_1 = x u_+ +_{u_+} b''_1 -_{u_-}
  u_-$ and $c_2 = x u_+ +_{u_+} b''_2 -_{u_-} u_-$.

  A similar argument can be used when $b_2$ is not an interval. This
  concludes the proof.
\end{proof}

\begin{theorem}[Confluence]
  \label{thm:church-rosser}
  If there are reactions $a \rea_\Phi b_1$ and $a \rea_\Phi b_2$, then
  there is a situation $c \in \dom \Phi$ such that $b_1 \rea_\Phi c$
  and $b_2 \rea_\Phi c$.
\end{theorem}
\begin{proof}
  We apply the induction steps outlined in
  Lemma~\ref{thm:inductive-confluence}.

  Since $b_1$ and $b_2$ are finite, there is certainly a time $t_0$ such
  that $b_1$ and $b_2$ are equal until $t_0$. By repeated application of
  the lemma we get the four sequences of reactions
  \begin{subequations}
    \begin{alignat}{2}
      a &\rea_\Phi c_{1, k},\qquad
      &b_1 &\rea_\Phi c_{1, k}, \\
      a &\rea_\Phi c_{2, k}, 
      &b_2 &\rea_\Phi c_{2, k},
    \end{alignat}
  \end{subequations}
  with $c_{1, k}$ and $c_{2, k}$ equal until time $k$. The only
  remaining question is whether this process stops after a finite
  number of steps.

  To see this this we note that if $\pr(b_1)^{(t)} = \pr(b_1)^{(t)} =
  \emptyset$ for all time steps $t > t_0$, then the same is true for
  $c_{1, k}$ and $c_{2, k}$, and for arbitrary $k$. This can be
  verified by following the constructions in
  Lemma~\ref{thm:slope-decomposition} and
  Lemma~\ref{thm:inductive-confluence}.
\end{proof}

\section{Summary}

In this chapter we have found a way to construct a reaction system
from a transition rule.

Separating intervals played an important role. They allowed us to
construct the set of \emph{achronal situations}; and an easily
recognisable subset of them, the ordered achronal sets, were shown to
have always a closure. We have therefore found a subset of situations
that generalises intervals but nevertheless consist of events at
different times.

To prove this we had to restrict the set of transition rules a bit
further, from interval-preserving to well-behaved rules. I expect that
this restriction is only temporary and later may be loosened to allow
for an extension of achronal sets to a larger class of transition
rules.

For the moment we have nevertheless the definition of a reaction
system that is usable for ``naturally occurring'' transition rules,
like Rule 54 in the next chapter. This \emph{local reaction system}
was introduced and shown to have useful properties. Since it has the
covering property, all information that can be found with help of the
closure operator can also be found with reactions. We are therefore no
longer dependent on processes to derive results on cellular automata.



\chapter{Rule 54}
\label{sec:rule-54}

Up to now we have worked with cellular automata only in an abstract
way. Now we will introduce a concrete cellular automaton which already
has a complex behaviour.

The aim of this chapter is then to demonstrate the concepts of the
previous chapters for an elementary cellular automaton, Rule 54. It
also shows how structures of intermediate complexity manifest in the
context of Flexible Time.

\section{Elementary Cellular Automata}

Rule 54 arises in the context of the \intro{elementary cellular
  automata}. We have seen in the introduction that they are the
one-dimensional cellular automata with radius 1 and $\Sigma = \{ 0,
1\}$ and that Stephen Wolfram \cite{Wolfram1983} has provided an
enumeration scheme for them.

In Wolfram's enumeration scheme we interpret the state set $\Sigma$ as
a set of integers. There is a number $s$ such that $\Sigma = \{ 0,
\dots, s - 1\}$, and $\Sigma$ can be seen as the set of digits for
base $s$. A sequence of such digits is then an integer. Then we can
view every state of the neighbourhood of a cell as a number with $2r +
1$ digits, the code number for the neighbourhood. If we then enumerate
the results of $\phi$ applied to every neighbourhood $w \in
\Sigma^{2r+1}$ by the code number of $w$, the transition rule itself
is another number under base $s$, this time with $s^{2r+1}$ digits.
This number is the code number for the function $\phi$.
\begin{definition}[Code numbers]
  \label{def:code-numbers}
  Let $\Sigma = \{0, \dots, s-1 \} \subseteq \N_0$, $r \in \N$, and
  let $\phi \colon \Sigma^{2r+1} \to \Sigma$ be a transition rule. For
  any sequence $w = \omega_0 \dots \omega_{2r} \in \Sigma^{2r+1}$, let
  $c(w) = \Sigma_{i=0}^{2r} \omega_i s^i$.
  Then the \intro{code number} for $\phi$ is
  \begin{equation}
    \label{eq:code-number}
    \sum_{u \in \Sigma^d}
    \phi(u)
    s^{c(u)}\,.
  \end{equation}
\end{definition}

\paragraph{Cellular Processes as Diagrams} We need to determine the
local reaction system for Rule 54. These computations involve some
cellular processes, and the easiest way to write them down is as a
rectangular diagram---especially since these cellular processes will
contain events from at most two different time steps.

Such a diagram may have the shape $\diag{&\tau \\ \sigma_0 & \sigma_1
  & \sigma_2}$. This specific diagram describes a cellular process in
which the cells in the states $\sigma_0$, $\sigma_1$ and $\sigma_2$
belong to time step $0$ and the cell in state $\tau$ belongs to time
step $1$. In such diagrams the leftmost event in the bottom line
always belongs to the space-time point $(0, 0)$, therefore the process
can be written in the set notation as $\{ [0, 0]\sigma_0, [0,
1]\sigma_1, [0, 2] \sigma_2, [1, 1]\tau \}$. 

\section{Basic Properties of Rule 54}
\label{sec:basic-prop-rule54}

I have chosen Rule 54 because it has some complex behaviour
\cite{Boccara1991, Ju'arezMart'inez2006a}, and it is a relatively
simple rule in which an ether appears. An example for ether generation
is Figure~\ref{fig:random_54}.

Rule 54 has the transition rule
\begin{equation}
  \label{eq:def54}
  \phi_{54}(s) =
  \left\{
    \begin{array}{r@{\quad}l}
      1 & \text{for $s \in \{(0,0,1), (1,0,0), (0,1,0), (1,0,1)\}$,}\\
      0 & \text{otherwise.}
    \end{array}
  \right.
\end{equation}
Note that $\phi_{54}$ is symmetric under the interchange of left and
right.

\begin{figure}[ht]
  \centering
  \ci{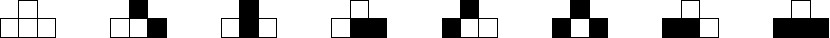}
  \caption{``Rule icon'' for  Rule 54.}
  \label{fig:54icon}
\end{figure}
The rule can be described in a diagram in Figure~\ref{fig:54icon}. The
diagram displays each of the eight possible 3-cell neighbourhoods
together with the next state of the central cell. This diagram has
been sometimes called the ``Rule Icon''\footnote{For example by
  \cite[p.~239]{Rowland2006} and in the ``Wolfram Atlas''---see
  \url{http://atlas.wolfram.com/01/01/54/} for Rule 54.}.

The description in~\eqref{eq:def54} is for a human reader (in contrast
to a computer) difficult to memorise. A simpler description is the
following slogan, which can be verified from Figure~\ref{fig:54icon}.
\begin{quote}
  ``$\phi_{54}(w) = 1$ if $w$ contains at least one $1$, except if the
  cells in state $1$ touch.''
\end{quote}
Here we say that two cells ``touch'' if they are direct neighbours.
Thus the two cells in state $1$ touch in the neighbourhood $(1, 1,
0)$, but not in the neighbourhood $(1, 0, 1)$.

\paragraph{Interval Preservation} Next we must check whether Rule 54
is interval-preserving. To do this, we must test for all intervals $w
\in \Sigma^k$ with $k \leq 3$ whether $\Delta \pr(w)$ is an interval
under the transition rule $\phi_{54}$. If this is true, then
$\phi_{54}$ is interval-preserving by
Theorem~\ref{thm:preserving-local}.

To do this we need a practical way to compute all the events
determined by an interval $w$. Among them are the events that can be
found directly by the transition rule, when applied to the intervals
of length $2r + 1$, together with those events that are determined by
smaller cellular processes. The transition rule implies that an
interval $w$ of length $2r + 1$ determines the event $[1, r] \phi(w)$.
We are now interested in all subsets of $\pr(w)$ that already
determine the event $[1, r] \phi(w)$. Since $\pr(w)$ is an interval,
every subset of $\pr(w)$ can be extended to an interval by adding
events. Therefore it is enough to search for the interval subsets of
$\pr(w)$ that determine the event $[1, r] \phi(w)$.

These intervals can be found by an application of
Definition~\ref{def:determined}: If there is a decomposition $w = x w'
y$ such that the value of $\phi(x w' y)$ is independent of the
contents of $x$ and $y$, then $\prx{x}{w'}$ already determines $[1, r]
\phi(w)$.\footnote{This means that the application of the transition
  rule must lead to the same result $\phi(x' w' y') = \phi(w)$ for all
  intervals $x'$, $y' \in \Sigma^*$ with $\abs{x'} = \abs{x}$ and
  $\abs{y'} = \abs{y}$.} This also means that the interval $\pr(w')$
determines the event $[1, r - \abs{x}] \phi(w)$. With this method we
can find all the intervals of length $\leq 2r + 1$ that determine an
event under the transition rule~$\phi$.

All this can then be expressed by a rule: \emph{If $\phi(x w y) =
  \sigma$ for all $x \in \Sigma^k$ and $y \in \Sigma^\ell$, then $w$
  determines the event $[1, r - k] \sigma$.} We will now find these
reactions for Rule 54. Rule 54 has the following cases where this rule
can be applied:
\begin{itemize}
\item There are two cases with $k = 1$ and $\ell = 0$, namely
  $\phi(001) = \phi(101) = 1$ and $\phi(011) = \phi(111) = 0$. The
  first equation shows that the interval $01$ determines the event
  $[1, 0] 1$ and the second that $11$ determines $[1, 0] 0$.

\item There are two cases with $k = 0$ and $\ell = 1$, namely
  $\phi(100) = \phi(101) = 1$ and $\phi(110) = \phi(111) = 0$. The
  first equation proves that the interval $10$ determines the event
  $[1, 1] 1$ and the second that $11$ determines $[1, 1] 0$.
\end{itemize}
With these arguments we have found four new rules to find a determined
event that belongs to an interval. With them it is now possible to
find the next state of the middle cell for all intervals of length $3$
that begin with $10$ or $11$ and for those that end with $01$ or $11$.
There remain the reactions for the neighbourhoods that cannot be
shortened in this way, namely $000$ and $010$. These neighbourhoods
determine the events $[1, 1] 0$ and $[1, 1] 1$, respectively. To the
other intervals we can apply one of the four new reactions to get the
state of the middle cell one time step later. Therefore we have now
the six cases where an interval determines an event,
\begin{subequations}
  \label{eq:new-determined}
  \def\x{\phantom1}
  \begin{align}
    000 &\text{ determines }  [1, 1] 0, & \diag{&0\\ 0&0&0} \\
     01 &\text{ determines }  [1, 0] 1, & \diag{1\\ 0&1} \\
    11  &\text{ determines }  [1, 0] 0, & \diag{0\\ 1&1} \\
    010 &\text{ determines }  [1, 1] 1, & \diag{&1\\ 0&1&0} \\
     10 &\text{ determines }  [1, 1] 1, & \diag{&1\\ 1&0&\x} \\
     11 &\text{ determines }  [1, 1] 0\,. & \diag{&0\\ 1&1&\x}
  \end{align}
\end{subequations}
With these rules we can compute the events determined by an interval.
At the right they are visualised with space-time diagrams. In them,
the bottom line contains the process of $w$, and on top of it there is
the event that is determined by it. Each of these diagrams, when
applied to an interval, gives us the identity of one event that is
determined by this interval.

No we can prove that $\phi_{54}$ is interval-preserving. For this we
take all intervals whose length is at most $2r + 1 = 3$ and apply
graphically all reactions to them that can be applied. We then get the
following diagrams; all of them have $\pr(w)$ as their bottom row and
$\Delta \pr(w)$ as the top row.
\begin{equation}
  \begin{array}[b]{*8c}
    \diag{\phantom 1\\ 0} &
    \diag{\phantom 1\\ 1} &
    &&
    \diag{\phantom 1\\ 0&0} &
    \diag{1\\ 0&1} &
    \diag{&1\\ 1&0} &
    \diag{0&0\\ 1&1} \\[3ex]
    \diag{&0\\ 0& 0& 0} &
    \diag{&1\\ 0& 0& 1} &
    \diag{1&1&1\\ 0& 1& 0} &
    \diag{&0&0\\ 0& 1& 1} &
    \diag{&1\\ 1& 0& 0} &
    \diag{&1\\ 1& 0& 1} &
    \diag{0&0&1\\ 1& 1& 0} &
    \diag{0&0&0\\ 1& 1& 1}
  \end{array}
\end{equation}
They show that $\phi_{54}$ is interval-preserving for the intervals of
length $\leq 3$; by Theorem~\ref{thm:preserving-local} it must then
preserve all intervals.

\paragraph{Characteristic Reactions} The diagrams
in~\eqref{eq:new-determined} allow us now to write down the
characteristic reactions for all those intervals that determine at
least one event. Their characteristic reactions are
\begin{subequations}
  \begin{align}
    \label{eq:char01}
    01  & \rea [1, 0] 1 [-1, 1],  \\
    \label{eq:char10}
    10  & \rea [1, 1] 1 [-1, 0],  \\
    \label{eq:char11}
    11  & \rea [1, 0] 00 [-1, 0], \\
    \label{eq:char000}
    000 & \rea [1, 1] 0 [-1, 1],  \\
    \label{eq:char111}
    010 & \rea [1, 0] 111 [-1, 0]\,.
  \end{align}
\end{subequations}
\begin{figure}[ht]
  \centering
  \ci{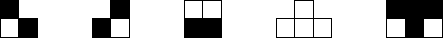}
  \caption{``Invariant Rule Icon'' for Rule 54.}
  \label{fig:54invariant-icon}
\end{figure}
Together, these reactions allow us to derive all the events that are
determined by a given cellular process. We can write them as an
alternative form of the rule icon of Figure~\ref{fig:54icon}, one that
in contrast to it does no longer depend explicitly on the radius. This
diagram is shown in Figure~\ref{fig:54invariant-icon}.

As we will see later, the minimal separating intervals of Rule 54 are
the intervals of length $2$. For the first three of them we have just
determined the characteristic reactions. The characteristic reaction
for the last interval, $00$, can now be determined according to
Definition~\ref{def:characteristic-separating}.

The characteristic reaction for the interval $a = 00$ must have the
form
\begin{equation}
  \label{eq:characteristic-interval}
  a \rea [1, i] \hat a [-1, j],
\end{equation}
and we must now determine $i$, $j$ and $\hat a$. This can be done with
help of reaction~\eqref{eq:char000}. If we extend $a$ to the left with
$0$, we get the reaction $0a \rea [1, 1] 0 [-1, 1]$, therefore we
must have $j = 1$. And if we extend $a$ to the right with $0$, we get
$a0 \rea [1, 1] 0 [-1, 1]$ and therefore $i = 1$. The only value
for $\hat a$ for which $\delta(a) = \delta([1, i] \hat a [-1, j])$
is $a = [0]$, therefore the characteristic reaction for $00$ is
\begin{equation}
  \label{eq:char00}
  00 \rea [1, 1] [-1, 1]\,.
\end{equation}

Next we can see that the intervals of length $1$ are not separating.
This is because we cannot construct a characteristic reaction for
them. For the ``interval'' consisting of a cell in state $0$, we would
have $i = 0$ in reaction~\eqref{eq:characteristic-interval} because of
the characteristic reaction~\eqref{eq:char01} for $01$ but $i = 1$
because of reaction~\eqref{eq:char00}. To verify that the cell in
state $1$ does not form a separating interval, note
that~\eqref{eq:char01} requires $j = 1$ while~\eqref{eq:char11}
requires $j = 0$.

\begin{figure}[ht]
  \centering
  \tabcolsep=1.5em
  \begin{tabular}{*4c}
    \ci{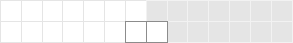} & \ci{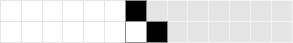} \\
    $00 \rea [1, 1] [-1, 1]$      & $01 \rea [1, 0] 1 \, [-1, 1]$
    \\[4ex]
    \ci{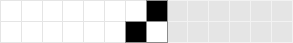} & \ci{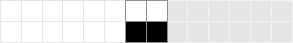} \\
    $10 \rea [1, 1] 1 \, [-1, 0]$ & $11 \rea [1, 1] [-1, 1]$
  \end{tabular}
  \caption{Separating intervals and their characteristic reactions for
    Rule 54.}
  \label{fig:separating}
\end{figure}
On the other hand, Lemma~\ref{thm:separating-exists} shows that the
intervals of length 2 are separating, so they must be the minimal
separating intervals. They and their characteristic reactions are
shown in Figure~\ref{fig:separating}.

\section{The Local Reaction System}

With the characteristic reactions for the minimal separating intervals
we can now determine the structure of the achronal situations for Rule
54.

\paragraph{Notation} According to Definition~\ref{def:achronal}, the
set of achronal situations is known when the generating slopes $\hat b
-_b b$ and $b +_b \hat b$ are known for all minimal separating
intervals $b$.

For concrete calculations, the repetition of $b$ in these terms
becomes however easily annoying. Therefore we will first introduce
another, related, notation. It is a variant of the slope operators
of Definition~\ref{def:slope-operators}.
\begin{definition}[Concrete Slope Operators]
  \label{def:concrete-boundaries}
  \idef{slope operator}
  Let $i \in \Z$. Then we write
  \begin{equation}
    \label{eq:osymbols}
    \om_i = [-1, -i] \qtext{and} \op_i = [1, -i]\,.
  \end{equation}
  If $r$ is the radius of the transition rule, we will use the
  abbreviations $\om$ for $\om_r$ and $\op$ for $\op_r$.
\end{definition}
This notation had been introduced in \cite{Redeker2010} and was
already used in \cite{Redeker2010a}.

\paragraph{Achronal Situations} Now we will derive the generating
slopes $\hat b -_b b$ and $b +_b \hat b$ from the characteristic
reactions of Figure~\ref{fig:separating}. They are listed in
Table~\ref{tab:steps54}. Its first column contains the characteristic
reactions, and the other columns contain the generating slopes derived
from them.
\begin{table}[t]
  \centering
  \caption{Characteristic reactions and generating slopes of Rule 54.}
  \label{tab:steps54}
  \def\r{$&${}\rea}
  \begin{tabular}[b]{r@{}l@{\qquad}lr}
    \toprule
    $b \r [1, i] \hat b [-1, j]$ & $b +_b \hat b$ & $\hat b -_b b$ \\
    \midrule
    $00 \r [1, 1] [-1, 1]$    & $00 \op$      & $\om 00$      \\
    $01 \r [1, 0] 1 [-1, 1]$  & $01 \op_2 1$  & $1 \om 01$    \\
    $10 \r [1, 1] 1 [-1, 0]$  & $10 \op 1$    & $1 \om_2 10$  \\
    $11 \r [1, 0] 00 [-1, 0]$ & $11 \op_2 00$ & $00 \om_2 11$ \\
    \bottomrule
  \end{tabular}
\end{table}

First we need expressions for the slope operators $+_b$ and $-_b$ in
terms of the new operators of
Definition~\ref{def:concrete-boundaries}. We assume here, as before,
that the characteristic reaction for every minimal separating interval
$b$ is $b \rea [1, i] \hat b [-1, j]$. When we then use the notation
of Definition~\ref{def:concrete-boundaries}, the two kinds of slope
operators are related by the equations
\begin{subequations}
  \begin{align}
    \label{eq:equivalent-operators}
    +_b &= [1, i - \abs{b}] = \oplus_{\abs{b} - i} \\
    -_b &= [-1, j - \abs{b}] = \ominus_{\abs{b} - j}
  \end{align}
\end{subequations}
Here we have used the fact that $b$ is a minimal separating interval
and that therefore $b_L = b_R = b$.

With the equations in~\eqref{eq:equivalent-operators} we now can
derive the entries is the second and third column of
Table~\ref{tab:steps54} from the characteristic reactions in the first
columns. This derivation consists of two steps. The first is finding
the values of $b$ and $\hat b$: They can be read of the characteristic
reactions in the first column. The second step consists of finding the
slope operators $+_b$ and $-_b$. I will now show this in detail for
the second column.

We see from the first column of Table~\ref{tab:steps54} that there are
two kinds of characteristic reactions, namely those where the reaction
product starts with $[1, 1]$ and those where it starts with $[1, 0]$.
In the first case there is $i = 1$ and therefore $+_b = \oplus_{2 - 1}
= \oplus$, and that is why in the second column of
Table~\ref{tab:steps54} the first and the third entry contains a
$\oplus$ operator. In the second case there is $i = 0$ and $+_b =
\oplus_{2 - 0} = \oplus_2$, and therefore the last second and fourth
entry in the second column in Table~\ref{tab:steps54} contain a
$\oplus_2$ operator.

The same way we can derive the third column of
Table~\ref{tab:steps54}.

\paragraph{Generating Reactions} Now, to complete the description of
the local reaction system for Rule 54, we need to find its generating
reactions. They are defined in Definition~\ref{def:slow-system} and
consist of two subsets: those that are associated to the minimal
separating intervals, and those that are found by extending a minimal
separating interval to the left or to the right.

\emph{(a)} The first subset consists of the reactions $b \rea_\Phi b
+_b \hat b -_b b$ and $\hat b -_b b +_b \hat b \rea_\Phi \hat b$ for
all minimal separating intervals $b$. We do already know that
\begin{subequations}
  \begin{alignat}{4}
    \label{eq:oplus-2}
    +_{00} &= +_{10} = \oplus, &\qquad
    +_{01} &= +_{11} = \oplus_2, \\
    \label{eq:ominus-2}
    -_{00} &= -_{01} = \ominus, &
    -_{10} &= -_{11} = \ominus_2
  \end{alignat}
\end{subequations}
and that
\begin{equation}
  \label{eq:hat-2}
  \widehat{00} = [0], \qquad
  \widehat{01} = \widehat{10} = 1, \qquad
  \widehat{11} = 00\,.
\end{equation}
With this we can calculate the reactions of the form $b \rea_\Phi b
+_b \hat b -_b b$ in the following way,
\begin{subequations}
  \begin{alignat}{2}
    00 &\rea_\Phi 00 +_{00} \widehat{00} -_{00} 00
    &&= 00 \oplus \ominus 00, \\
    01 &\rea_\Phi 01 +_{01} \widehat{01} -_{01} 01
    &&= 01 \oplus_2 1 \ominus 01, \\
    10 &\rea_\Phi 10 +_{10} \widehat{10} -_{10} 10
    &&= 10 \oplus 1 \ominus_2 10, \\
    11 &\rea_\Phi 11 +_{11} \widehat{11} -_{11} 11
    &&= 11 \oplus_2 00 \ominus_2 11,
  \end{alignat}
\end{subequations}
and the reactions of the form $\hat b -_b b +_b \hat b
\rea_\Phi \hat b$ in the following way,
\begin{subequations}
  \begin{align}
    \ominus 00 \oplus =
    \widehat{00} -_{00} 00 +_{00} \widehat{00}
    &\rea_\Phi \widehat{00} = [0], \\
    1 \ominus 01 \oplus_2 1 =
    \widehat{01} -_{01} 01 +_{01} \widehat{01}
    &\rea_\Phi \widehat{01} = 1, \\
    1 \ominus_2 10 \oplus 1 =
    \widehat{10} -_{10} 10 +_{10} \widehat{10}
    &\rea_\Phi \widehat{10} = 1, \\
    00 \ominus_2 11 \oplus_2 00 =
    \widehat{11} -_{11} 11 +_{11} \widehat{11}
    &\rea_\Phi \widehat{11} = 11\,.
  \end{align}
\end{subequations}
The results of these two sets of calculations are collected in the
left and right bottom fields of Table~\ref{tab:generator54},
respectively.

\emph{(b)} The second subset of the reactions in
Definition~\ref{def:slow-system} consists of reactions of the form
$\sigma b +_b \hat b \rea_\Phi (\sigma b)_L +_{\sigma b}
\widehat{\sigma b}$ and $\hat b -_b b\sigma \rea_\Phi
\widehat{b\sigma} -_{b\sigma} (b\sigma)_R$ with $\sigma \in \Sigma$,
where $b$ is a left or right minimal separating interval,
respectively. We will now concentrate on the second type of reactions,
which is sufficient because Rule 54 is symmetric.

To compute the reactions $\hat b -_b b\sigma \rea_\Phi
\widehat{b\sigma} -_{b\sigma} (b\sigma)_R$ for all right minimal
separating intervals $b$ we need to know $\widehat{b \sigma}$, $(b
\sigma)_R$ and $-_{b \sigma}$ for every right minimal interval $b$ and
every $\sigma \in \Sigma$. And for this we first need to know the set
of right minimal intervals for Rule 54.

The easiest way to do it is to start in greater generality and to
determine the values of $b_L$ and $b_R$ for every separating interval
$b$. In case of Rule 54 this is simple: Since every interval of length
$2$ is a minimal separating interval, $b_L$ consists of the two
leftmost events in $b$ and $b_R$ if the two rightmost events in $b$.
(If $b$ is separating, it must contain a minimal separating interval
and is therefore at least two cells long.) For this reason the set of
\emph{left} and \emph{right minimal separating intervals} under Rule
54 is both times $\Sigma^2$.

Next we need the characteristic reactions for all elements of
$\Sigma^3$. For them we need, in turn, to know the cells determined by
all the intervals in $\Sigma^3$. They can be determined
from~\eqref{eq:new-determined} and are
\begin{equation}
  \begin{array}[b]{*8c}
    \diag{&0\\ 0&0&0} &
    \diag{&1\\ 0&0&1} &
    \diag{1&1&1\\ 0&1&0} &
    \diag{1&0&0\\ 0&1&1} &
    \diag{&1\\ 1&0&0} &
    \diag{&1\\ 1&0&1} &
    \diag{0&0&1\\ 1&1&0} &
    \diag{0&0&0\\ 1&1&1}\,.
  \end{array}
\end{equation}
These diagrams are also a short description of the connection between
$b$ and $\hat b$ for all $b \in \Sigma^3$. We can derive the values of
$\hat b$ for $b \in \Sigma^3$ directly from them,
\begin{equation}
  \begin{aligned}[b]
    \widehat{000} &= 0, & \widehat{010} &= 111, &
    \widehat{100} &= 1, & \widehat{110} &= 001, \\
    \widehat{001} &= 1, & \widehat{011} &= 100, &
    \widehat{101} &= 1, & \widehat{111} &= 000\,.
  \end{aligned}
\end{equation}
The values of $\hat b$ for all $b \in \Sigma^2$ are already listed
in~\eqref{eq:hat-2}. To compute $(b \sigma)_R$ and $-_{b \sigma}$ we
use the relations
\begin{align}
  (\sigma_1 \sigma_2 \sigma_3)_R &= \sigma_2 \sigma_3, &
  -_{\sigma_1 \sigma_2 \sigma_3} &= -_{\sigma_2 \sigma_3}
\end{align}
for all $\sigma_1$, $\sigma_2$, $\sigma_3 \in \Sigma$; for the second
relation the values of $-_{\sigma_2 \sigma_3}$ can be found
in~\eqref{eq:ominus-2}. With this data we can now calculate the
reactions of the type $\hat b -_b b\sigma \rea_\Phi \widehat{b\sigma}
-_{b\sigma} (b\sigma)_R$ in the following way,
\begin{subequations}
  \begin{align}
    \ominus 000 =
    \widehat{00} -_{000} 000 &
    \rea_\Phi \widehat{000} -_{000} (000)_R
    = 0 \ominus 00, \\
    \ominus 001 =
    \widehat{00} -_{001} 001 &
    \rea_\Phi \widehat{001} -_{001} (001)_R
    = 1 \ominus 01, \\
    1 \ominus 010 =
    \widehat{01} -_{010} 010 &
    \rea_\Phi \widehat{010} -_{010} (010)_R
    = 111 \ominus_2 10, \\
    1 \ominus 011 =
    \widehat{01} -_{011} 011 &
    \rea_\Phi \widehat{011} -_{011} (011)_R
    = 100 \ominus_2 11, \\
    1 \ominus_2 100 =
    \widehat{10} -_{100} 100 &
    \rea_\Phi \widehat{100} -_{100} (100)_R
    = 1 \ominus 00, \\
    1 \ominus_2 101 =
    \widehat{10} -_{101} 101 &
    \rea_\Phi \widehat{101} -_{101} (101)_R
    = 1 \ominus 01, \\
    00 \ominus_2 110 =
    \widehat{11} -_{110} 110 &
    \rea_\Phi \widehat{110} -_{110} (110)_R
    = 001 \ominus_2 10, \\
    00 \ominus_2 111 =
    \widehat{11} -_{111} 111 &
    \rea_\Phi \widehat{111} -_{111} (111)_R
    = 000 \ominus_2 11\,.
  \end{align}
\end{subequations}
The reactions that are found in this calculation are shown in
Table~\ref{tab:generator-diagrams}, both as formulas and as diagrams.
\begin{table}[t]
  \centering
  \caption{Diagrams of the generator reactions for Rule 54.}
  \tabcolsep=1.5em
  \def\gen#1#2{#2 &\ci{gen54_#1}}
  \def\r{$&${}\rea}
  \begin{tabular}{r@{}lc}
    \toprule
    \multicolumn2c{Reactions} & Diagrams \\
    \midrule
    \gen{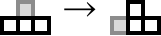}{$\om 000 \r 0 \om 00$} \\
    \gen{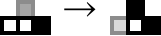}{$\om 001 \r 1 \om 01$} \\
    \gen{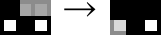}{$1 \om 010 \r 111 \om_2 10$} \\
    \gen{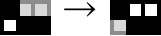}{$1 \om 011 \r 100 \om_2 11$} \\
    \gen{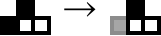}{$1 \om_2 100 \r 1 \om 00$} \\
    \gen{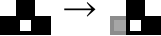}{$1 \om_2 101 \r 1 \om 01$} \\
    \gen{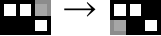}{$00 \om_2 110 \r 001 \om_2 10$} \\
    \gen{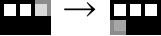}{$00 \om_2 111 \r 000 \om_2 11$} \\
    \bottomrule
  \end{tabular}
  \label{tab:generator-diagrams}
\end{table}
Together with the reactions that are their left-to right mirror images
they form the upper part of the ``Reactions'' section in
Table~\ref{tab:generator54}.

\begin{table}[t]
  \caption{Generator reactions of the local reaction system for Rule 54.}
  \label{tab:generator54}
  \def\r{$&${}\rea}
  \tabcolsep=0pt
  \begin{tabular*}{\textwidth}{lrl@{\quad}rl}
    \toprule
    Separating \rlap{Intervals:}
    && $00$, $01$, $10$, $11$. \\
    \midrule
    Generating\rlap{ Slopes:} \\
    & $\om 00$, $1 \om 01$&,   $1 \om_2 10$, $00 \om_2 11$,
    & $00 \op$, $01 \op_2 1$&, $10 \op 1$, $11 \op_2 00$. \\
    \midrule
    Reactions:
    & $   \om 000   \r 0 \om 00$     & $000 \op \r 00 \op 0$ \\
    & $   \om 001   \r 1 \om 01$     & $100 \op \r 10 \op 1$ \\
    & $ 1 \om 010   \r 111 \om_2 10$ & $010 \op 1 \r 01 \op_2 111$ \\
    & $ 1 \om 011   \r 100 \om_2 11$ & $110 \op 1 \r 11 \op_2 001$ \\
    & $ 1 \om_2 100 \r 1 \om 00 $    & $001 \op_2 1 \r 00 \op 1$ \\
    & $ 1 \om_2 101 \r 1 \om 01 $    & $101 \op_2 1 \r 10 \op 1$ \\
    & $00 \om_2 110 \r 001 \om_2 10$ & $011 \op_2 00 \r 01 \op_2 100$ \\
    & $00 \om_2 111 \r 000 \om_2 11$ & $111 \op_2 00 \r 11 \op_2 000$ \\[1.2ex]

    & $ 00 \r 00 \opom 00$          & $  \om 00 \op     \r [0]$    \\
    & $ 01 \r 01 \op_2 1 \om 01$    & $1 \om 01 \op_2 1 \r 1$      \\
    & $ 10 \r 10 \op 1 \om_2 10$    & $1 \om_2 10 \op 1 \r 1$      \\
    & $ 11 \r 11 \op_2 00 \om_2 11$ & $00 \om_2 11 \op_2 00 \r 11$ \\
    \bottomrule
  \end{tabular*}
\end{table}
This completes the calculation of the local reaction system for Rule
54. The result is a new form of the transition rule $\phi_{54}$.

\section{Understanding Rule 54 Better}

While $\phi_{54}$ describes how a cell's neighbourhood influences its
state in the next time step, each reaction in the local reaction
system describes the relation between a separating interval $\pi$ and
the interval $\Delta \pi$ that is determined by it. The generating
slopes in Table~\ref{tab:generator54} describe the relations between
the boundaries of $\pi$ and $\Delta \pi$: The slope $00 \oplus$ means
that if $\pi$ begins with $00$, then $\Delta \pi$ reaches one cell to
the right of the left end of $\pi$; the slope $10 \oplus 1$ tells that
if $\pi$ begins with $10$, then the left end of $\Delta \pi$ reaches
to the same position, but its leftmost event must be in state $1$, and
so on.

Even this localised knowledge helps us to understand the behaviour of
Rule 54 better. For an example we use the task of finding the closure
of an interval, something that we had already begun in
Figure~\ref{fig:dependent}. We can now express the closure of an
interval with a reaction $u \rea_\Phi a_+ a_-$, where $a_+$ is a
positive and~$a_-$ a negative slope. The reaction system $\Phi$ has
been constructed in such a way that there is a reaction in which the
situations $a_+$ and $a_-$ form the boundaries of the triangle, which
we will now assume. Then the process of $a_+$ consists of the leftmost
separating intervals of each time slice of $\cl \pr(u)$. The analysis
of the previous paragraph then helps us to understand better the
ragged boundaries of the closure in Figure~\ref{fig:dependent}.

Now let us add an event to the left side of the interval at the base
of Figure~\ref{fig:dependent}. Then its closure grows too. The kind of
growth, and how it depends on the added event, tells us how a change
in the initial configuration is propagated to later time steps.

For Rule 54, this is expressed by the reactions in the top half of the
``Reactions'' section of Table~\ref{tab:generator54}. The reactions at
the left side of the table show what happens when a cell is added to
the right, and those at the right side of the table show what happens
when a cell is added to the left.

The influence of the added event varies greatly depending on the
states of the cells at the end of the original interval. We see from
one pair of reactions, $010 \op 1 \rea 01 \op_2 111$ and $110 \op 1
\rea 11 \op_2 001$, that when the left side of the original interval
is $00$, the added event adds two events in the next time step; on the
other hand, another pair, $001 \op_2 1 \rea 00 \op 1$ and $101 \op_2 1
\rea 10 \op 1$, proves that the closure may also stay unchanged. (It
is peculiar to Rule 54 that the state of the added event has no
influence on the number of cells that are added in the next time step,
only on their states.)

We have therefore found for each cellular automaton a specific pattern
of influence, described by the generators of the local reaction
system.

\section{Summary}

In this chapter we have seen how the local reaction system is computed
for a concrete rule.

In calculations with a concrete system, brevity is an advantage and
redundancy is annoying. Therefore we used in this chapter the symbols
$\ominus_k$ and $\oplus_k$ instead of $-_a$ and $+_a$ for the display
of the resulting reaction system. In spite of this the computation of
the reaction system may appear to be quite long and complex, with all
the explanations given. If one leaves them out, it is however possible
to do the whole work described here on a single piece of paper.

Nevertheless the resulting system in Table~\ref{tab:generator54} looks
somewhat voluminous when compared with the original description of
Rule 54 in~\eqref{eq:def54}. This is caused, among other things, by
the requirement that a local reaction system covers the full closure
of each of its situations. If we drop this requirement, then we can
create for special purposes reaction systems that are easier to
describe and more powerful. One of them will be constructed in the
next chapter.

The advantage of the large size of Table~\ref{tab:generator54} is
however that it provides additional information about the way in which
information travels in the cellular automaton. This was not directly
visible from the transition rule.

\begin{subappendices}
  \newpage

  \section{Appendix: Rule 110}

  Here I will give another example and construct the local reaction
  system for another elementary cellular automaton, Rule 110. The
  derivation will be much more sketchy, but it should also show how
  the calculation of a concrete local reaction system can be done in a
  relatively small space.

  \paragraph{Nature of the Rule} We will use as the initial
  description of the rule an icon similar to that of
  Figure~\ref{fig:54icon}. We see especially that Rule 110 is an
  asymmetric rule, in contrast to Rule 54.
  \begin{figure}[ht]
    \centering
    \ci{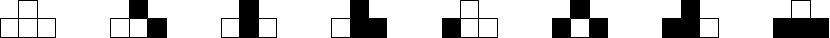}
    \caption{``Rule icon'' for  Rule 110.}
    \label{fig:110icon}
  \end{figure}

  To understand Rule 110 better, we will now find a slogan for it, as
  we had done for Rule 54. This time the slogan becomes especially
  simple if we take the states $0$ and $1$ as Boolean values. Then we
  can write,
  \begin{quote}
    ``$\phi_{110}(\sigma_{-1}, \sigma_0, \sigma_1) = \sigma _0
    \mathbin\mathbf{xor} \sigma_1$, except that $\phi_{110}(0, 1, 1) =
    1$.''
  \end{quote}

  \paragraph{Graphical Evolution} With Figure~\ref{fig:110icon} we
  will now search for the cases where less than three cells determine
  the state of the central cell in the next time step. From the slogan
  we know that the value of this cell depends only on the interval
  consisting of the central cell and its neighbour, except when that
  interval is $11$. For the other cases we can write this as the
  diagrams $\diag{0\\ 0&0}$, $\diag{1\\ 0&1}$ and $\diag{1\\ 1&0}$, as
  we have done in~\eqref{eq:new-determined}. There is also a fourth
  case when two adjacent cells determine the next state of a cell: In
  the interval $01$, the next state of the right cell is always $1$.
  Thus we get the following diagrams,
  \begin{equation*}
    \begin{array}[b]{*4c}
      \diag{0\\ 0&0} &
      \diag{1\\ 0&1} &
      \diag{1\\ 1&0} &
      \diag{&1\\ 0&1}\,.
    \end{array}
  \end{equation*}
  We can also see from Figure~\ref{fig:110icon} that there are no
  shorter intervals that determine an event.

  To get all the diagrams that are needed for a graphical evolution we
  have to add the diagram for the interval $111$, because the next
  state of its central cell cannot be derived otherwise; then we get
  \begin{equation}
    \label{eq:evolution-110}
    \begin{array}[b]{*6c}
      \diag{0\\ 0&0} &
      \diag{1\\ 0&1} &
      \diag{1\\ 1&0} &
      \diag{&1\\ 0&1} &
      \diag{&0\\ 1&1&1}\,.
    \end{array}
  \end{equation}
  These diagrams are then enough to find for every cellular process
  the events that are determined by it.

  \paragraph{Minimal Intervals} We have seen in
  Table~\ref{tab:interval-preserving-2} that Rule 110 is
  interval-preserving, and from Table~\ref{tab:separating} that its
  minimal separating intervals are $0$ and $11$. For the theory we
  will however need also the intervals that can occur as rightmost or
  leftmost minimal intervals. There is an additional leftmost minimal
  interval, $10$, and another rightmost minimal interval, $01$.

  \paragraph{Characteristic Reactions} For these intervals we now
  determine the events that are determined by them, using the diagrams
  of~\eqref{eq:evolution-110}. We get another list of diagrams,
  \begin{equation}
    \begin{array}[b]{*4c}
      \diag{&\phantom 1\\ 0} &
      \diag{&\phantom 1\\ 1&1} &
      \diag{1&1\\ 0&1} &
      \diag{1\\ 1&0}\,.
    \end{array}
  \end{equation}
  The characteristic reactions for the minimally separating intervals
  are then
  \begin{subequations}
    \label{eq:characteristic-110}
    \begin{align}
      0 &\rea [1, 0] [-1, 1], &
      11 &\rea [1, 1] [-1, 1], \\
      01 &\rea [1, 0] 11 [-1, 0], &
      10 &\rea [1, 0] 1 [-1, 1]\,.
    \end{align}
  \end{subequations}
  We find the coefficients in the first reaction once we realise that
  the next state of a cell in state $0$ is always the state of the
  cell at its right. Therefore, the boundary between the cells that
  are determined by the left side and those determined by the right
  side is at the left of the cell in state $0$, and this is reflected
  in the reaction. For the second reaction we must note that the next
  state for the left cell in the interval $11$ depends on information
  at the left, and the state of the right cell depends on information
  from the right. The last two reactions
  in~\eqref{eq:characteristic-110} can be read directly from the
  diagrams.

  \paragraph{Generating Slopes} From the characteristic reactions we
  derive the generating slopes. For the positive slopes we use the
  left coefficients of the characteristic reactions that belong to the
  left minimal separating intervals. We then get the situations $0
  \oplus$, $11 \oplus$ and $01 \oplus_2 1$.

  For the negative slopes we use the rightmost minimal separating
  intervals and the coefficients of the right side of the
  characteristic reactions and get $\ominus_0 0$, $\ominus 11$ and $11
  \ominus_2 01$.

  \paragraph{Reactions} The generating reactions can also be derived
  in a graphical manner, especially the slope reactions. We will give
  here only a few examples.

  \def\b#1{\colorbox{gray!45}{$\scriptstyle #1$}}

  One example is the reaction $00 \oplus \rea 0 \oplus 0$. First we
  write down the diagram for the input situation, $\diag{\phantom 1\\
    0&0}$. This we extend with help of the rule for graphical
  evolution and get $\diag{0&\\ 0&0}$. We now mark on this diagram the
  rightmost positive slope. Then we get $\diag{\b0\\ \b0&0}$, and when
  we translate the sequence of marked events back into a positive
  slope, it is $0 \oplus 0$.

  In a similar way the reaction $010 \oplus_2 1 \rea 0 \oplus 11$ is
  derived. Here the diagram for the initial situation is
  $\diag{&1\\0&1&0}$. We extend it to $\diag{1&1\\0&1&0}$ and mark in
  it the rightmost positive slope. The result is
  $\diag{\b1&\b1\\\b0&1&0}$, and the marked events belong to the situation
  $0 \oplus 11$.

  The reactions in the bottom of Table~\ref{tab:generator110} are
  derived more easily algebraically. For the bottom left block in that
  table we need to find reactions of the form $u \rea u +_u \hat u -_u
  u$, as in~\eqref{eq:slow-create}, where $u$ is a minimal separating
  interval. In the case of the interval $11$, we must therefore find
  generating slopes $u +_u \hat u$ and $u +_u \hat u$, with $u = 11$.
  We have already seen that these generating slopes are $11 \oplus$
  and $\ominus 11$. This leads to the reaction $11 \rea 11 \oplus
  \ominus 11$.

  For the bottom right block we need to find in the same way reactions
  of the form $\hat u -_u u +_u \hat u \rea_\Phi \hat u$, as in
  \ref{eq:slow-destroy}, again for the minimal blocking intervals. In
  the case of the interval $11$ we get the reaction $\ominus 11 \oplus
  \rea [0]$, since $\widehat{11} = [0]$.

  \begin{table}[t]
    \caption{Generator reactions of the local reaction system for Rule 110.}
    \label{tab:generator110}
    \def\r{$&${}\rea}
    \tabcolsep=0pt
    \begin{tabular*}{\textwidth}{lrl@{\qquad}rl}
      \toprule
      Separating Intervals: \\
      \qquad Rightmost: && $0$, $01$, $11$. \\
      \qquad Leftmost:  &&&& $0$, $10$ $11$. \\
      \midrule
      Generating Slopes: 
      & $\ominus_0 0$&, $\ominus 11$, $11 \ominus_2 01$,
      & $0 \oplus$&, $11 \oplus$, $10 \oplus_2 1$. \\
      \midrule
      Reactions:
      & $   \om_0 00  \r 0  \om_0 0$  & $00 \op \r 0 \op 0$       \\
      & $   \om_0 01  \r 11 \om_0 01$ & $10 \op \r 10 \op_2 1$    \\
      & $11 \om_2 010 \r 11 \om_0 0$  & $010 \op_2 1 \r 0 \op 11$ \\
      & $11 \om_2 011 \r 11 \om 11$   & $110 \op_2 1 \r 11 \op 1$ \\
      & $   \om 110   \r 1  \om_0 0$  & $011 \op \r 0 \op 11$     \\
      & $   \om 111   \r 0  \om 11$   & $111 \op \r 11 \op 0$     \\[1.2ex]

      & $ 0 \r 0 \op\om_0 0$  & $ \om_0 0 \op \r [0]$ \\
      & $ 11 \r 11 \op\om 11$ & $ \om  11 \op \r [0]$ \\
      \bottomrule
    \end{tabular*}
  \end{table}

  \paragraph{Results} All these results are summarised in
  Table~\ref{tab:generator110}. It is clearly visible that this local
  reaction system has a completely different structure than that of
  Rule 54. But it is not yet clear what exactly this difference means.
\end{subappendices}


\chapter{The Ether}
\label{cha:ether}

Now we will use the reaction system we found in the previous chapter
to describe ether formation under Rule 54. As a first step towards
this goal we must find a description of the ether in terms of
situations and reactions. We will take a more general point of view
and explain how periodic structures in a cellular automaton are
described in the formalism of Flexible Time.

First we introduce some concepts for simple periodic structures that
occur in one-dimensional cellular automata. We will define them in
general; then we introduce terms for the special forms they have under
Rule 54. At the end of the chapter we will use them to describe why
the ether under Rule 54 occurs spontaneously and why it is stable.

\section{Triangles}

As yest we know only the generating reactions of a local reaction
system. All of them involve only a small number of cellular events. In
order to understand the large-scale behaviour of a cellular automaton
we need then to find reactions that involve a larger number of events.
And in order to find general laws for the behaviour of the cellular
automaton that can be expressed with a theorem or a formula, we need
to find sets of situations that all behave in a similar way.

\paragraph{Families of Situations} This section is about the simplest
form of such a general law, namely about the evolution of repeated
patterns in the initial configuration of a cellular automaton. In case
of Rule 54 we are interested in sequences of cells in the same state
together with a boundary of a single cell in the opposite state, such
as $011110$ and $1000001$. We can then decompose the initial
configurations into such sequences, which overlap at their boundaries.
A cell sequence $\dots 100011110 \dots$ can then be decomposed into
$10001$ and $011110$, and when it is part of an interval situation it
can be written as the overlapping product $10001 \ovl{01} 011110$. The
boundary cells in the opposite state are included here because they
make the decomposition uniquely determined, and also because the
reactions that originate from them lead also to situations that are
useful in the model for ether formation at the end of this chapter.
For a different kind of problem another decomposition might be more
useful.

In the following definition this kind of decomposition is formalised
and generalised, as much as we can do without making it difficult to
handle. We especially drop the requirement that the repeated pattern
must always be an interval: this would be unimportant for most
calculations with situations.

\begin{definition}[Family of Situations]
  Let $R$ be a reaction system and $a$, $x$, $b \in \dom R$ be
  situations. Then the set
  \begin{equation}
    \label{eq:interval-basic}
    \set{a x^k b \colon k \geq 0}
  \end{equation}
  is a \introthree{family of situations} in $R$.
\end{definition}
The representation~\eqref{eq:interval-basic} is not the only one for a
family of situations. For all constants $n$, $k_0 \in \N_0$ we have
the equivalences
\begin{align}
  \set{a x^k b \colon k \geq k_0}
  &= \set{(a x^{k_0}) x^k b \colon k \geq 0}, \\
  \set{a x^{n k + k_0} b \colon k \geq 0}
  &= \set{(a x^{k_0}) (x^n)^k b \colon k \geq 0},
\end{align}
therefore the left sides of these equations are also valid
representations of situation families. Now we make use of the first of
these equations and introduce names for the two families of initial
intervals described above:
\begin{equation}
  \label{eq:intervals-54}
  F_0 = \set{ 10^k1 \colon k \geq 1}
  \qqtext{and}
  F_1 = \set{ 01^k0 \colon k \geq 1}\,.
\end{equation}

\paragraph{Layers} Our goal was to find general laws for the evolution
of situation families like $F_0$ and $F_1$. We will now find the
reactions that describe the evolution of the elements of such a
situation family over a single time step. Most of these reactions,
namely those that start from an element of a situation family that is
larger than a certain minimal size, have a similar form.

This general form for a set of reactions is given in the next
definition. The elements of these \emph{layer reactions} will be used
as building blocks for other reactions that extend over several time
steps, therefore their name.
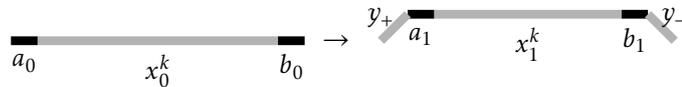
\begin{figure}[ht]
  \centering
  \begin{tikzpicture}[
    direct/.style={line width=.75ex},
    somewhere/.style={direct, draw=shaded, decorate},
    decoration={random steps, amplitude=1.5, segment length=5}]
    \matrix [cells=midway] {
      \path   (0,0) coordinate (a0) ++(11,0) coordinate (b0')
              (a0)  ++(1,0) coordinate (a0')
              (b0') ++(-1,0) coordinate (b0);
      \draw[direct]
              (a0) -- node[below] {$a_0$} (a0')
              (b0) -- node[below] {$b_0$} (b0');
      \draw[direct,draw=shaded]
              (a0') -- node[below] {$x_0^k$} (b0);
      & \node {$\rea$}; &
      \path   (0,0) coordinate (yplus) ++(11,0) coordinate (yminus')
              (yplus)   ++(1,1)  coordinate (a1)  ++(1,0)  coordinate (a1')
              (yminus') ++(-1,1) coordinate (b1') ++(-1,0) coordinate (b1);
      \draw[direct]
              (a1) -- node[below] {$a_1$} (a1')
              (b1) -- node[below] {$b_1$} (b1');
      \draw[direct, draw=shaded]
              (yplus) -- node[above, pos=0] {$y_+$} (a1)
              (b1')   -- node[above, pos=1] {$y_-$} (yminus')
              (a1')   -- node[below] {$x_1^k$} (b1);
            \\
    };
  \end{tikzpicture}
  \caption{A layer reaction.}
  \label{fig:generic-layer}
\end{figure}
\begin{definition}[Layer Reactions]
  \label{def:layers}
  Let $R$ be a reaction system and let $A_0 = \set{a_0 x_0^k b_0
    \colon k \geq 0}$ and $A_1 = \set{a_1 x_1^k b_1 \colon k \geq 0}$
  be two families of situations in $\dom R$. A set of reactions
  (Figure~\ref{fig:generic-layer})
  \begin{equation}
    \label{eq:generic-layer}
    \set{a_0 x_0^k b_0 \rea_R y_+ a_1 x_1^k b_1 y_-
      \colon k \geq 0},
  \end{equation}
  with $y_+$, $y_- \in \dom R$ is then a \indextwoshort{layer
    reaction}\introthree{family of layer reactions} from $A_0$ to
  $A_1$.
\end{definition}
In order to make the definition not unnecessarily specific, the
requirement that the situation families $A_0$ and $A_1$ consist of
intervals is not part of it. In the case that $A_0$ and $A_1$ are
actually families of intervals, we have usually $y_+ \in \dom R_+$ and
$y_- \in \dom R_-$. Then $y_+$ represents a step into the future and
$y_-$ the corresponding movement back to the past.

The following lemmas can be used to construct a family of layer
reactions from simpler reactions. We will use them to find layer
reactions for the interval families $F_0$ and $F_1$.
\begin{lemma}[Repeatable Reactions]
  \label{thm:existence-layers}
  Let $R$ be a reaction system. Then,
  \begin{subequations}
    \begin{align}
      \label{eq:layer-left}
      \text{if}\qquad
      a x &\rea_R y a,
      \qqtext{then}
      a x^k \rea_R y^k a
      &&\text{for all $k \geq 0$,} \\
      \label{eq:layer-right}
      \text{if}\qquad
      x a &\rea_R a y,
      \qqtext{then}
      x^k a \rea_R a y^k
      &&\text{for all $k \geq 0$.}
    \end{align}
  \end{subequations}
\end{lemma}
\begin{proof}
  The proof is by induction. The first equation is trivially true for
  $k = 0$. Now assume that $k \geq 0$ and $a x^k \rea_R x^k a$. Then
  there is a chain of reactions $a x^{k+1} = a x^k x \rea_R y^k a x
  \rea_R y^k y a = y^{k+1} a$. Therefore we have by induction $a x^k
  \rea y^k a$ for every $k \geq 0$. The second equation is proved in
  the same way.
\end{proof}

\begin{lemma}[Generators of Layer Reactions]
  \label{thm:layer-generation}
  Let $R$ be a reaction system which contains the reactions
  \begin{equation}
    a \rea_R a' c, \qquad
    c x \rea_R y c  \qqtext{and}
    c b \rea_R b'\,.
  \end{equation}
  Then $R$ has the family of layer reactions $\set{a x^k b \rea_R a'
    y^k b' \colon k \geq 0}$.
\end{lemma}
\begin{proof}
  By Lemma~\ref{thm:existence-layers}, applied to $c x \rea_R y c$, we
  have for every $k \geq 0$ a reaction $c x^k \rea_R y^k c$. We have
  then the chain of reactions $a x^k b \rea_R a' c x^k b \rea_R a' y^k
  c b \rea_R a' y^k b'$, which proves the lemma.
\end{proof}

The layer reactions for the interval families $F_0$ and $F_1$ are then
\begin{subequations}
  \label{eq:rule54-layers}
  \begin{alignat}{3}
    \label{eq:rule54-layer-0}
    L_0 &= \set{1 0^{k+2}1 &&\rea_\Phi 10 \op 1 0^k 1 \om 01
      &&\colon k\geq 0}, \\
    \label{eq:rule54-layer-1}
    L_1 &= \set{01^k0 &&\rea_\Phi 01 \op_2 1 0^k 1 \om_2 10
      &&\colon k\geq 2}\,.
  \end{alignat}
\end{subequations}
In the terminology of Definition~\ref{def:layers}, the set $L_0$ is a
family of layer reactions from $F_0$ to $F_0 \cup \{ 11 \}$, and $L_1$
is a family of layer reactions from $F_1$ to $F_0$. The set $F_0 \cup
\{ 11 \}$ is indeed a situation family: It is equal to $\set{1 0^k 1
  \colon k \geq 0}$.

For the proof of the first formula we use the reactions
\begin{equation}
  1 0^2 \rea_\Phi 10 \oplus 1 (\ominus 00), \quad
  (\ominus 00) 0 \rea_\Phi 0 (\ominus 00), \quad
  (\ominus 00) 1 \rea_\Phi 1 \ominus 01.
\end{equation}
The common part $c = \ominus 00$ of these reaction is put in
parentheses for better legibility. Then we can see that $a = 1 0^2$,
$a' = 10 \oplus 1$, $x = y = 0$, $b = 1$ and $b' = 1 \ominus 01$ in
the terminology of Lemma~\ref{thm:layer-generation}. Therefore there
is a family of layer reactions $\set{1 0^2 0^k 1 \rea_\Phi 10 \oplus 1
  0^k 1 \ominus 01 \colon k \geq 0}$ under Rule 54, the same as
in~\eqref{eq:rule54-layer-0}.

The second formula is derived from the reactions
\begin{multline}
  01^2 \rea_\Phi 01 \oplus_2 1 (00 \ominus_2 11), \quad
  (00 \ominus_2 11) 1 \rea_\Phi 0 (00 \ominus_2 11), \\
  (00 \ominus_2 11) 0 \rea_\Phi 0^2 1 \ominus_2 10.
\end{multline}
The common part of these reactions is once again put into parentheses.
We can see from these reactions that $a = 01^2$, $a' = 01 \oplus_2 1$,
$x = 1$, $y = b = 0$ and $b' = 0^2 1 \ominus_2 10$ in the terminology
of Lemma~\ref{thm:layer-generation}. Therefore there is a family of
reactions $\set{01^2 1^k 0 \rea_\Phi 01 \oplus_2 1 0^k 0^2 1 \ominus_2
  10 \colon k \geq 0}$ under Rule 54, the same as
in~\eqref{eq:rule54-layer-1}.

\paragraph{Triangles} As a final step in this subproject of finding
general laws for the evolution of finitely often repeated patterns in
the initial configuration, we now trace their evolution over as many
time steps as possible. This is done with \emph{triangle reactions}.

\begin{figure}[ht]
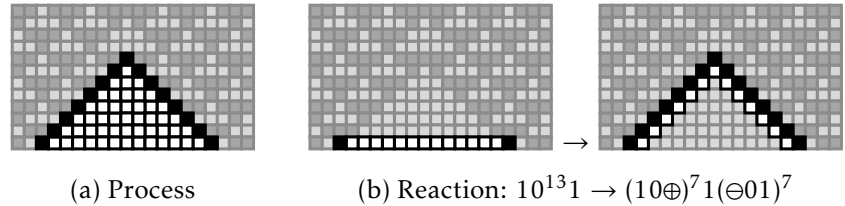

  \centering
  \tabcolsep1em
  \def\arraystretch{1.5}
  \begin{tabular}{cc}
    \ci{triangle} & \ci{triangle_reaction} \\
    (a) Process &
    (b) Reaction: $1 0^{13} 1 \rea (10\op)^7 1 (\om 01)^7$
  \end{tabular}
  \caption{Triangle process and triangle reaction under Rule 54.}
  \label{fig:triangles}
\end{figure}

An example for a triangle reaction under Rule 54 is shown in
Figure~\ref{fig:triangles}. It is a copy of Figure~\ref{fig:slopes},
but now it shows the reaction in the new notation, as $1 0^{13} 1
\rea_\Phi (10\op)^7 1 (\om 01)^7$. The evolution diagram at the left
side shows the closure of the process for initial situation $1 0^{13}
1$. The two diagrams at the right side show the processes of the input
and the result situation of the triangle reaction: they are the edges
of the triangle process at the right. If we view the triangle as a
temporal process, then its input situation shows two ``particles'',
the interval situations $10$ and $01$, that are positioned at the
boundaries of a sequence of $11$ cells in state $0$; the result
situation shows how the particles move towards each other until they
collide. This similarity to the collision of macroscopic particles
makes triangle reactions a promising tool for understanding the
behaviour of cellular automata.\footnote{In the typical behaviour of
  Rule 54, the structures $10$ and $01$ are very short-lived
  \cite{Boccara1991}. For it and other cellular automata, the word
  ``particle'' means therefore generally larger, longer-lived
  structures. Nevertheless, the interpretation given here looks like a
  useful generalisation.}

\begin{figure}[ht]
  \centering
  \begin{tikzpicture}[
    direct/.style={line width=.75ex},
    somewhere/.style={direct, draw=shaded, decorate},
    decoration={random steps, amplitude=1.5, segment length=5}]
    \matrix [cells=midway] {
      \path   (0,0) coordinate (a) ++(11,0) coordinate (b')
              (a)  ++(1,0) coordinate (a')
              (b') ++(-1,0) coordinate (b);
      \draw[direct]
              (a) -- node[below] {$a$} (a')
              (b) -- node[below] {$b$} (b');
      \draw[direct,draw=shaded]
              (a') -- node[below] {$x^k$} (b);
      & \node {$\rea$}; &
      \path   (0,0) coordinate (yplus) ++(11,0) coordinate (yminus')
              let \p1 = (yminus'), \n1 = {(\x1 - 2.5ex) / 2} in
              (yplus)   ++( \n1,\n1) coordinate (c)
              (yminus') ++(-\n1,\n1) coordinate (c');
      \draw[direct]
              (c) -- node[below] {$c$} (c');
      \draw[direct, draw=shaded]
              (yplus) -- node[above] {$y_+^k$} (c)
              (c')    -- node[above] {$y_-^k$} (yminus');
            \\
    };
  \end{tikzpicture}
  \caption{Generic form of a triangle reaction.}
  \label{fig:triangle}
\end{figure}
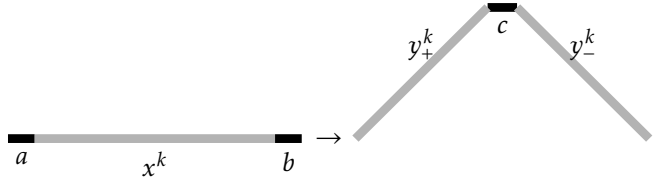

The following definition of triangle reactions harmonises with the
definition of layer reactions. As before, the definition ignores the
temporal aspect of the triangle reactions: In the cases that interest
us here most, the situations $a$, $x$ and $b$ are intervals, while
$y_+$ is an element of $\dom R_+$ and $y_-$ of $\dom R_-$.
\begin{definition}[Triangle Reactions]
  \label{def:triangle}
  Let $R$ be a reaction system. Let $A = \set{a x^k b \colon k \geq 0}
  \subseteq \dom R$ be a family of situations. A \indextwoshort{triangle
    reaction}\introthree{family of triangle reactions} for $A$ is a
  set of reactions (Figure~\ref{fig:triangle}),
  \begin{equation}
    \label{eq:triangle-reaction}
    \set{a x^k b \rea_R y_+^k c y_-^k \colon k \geq 0}\,.
  \end{equation}
\end{definition}

As before with layers, there are many equivalent forms to represent
triangle reactions. We are especially interested in the apparently
more general form
\begin{equation}
  \label{triangle-generalised}
  \set{a x^{kn + i} b \rea_R
    y_+^{k+k_0} c y_-^{k+k_0} \colon k \geq 0}
\end{equation}
with $n$, $i$ and $k_0 \in \N_0$. This set of reactions can be brought
into the same form as~\eqref{eq:triangle-reaction} by writing it as
$\set{(a x^{i}) (x^n)^k b \rea_R y_+^k \,(y_+^{k_0} c y_-^{k_0})\,
  y_-^k \colon k \geq 0}$ and so be shown to be equivalent to it. We
can then bring the reaction of Figure~\ref{fig:triangles} into the
form~\eqref{triangle-generalised} by writing it as $1 0^{2 \times 6
  + 1} 1 \rea_\Phi (10 \oplus)^{6+1} 1 (\ominus 01)^{6+1}$.

Triangle reactions can be derived from layer reactions. If $A = \set{a
  x^k b \colon k \geq 1}$ is a family of situations and $L$ is a
family of layer reactions from $A$ to $A \cup \{ ab \}$, then there is
a family of triangle reactions for $A$. The following lemma shows why.
(The requirement in the lemma that there is a reaction $a b \rea _Rc$
is no restriction, since one can always set $c = a b$.)
\begin{lemma}[Generators of Triangle Reactions]
  \label{thm:triangle-reactions}
  Let $R$ be a reaction system. If there is a reaction $a b \rea _Rc$
  and a family of layer reactions
  \begin{equation}
    L = \set{ a x^{k+1} b \rea_R y_+ a x^k b y_- \colon k \geq 0},
  \end{equation}
  then there exists in $R$ a family of triangle reactions
  \begin{equation}
    \set{a x^k b \rea_R y_+^k c y_-^k \colon k \geq 0}\,.
  \end{equation}
\end{lemma}
\begin{proof}
  Let a $k \geq 0$ be given. We can then apply for $k$ times one of
  the reactions of $L$, first to the situation $a x^k b$, and then
  always to the result of the previous reaction. Then we have found a
  reaction $a x^k b \rea_R y_+^k a b y_-^k$. Next we apply the
  reaction $a b \rea _Rc$ to the result of this reaction and get $a
  x^k b \rea_R y_+^k c y_-^k$.
\end{proof}

An interesting phenomenon occurs when the layer reaction consumes more
that one copy of the repeated pattern $x$. Then the one family of
layer reactions splits into several sets of triangle reactions. This
occurs especially in the family of layer reactions $L_0$ for Rule 54.
\begin{lemma}[Multiple Triangle Reactions]
  \label{thm:triangle-reactions-multi}
  Let $R$ be a reaction system that contains a family of layer
  reactions,
  \begin{equation}
    L = \set{a x^{k+n} b \rea_R y_+ a x^k b y_- \colon k \geq 0},
  \end{equation}
  and the reactions $a x^{i} b \rea_R c_i$ for $i = 0$, \dots, $n -
  1$. Then $R$ contains for every value of $i$ a family of triangle
  reactions,
  \begin{equation}
    \label{eq:triangle-reactions-multi}
    \set{a x^{kn + i} b \rea_R y_+^k c_i y_-^k \colon k \geq 0}\,.
  \end{equation}
\end{lemma}
\begin{proof}
  We will write the terms $x^k$ and $x^{k+n}$ in differing ways,
  depending on the value of $k$. If there is a $j \in \N_0$ such that
  $k = jn + i$ and $i < n$, then $x^k = (x^n)^j$ and $x^{k+n} = x^i
  (x^n)^j$. Therefore $L$ is the disjoint union of $n$ families of
  layer reactions,
  \begin{equation}
    L_i = \set{(a x^i) (x^n)^j b \rea_R y_+ a (x^n)^j b y_-
      \colon j \geq 0}
    \qquad\text{for $i = 0$, \dots, $n - 1$}.
  \end{equation}
  Then we can apply the previous lemma to each of the reactions
  families $L_i$. We get for every value of $i$ a family of triangle
  reactions, $\set{(a x^i) (x^n)^k b \rea_R y_+^k c_i y_-^k \colon j
    \geq 0}$, the same as in~\eqref{eq:triangle-reactions-multi}.
\end{proof}

Now we will derive triangles and the triangle reaction for Rule 54.
Among the two families of layer reactions only $L_0$ fulfils the
requirement of Lemma~\ref{thm:triangle-reactions-multi}. So we set
$L_0 = \set{1 0^{k+2}1 \rea_\Phi 10 \op 1 0^k 1 \om 01 \colon k\geq
  0}$ for $L$ and choose as the finishing reactions $a x^{i} b \rea_R
c_i$ the two reactions $11 \rea_\Phi 11$ and $101 \rea_\Phi 10 \oplus
1 \ominus 01$. Then we can apply the lemma and get the following
families of triangle reactions,
\begin{subequations}
  \label{eq:triangle-raw}
  \begin{alignat}{3}
    T_{00} &= \set{1 0^{2k} 1 && \rea_\Phi (10 \oplus)^k 11 (\ominus 01)^k
      &&\colon k \geq 0}, \\
    \label{eq:00-block-odd}
    T_{01} &= \set{1 0^{2k + 1} 1 && \rea_\Phi (10 \oplus)^k (10 \oplus
      1 \ominus 01) (\ominus 01)^k &&\colon k \geq 0}\,.
  \end{alignat}
\end{subequations}
We can view these two families of reactions as an improved version of
the layer reactions $L_0$, and as their replacement. For $L_1$ no such
replacement has been found, therefore the general laws for Rule 54
that we have found with the methods of this section are $T_{00}$,
$T_{01}$ and $L_1$. They are listed in Table~\ref{tab:large-scale}.
\begin{table}[t]
  \centering
  \caption{Large-scale reactions under Rule 54}
  \label{tab:large-scale}
  \def\a{$&${}}               
  \begin{tabular}{c*3{l@{}}}
    \toprule
    Name & \multicolumn3l{Family of reactions} \\
    \midrule
      $T_{00}$ & $\set{1 0^{2k} 1
        \a \rea_\Phi (10 \oplus)^k 11 (\ominus 01)^k
        \a \colon k \geq 0}$ \\
      $T_{01}$ & $\set{1 0^{2k+1} 1
        \a \rea_\Phi (10 \oplus)^{k+1} 1 (\ominus 01)^{k+1}
        \a \colon k \geq 0}$ \\
      $L_1$ & $\set{0 1^k 0
        \a \rea_\Phi 01 \op_2 1 0^k 1 \om_2 10
        \a \colon k\geq 2}$ \\
      \bottomrule
  \end{tabular}
\end{table}

\section{What Is the Ether?}

Now we can try to understand ether formation under Rule 54. For this
we need to express explicitly what the ether is. We will do this first
in terms of configurations and then in terms of situations and
reactions.

In Figure~\ref{fig:random_54} at the beginning we have seen an example
of ether formation. In it we can now recognise how, when starting from
a random initial configuration, large regions that consist of a
regular pattern begin to form. We see that this pattern consists at
alternating times of a configuration in which blocks of three cells in
state $0$ are separated by one cell in state $1$, and of a
configuration in which three cells in state $1$ are separated by one
cell in state $0$. After two time steps the same patterns arise again,
but but shifted to the side by a distance of two cells, such that a
true repetition only occurs after $4$ time steps. We will speak of the
two configurations as the \introx[phase]{phases} of the ether, in
analogy to the usage of Martínez et al.\ \cite{Ju'arezMart'inez2008}
for Rule 110.

We also see that there are several such ether regions in
Figure~\ref{fig:random_54}. They are separated by larger structures
between them, or sometimes just by phase differences. This is a
typical phenomenon in ether formation: when starting from a random
configuration one almost never gets a ``pure'', or empty ether.
Instead one gets large regions with a regular pattern that are
separated by disturbances. Over time the ether regions coalesce and
the distance between the disturbances increases. This has already been
noted by Boccara, Nasser and Roger \cite{Boccara1991}. We will
therefore first describe an empty ether and then turn to the more
realistic case of an ether with disturbances.

\paragraph{The Empty Ether} A configuration of empty ether then consists
of an infinite repetition of one of the patterns $0^3 1$ and $1^3 0$.
In contrast to configurations, situations are always finite. We could
now represent a finite part of such an ether configuration by a
situation of the form $(0^3 1)^k$ or $(1^3 0)^k$, depending on the
phase. We choose however the family of situations $\set{ 1 (0^3 1)^k
  \colon k \geq 0}$ as our standard representative of an ether
configuration. The reason for the choice of this phase and for adding
a $1$ at the left is that then the situation $1 0^3 1$ is an element
of the family and that this situation is the input of one of the
triangle reactions of the family $T_{01}$, namely $1 0^3 1 \rea_\Phi
(10 \oplus)^2 1 (\ominus 01)^2$.

In order to get a notation in which no overlapping situations occur we
will write this reaction in a slightly different form,\footnote{The
  right site of the reaction will be later decomposed into the
  situations $1$, $(0 \oplus 1)^2$ and $(\ominus 01)^2$. These
  situations do not overlap, even if two of their \emph{processes},
  namely $\prx{1}{(0 \oplus 1)^2}$ and $\prx{1 (0 \oplus
    1)^2}{(\ominus 01)^2}$, do overlap.}
\begin{equation}
  \label{eq:ether-triangle}
  1 0^3 1 \rea_\Phi 1 (0 \oplus 1)^2 (\ominus 01)^2\,.
\end{equation}
The next step to construct a description of the empty ether is then to
find reactions similar to~\eqref{eq:ether-triangle} for all intervals
of the form $1 (0^3 1)^k$. To do this we need the following lemma,
which provides a kind of converse to the
reaction~\eqref{eq:ether-triangle}.

\begin{lemma}[Converse Triangle Reaction]
  \label{thm:converse-triangle}
  \begin{equation}
    \label{eq:converse-triangle}
    1 (\ominus 01)^2 (0 \oplus 1)^2 \rea_\Phi 1 0^3 1\,.
  \end{equation}
\end{lemma}
\begin{proof}
  First we show that $1 \ominus 010 \oplus 1 \rea_\Phi 1^3$. This is
  true because
  \begin{equation}
    \label{eq:converse-1}
    \underline{1 \ominus 010} \oplus 1
    \rea_\Phi 11 \underline{1 \ominus_2 10 \oplus 1}
    \rea_\Phi 111;
  \end{equation}
  the terms that change in the next reaction step are underlined. We
  have then also shown that $1 (\ominus 01)^2 (0 \oplus 1)^2 \rea_\Phi
  1 \ominus 0 1^3 1 \oplus 1$ and only must verify that the reaction
  product reacts to $1 0^3 1$:
  \begin{align}
    \label{eq:converse-2}
    \underline{1 \ominus 01} 1 \underline{10 \oplus 1}
    &\rea_\Phi 1 \underline{00 \ominus_2 111} \oplus_2 001 \\
    &\rea_\Phi 10 \underline{00 \ominus_2 11 \oplus_2 00} 1
    \rea_\Phi 10001\,.
  \end{align}
  This proves the lemma.
\end{proof}

We now combine~\eqref{eq:converse-triangle}
and~\eqref{eq:ether-triangle} to a reaction in which the interval $1
0^3 1$ does no longer occur,
\begin{align}
  \label{eq:ether-square}
  1 (\ominus 01)^2 (0 \oplus 1)^2
  \rea_\Phi  1 (0 \oplus 1)^2 (\ominus 01)^2\,.
\end{align}
The reaction~\eqref{eq:ether-triangle} and this reaction are the
building blocks for the larger reactions that originate from the $1
(0^3 1)^k$ terms. In order to express them better we introduce the
abbreviations
\begin{equation}
  \label{eq:epsilon-def}
  \epsilon_- = \ominus 01
  \qqtext{and}
  \epsilon_+ = 0 \oplus 1\,.
\end{equation}
Note that these terms are not achronal. In the formalism they can
therefore appear only as factors of situations in $\dom \Phi$, not as
terms in their own right.

With the abbreviations of~\eqref{eq:epsilon-def} we will now summarise
the reactions that characterise the empty ether.

\begin{definition}[Basic Ether Reactions]
  \label{def:basic-ether}
  The \intro{basic ether reactions} for Rule 54 are
  \begin{equation}
    \label{eq:basic-ether}
    1 0^3 1 \rea_\Phi 1 \epsilon_+^2 \epsilon_-^2
    \qqtext{and}
    1 \epsilon_+^2 \epsilon_-^2 \rea_\Phi  1\epsilon_+^2 \epsilon_-^2\,.
  \end{equation}
\end{definition}

\begin{figure}[ht]
  \centering
  \def\arraystretch{1.1}
  \begin{tabular}{c}
    \ci{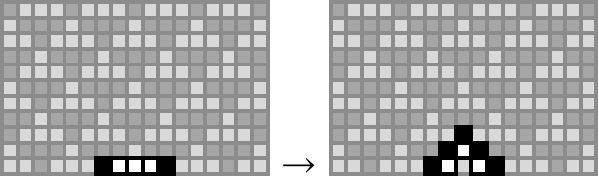} \\
    $1 0^3 1 \rea 1 \epsilon_+^2 \epsilon_-^2$ \\[2ex]
    \ci{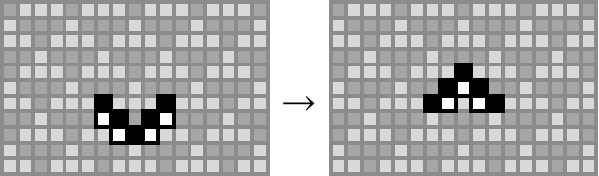} \\
    $1 \epsilon_-^2 \epsilon_+^2 \rea 1 \epsilon_+^2 \epsilon_-^2$
  \end{tabular}
  \caption{Basic ether reactions on a background of empty ether.}
  \label{fig:basic-ether}
\end{figure}

The basic ether reactions are shown in Figure~\ref{fig:basic-ether}.
The cells of the ether situations are shown in black and white, and
the background in grey tones is a finite part of the empty ether.

From these two reactions we will now derive two sets of reactions that
involve larger parts of the empty ether. The set one describes the
evolution of larger segments of the initial configuration, and the
second one is a description of the evolution of a part of the empty
ether.

\begin{theorem}[Ether Reactions]
  \label{thm:ether-reactions}
  \begin{subequations}
    \label{eq:ether-reactions}
    \begin{align}
      \label{eq:ether-init}
      1 (0^3 1)^k & \rea_\Phi 1 \epsilon_+^{2k} \epsilon_-^{2k}
      &&\text{for $k \geq 0$,}\\
      \label{eq:ether-later}
      1 \epsilon_-^{2k} \epsilon_+^{2\ell}
      & \rea_\Phi 1 \epsilon_+^{2\ell} \epsilon_-^{2k}
      &&\text{for $k$, $\ell \geq 0$.}
    \end{align}
  \end{subequations}
\end{theorem}
\begin{proof}
  We need for this proof a reaction that is derived from the reaction
  at the right side of~\eqref{eq:basic-ether} with the help of
  Lemma~\ref{thm:existence-layers}.

  In order to apply this lemma to the reaction $1 \epsilon_-^2
  \epsilon_+^2 \rea_\Phi 1 \epsilon_+^2 \epsilon_-^2$ we need to write
  it in the form $x a \rea_\Phi a y$. We do this by resolving the
  $\epsilon_-$ terms at the left side of the reaction. Then we get $1
  \ominus 01 \ominus 0 (1 \epsilon_+^2) \rea_\Phi (1 \epsilon_+^2)
  \epsilon_-^2$, with the repeated factor $a$ put in parentheses to
  emphasise it. Now we can apply~\eqref{eq:layer-right} and get a
  reaction $(1 \ominus 01 \ominus 0)^k 1 \epsilon_+^2 \rea_\Phi 1
  \epsilon_+^2 (\epsilon_-^2)^k$ for every $k \geq 0$. Then we rewrite
  the left side of this reaction with $\epsilon_-$ terms. The result
  is
  \begin{align}
    \label{eq:ether-immediate}
    1 \epsilon_-^{2k} \epsilon_+^2
    &\rea_\Phi 1 \epsilon_+^2 \epsilon_-^{2k}
    &&\text{for all $k \geq 0$.}
  \end{align}
  We use it to prove the reactions in~\eqref{eq:ether-reactions} by
  induction over $k$. For both families of reactions the case $k = 0$
  is trivially true, and for $k \geq 1$ we get
  \begin{subequations}
    \begin{align}
      1 (0^3 1)^k
      &\rea_\Phi 1 (0^3 1)^{k-1} \epsilon_+^2 \epsilon_-^2 \notag \\
      &\rea_\Phi 1 \epsilon_+^{2(k-1)} \epsilon_-^{2(k-1)}
      \epsilon_+^2 \epsilon_-^2
      &&\text{by induction,} \notag \\
      &\rea_\Phi 1 \epsilon_+^{2(k-1)} \epsilon_+^2
      \epsilon_-^{2(k-1)} \epsilon_-^2
      &&\text{with~\eqref{eq:ether-immediate}} \notag \\
      &= \epsilon_+^{2k} \epsilon_-^{2k}, \\
      1\epsilon_-^{2k} \epsilon_+^{2\ell}
      &\rea_\Phi 1\epsilon_-^2 \epsilon_+^{2\ell} \epsilon_-^{2(k-1)}
      &&\text{by induction,} \notag \\
      &\rea_\Phi 1\epsilon_+^{2\ell} \epsilon_-^2 \epsilon_-^{2k-1}
      &&\text{with~\eqref{eq:ether-immediate}.} \notag \\
      &= 1\epsilon_+^{2\ell} \epsilon_-^{2k}\,.
    \end{align}
  \end{subequations}
  This then proves the theorem.
\end{proof}

The reactions of Theorem~\ref{thm:ether-reactions} now serve as an
inspiration for the way in which one can express the disturbed ether
that arises from a random initial configuration in terms of situations
and reactions.

\paragraph{Random Initial Situations} We need a method to express with
the help of situations the behaviour of random initial configurations.
First we define what we mean by a \introtwo{random initial
  configuration}. Here we assume that the states of a configuration $c
\in \Sigma^\Z$ are chosen at random in such a way that the state of
every cell is equal to $1$ with probability $p_1$ and that the states
of all cells are independent of each other. We will always exclude the
trivial cases $p_1 = 0$ and $p_1 = 1$.

In the language of probability theory \cite[Chapter 1]{Grimmett1989}
we have then performed a \introtwo{random experiment}. The random
choice of a configuration is represented by a \introtwo{random
  variable} $C$ with values in $\Sigma^\Z$. As it is usual in
probability theory, this and other random variables are written in
capital letters. The configuration $c$ mentioned above is then a
possible \intro{outcome} of the experiment. The probabilities that
define the random experiment will however refer to whole sets of
outcomes. These sets are called in probability theory ``events'', but
here they will be called \indextwoshort{stochastic
  event}\emph{stochastic events}, to avoid confusion with the already
existing use of the word ``event'' in the context of cellular
processes.

First we must now specify the probabilities for the outcomes of $C$
and make precise the informal definition given above. We can express
the random choice of $C$ by saying that for all $x \in \Z$,
\begin{equation}
  \label{eq:measure}
  P(C(x) = 1) = p_1,
\end{equation}
In a more formal way of speaking, this equation assigns the
probability $p_1$ to the set of outcomes, or stochastic event, $\set{c
  \in \Sigma^\Z \colon c(x) = 1}$. Other expressions with random
variables are understood in a similar way.

From~\eqref{eq:measure} we can derive the probailities for other
stochastic events. These are the events that belong to the
$\sigma$-field \cite[p.~19]{Grimmett1989} that is generated by the
sets $\set{c \in \Sigma^\Z \colon c(x) = 1}$. We will however not use
this $\sigma$-field explicitly and work with probabilities in a less
formal way.

For the work with situations we can only use finite pieces of the
initial configuration $C$. So we define now the sequence $(U_n)_{n
  \geq 0}$ of \indextwo{random interval}\emph{random intervals}. Each
$U_n$ is a random variable with values in $\Sigma^n$ and defined by
the relation
\begin{align}
  \label{eq:random-intervals}
  U_n(c) = C(-n + 1) \dots C(1) C(0)\,.
\end{align}
With this definition we have a growing sequence of random cellular
processes,
\begin{equation}
  \label{eq:random-inclusion}
  \dots \supset \pr([-n] U_n) \supset
  \dots \supset \pr([-1] U_{1}) \supset \pr([0]U_0) = \emptyset,
\end{equation}
that ultimately incorporate all cells of $\Sigma^Z$ with indices $\leq
0$. For each $n$, the closure of $\pr([-n]U_n)$ then represents a view
into the evolution of an element of $C$. The larger $n$ becomes, the
larger the window on $C$ becomes. All these windows contain only
events with non-positive space coordinates, but this does not matter,
since the probability distribution for $C$ is invariant under
left-to-right shifts.

Later we will need to do induction proofs over the length of the
intervals $U_n$. In them we need to express the sequence $(U_n)_{n
  \geq 0}$ in a recursive way. To do this we now imagine the
construction of the $U_n$ as a stochastic process in which the random
interval $U_{n-1}$ is extended by an event that is in state $1$ with
probability $p_1$ and in state $0$ with probability $1 - p_1$. We can
then express the values of the $P(U_n)$ in the language of conditional
probability with the equations
\begin{equation}
  \label{eq:random-recursion}
  P(U_n = \sigma u \mid U_{n-1} = u) =
  \begin{cases}
    p_1 & \text{if $\sigma = 1$,} \\
    (1 - p_1) & \text{if $\sigma = 0$,}
  \end{cases}
\end{equation}
valid for all $n \geq 1$ and $u \in \Sigma^n$. The starting point of
this recursion is the trivial case of $n = 0$, with $P(U_0 = [0]) =
1$.

\paragraph{Ether Fragments} We also need a way to express the fact
that the closure of $\pr(U_n)$ contains fragments of the empty ether.
To do this we will first consider reactions of the form
\begin{align}
  \label{eq:irregular-triangle}
  U_n \rea a_+ a_-
\end{align}
with $a_+ \in \dom \Phi_+$ and $a_- \in \dom \Phi_-$. They are a
generalisation of the triangle reactions~\eqref{eq:ether-init} for the
empty ether. Since however $U_n$ has in general not the form $1 (0^3
1)^k$, the situation $a_+$ will in general not consist entirely of
$\epsilon_+$ terms and $a_-$ not only of $\epsilon_-$ terms. Our goal
in this chapter then is to prove that for large $n$, the closure of
$\pr(U_n)$ nevertheless contains pieces of the ether.

We will therefore look for reactions of the
form~\eqref{eq:irregular-triangle} in which $a_+$ contains factors of
the form $1 \epsilon_+^2$. The exponent of $2$ in $\epsilon_+^2$
occurs because all the ether reactions in~\eqref{eq:ether-reactions}
involve only even numbers as exponents for $\epsilon_+$ and
$\epsilon_-$.
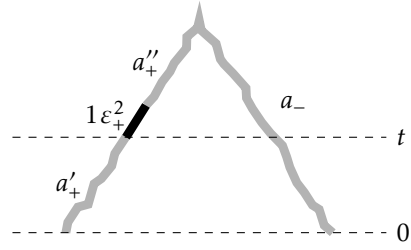
\begin{figure}[ht]
  \centering
  \begin{tikzpicture}[
    direct/.style={line width=.75ex},
    somewhere/.style={direct, draw=shaded, decorate},
    decoration={random steps, amplitude=1.5, segment length=5}]
    \path let \n{x} = 5, \n{y} = {1.6 * \n{x}} in
       (2,0) coordinate (a'plus)
       ++(\n{x}, \n{y}) coordinate (aminus)
       ++(\n{x}, -\n{y}) coordinate (aminus')
       ++(2, 0) coordinate (end)
       ($(a'plus)!0.45!(aminus)$) coordinate (a'plus')
       ($(a'plus)!0.6!(aminus)$) coordinate (a''plus);
    \draw[somewhere]
       (a'plus)  -- node [left] {$a'_+$} (a'plus')
       (a''plus) -- node [left] {$a''_+$} (aminus)
                 -- node [above right] {$a_-$} (aminus');
    \draw[direct]
       (a'plus') -- node[left, pos=.7] {$1 \epsilon_+^2$} (a''plus);
    \draw[dashed] let \p{right} = (end), \p{t} = (a'plus') in
       (0, 0)     -- (\x{right}, 0)     node [right] {$0$}
       (0, \y{t}) -- (\x{right}, \y{t}) node [right] {$t$};
  \end{tikzpicture}
  \caption{Location of the ether fragments
    of~\eqref{eq:ether-fragment} in the situation $a_+ a_-$.}
  \label{fig:ether-fragment}
\end{figure}
We will say that an \intro{ether fragment} occurs at time $t$ if
\begin{equation}
  \label{eq:ether-fragment}
  a_+ = a'_+ 1 \epsilon_+^2 a''_+
  \qqtext{and}
  \delta(a'_+)_T = t\,.
\end{equation}
The situation $a_+ a_-$ serves here as a probe into the closure of
$\pr(U_n)$. We can then use it as a means to express ether formation
in such a way that it can be proved with the help of situations and
reactions.

We could also have looked for reactions in which $a_-$ contains
factors of the form $\epsilon_+^2 1$. The results would be equivalent
because Rule 54 is symmetric under exchange of left and right. The
advantage of the definition we use here is that with it we can express
more easily the time at which an ether fragment occurs.

\paragraph{Reactions and Probability} Now we will consider arbitrary
reactions of the form $U_n \rea_\Phi a$ for a given situation $a$.
Whether such a reaction is possible depends on the value of $U_n$. We
will write the probability of such a reaction as $P(U_n \rea_\Phi a)$;
it is the probability that $U_n$ belongs to the set of situations $u
\in \Sigma^n$ for which there is a reaction with result $a$. So we can
write
\begin{equation}
  \label{eq:reaction-probability}
  P(U_n \rea_\Phi a) =
  P(\exists u \in \Sigma^n \colon u \rea_\Phi a)\,.
\end{equation}

A generalisation of this is the case where a random interval may react
to a whole set of possible outcomes. When $\mathcal{A} \subseteq
\mathcal{S}$ is a set of situations, we will write
\begin{equation}
  \label{eq:set-probability}
  P(U_n \rea_\Phi \mathcal{A}) =
  P(\exists u \in \Sigma^n, a \in \mathcal{A} \colon u \rea_\Phi a)\,.
\end{equation}
This definition counts the probability that the random situation $U_n$
can react to at least one element of $\mathcal{A}$. So it is
meaningful even if there is more than one possible reaction result for
a certain value of $u$. We have therefore always $P(U_n \rea_\Phi
\mathcal{S}) = 1$.

\paragraph{Notation for Fragments} The definitions of the last
paragraphs are aimed at a special kind of situations that are needed
in the following proof. We need to express that a certain situation
$f$, the ``fragment'', occurs at one or more specified time steps. The
set of situations that have in common a certain fragment $f \in
\mathcal{S}$, beginning at one of the time steps $t_1$, \dots, $t_n$,
is then written
\begin{multline}
  \label{eq:fragments}
  \mathcal{F}(f, t_1, \dots, t_n) =
  \set{ a_+ f a' \colon
    a_+ \in \dom \Phi_+, a' \in \dom \Phi, \\
    \delta(a_+)_T \in \{ t_1, \dots t_n \}}\,.
\end{multline}
Among these sets of situations, the most important one is that in
which $f$ is an ether fragment; it is called
\begin{equation}
  \label{eq:R-slope}
  \mathcal{E}(t_1, \dots, t_n) =
  \mathcal{F}(1 \epsilon_+^2, t_1, \dots, t_n)\,.
\end{equation}

\section{The Ether Is Inevitable}

The following theorem is a way to express the necessity of ether
formation in a weak form. It shows that for a random initial
configuration, ether fragments can be found arbitrarily far in the
future. Ether fragments never vanish totally, and we can view that as
a sign (but not a proof) that the ether, too, persists.

\begin{theorem}[Ether Formation]
  \label{thm:ether-formation}
  Let $\epsilon > 0$ be a probability. The random intervals $(U_k)_{k
    \geq 0}$ are defined as in~\eqref{eq:random-intervals}. Then for
  every $t \in \N$ there is an $n \in \N$ such that
  \begin{equation}
    \label{eq:ether-formation}
    P(U_n \rea_\Phi \mathcal{E}(t, t + 1, t + 2)) \geq 1 - \epsilon\,.
  \end{equation}
\end{theorem}

Since $U_n$ is part of the random initial configuration $C$, the
theorem shows that every given sequence of three time steps contains
with probability $1$ the starting time of an ether fragment. Moreover,
$C$ consists of infinitely many intervals of length $n$ and these
intervals are stochastically independent of each other. Therefore the
theorem also shows that during every sequence of three steps time,
infinitely many ether fragments begin.

The theorem will now be proved with help of several lemmas. Among
them, Lemma~\ref{thm:fragment-formation} shows how ether fragments
arise from the initial configuration, and
Lemma~\ref{thm:fragment-propagation} shows how they propagate to later
times.

\begin{lemma}[Formation of Ether Fragments]
  \label{thm:fragment-formation}
  Let $(U_k)_{k \geq 0}$ be the sequence of random intervals
  from~\eqref{eq:random-intervals}. Let $n \in \N_0$ and let $u \in
  \Sigma^n$ be an interval. Then for every $\epsilon > 0$ there is an
  integer $m > n$ such that
  \begin{equation}
    \label{eq:fragment-formation}
    P(U_m \rea_\Phi \mathcal{E}(0) \mid U_n = u)
    \geq 1 - \epsilon\,.
  \end{equation}
\end{lemma}

\begin{proof}
  When extending a random interval $U_k$ by $5$ events to $U_{k+5}$,
  the probability that $U_{k+5}$ begins with $1 0^3 1$ is
  \begin{equation}
    \label{eq:extension-by-5}
    P(U_{k+5} = 1 0^3 1 U_k) = p_1^2 (1 - p_1)^3,
  \end{equation}
  independent of the value of $U_k$. We will call this probability
  $p_{10^31}$. Then $p_{10^31} > 0$ because we have required earlier
  that $p_1 \notin \{0, 1\}$.

  Let now $k_0 \geq 0$ be an integer. Then the probability that one of
  the random intervals $U_{n + 5k}$ with $k \leq k_0$ begins with $1
  0^3 1$ is $(1 - p_{10^31})^{k_0}$. These probabilities are
  independent of each other because each of them only depends on the
  states of the cells at the locations $-(n + 5k) + 1$, \dots, $-(n +
  5k) + 5$, and those intervals do not overlap.\footnote{Note that the
    intervals $U_k$ grow to the left, as described in
    \eqref{eq:random-intervals}.}

  Since $1 - p_{10^31} < 1$, we can find a $k_0$ such that $(1 -
  p_{10^31})^{k_0} < \epsilon$. Now we can set $n_0 = n + 5 k_0$. Then
  with a probability greater than $1 - \epsilon$ there is a $m \leq
  n_0$ such that $U_m$ begins with $1 0^3 1$. With such an $m$ we also
  have a reaction $U_m \rea_\Phi 1 \epsilon_+^2 \epsilon_-^2 U_{m-5}$.
\end{proof}

So we see that as $n$ approaches infinity, there is always an ether
fragment $1 \epsilon_+^2$ at time $0$ that is generated by $U_n$. Next
we will prove that these fragments are almost always copied by other
reactions to space-time locations at later times.

For this proof we will need reactions that create a situation with an
interval of a given minimal length and at a specified time as a factor.
Furthermore, initial and final situation of the reaction must be
positive slopes. This means that the required reaction must be a part
of the positive slope reaction system~$\Phi_+$.

The following lemma then tells us how we can find reactions that
create such an interval from a given situation $a_+ \in \dom \Phi_+$.
When the conditions of the lemma are met, we will say that every $u$
\introx[enforce]{enforces} $v$.

\begin{lemma}[Enforcement of Intervals]
  \label{thm:propagate-interval}
  Let $a_+ \in \dom \Phi_+$ be chosen arbitrarily. Then for every
  $\ell > 0$ there is a $k > 0$ such that for all $u \in \Sigma^*$
  with $\abs{u} \geq k$ there is a reaction of the form
  \begin{equation}
    \label{eq:propagate-interval}
    u a_+ \rea_{\Phi_+} b_+ v
  \end{equation}
  with $b_+ \in \dom \Phi_+$ and $v \in \Sigma^\ell$.
\end{lemma}

\begin{proof}
  We first consider the case where $\ell = 1$ and $\delta(a_+)_T = 1$.
  Because $a_+$ is achronal we must then have a decomposition $a_+ =
  u_1 s_+ u_2$, with $u_1$, $u_2 \in \Sigma^*$ and $s_+$ an element of
  the set of positive generating slopes for Rule 54. They are found in
  Table~\ref{tab:steps54}, and we will refer to them as the set
  \begin{equation}
    \label{eq:positive-steps}
    S_+ =
    \{ 00 \oplus, 01 \oplus_2 1, 10 \oplus 1, 11 \oplus_2 00 \}\,.
  \end{equation}
  The worst possible case for the proof occurs when both $u_1$ and
  $u_2$ are the empty situation. So we may assume without loss of
  generality that $a_+ \in S_+$. Now, by checking the generating
  reactions in Table~\ref{tab:generator54}, we see that as long as
  $s_+ \neq 01 \oplus_2 1$, they always have the form $\sigma s_+
  \rea_{\Phi_+} s'_+ v_1$, with $\sigma \in \Sigma$, $v_1 \in
  \Sigma^*$ and $\abs{v} \geq 1$. The two remaining generating
  reactions, those with $s_+ = 01 \oplus_2 1$, have the form $\sigma
  s_+ \rea_{\Phi_+} s'_+$, with $s'_+ \neq s_+$. Now we consider the
  chain of two generating reaction that starts from $\sigma_1 \sigma_2
  s_+ $, with $\sigma_1$, $\sigma_2 \in \Sigma$ and $v_1$, $v_2 \in
  \Sigma^*$.
  \begin{equation}
    \label{eq:enforce-chain}
    \sigma_1 \sigma_2 s_+
    \rea_{\Phi_+} \sigma_1 s'_+ v_1
    \rea_{\Phi_+} s''_+ v_2 v_1\,.
  \end{equation}
  We have just seen that if $\abs{v_1} = 0$, then $\abs{v_2} = 1$.
  Therefore $\abs{v_1 v_2} \geq 1$ for all $s_+ \in S_+$.

  This means that for $\delta(a_+)_T = 1$ and $\ell = 1$ we always can
  set $k = 2$. By induction we see then that for $\delta(a_+)_T = 1$
  and arbitrary $\ell$, a value of $k = 2 \ell$ is enough. When we
  also drop the condition that $\delta(a_+)_T = 1$, a value of $k = 2
  \ell \delta(a_+)_T$ is enough for the proposition of the lemma.
\end{proof}

The following two reactions are not derived in a completely formal
mode; they can instead be derived from the illustrations. Together
they will show how an ether fragment causes another ether fragment to
occur at a later time.

\begin{lemma}[Destruction of an Ether Fragment]
  \label{thm:fragment-destruction}
  For every $\sigma \in \Sigma$ there is a reaction of the form
  \begin{align}
    \label{eq:ether-1^3}
    \sigma 1 \epsilon_+^2 &\rea_\Phi a_+ 1^3 a'_+
  \end{align}
  with $a_+$, $a'_+ \in \dom \Phi_+$ and $\delta(a_+)_T \in \{ 1, 2
  \}$.
\end{lemma}

\begin{figure}[ht]
  \centering
  \tabcolsep1em
  \def\arraystretch{1.5}
  \begin{tabular}{cc}
    \ci{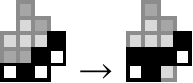} & \ci{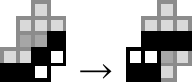} \\
    (a) $01 \epsilon_+^2 \rea_\Phi 01 \oplus_2 1^3 \epsilon_+$ &
    (b) $11 \epsilon_+^2 \rea_\Phi 11 \oplus_2 00 \oplus 1^3$
  \end{tabular}
  \caption{Destruction of ether fragments.}
  \label{fig:fragment-destruction}
\end{figure}
\begin{proof}
  Depending on the value of $\sigma$, one of the following two
  reactions can be applied to the left side of the
  reaction~\eqref{eq:ether-1^3},
  \begin{equation}
    \label{eq:ether-destruction}
    01 \epsilon_+^2 \rea_\Phi 01 \oplus_2 1^3 \epsilon_+
    \qqtext{or}
    11 \epsilon_+^2 \rea_\Phi 11 \oplus_2 00 \oplus 1^3\,.
  \end{equation}
  An example for these reactions can be seen in
  Figure~\ref{fig:fragment-destruction}. Now we can set either $a_+ =
  01 \oplus_2$ and $a'_+ = \epsilon_+$ or $a'_+ = 11 \oplus_2 00
  \oplus$ and $a'_+ = [0]$.
\end{proof}

\begin{lemma}[Creation of an Ether Fragment]
  \label{thm:fragment-creation}
  For every $j \in \N_0$ there is a reaction of the form
  \begin{align}
    \label{eq:1^3-ether}
    0 1^{2j+3} \rea_\Phi a_+ 1 \epsilon_+^{j+2} a,
  \end{align}
  with $a_+ \in \dom \Phi_+$, $a \in \dom \Phi$ and $\delta(a_+)_T =
  1$.
\end{lemma}
\begin{proof}
  \begin{figure}[ht]
    \centering
    \tabcolsep1em
    \def\arraystretch{1.5}
    \begin{tabular}{c}
      \ci{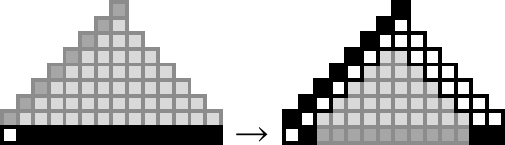} \\
      $0 1^{2j +3} \rea_\Phi 01 \oplus_2 1 \epsilon_+^{j+2}
      \ominus_2 10 (\ominus 00)^{j+1} \ominus_2 11$
    \end{tabular}
    \caption{Creation of new ether fragments.}
    \label{fig:fragment-creation}
  \end{figure}
  We use the following family of reactions,
  \begin{equation}
    \label{eq:ether-recreation}
    \set{ 0 1^{2j +3} \rea_\Phi 01 \oplus_2 1 \epsilon_+^{j+2}
    \ominus_2 10 (\ominus 00)^{j+1} \ominus_2 11
    \colon j \geq 0}\,.
  \end{equation}
  It can be derived in a way similar to that of the triangle reactions
  in Table~\ref{tab:large-scale}. An example for these reactions is
  shown in Figure~\ref{fig:fragment-creation}. We can then set $b_+ =
  01 \oplus_2$ and $b = \ominus_2 10 (\ominus 00)^{j+1} \ominus_2 11$.
\end{proof}

These two types of reactions then play a role in the following lemma.
Its proof uses the fact that the existence of an ether fragment $1
\epsilon_+^2$ at a given time causes the existence of a fragment of
another type, which then causes the existence of another ether
fragment at still another time. This second type of fragment may be
any situation of form $0 1^{2j + 3}$ with $j \geq 0$. For this kind of
fragment we need another set of situations analogous to
$\mathcal{E}(t_1, \dots, t_n)$. We will therefore write
\begin{equation}
  \label{eq:ones-set}
  \mathcal{E}_1(t_1, \dots, t_n) =
  \bigcup_{j \geq 0} \mathcal{F}(0 1^{2j + 3}, t_1, \dots, t_n)\,.
\end{equation}
for the set of intermediate fragments that occur in the propagation of
ether fragments to later times.

\begin{lemma}[Propagation of Ether Fragments]
  \label{thm:fragment-propagation}
  Let $(U_n)_{n \geq 0}$ be the sequence of random intervals
  in~\eqref{eq:random-intervals}. Assume that there is a reaction $u
  \rea_\Phi a$ with $u \in \Sigma^n$ and $a \in \mathcal{E}(t)$. Then
  for every $\epsilon > 0$ there is a number $m \geq n$ for which
  \begin{equation}
    \label{eq:fragment-propagation}
    P(U_m \rea_\Phi \mathcal{E}(t + 2, t + 3) \mid U_n = u)
    \geq 1 - \epsilon\,.
  \end{equation}
  The number $m$ depends only on $\epsilon$ and $t$, not on $a$.
\end{lemma}

\begin{proof}
  \begin{figure}[ht]
    \centering
    \begin{tikzpicture}[
      direct/.style={line width=.75ex},
      somewhere/.style={direct, draw=shaded, decorate},
      decoration={random steps, amplitude=1.5, segment length=5}]
      \path let \n{x} = 5, \n{y} = {1.6 * \n{x}} in
         (2,0) coordinate (aplus)
         ++(\n{x}, \n{y}) coordinate (a'max)
         ++(\n{x}, -\n{y}) coordinate (a'>)
         ++(2, 0) coordinate (end)
         ($(aplus)!0.45!(a'max)$) coordinate (aplus>)
         ($(aplus)!0.6!(a'max)$) coordinate (a');
      \draw[somewhere]
         (aplus) -- node [left] {$a_+$} (aplus>)
         (a')    -- (a'max)
                 -- node [right, pos=0.2] {$a'$} (a'>);
      \draw[direct]
         (aplus>)   -- node[left, pos=.7] {$1 \epsilon_+^2$} (a');
      \draw[dashed] let \p{right} = (end), \p{t} = (aplus>) in
         (0, 0)     -- (\x{right}, 0)     node [right] {$0$}
         (0, \y{t}) -- (\x{right}, \y{t}) node [right] {$t$};
    \end{tikzpicture}
    \caption{Decomposition of $a$ in \eqref{eq:a-decompose}.}
    \label{fig:a-decompose}
  \end{figure}
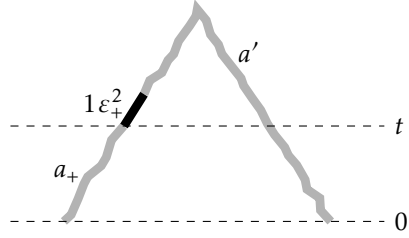
  Since $a$ is an element of $\mathcal{E}(t)$, it has a representation
  (Figure~\ref{fig:a-decompose})
  \begin{equation}
    \label{eq:a-decompose}
    a = a_+ 1 \epsilon_+^2 a'
  \end{equation}
  with $a_+ \in \dom \Phi_+$, $a' \in \dom \Phi$ and $\delta(a_+)_T =
  t$. We now will show that by putting a random interval $v$ at the
  left of $a$ we will get a situation that reacts to an element of
  $\mathcal{E}(t + 2, t + 3)$, with the probability that this happens
  growing arbitrarily large as the length of $v$ becomes arbitrarily
  large.

  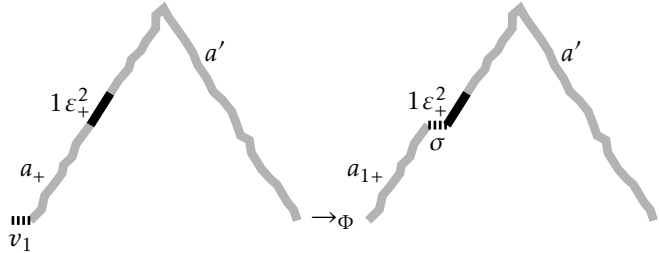
\begin{figure}[ht]
    \centering
    \begin{tikzpicture}[
      direct/.style={line width=.75ex},
      zebra/.style={direct, draw=black, dash pattern=on 1pt off 1pt},
      somewhere/.style={direct, draw=shaded, decorate},
      decoration={random steps, amplitude=1.5, segment length=5}]
      \def\backgroundpath{
      \pgfmathsetseed{999}
      \path let \n{x} = 5, \n{y} = {1.6 * \n{x}}, \n{v1} = {0.7} in
           (0,0)           coordinate (a1plus)
         ++(\n{v1},0)      coordinate (aplus)
         ++(\n{x},  \n{y}) coordinate (a'max)
         ++(\n{x}, -\n{y}) coordinate (a'>)
         ($(aplus)!0.45!(a'max)$) coordinate (aplus>)
             ++(-\n{v1},0) coordinate (a1plus>)
         ($(aplus)!0.6!(a'max)$)  coordinate (a');
       }
      \matrix [cells=midway] {
      \backgroundpath
      \draw[somewhere]
         (aplus) -- node [left] {$a_+$} (aplus>)
         (a')    -- (a'max)
                 -- node [right, pos=0.2] {$a'$} (a'>);
      \draw[zebra]
         (a1plus) -- node [below] {$v_1$} (aplus);
      \draw[direct]
         (aplus>) -- node[left, pos=.7] {$1 \epsilon_+^2$} (a');
      & \node {$\rea_\Phi$}; &
      \backgroundpath
      \draw[somewhere]
         (a1plus) -- node [left] {\llap{$a_{1+}$}} (a1plus>)
         (a')    -- (a'max)
                 -- node [right, pos=0.2] {$a'$} (a'>);
      \draw[direct]
         (aplus>)  -- node[left, near end] {$1 \epsilon_+^2$} (a');
      \draw[zebra]
         (a1plus>) -- node[below] {$\sigma$} (aplus>);
      \\ };
    \end{tikzpicture}
    \caption{Reaction \eqref{eq:enforce-event} puts an event
      $\sigma$ to the left of the ether fragment.}
    \label{fig:enforce-event}
  \end{figure}
  We know from Lemma~\ref{thm:propagate-interval} that we can find a
  number $m_1$ such that for all $v_1 \in \Sigma^{m_1}$ there is a
  reaction $v_1 a_+ \rea_\Phi a_{1+} \sigma$ with $a_{1+} \in \dom
  \Phi_+$ and $\sigma \in \Sigma$, independent of the value of $v_1$.
  The values of $a_{1+}$ and $\sigma$ depend on the value of $u$. This
  also means that for every $a_+$ there is a reaction
  (Figure~\ref{fig:enforce-event})
  \begin{equation}
    \label{eq:enforce-event}
    v_1 a_+ 1 \epsilon_+^2 a'
    \rea_\Phi a_{1+} \sigma 1 \epsilon_+^2  a'\,.
  \end{equation}
  This reaction can now be extended to a reaction that modifies the
  factor $1 \epsilon_+^2$ in the result term. In order to find this
  reaction we must take both possible values for $\sigma$ into
  consideration and need to understand the reactions that start at $01
  \epsilon^2$ and $11 \epsilon^2$. From
  Lemma~\ref{thm:fragment-destruction} we know that of all $\sigma \in
  \Sigma$ there is a reaction $0 \sigma \epsilon_+^2 \rea_\Phi x_+ 1^3
  a'_{2+}$, with $x_+$, $a'_{2+} \in \dom \Phi_+$ and $\delta(x_+)_T
  \in \{ 1, 2 \}$. One of these two reactions can then always be
  applied to the result of~\eqref{eq:enforce-event}. The result is a
  new reaction, $v_1 a_+ 1 \epsilon_+^2 \rea_\Phi a_{1+} x_+ 1^3
  a'_{2+}$. Now we introduce a new situation, $a_{2+} = a_{1+} x_+$
  and have then found that for every $v_1 \in \Sigma^{m_1}$ there is a
  reaction
  \begin{equation}
    \label{eq:enforce-1^3}
    v_1 a_+ 1 \epsilon_+^2  a'
    \rea_\Phi a_{2+} 1^3  a'
  \end{equation}
  with $\delta(a_{2+})_T \in \{ t + 1, t + 2 \}$.

  For the next step of the proof we need to find a reaction that
  transforms the result of~\eqref{eq:enforce-1^3} into a situation
  with $1 0^{2j+3}$ as factor, where $j \geq 0$. It would be nice if
  there were a reaction of the form $a_{2+} 1^3 a' \rea_\Phi a_{3+} 0
  1^{2j + 3} a''$, because then we could apply one the reactions of
  Lemma~\ref{thm:fragment-creation} to the term $0 1^{2j + 3}$ and get
  another situation that contains an ether fragment. However we will
  prove instead a weaker form of this statement: We will show that
  there is a number $m_2 \geq 0$ such that the probability for a
  reaction $v_2 a_{2+} 1^3 a' \rea_\Phi a_{3+} 0 1^{2j + 3} a''$, with
  a random interval $v_2 \in \Sigma^{m_2}$, is arbitrarily close to
  $1$. The interval $v_2$ is then another fragment of the random
  initial configuration.

  We begin with another application of
  Lemma~\ref{thm:propagate-interval}. It shows that by choosing an
  appropriate minimal length for the interval $v_2$ we can ensure that
  there is always a reaction $v_2 a_{2+} 1^3 \rea_\Phi x_+ w$, with
  $x_+ \in \dom \Phi_+$ and $w \in \Sigma^*$, in which $w$ is
  arbitrarily long. We now multiply $a'$ to the right of this reaction
  and get $v_2 a_{2+} 1^3 a' \rea_\Phi x_+ w a'$. The result of this
  reaction will now written in a different form, depending on the
  value of $w$. If $w$ does not consist entirely of events with state
  $1$, we can write it as $w = w' 0 1^\ell$. Here we must have $\ell
  \geq 3$, because $w$ is an extension of the interval $1^3$. Now we
  can set $j = \Floor{\frac{\ell - 3}{2}}$. Then we have either $\ell
  = 2j + 3$ or $\ell = 2 j + 4$. In the first case we have $w = w' 0
  1^{2j + 3}$, and we can set $a_{3+} = x_+ w'$ and $a'' = a'$; in the
  second case we have $w = w' 0 1^{2j + 3} 1$ and can set $a_{3+} =
  x_+ w'$ and $a'' = 1 a'$. The reaction becomes in both cases $v_2
  a_{2+} 1^3 a' \rea_\Phi a_{3+} 0 1^{2j + 3} a''$, as required.
  \begin{figure}[ht]
    \centering
    \begin{tikzpicture}[
      direct/.style={line width=.75ex},
      zebra/.style={direct, draw=black, dash pattern=on 1pt off 1pt},
      somewhere/.style={direct, draw=shaded, decorate},
      decoration={random steps, amplitude=1.5, segment length=5}]
      \def\backgroundpath{
      \pgfmathsetseed{999}
      \path let \n{x} = 5, \n{y} = {1.6 * \n{x}}, \n{v1} = {2.5} in
           (0,0)           coordinate (a1plus)
         ++(\n{v1},0)      coordinate (aplus)
         ++(\n{x},  \n{y}) coordinate (a'max)
         ++(\n{x}, -\n{y}) coordinate (a'>)
         ($(aplus)!0.45!(a'max)$) coordinate (aplus>)
             ++(-\n{v1},0) coordinate (a1plus>)
         ($(aplus)!0.50!(a'max)$)  coordinate (a'')
             ++(-\n{v1},0) coordinate (a3plus>)
         ($(aplus)!0.6!(a'max)$)  coordinate (a');
       }
      \matrix [cells=midway] {
      \backgroundpath
      \draw[somewhere]
         (aplus) -- node [left] {$a_+$} (aplus>)
         (a')    -- (a'max)
                 -- node [right, pos=0.2] {$a'$} (a'>);
      \draw[zebra]
         (a1plus) -- node [below] {$v_2 v_1$} (aplus);
      \draw[direct]
         (aplus>) -- node[left, pos=.7] {$1 \epsilon_+^2$} (a');
      & \node {$\rea_\Phi$}; &
      \backgroundpath
      \draw[somewhere]
         (a1plus) -- node [left] {\llap{$a_{3+}$}} (a3plus>)
         (a'') -- (a'max)
               -- node [right, pos=0.2] {$a''$} (a'>);
      \draw[direct]
         (a3plus>) -- node[below, near end] {$0 1^{2j+3}$} (a'');
      \\ };
    \end{tikzpicture}
    \caption{Reaction \eqref{eq:enforce-intermediate}. The interval $0
      1^{2j+3}$ occurs later than the ether fragment.}
    \label{fig:enforce-intermediate}
  \end{figure}
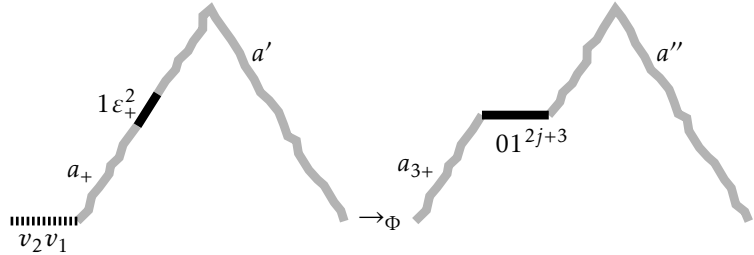
  When we now multiply both sides of~\eqref{eq:enforce-1^3} from the
  right with $v_2$ and apply the previous reaction to its result, we
  get (Figure~\ref{fig:enforce-intermediate})
  \begin{equation}
    \label{eq:enforce-intermediate}
    v_2 v_1 a = v_2 v_1 a_+ 1 \epsilon_+^2 a'
    \rea _\Phi a_{3+} 0 1^{2j + 3} a''\,.
  \end{equation}
  We have in this reaction $\delta(a_{3+})_T \in \{ t + 1, t + 2 \}$,
  because $\delta(a_{3+})_T = \delta(a_{2+})_T$.

  The proof for the existence of this reaction is however only valid
  if $w$ does not consist entirely of ones. Next we must therefore
  show that the probability that $w$ contains a zero can be made
  arbitrarily large by choosing $m_2$ large enough. For the proof we
  use the fact that the events of $pr_{x_+}(w)$ in the reaction $v_2
  a_{2+} 1^3 \rea_\Phi x_+ w$ above must occur either at the time step
  $t + 1$ or $t + 2$. We must therefore show that with a high
  probability there is an event with state $0$ at these time steps.
  For this we use the fact that if the random interval $v_2$ becomes
  long enough, the probability that it contains a given interval as
  its factor becomes arbitrarily close to $1$. (This can be proved in
  a similar way as Theorem~\ref{thm:fragment-formation}.) In the
  current proof we use the factor $1 0^{2t + 3} 1$ because its closure
  is a triangle process, and it contains zeros at all time steps from
  $0$ to $t + 2$. Therefore, if $m_2$ is large enough, the probability
  that a random $v_2$ contains the interval $1 0^{2t + 3} 1$ becomes
  arbitrarily close to $1$, and when $v_2$ contains such an interval,
  there is a cell in state $0$ in the time steps $t + 1$ and $t + 2$.
  Since the interval $v_2$ is at the left of the situation $a_+ 1^3$,
  a cell in state $0$ must therefore occur in~$w$.

  The result of reaction~\eqref{eq:enforce-intermediate} is always an
  element of $\mathcal{E}_1(t + 1, t + 2)$. We have therefore shown
  that for every $\epsilon > 0$ there is a number $m = m_2 + m_1 + n$
  such that we have the probability
  \begin{equation}
    \label{eq:intermediate-step}
    P(U_m \rea_\Phi \mathcal{E}_1(t + 1, t + 2) \mid U_n = u)
    \geq 1 - \epsilon\,.
  \end{equation}

  As a final step we now prove that if there is a reaction $v
  \rea_\Phi b$ with $v \in \Sigma^m$ and $b \in \mathcal{E}_1(t + 1, t
  + 2)$, then there is a reaction from $b$ to an element of
  $\mathcal{E}(t + 2, t + 3)$.
    \begin{figure}[ht]
    \centering
    \begin{tikzpicture}[
      direct/.style={line width=.75ex},
      zebra/.style={direct, draw=black, dash pattern=on 1pt off 1pt},
      somewhere/.style={direct, draw=shaded, decorate},
      decoration={random steps, amplitude=1.5, segment length=5}]
      \def\backgroundpath{
      \pgfmathsetseed{999}
      \path let \n{x} = 5, \n{y} = {1.6 * \n{x}}, \n{v1} = {2.5} in
           (0,0)           coordinate (a1plus)
         ++(\n{v1},0)      coordinate (aplus)
         ++(\n{x},  \n{y}) coordinate (a'max)
             +(-\n{v1},0) coordinate (shiftmax)
         ++(\n{x}, -\n{y}) coordinate (a'>)
         ($(aplus)!0.55!(a'max)$) coordinate (aplus>)
             ++(-\n{v1},0) coordinate (a1plus>)
         ($(aplus)!0.50!(a'max)$)  coordinate (a'')
             ++(-\n{v1},0) coordinate (a3plus>)
         ($(aplus)!0.7!(a'max)$)  coordinate (a');
       }
      \matrix [cells=midway] {
      \backgroundpath
      \draw[somewhere]
         (a1plus) -- node [left] {$b_+$} (a3plus>)
         (a'') -- (a'max)
               -- node [right, pos=0.2] {$b'$} (a'>);
      \draw[direct]
         (a3plus>) -- node[below, near end] {$0 1^{2j+3}$} (a'');
      & \node {$\rea_\Phi$}; &
      \backgroundpath
      \coordinate (b1plus>) at ($(a1plus)!0.58!(shiftmax)$);
      \coordinate (y)       at ($(a1plus)!0.73!(shiftmax)$);
      \draw[somewhere]
         (a1plus) -- node [left] {\llap{$b_{1+}$}} (b1plus>)
         (y) -- node [right, near start] {$y$} (a'')
             -- (a'max)
             -- node [right, pos=0.2] {$b'$} (a'>);
      \draw[direct]
         (b1plus>) -- node[left] {$1 \epsilon_+^2$} (y);
      \\ };
    \end{tikzpicture}
    \caption{Reaction \eqref{eq:fragment-regenerate}. A new ether
      fragment is generated from the interval $0 1^{2j + 3}$.}
    \label{fig:fragment-regenerate}
  \end{figure}
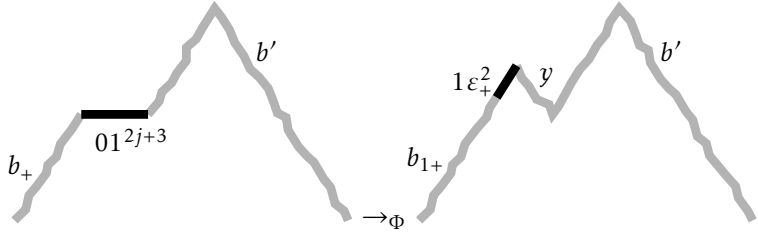
  For this we do the same with $b$ as we did before with $a$ and write
  it as $b = b_+ 0 1^{2j + 3} b'$. We have already seen in
  Lemma~\ref{thm:fragment-creation} that there is a reaction $0 1^{2j
    + 3} \rea_\Phi y_+ 1 \epsilon_2^{j + 2} y$, with $\delta(y_+)_T =
  1$. We now set $b_{1+} = b_+ y_+$. Then we have found a reaction
  (Figure~\ref{fig:fragment-regenerate})
  \begin{equation}
    \label{eq:fragment-regenerate}
    b_+ 0 1^{2j + 3}  b'
    \rea_\Phi b_{1+} 1 \epsilon_+^{j+2} y b'
  \end{equation}
  with $\delta(b_{1+})_T \in \{ t + 2, t + 3 \}$, because
  $\delta(b_{1+})_T = \delta(b_+)_T + 1$.

  Therefore the right side of the
  reaction~\eqref{eq:fragment-regenerate} is always an element of
  $\mathcal{E}(t + 2, t + 3)$. Putting all these steps together we see
  therefore that we have found for every $\epsilon > u$ a probability
  \begin{equation}
    P(U_m \rea_\Phi \mathcal{E}(t + 2, t + 3) \mid U_n \rea_\Phi a)
    \geq 1 - \epsilon,
  \end{equation}
  as stated in the lemma.
\end{proof}

With the reactions in Lemma~\ref{thm:fragment-propagation} we see that
the ether fragments move to the future, and with
Lemma~\ref{thm:propagate-interval} we see that ether fragments are
generated from random initial configurations. Together the lemmas show
that ether fragments continue to exist for arbitrarily long times.
This is now described in more detail in the following proof.

\begin{proof}[Proof of Theorem~\ref{thm:ether-formation}]
  The proof is done by induction.

  We can see from Lemma~\ref{thm:fragment-formation} that the theorem
  is true for $t = 0$: The lemma shows that there is a $m \geq 0$ such
  that $P(U_m \rea_\Phi \mathcal{E}(0)) \geq 1 - \epsilon$. Therefore
  we have $P(U_m \rea_\Phi \mathcal{E}(0, 1, 2)) \geq P(U_m \rea_\Phi
  \mathcal{E}(0) \geq 1 - \epsilon)$.

  In the main part of the induction we assume that the theorem is true
  for a given time $t \geq 0$. With this we mean that for every
  probability $\epsilon_1 \geq 0$ there is a number $n$ such that
  $P(U_n \rea_\Phi \mathcal{E}(t, t + 1, t + 2)) \geq 1 - \epsilon_1$.
  We then need to show that for every probability $\epsilon > 0$ there
  is a length $m \geq n$ such that
  \begin{equation}
    P(U_m \rea_\Phi \mathcal{E}(t + 1, t + 2, t + 3))
    \geq 1 - \epsilon\,.
  \end{equation}
  In order to do this we split the reaction $U_m \rea_\Phi
  \mathcal{E}(t + 1, t + 2, t + 3)$ into two subreactions. The first
  one is derived from the reaction $U_n \rea_\Phi \mathcal{E}(t, t +
  1, t + 2)$. This reaction exists by the induction assumption, and we
  can make its probability arbitrarily high by choosing the right $n$,
  but its input interval $U_n$ is by definition shorter than $U_m$.
  Nevertheless, a similar reaction starting from $U_m$ does also
  exist. It will be our first subreaction. In the second subreaction,
  the result of the first subreaction then reacts to an element of
  $\mathcal{E}(t + 1, t + 2, t + 3)$, and we will show that we can
  make its probability as large as we want by making $m$ large enough.

  To combine these subreactions we need to split the set of possible
  input intervals for the first reaction, $U_n \rea_\Phi
  \mathcal{E}(t, t + 1, t + 2)$, into two disjoint sets. For this we
  introduce the sets $\mathcal{U}_0$ and $\mathcal{U}_1 \subseteq
  \Sigma^n$, satisfying the requirement that for every reaction $u
  \rea_\Phi a$ with $u \in \Sigma^n$ and $a \in \mathcal{E}(t, t + 1,
  t + 2)$ we have either $u \in \mathcal{U}_0$ and $a \in
  \mathcal{E}(t)$, or $u \in \mathcal{U}_1$ and $a \in \mathcal{E}(t +
  1, t + 2)$. If $u$ satisfies both conditions, then we may choose
  arbitrarily $u \in \mathcal{U}_0$ or $u \in \mathcal{U}_1$. The same
  is true if there is no such reaction $u \rea_\Phi a$ for a given
  $u$.

  Then we can write the probability for the first subreaction as a sum
  of the probability of two independent stochastic events, namely as
  \begin{equation}
    \label{eq:fragment-decomposition}
    P(U_n \rea_\Phi \mathcal{E}(t, t + 1, t + 2)) =
    P(U_n \in \mathcal{U}_0) + P(U_n \in \mathcal{U}_1)\,.
  \end{equation}
  The requirements for the reactions from the sets $\mathcal{U}_0$ and
  $\mathcal{U}_1$ are expressed by the (trivial) probabilities
  \begin{subequations}
    \label{eq:U-intervals}
    \begin{align}
      \label{eq:U0-intervals}
      P(U_n &\rea_\Phi \mathcal{E}(t)
      \mid U_n \in \mathcal{U}_0) = 1, \\
      \label{eq:U1-intervals}
      P(U_n &\rea_\Phi \mathcal{E}(t + 1, t + 2)
      \mid U_n \in \mathcal{U}_1) = 1\,.
    \end{align}
  \end{subequations}
  The equations stay true if we replace the leftmost $U_n$ in them
  with a larger random interval $U_m$, where $m \geq n$. We will use
  this now for the second subreaction.

  The results~\eqref{eq:U-intervals} are the input situations for the
  second subreaction, so it has two cases as well. We need to show
  that in both cases there is a reaction into the set $\mathcal{E}(t +
  1, t + 2, t + 3)$. If $u \in \mathcal{U}_0$, we use
  Lemma~\ref{thm:fragment-propagation}. It applies to situation $u \in
  \Sigma^n$ for which there is a reaction $u \rea_\Phi a \in
  \mathcal{E}(t)$. This property is of course equivalent to the
  condition that $u \in \mathcal{U}_0$. The lemma then states that for
  every $\epsilon_2 > 0$ there is a number $m \geq n$ such that $P(U_m
  \rea_\Phi \mathcal{E}(t + 2, t + 3) \mid U_n = u) \geq 1 -
  \epsilon_2$. We now collect the probabilities for all $u \in
  \mathcal{U}_0$ and get the following reaction,
  \begin{subequations}
    \label{eq:common-base}
    \begin{equation}
      P(U_m \rea_\Phi \mathcal{E}(t + 2, t + 3)
      \mid U_n \in \mathcal{U}_0) \geq 1 - \epsilon_2\,.
    \end{equation}
    If $u \in \mathcal{U}_1$, then we can use a lengthened form of
    reaction~\eqref{eq:U1-intervals},
    \begin{equation}
      P(U_m \rea_\Phi \mathcal{E}(t + 1, t + 2) \mid U_n \in
      \mathcal{U}_1) = 1\,.
    \end{equation}
  \end{subequations}
  The results of both reactions are elements of $\mathcal{E}(t + 1, t
  + 2, t + 3)$, as was required.

  Now we can perform the complete induction step. The following
  computation begins by splitting the probability for the reaction
  $U_m \rea_\Phi \mathcal{E}(t + 1, t + 2, t + 3)$ into two cases,
  depending on whether the right end of $U_m$, i.\,e.\ the interval
  $U_n$, is an element of $\mathcal{U}_0$ or $\mathcal{U}_1$. Then the
  equations~\eqref{eq:common-base} are used to get estimates for these
  probabilities, and later, at the penultimate step, the two cases are
  unified again with the help of~\eqref{eq:fragment-decomposition}.
  \begin{multline}
    P(U_m \rea_\Phi \mathcal{E}(t + 1, t + 2, t + 3)) \\
    \begin{aligned}[b]
      &\geq P(U_m \rea_\Phi \mathcal{E}(t + 1, t + 2, t + 3)
      \mid U_n \in \mathcal{U}_0) P(U_n \in \mathcal{U}_0) \\
      &\quad + P(U_m \rea_\Phi \mathcal{E}(t + 1, t + 2, t + 3)
      \mid U_n \in \mathcal{U}_1) P(U_n \in \mathcal{U}_1) \\
      &\geq P(U_m \rea_\Phi \mathcal{E}(t + 1, t + 2)
      \mid U_n \in \mathcal{U}_0) P(U_n \in \mathcal{U}_0) \\
      &\quad + P(U_m \rea_\Phi \mathcal{E}(t + 2, t + 3)
      \mid U_n \in \mathcal{U}_1) P(U_n \in \mathcal{U}_1) \\
      &\geq (1 - \epsilon_2) P(U_n \in \mathcal{U}_0)
      + P(U_n \in \mathcal{U}_1) \\
      &\geq (1 - \epsilon_2)
      (P(U_n \in \mathcal{U}_0) + P(U_n \in \mathcal{U}_1)) \\
      &= (1 - \epsilon_2) P(U_n \rea_\Phi \mathcal{E}(t, t + 1, t + 2)) \\
      &\geq (1 - \epsilon_2) (1 - \epsilon_1)\,.
    \end{aligned}
  \end{multline}
  Therefore the probability for this reaction can be made greater than
  $1 - \epsilon$ by making both $\epsilon_1$ and $\epsilon_2$ small
  enough.
\end{proof}

\section{Generalisation to Other Rules}
\label{sec:gener-other-rules}

While the arguments in this chapter were tailored specifically for the
use with Rule 54, it was with the hope that they would lead to ideas
that are useful for the understanding of ether formation under other
rules, e.\,g.\ Rule 110.

To find these new ideas we must generalise the proofs and definitions
of this chapter. For some of them it is easy to see how to generalise
them, but in other places new ideas are needed. I will now describe
the necessary changes in more detail.

First we need a characterisation of the ether in the relevant cellular
automaton. The definition of the ether reaction
(Definition~\ref{def:basic-ether}) would be generalised to a pair of
reactions
\begin{equation}
  \label{eq:generalised-ether}
  e_0 v \rea_\Phi e_0 e_+ e_-
  \qqtext{and}
  e_0 e_- e_+ \rea_\Phi e_0 e_+ e_-,
\end{equation}
with $e_0$, $v \in \Sigma^*$, $e_+ \in \dom \Phi_+$ and $e_- \in \dom
\Phi_-$. Under Rule 54 we had $e_0 = 1$, $v = 0^3 1$, $e_+ =
\epsilon_+^2$ and $e_- = \epsilon_-^2$. This can be done in every
one-dimensional cellular automaton with an ether. The situation $e_0
e_+$ then plays the role of the ether fragment.

Then we can construct sets of situations $\mathcal{E}'(t_1, \dots,
t_n) = \mathcal{F}(e_0 e_+, t_1, \dots, t_n)$, in analogy to the sets
$\mathcal{E}(t_1, \dots, t_n)$ in~\eqref{eq:R-slope}.
Lemma~\ref{thm:fragment-formation}, which proves the generation of an
ether fragment from a large enough random interval, can be extended to
$\mathcal{E}'$: The only requirement for its proof is the existence of
a reaction $e_0 v \rea \mathcal{E}'(0)$, but this is the first
reaction in~\eqref{eq:generalised-ether}. Then, if we have an
equivalent to Lemma~\ref{thm:fragment-propagation}, we can prove ether
formation in essentially the same way as here, by showing that if
there is an ether fragment at time $t$, there is always a starting
time $t' > t$ at which another ether fragment exists with an
arbitrarily high probability.

Lemma~\ref{thm:fragment-propagation}, however, is proved with the help
of the Lemmas~\ref{thm:fragment-destruction}
and~\ref{thm:fragment-creation}, which are highly specific to Rule 54.
It is not clear whether an equivalent to these lemmas exists for other
cellular automata with an ether, and if it exists, how to find it in a
systematic way. In Rule 54, they were the result of some experimenting
and an already well-developed understanding of the way this rule
works. This means that an expert for, say, Rule 110 could find an
equivalent to Lemma~\ref{thm:fragment-destruction}
and~\ref{thm:fragment-creation} after a similar amount of
experimenting, but there is no recipe for a proof of ether formation
if the behaviour of a rule is not yet well understood---even if one
has already seen that it has an ether.

\section{Summary}

In this chapter we have begun to define concepts with which one can
express larger structures in a cellular automaton. There is a small
theory of regular structures like triangles. We have seen how one can
derive families of reactions.

Then we turned our view to the ether. We first found a way to express
the empty ether with situations and reactions. The ether reactions
motivated the definition of ether fragments, the situations $1
\epsilon_+^2$. The existence of ether fragments at arbitrary times was
defined as a way to express the existence of ether when the initial
configuration was chosen at random.

We had to invent an extension of the calculus of Flexible Time that
can handle probabilities. This was done in an incomplete way, just
enough to get the proofs done. Nevertheless a basic principle did
appear: in a random reaction the input had to be a random variable,
while the result had to be a set of situations. With these extensions
we expressed how ether fragments were generated in the initial
configuration and how an ether fragment that was positioned at a given
time $t$ caused the existence of an ether fragment at a later time.
This then lead to a proof that ether fragments must exist at all time
steps when the cellular automaton started from a random initial
configuration.

Finally we considered the question whether the proof that we had done
for Rule 54 could be generalised to other transition rules. The answer
was that it could be only partially, and that filling the gaps would
require knowledge about the specific rule involved.


\chapter{Conclusions}

This work was about finding a way to speak about cellular automata in
terms of ``traditional mathematics'', as I have called it in the
introduction. The main results of this thesis are therefore concepts,
not theorems.

\paragraph{Reaction Systems} The idea that started my work on cellular
automata was that of \emph{situations} and \emph{reaction}, first
understood only in an intuitive way and for concrete cellular
automata. There was much choice in the way the situations were
defined, and it was resolved by trial and error for a specific
transition rule. My article \cite{Redeker2010} was a product of this
phase. Already then I tried to find definitions that applied to a
large number of cellular automata, even if the work was done for one
specific transition rule. This was so because only when a definition
applies to a large number of cases, the methods developed for it can
also be used to explore the behaviour of unknown cellular automata.

In this work I have therefore searched for principles on which I could
base the definition of a reaction system that are valid for a large
number of transition rules. It was still possible to exclude certain
rules from consideration if their behaviour was difficult to express
with situations. One of the first results was the restriction to
\emph{interval-preserving} rules. The reason for this decision was
that intervals are especially easy to express with situations. We have
also seen that the closure of an interval has a simple structure and
that the possibilities for their left-to-right arrangement are
limited. Only the restriction to interval-preserving rules made the
theoretical understanding of cellular processes and their closures
possible.

The question which kinds of situations to choose was then resolved by
finding the concept of \emph{separating intervals}. A separating
interval forms a boundary between the cells left of it and the cells
right of it. If two separating intervals form the left and right ends
of an interval process $\pi$, then the two intervals already determine
which points are determined by $\pi$, even if the rest of $\pi$ is
unknown. This occurs especially when the events between the separating
intervals belong to the closure of a larger process. This then occurs
in \emph{achronal situations}: They contain separating intervals at
strategic places, so that it is possible to limit the extension of
their closure. Thanks to this construction we have a guarantee that
the closure exists at all. The definition of achronal situations and
the proof that ordered achronal situations have a closure is another
important progress in comparison to \cite{Redeker2010}.

Separating intervals are also interesting in a more general context.
They tell about the information transmission in a cellular automaton.
The set of separating intervals for a transition rule is a better
measure for the speed of information transmission than the radius of
the transition rule, another a measure for the speed of information
transmission. The radius does however only measure of the maximal
possible speed. In contrast to this, the set of separating intervals
is an invariant of the cellular automaton that is radius-invariant in
the sense of Definition~\ref{def:radius-invariance}.

\paragraph{Ether Formation} The sections about ether formation were
intended as a reality check for the formalism of Flexible Time. They
are a test whether the formalism has been developed far enough to find
solutions for a natural-looking question about cellular automata,
i.\,e.\ a question that did not occur as part of the development of
the formalism. The question of ether formation was such a problem.

To become answerable the ether problem had to be reduced to a very
simple question. We reduced the question of ether formation from the
general case to that of Rule 54, and then characterised the ether by
the ether fragments $1 \epsilon_+^2$. The result was a theorem that
only showed that such ether fragments exist at arbitrary times when
starting from a random initial configuration, not that they become
more common with time. The latter fact is clearly visible from
computer simulations.

A more positive result is that the proof of the Ether Formation
Theorem~\ref{thm:ether-formation} is valid for all probability
distributions in which both cells in state $0$ and in state $1$ can
occur in the initial configuration. This is more general than the
empirical results, which usually refer to the case that zeros and ones
initially occur with equal probability.

Another positive result of the approach with Flexible Time is that it
can express the mechanism with which the ether is created and
preserved. We have seen that the initial configuration creates with a
high probability an ether fragment $1 \epsilon_+^2$, which then causes
the creation of an interval $0 1^{2j + 3}$ at a later time, which in
turn reacts to another ether fragment at a still later time: All this
was expressed with help of the sets $\mathcal{E}$ and $\mathcal{E}_1$.

The mechanism of Lemma~\ref{thm:fragment-propagation} then lets us
formulate one cause of ether formation under Rule 54. This is the
connection of structure formation with loss of information about the
initial state. Structure formation in a dynamical system can always be
expressed as a case of information loss: Several initial
configurations must evolve to the same, more ordered, later
configuration. Under Rule 54, information loss occurs during the
propagation of ether fragments: In the reactions of
Lemma~\ref{thm:fragment-destruction}, the two situations $01
\epsilon_+^2$ and $11 \epsilon_+^2$ react to a situation that contains
the term $1^3$. This idea has a chance to be also the cause of ether
formation in other cellular automata.

\paragraph{The Results in Context} This work has therefore shown that
there is an essentially two-dimensional approach to structure
formation in cellular automata and that it can prove nontrivial facts
about a cellular automaton. Which place does then the formalism of
Flexible Time take among the approaches to understand the behaviour of
one-dimensional cellular automata, especially among those described in
Section~\ref{sec:struct-cell-autom}?

In its two-dimensionality it is similar especially to the work of
Ollinger and Richard \cite{Ollinger2007,Richard2008}, which was
however mainly applied to model a network of colliding particles.
While this kind of research is also possible with Flexible Time
\cite{Redeker2010a}, it has not been done in detail.

In its formalisation of the ether, the approach of this thesis
diverges from most other works in that it concentrates directly on
fragments of the ether. Most other works, not just Ollinger and
Richard, describe in great details the particles that move through the
ether. This kind of approach could also become the basis of a proof of
ether formation: an analysis how the particles interact and how they
gradually destroy each other. (The destruction of gliders does occur
under Rule 54 and 110 and was shown experimentally by Boccara
\textit{et al.} \cite{Boccara1991} and by Li and Nordahl
\cite{Li1992}.) The number of particles and their possible
interactions can become however quite large, as it does under Rule 110
\cite{Ju'arezMart'inez2001}, and it would be tedious to use them all
in a formal proof.

Among the works about the behaviour of random initial configurations,
many concentrate on the behaviour of the defects that occur between
domains \cite{Eloranta1993,Jen1990}. It has been shown that under many
conditions such a defect performs a random walk. In our proof for
ether formation we also have a moving ``particle'', namely the ether
fragment, but its position cannot be directly identified from an
evolution diagram. Nevertheless it performs a kind of random walk,
driven by the states of the cells in the initial configuration. With
Flexible Time we have therefore another tool with which one can trace
the influence of a random initial configuration to later time step.
(The tool however has not yet been developed very far.)

There is also some similarity to the ``grouping'' approach
\cite{Delorme2011a,Delorme2011}, since Flexible Time also groups
several cells to a greater entity. With Flexible Time there is however
much more freedom in the choice of the situations, and as a result
this approach does not automatically provide tools to put cellular
automata into groups according to their behaviour.

Unintentionally, Flexible Time may however lead to new ideas for the
classification of cellular automata. We could now classify them by
their separating intervals (Table~\ref{tab:separating}) or by the
pattern of their generating reactions (Table~\ref{tab:generator54}).
What this classification means is not yet clear, but it must have
something to do with the information transmission in the cellular
automata.

\section{Ideas for Further Research}

Finally I list here a small number of ideas for further research.
Their purpose is to extend the system of Flexible Time and also to
apply its ideas to other domains.

\paragraph{Separating Intervals} We have defined separating intervals
as a purely technical tool. It is not yet clear whether they have an
intrinsic meaning, except that they are somewhat related to signal
transmission. A possible starting point for further research is
therefore the question whether transition rules with the same set of
separating intervals have something in common. Which properties of a
transition rule can be derived from knowing its separating intervals?

Another starting point to find out more about separating intervals is
the growing body of research about cellular automata with memory
\cite{Alonzo-Sanz2009}. Does the addition of memory change the
separating intervals of cellular automata, and if so, in which way?

\paragraph{Explicit Probabilities} Another idea for later work is the
search for good explicit probabilities for ether propagation. The
reactions only show that the ether fragments survive over time, but
they do not give a good estimate about their density. A description of
the ether propagation that was more detailed will be needed to get
better estimates. With it, there is a chance to find a proof that the
density of ether fragments actually grows over time.

\paragraph{Generalising the Way to Find the Ether} Some ideas for this
were already outlined in Section~\ref{sec:gener-other-rules}. The
current argument for the ether required many \emph{ad hoc}
constructions. An example are the ether fragments, which were only
found after studying the evolution of configurations under Rule 54 for
a long time. There was nothing systematic in their construction.
Another example is the structure of the proof for
Lemma~\ref{thm:fragment-propagation}. All relied on phenomenology.
Nevertheless the current proof may contain the ideas that can be
generalised to a more systematic proof. This in turn will of course
require a further development of the formalism in order to make it
more streamlined and easier to use.

\paragraph{Analysis of Particle Interactions} The ether in Rule 54 is
the medium in which particles move. There has been a large amount of
research about the particles and their interactions under Rule 54
\cite{Boccara1991,Martin2000,Ju'arezMart'inez2006a}, Rule 110
\cite{Cook2004,Ju'arezMart'inez2001,Ju'arezMart'inez2003,Ju'arezMart'inez2006,Ju'arezMart'inez2008,Ju'arezMart'inez2007,Ollinger2007,Richard2008}
and other rules \cite{Letourneau2010,Letourneau2010a}. My own earlier
paper \cite{Redeker2010a} contains the beginnings of an analysis of
the particle interactions under Rule 54. This line of research was
left incomplete because the structure of the reaction system for Rule
54 was not yet clear. Now it could be continued, with the hope for a
reasonably simple algebra of particle interactions for Rule 54.

A good knowledge of such particle interactions could make another
proof for ether formation possible. It has already been noted by
Boccara, Nasser and Roger \cite{Boccara1991} that in the typical
evolution of a random initial configuration at a very early time
configurations arise that consist of small regions of ether, with
particles between them. The particles then interact and slowly destroy
each other, such that the ether between them grows. All this has been
found in computer simulations, but not proved. A good understanding of
particle interactions would therefore allow to understand this process
in detail and provide a quantitative estimate for the speed with which
the ether grows.

\paragraph{A Generic Case of Self-Organisation} Finally an idea for a
larger project. It is inspired by a paper by Boccara and Roger
\cite{Boccara1991a}, in which the authors describe a whole class of
self-organising rules. They are generated from \emph{totalistic}
rules, in which the states of the cells are numbers and the next state
of a cell only depends on the sum of the states of the cells in the
neighbourhood. The authors have found a transformation that transforms
an arbitrary totalistic rule into a rule that shows a certain amount
of self-organisation. In it a rule $\phi$ of radius $r$ is transformed
into a rule $\phi'$ of radius $n r$ which works on initial
configurations in which every block of $n$ cells have the same state.
It is required that if every cell in the initial configuration of
$\phi$ is expanded to $n$ cells, then the evolution of this
configuration under $\phi'$ corresponds exactly to the evolution of
the original configuration under $\phi$. If both $\phi$ and $\phi'$
are totalistic, then $\phi'$ is uniquely determined by $\phi$. Now if
the initial configuration is arbitrary, then in the following time
steps in the evolution under $\phi'$ the cells begin to organise in
blocks of length $n$, again with defects between the blocks that move
randomly and sometimes annihilate.

The understanding of this kind of pattern formation, in the same or a
different way as we have done this here for Rule 54, would lead to the
understanding of the behaviour of a whole class of cellular automata,
not just one. It is therefore very interesting and would also allow
the formalism to grow.


\backmatter
\bibliography{../references}
\printindex

\end{document}